\documentclass{article}

\PassOptionsToPackage{numbers, compress}{natbib}

 \usepackage[creativeai, final]{neurips_2025}

\usepackage[utf8]{inputenc} 
\usepackage[T1]{fontenc}    
\usepackage{hyperref}       
\usepackage{xurl}            
\usepackage{booktabs}       
\usepackage{amsfonts}       
\usepackage{nicefrac}       
\usepackage{microtype}      
\usepackage{xcolor}         

\usepackage{graphicx}
\usepackage{subfigure}
\usepackage{amsmath}
\usepackage[capitalize,noabbrev]{cleveref}
\allowdisplaybreaks 
\usepackage{titletoc} 
\usepackage[export]{adjustbox} 
\usepackage{array} 

\newcommand{\spice}{\textsc{SPICE}}

\newcommand{\stepsgrid}[1]{%
  \setlength{\tabcolsep}{2pt}%
  \renewcommand{\arraystretch}{1.0}%
  \begin{tabular}{@{}cccccc@{}}
    \scriptsize Step 1 & \scriptsize Step 2 & \scriptsize Step 3 &
    \scriptsize Step 4 & \scriptsize Step 5 & \scriptsize Step 6 \\
    \includegraphics[width=0.155\linewidth]{figures/jpg/iterative-editing-#1-step-01.jpg} &
    \includegraphics[width=0.155\linewidth]{figures/jpg/iterative-editing-#1-step-02.jpg} &
    \includegraphics[width=0.155\linewidth]{figures/jpg/iterative-editing-#1-step-03.jpg} &
    \includegraphics[width=0.155\linewidth]{figures/jpg/iterative-editing-#1-step-04.jpg} &
    \includegraphics[width=0.155\linewidth]{figures/jpg/iterative-editing-#1-step-05.jpg} &
    \includegraphics[width=0.155\linewidth]{figures/jpg/iterative-editing-#1-step-06.jpg} \\
    \scriptsize Step 7 & \scriptsize Step 8 & \scriptsize Step 9 &
    \scriptsize Step 10 & \scriptsize Step 11 & \scriptsize Step 12 \\
    \includegraphics[width=0.155\linewidth]{figures/jpg/iterative-editing-#1-step-07.jpg} &
    \includegraphics[width=0.155\linewidth]{figures/jpg/iterative-editing-#1-step-08.jpg} &
    \includegraphics[width=0.155\linewidth]{figures/jpg/iterative-editing-#1-step-09.jpg} &
    \includegraphics[width=0.155\linewidth]{figures/jpg/iterative-editing-#1-step-10.jpg} &
    \includegraphics[width=0.155\linewidth]{figures/jpg/iterative-editing-#1-step-11.jpg} &
    \includegraphics[width=0.155\linewidth]{figures/jpg/iterative-editing-#1-step-12.jpg} \\
  \end{tabular}%
}

\title{SPICE: A Synergistic, Precise, Iterative, and Customizable Image Editing Workflow}

\author{%
  Kenan Tang\thanks{Equal contribution} \\
  University of California, Santa Barbara\\
  \texttt{kenantang@ucsb.edu} \\
  \And
  Yanhong Li$^*$ \\
  Allen Institute for AI \\
  \texttt{yanhongl@allenai.org} \\
  \AND
  Yao Qin \\
  University of California, Santa Barbara\\
  \texttt{yaoqin@ucsb.edu} \\
}

\begin{document}

\maketitle

\setcounter{footnote}{0}

\begin{abstract}
  Prompt-based models have demonstrated impressive prompt-following capability at image editing tasks. However, the models still struggle with following detailed editing prompts or performing local edits. Specifically, global image quality often deteriorates immediately after a single editing step. To address these challenges, we introduce \spice, a \emph{training-free} workflow that accepts arbitrary resolutions and aspect ratios, accurately follows user requirements, and consistently improves image quality during more than 100 editing steps, while keeping the unedited regions intact. By synergizing the strengths of a base diffusion model and a Canny edge ControlNet model, \spice~robustly handles free-form editing instructions from the user. On a challenging realistic image-editing dataset, \spice~quantitatively outperforms state-of-the-art baselines and is consistently preferred by human annotators. We release the workflow implementation for popular diffusion model Web UIs to support further research and artistic exploration.\footnote{\url{https://github.com/kenantang/spice}}
\end{abstract}

\section{Introduction}

\begin{figure*}[ht]
\begin{center}
\scriptsize
\begin{tabular}
{c@{\hspace{0.3em}}c@{\hspace{0.3em}}c@{\hspace{0.3em}}c@{\hspace{0.3em}}c@{\hspace{0.3em}}c@{\hspace{0.3em}}c@{\hspace{0.3em}}c}
Original Image & Open the Door & Remove a Jar & Add an Apple & Replace a Glass & Add a Word & Change a Color & Fix the Structure \\
\includegraphics[width=0.115\textwidth]{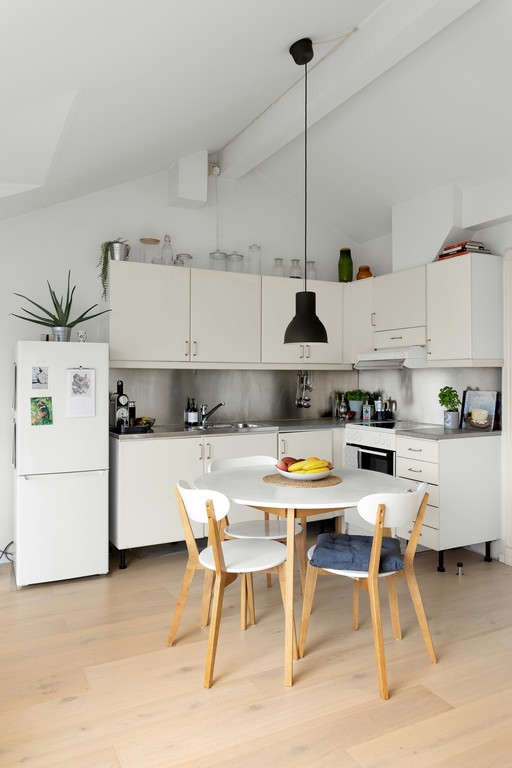} &
\includegraphics[width=0.115\textwidth]{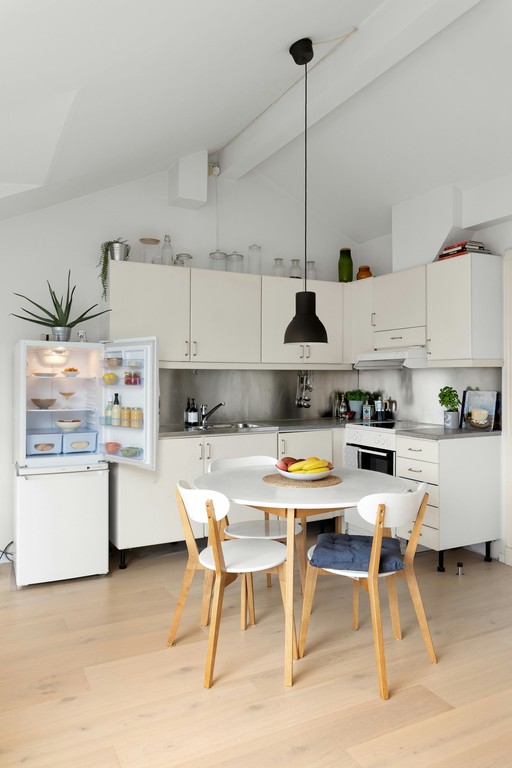} &
\includegraphics[width=0.115\textwidth]{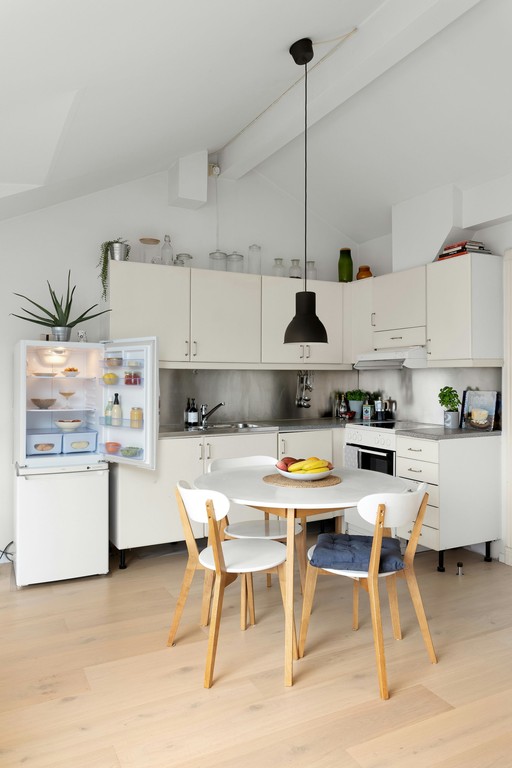} &
\includegraphics[width=0.115\textwidth]{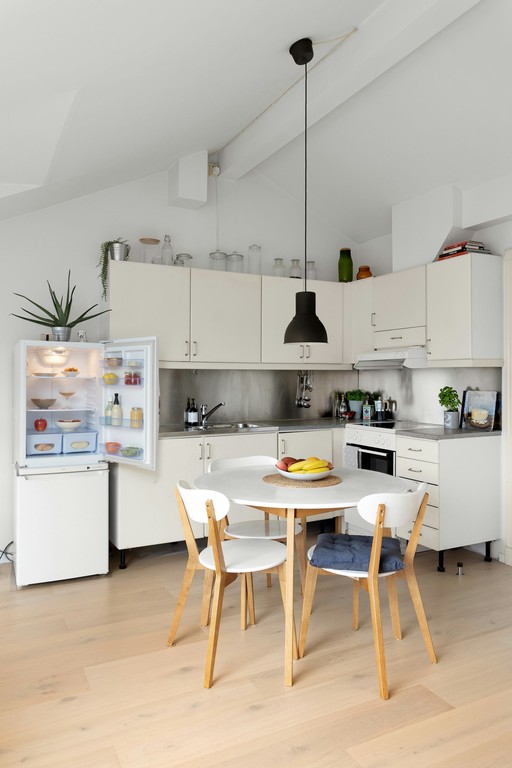} &
\includegraphics[width=0.115\textwidth]{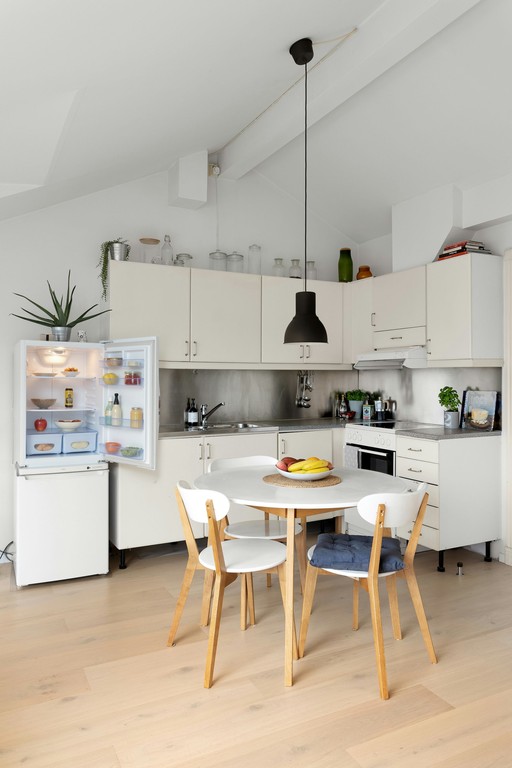} &
\includegraphics[width=0.115\textwidth]{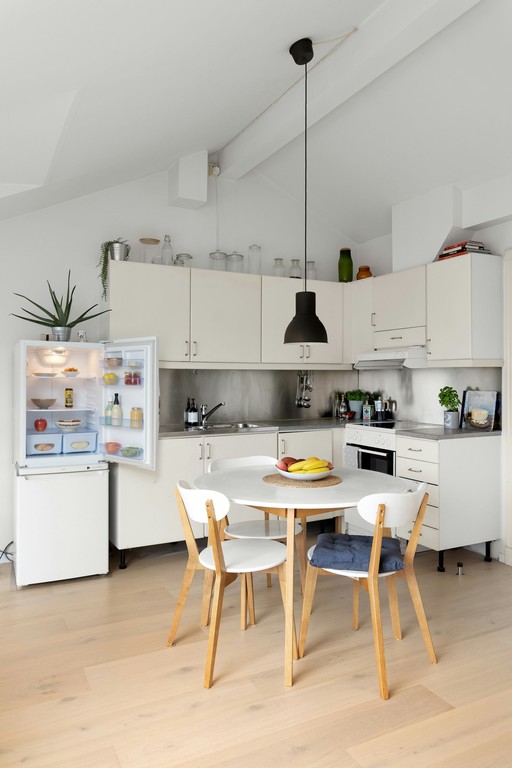} &
\includegraphics[width=0.115\textwidth]{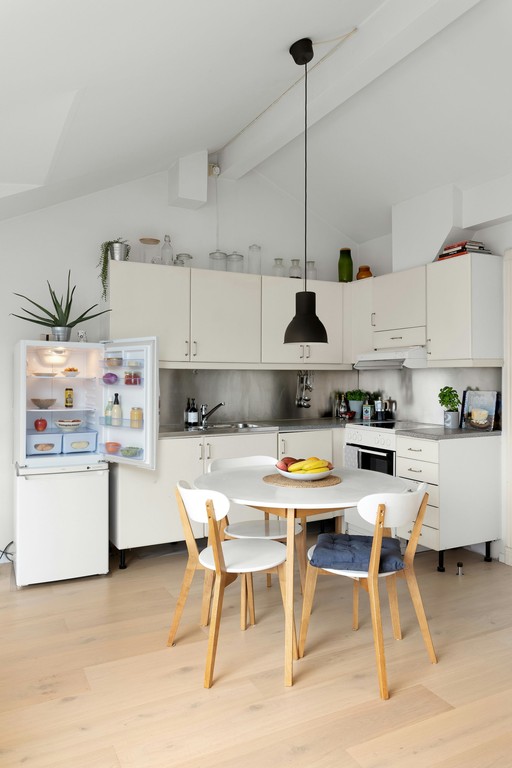} &
\includegraphics[width=0.115\textwidth]{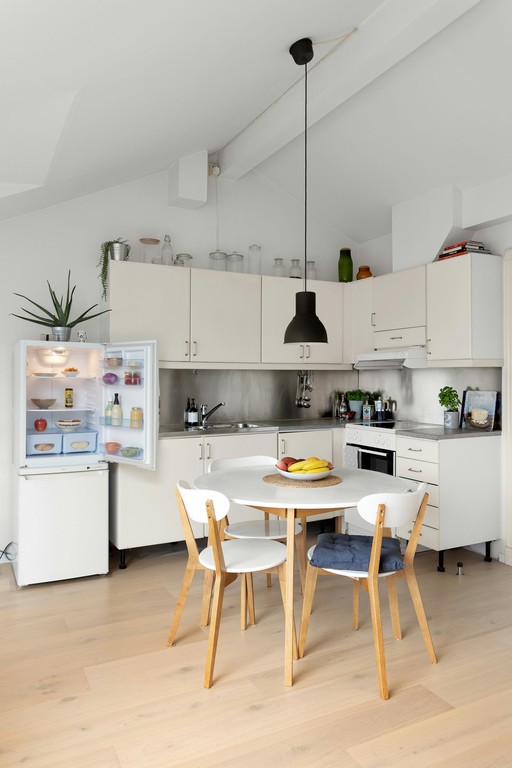} \\
\includegraphics[width=0.115\textwidth]{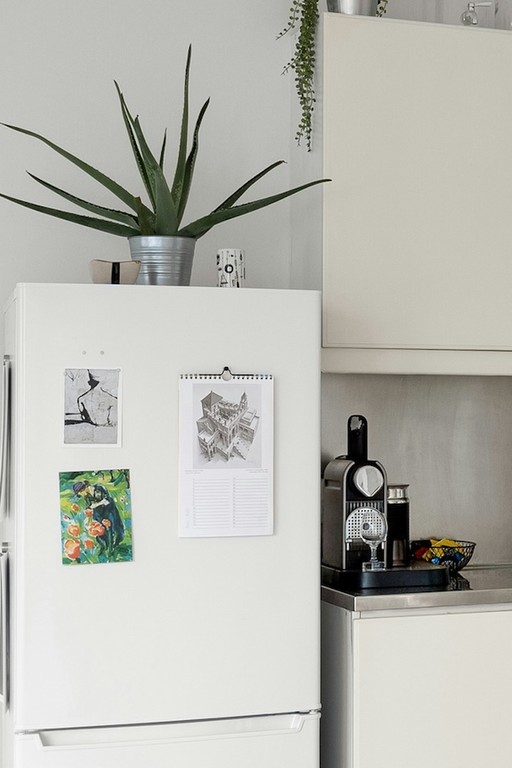} &
\includegraphics[width=0.115\textwidth]{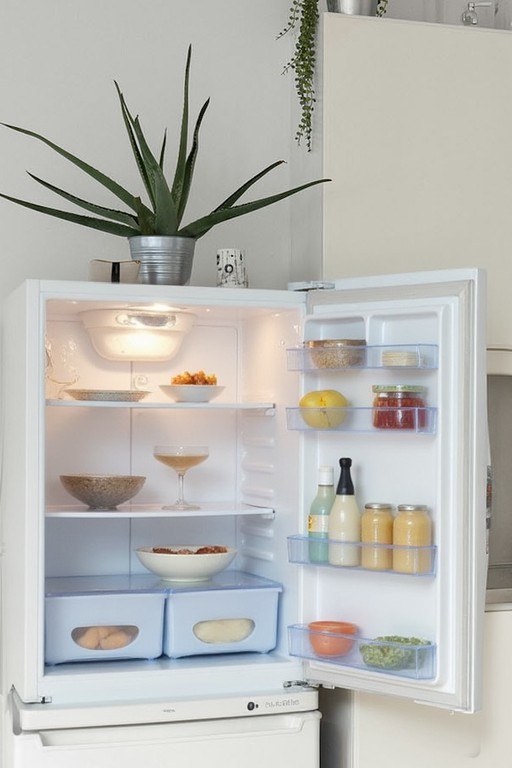} &
\includegraphics[width=0.115\textwidth]{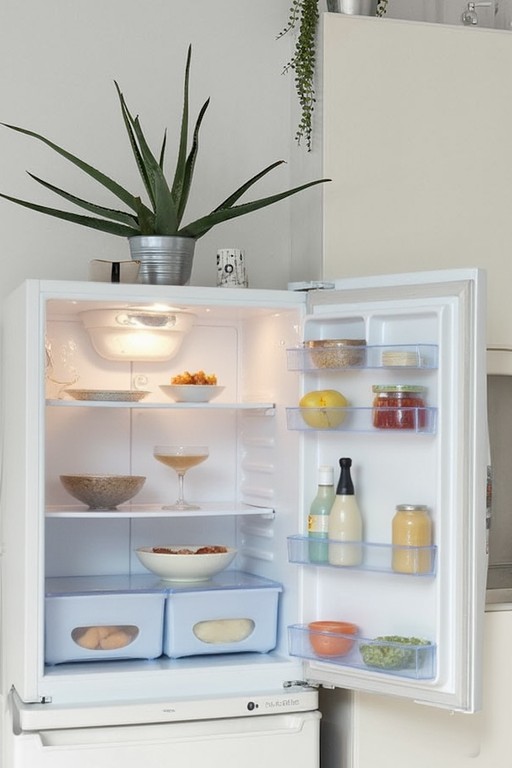} &
\includegraphics[width=0.115\textwidth]{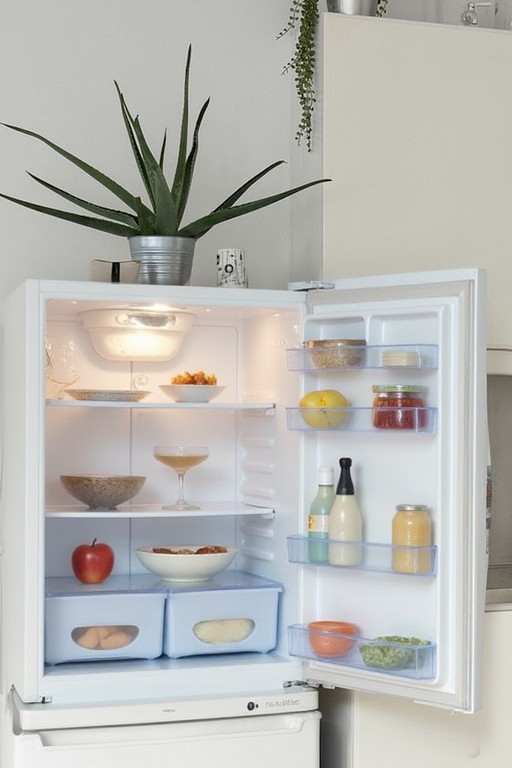} &
\includegraphics[width=0.115\textwidth]{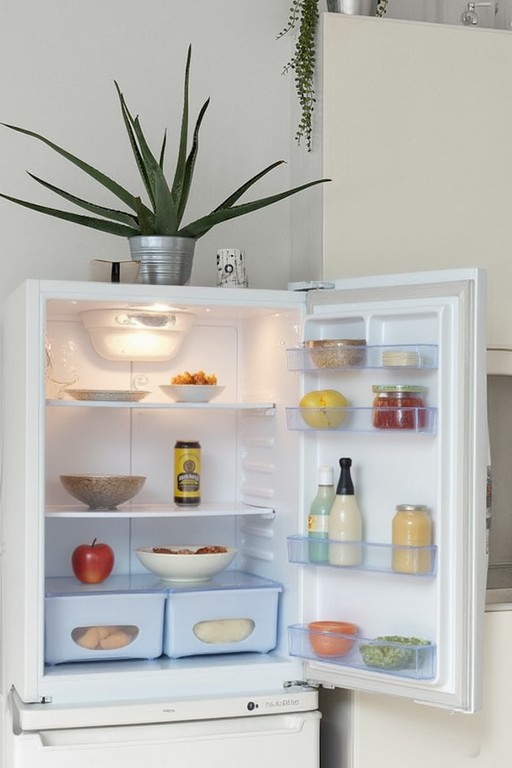} &
\includegraphics[width=0.115\textwidth]{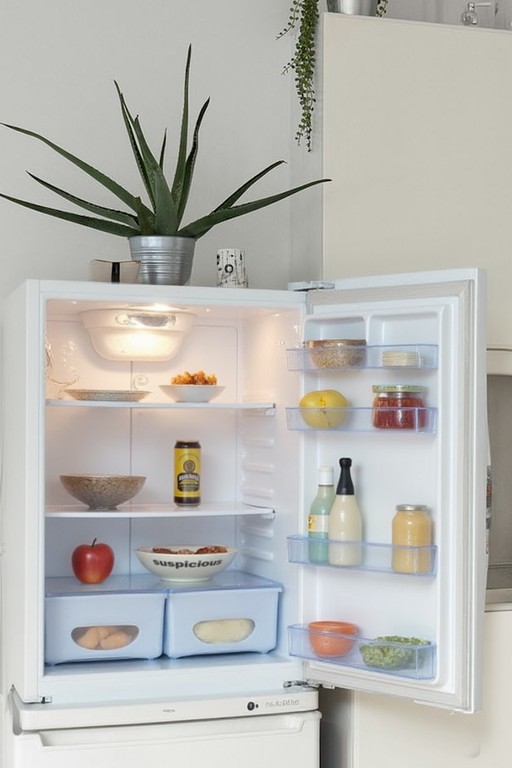} &
\includegraphics[width=0.115\textwidth]{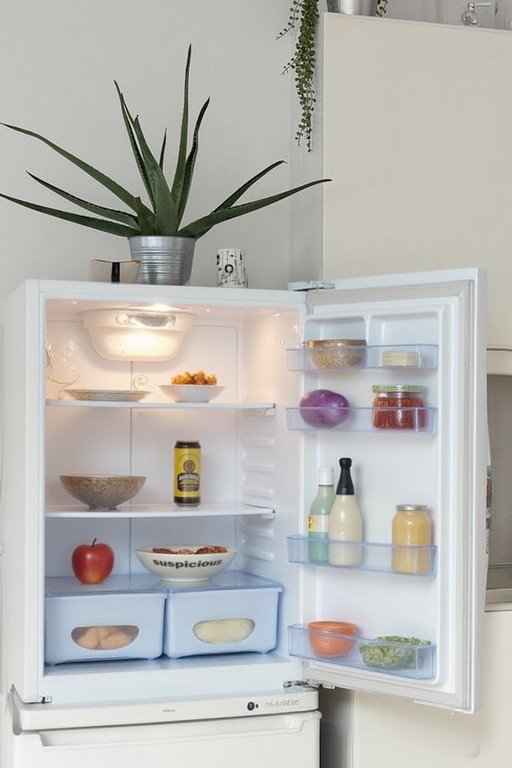} &
\includegraphics[width=0.115\textwidth]{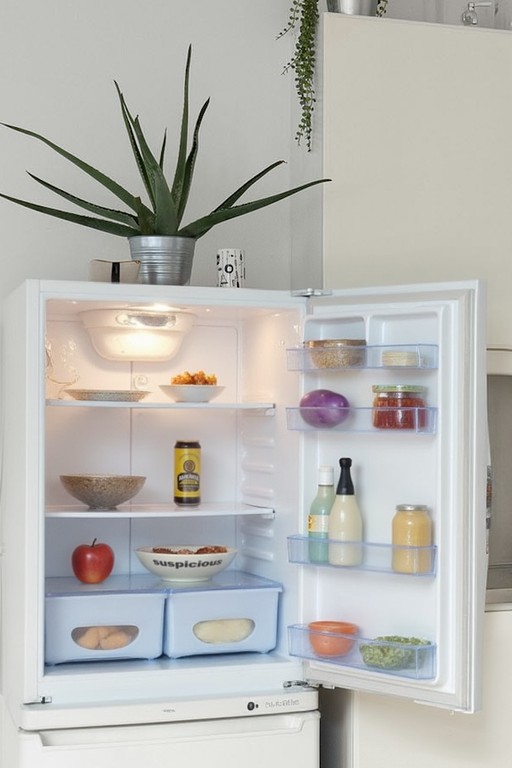} \\
\scriptsize\textsc{Step 0} & \scriptsize\textsc{Step 1} & \scriptsize\textsc{Step 2} & \scriptsize\textsc{Step 3} & \scriptsize\textsc{Step 4} & \scriptsize\textsc{Step 5} & \scriptsize\textsc{Step 6} & \scriptsize\textsc{Steps 7, 8, 9}  \\
\end{tabular}

\caption{\textbf{\spice~enables a user to edit the image exactly as they want, and image details outside the edited region are strictly intact after many editing steps.} The first row shows the full image of a 3000$\times$2000 resolution. The second row shows a 900$\times$600 region enlarged for better visibility. In this example, a user uses 9 editing steps to perform various edits, including structure change, object removal, object addition, object replacement, text addition, color change, and detail fixes. Steps 7, 8, and 9 together fix the fridge structure. The labels above the first row are the abbreviated version of the true editing instructions (more details in \cref{app:hints-and-prompts-fridge}). For example, in the ``Add a Word'' column, the user wants to add the specific word ``suspicious'' to the white bowl. The prompt is ``An open fridge with food in it. A bowl with a word `suspicious' on it.'' The result aligns with the user's requirement.}
\label{fig:advantages}
\end{center}

\end{figure*}

Image editing is the task of changing the content of an image according to a user's requirements. An example is adding an apple to a specific location on an image (\cref{fig:advantages}).\footnote{All figures are provided in high resolution. Readers are encouraged to zoom in to examine the details.} A powerful image editing tool is vital for many applications from creative design to scientific research, including photo editing \cite{mechrez2018saliency}, video editing \cite{videoshop}, data augmentation \cite{hirota-etal-2024-resampled}, and benchmark construction \cite{wu-etal-2024-autohallusion}. 

Existing vision-language models have achieved initial success on image editing \cite{brooks2023instructpix2pix, zhang2024magicbrush}. Such models take in an original image and the user's editing prompt as the input and output the edited image, sometimes with a binary mask as additional input \cite{wang2023imagen, zhao2024ultraedit}.
However, for advanced artistic purposes that require more than one editing step, existing methods are disqualified by the following limitations. First, pixels outside the mask deteriorate after editing. Second, the user cannot specify the precise size and location of an added object by the mask. 
Third, models struggle with unusual editing tasks, such as adding a backpack to a bench facing away from the viewer (\cref{fig:benchmark-results-main-text}). 
Taken together, these limitations prohibit iterative editing, as image degradation inevitably accumulates.\footnote{\url{https://www.reddit.com/r/ChatGPT/comments/1kbj71z/i_tried_the_create_the_exact_replica_of_this/}}

\begin{figure*}[t]
\hypertarget{figure2}{}
\begin{center}
\includegraphics[width=\textwidth]{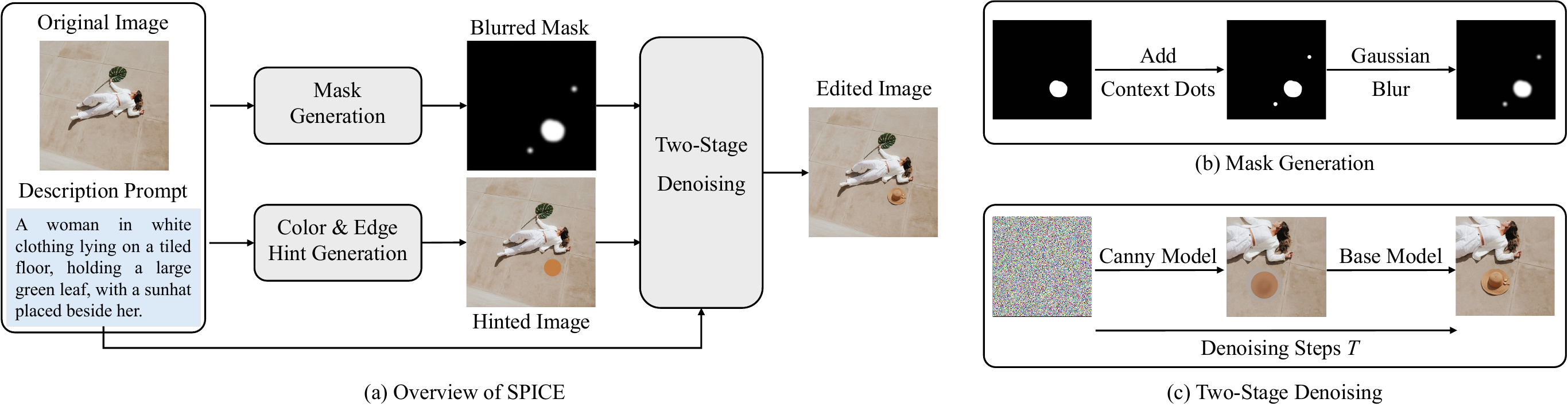}
\caption{\textbf{By sketching a binary mask with context dots and a color \& edge hint, users can effortlessly achieve realistic edits with \spice.} Subfigure (a) shows the overview of our workflow, while Subfigures (b) and (c) show the internal steps. In this example, the user requires a sunhat to be added next to the woman. First, the user sketches both a mask with context dots and a hinted image containing color and edge hints. The mask is automatically blurred after being sketched. Then, during the two-stage denoising step, the Canny and base models perform the early and late denoising steps, respectively. See \cref{fig:simple-editing} for more examples of masks and hints.}
\label{fig:methods}
\end{center}

\end{figure*}

To overcome these limitations, we propose \textbf{\spice}, a \emph{training-free} workflow consisting of 3 steps, namely mask generation, color and edge hint generation, and two-stage denoising (\cref{fig:methods}). 

First, in the mask generation step, the user provides a mask to define the editing region and the context size, localizing the modifications while using essential contextual information from the original image. \spice~strictly constrains the editing region to the user-provided mask, preventing deterioration. 

Second, by accepting color and edge hints as a hinted image, \spice~allows the user to provide arbitrarily detailed or simplified image-space information to the model. This resolves the issue that textual prompts cannot specify precise sizes and locations \cite{brooks2023instructpix2pix, zhang2024magicbrush}.

Finally, \spice~uses a two-stage denoising process, where a Canny edge ControlNet model \cite{zhang2023adding} integrates image-space hints in the \emph{early} denoising steps, and a base diffusion model refines and diversifies the output in the \emph{later} steps. By synergizing the complementary strengths of both models, the two-stage denoising process enables \emph{precise} and \emph{customizable} editing, in which the model faithfully follows the user's requirements on properties of the edited object.
Beyond image editing, the superior faithfulness also helps users to overcome fundamental limitations of AI-generated artwork, such as predominantly portraying a single character or repeatedly erring on details.

By overcoming the limitations, \spice~consistently achieves success in iterative editing tasks of \emph{more than 100 editing steps} (\cref{sec:iterative-editing}). For the first time, our workflow enables scaling test-time compute \cite{snell2024scaling} in image generation, where the user can decide how each unit of additional compute time contributes to the final image quality. This user-friendly design distinguishes our workflow from contemporary test-time scaling approaches that rely on automatic search algorithms \cite{ma2025inference, singhal2025general}.

Despite the strong capability, our workflow can be easily integrated into all popular diffusion model Web UIs, such as Stable Diffusion Web UI Automatic1111, ComfyUI, and Stable Diffusion Web UI Forge.
Unlike existing tools such as ADetailer\footnote{\url{https://github.com/Bing-su/adetailer}} or Regional Prompter,\footnote{\url{https://github.com/hako-mikan/sd-webui-regional-prompter}} which offer an overwhelming number of hyperparameters that can be difficult to navigate, \spice~provides a much smaller set of hyperparameters for ease-of-use.
Moreover, \spice~introduces minimal computational overhead in each editing step, making editing as efficient as generating an image from text with the same model and sampling hyperparameters. This extremely low overhead allows \spice~to run on consumer GPUs (e.g., a single NVIDIA GeForce RTX 4090), unlocking iterative editing for more users.
Combining strong capabilities with low implementation and computational cost, our workflow provides a powerful yet accessible tool for researchers and non-researchers alike.

\section{Methods}

\spice~is based on inpainting \cite{lugmayr2022repaint}, an operation that uses a diffusion model to replace a masked region on an image, conditioned on a prompt that describes the new image. Note that this prompt differs from the editing prompt, which uses a verb (e.g., add or remove) to describe the editing operation \cite{zhang2024magicbrush}. For example, when the editing prompt is ``add an apple in the fridge'', the description prompt will be ``an apple in a fridge''. In this section, we introduce the three key steps of our workflow, namely mask generation, color and edge hint generation and two-stage denoising (\cref{fig:methods}).
\subsection{Mask Generation}
\label{sec:mask-generation}
\paragraph{Context Selection.}
We denote the original image to be edited as \( I_T \in [0, 1]^{H \times W \times C} \), where \( T \) is the index of the editing step, \( H \) and \( W \) are the image height and width, and \( C = 3 \) represents the RGB color channels. Other than a prompt $p$, traditional inpainting methods\footnote{\url{https://github.com/AUTOMATIC1111/stable-diffusion-webui}} require the user to provide a binary mask \( M \in \{0, 1\}^{H \times W} \), where 1 indicates the region to be edited. Given the mask and the prompt, the inpainting operation uses the following steps to generate an edited image. First, a bounding box of the region to be edited is calculated, and the bounding box is extended either vertically or horizontally to ensure its aspect ratio matches a user-specified resolution supported by the diffusion model, such as 1216$\times$832 \cite{podell2024sdxl}. The user-specified resolution is usually larger than the extended bounding box. Next, pixels on $I$ and $M$ within the extended bounding box are upsampled to the user-specified resolution. Then, the diffusion model generates a new image from latent noise. The generation process is conditioned on the description prompt and the upsampled pixels from $I$ and $M$.
Finally, the output is downsampled to the resolution of the extended bounding box, and the inpainted region on $I_T$ is replaced by the downsampled output, resulting in $I_{T+1}$.

However, since these traditional methods generate the inpainted region without contextual information outside the extended bounding box, the output often appears unnatural, with inconsistent lighting or color compared to the rest of the image. A naive solution is the ``whole image'' mode, where the entire image is used as context, and the extended bounding box is not used. However, this introduces two major problems: (1) poor performance on small objects (e.g., deformed fingers or scrambled patterns), and (2) distortion caused by resizing the whole image to a model-supported resolution when aspect ratios of the two differ.

To address these issues, we introduce \textit{context dots}, a pair of dots at opposite corners of the desired bounding box. Context dots ensure that the extended bounding box includes sufficient context for generation. The user directly adds context dots to the original mask, resulting in a context mask \( M_\text{context} \in \{0, 1\}^{H \times W} \), as shown in Figure \hyperlink{figure2}{2(b)}. This simple enhancement offers three key advantages: (1) the user can exclude image areas that interfere with the inpainting process, (2) the user can specify a  resolution between that of the inpainted region and that of the full image, balancing local details and global context (\cref{sec:customizable-editing}), and (3) context dots minimally affect surrounding pixels, limiting changes to only the desired editing region.
\paragraph{Soft Inpainting.}

On inpainted images, there is usually an unwanted sharp boundary between the inpainted region and the remaining parts of the image. In more severe cases, an inpainted object can be incomplete. The reason is that even with a context, diffusion models are not robust enough to generate pixels that seamlessly blend with existing ones.  To mitigate these artifacts, we adopt Differential Diffusion \cite{levin2023differential}, a method that allows the diffusion model to be conditioned by continuous mask values in [0, 1] during generation. We provide the continuous values as a soft mask $M_\text{soft} \in [0, 1]^{H\times W}$ by applying a simple Gaussian blur to $M_\text{context}$ with a kernel size of a few pixels, as shown in Figure \hyperlink{figure2}{2(b)}. This procedure, together with thresholds for blending the original and inpainted image, is named as Soft Inpainting in popular Web UIs. We use Soft Inpainting when it is available in a Web UI or implement our own simplified version (without thresholds) when it is not available.

\subsection{Color and Edge Hint Generation}

In many existing image editing systems \cite{croitoru2023diffusion, yang2023diffusion, huang2024diffusion}, users typically specify desired content through a textual prompt. However, words alone cannot easily describe certain details, such as asymmetric apparel designs or intricate color patterns. Instead, these subtle details can be effectively conveyed with an additional visual hint in the image space.
To this end, \spice~allows a user to first create a rough sketch or color layout using standard editing software (e.g., Krita or Adobe Photoshop). The resultant image, denoted $I_\text{hinted} \in [0,1]^{H \times W \times C}$, then replaces the original image as the inpainting input. To incorporate information from this hinted image, a denoising strength hyperparameter in [0, 1] specifies how much the original pixels on the image should be changed.\footnote{The hyperparameter does not work as a simple linear blending coefficient at the post-processing stage, and its exact implementation differs for various algorithms. To avoid ambiguity, we refrain from providing a general equation here. Interested readers can refer to the source code of popular Web UIs for the equations. } At a moderate denoising strength (e.g., 0.5 to 0.7), the inpainting model produces pixels that remain close to the user’s sketched hints, preserving intended colors, shapes, or patterns. Meanwhile, the pixels still diverges enough from the sketch, forming realistic objects in the end.

Note that the user can provide any form of color and edge hints, including sketches, reference images pasted in a collage style, or even another region of the original image (such as using the Clone Stamp Tool in PhotoShop). This flexibility is an improvement over existing methods, such as MagicQuill \cite{liu2024magicquill} that uses downsampled 32$\times$32 color blocks for guidance and thus loses high-frequency details. Hence, \spice~allows a user to perform image editing from any point on the human-model collaboration spectrum. At one end, the user fully edit the image by drawing out every detail, without the help of diffusion models. At the other end, the user fully delegates the model to edit the image. Usually, the user can get decent editing results by staying on the end where the user input is minimal (\cref{sec:benchmark-results}). We will also discuss how a small set of hyperparameters enable the user to move freely along this spectrum (\cref{sec:customizable-editing}).

\subsection{Two-Stage Denoising}

Image editing methods \cite{lugmayr2022repaint} typically rely on a single diffusion model to perform all denoising steps when generating an image from latent noise. However, different diffusion models have complementary strengths. On the one hand, a general-purpose text-to-image base model, such as Flux.1 [dev],\footnote{\url{https://huggingface.co/black-forest-labs/FLUX.1-dev}} excels at generating rich variations in its output but can only be conditioned on textual prompts.
On the other hand, Flux.1 [dev] Canny,\footnote{\url{https://huggingface.co/black-forest-labs/FLUX.1-Canny-dev}} a model derived from Flux.1 [dev] that contains a Canny edge ControlNet (i.e., a Canny model), can be conditioned on Canny edge information \cite{zhang2023adding} but sacrifices variability.

To synergize the strengths, we propose a two-stage denoising process. Specifically, we use a Canny model $f_\text{Canny}$ during \emph{early} denoising steps to incorporate image-space hints. Starting from latent noise $z_0$, the Canny model can condition its generation on the Canny edge information $E_\text{hinted}$ (extracted from the hinted image $I_\text{hinted}$) within the extended bounding box, so that the edge hints can be followed. After the early steps, we get an intermediate latent image
\begin{equation}
    z_\text{Canny} = f_\text{Canny}(I_\text{hinted}, p, E_\text{hinted}, M_\text{soft}, z_0).
\end{equation}
In the \emph{late} denoising steps, we use a base model $f_\text{base}$ to generate diverse content from $z_\text{Canny}$, achieving realism and sophistication despite the simplicity of color and edge hints. This can be formulated as 
\begin{equation}
    z_\text{base} = f_\text{base}(I_\text{hinted}, p, M_\text{soft}, z_\text{Canny}).
\end{equation}
The final latent image $z_\text{base}$ will be decoded into the edited RGB image. This two-stage denoising process ensures that the denoising process benefits from the strengths of both models. Meanwhile, by simply adjusting the proportion of denoising steps assigned to each model, users can intuitively balance variability and controllability.

Empirically, we find that for a wide range of image editing tasks, Canny Edge ControlNet models outperforms other ControlNet variants (e.g., depth or pose) in the first denoising stage. Furthermore, Canny models are more widely available than other ControlNet models. Therefore, we only use Canny models but not other ControlNet models in our workflow.

\section{Experiments}

We comprehensively evaluate \spice~on challenging editing tasks. Specifically, we stress test the three features of \spice, namely precise (\cref{sec:precise-editing}), iterative (\cref{sec:iterative-editing}), and customizable editing (\cref{sec:customizable-editing}). We also provide ablation studies to justify each step in \spice. Due to page limitations, full details and more results are deferred to \cref{app:overview-of-results}. In this section, we highlight the results that demonstrate the unique strengths of \spice~over state-of-the-art baselines.

\subsection{Benchmark Results}

\begin{figure*}[htbp]
\begin{center}
\small
\begin{tabular}{c@{\hspace{0.3em}}c@{\hspace{0.3em}}c@{\hspace{0.3em}}c@{\hspace{0.3em}}c@{\hspace{1em}}c@{\hspace{0.3em}}c@{\hspace{0.3em}}c@{\hspace{0.3em}}c@{\hspace{0.3em}}c}
    \scriptsize\textsc{Original} & \scriptsize\textsc{IP2P} & \scriptsize\textsc{MB} & \scriptsize\textsc{UE} & \scriptsize\textsc{\textbf{Ours}} & \scriptsize\textsc{Original} & \scriptsize\textsc{IP2P} & \scriptsize\textsc{MB} & \scriptsize\textsc{UE} & \scriptsize\textsc{\textbf{Ours}} \\
    
    \includegraphics[width=0.09\textwidth]{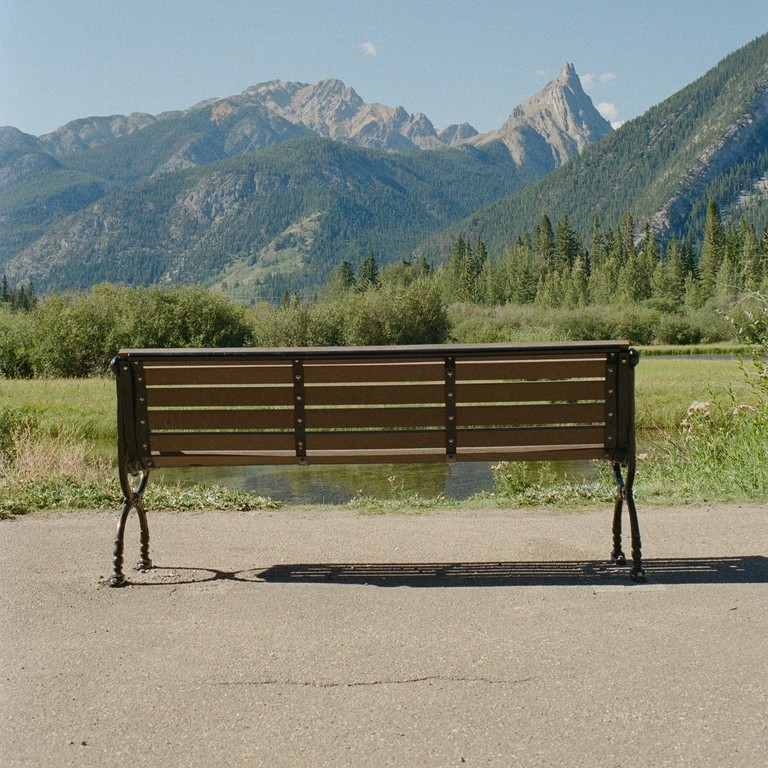} &
    \includegraphics[width=0.09\textwidth]{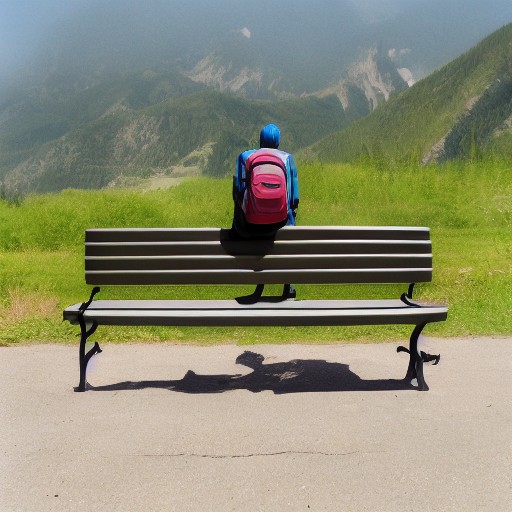} &
    \includegraphics[width=0.09\textwidth]{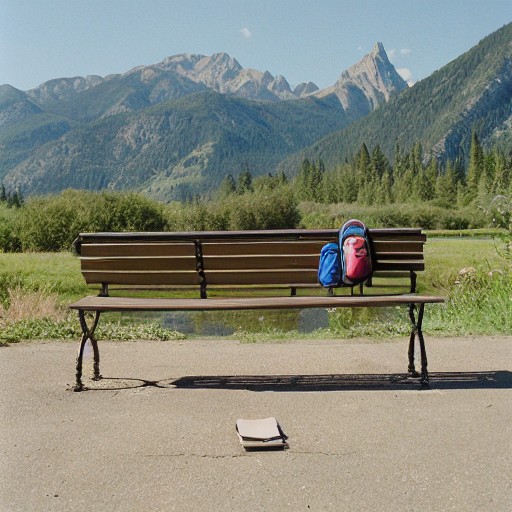} &
    \includegraphics[width=0.09\textwidth]{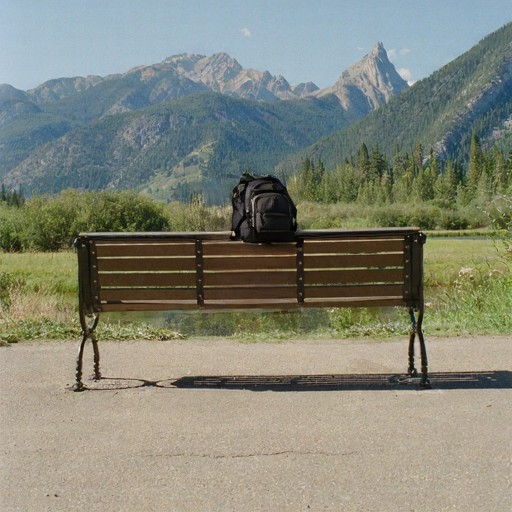} &
    \includegraphics[width=0.09\textwidth]{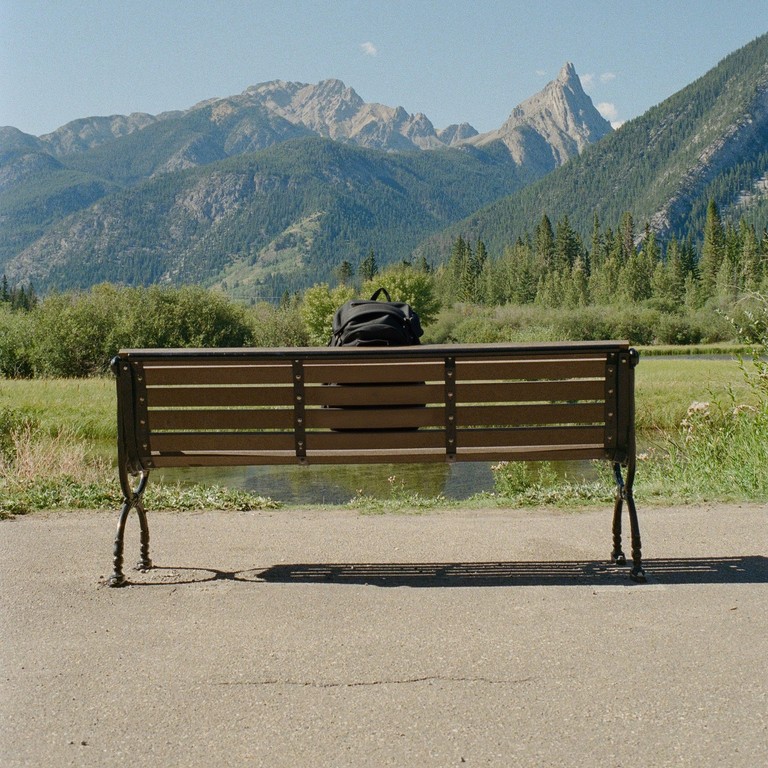} &
    \includegraphics[width=0.09\textwidth]{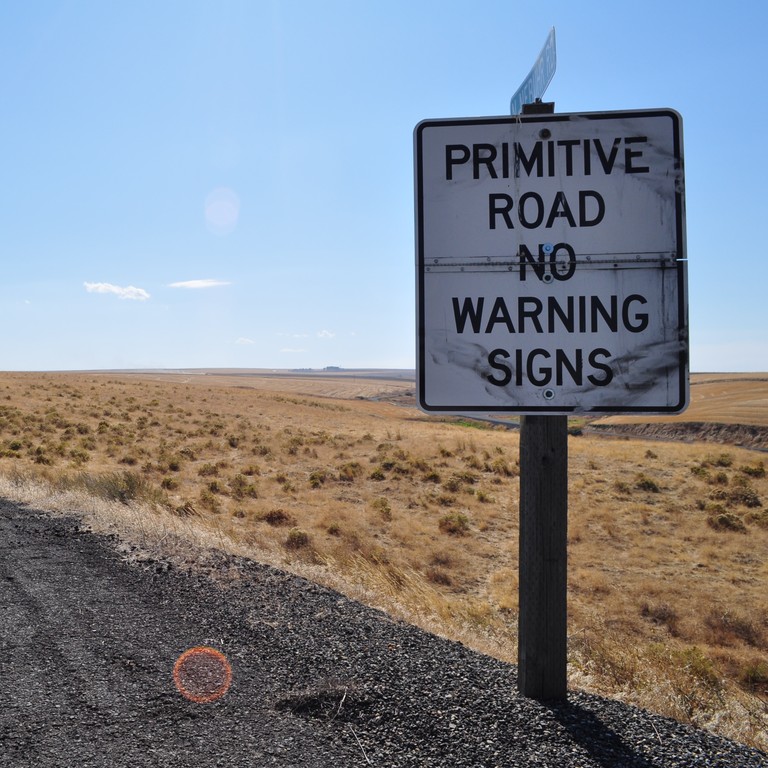} &
    \includegraphics[width=0.09\textwidth]{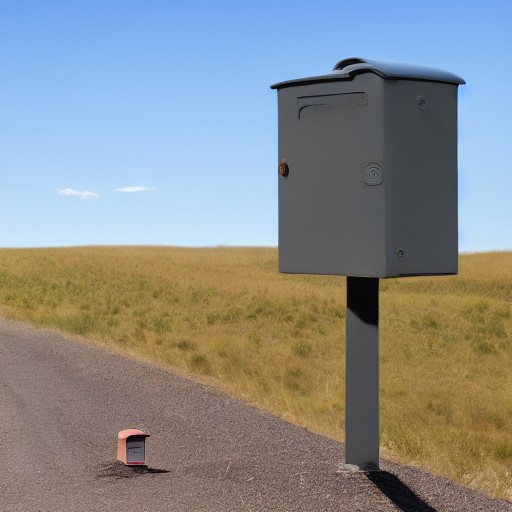} &
    \includegraphics[width=0.09\textwidth]{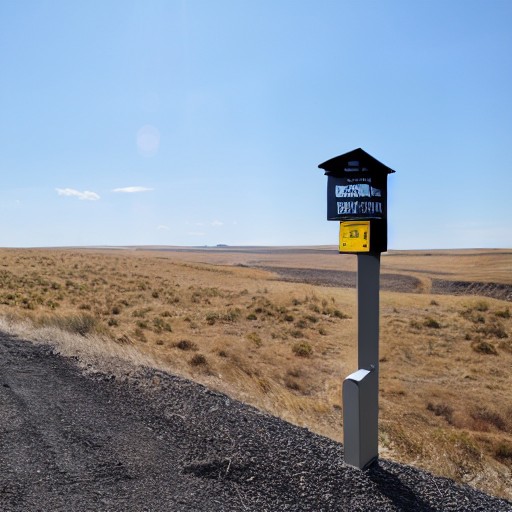} &
    \includegraphics[width=0.09\textwidth]{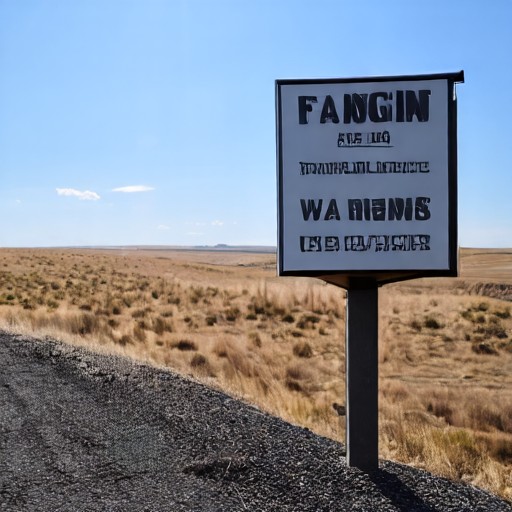} &
    \includegraphics[width=0.09\textwidth]{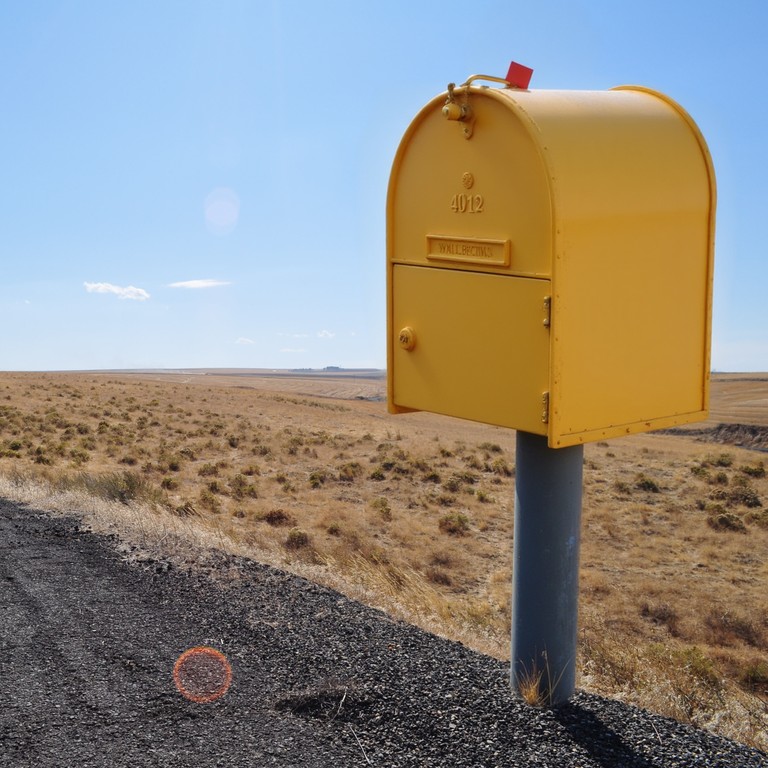} \\
    \multicolumn{5}{c}{\scriptsize \textbf{Object Addition:} Add a backpack placed on the bench.} & \multicolumn{5}{c}{ \scriptsize\textbf{Object Replacement:} Replace the road sign with a mailbox.} \\
    [0.5em]
    \includegraphics[width=0.09\textwidth]{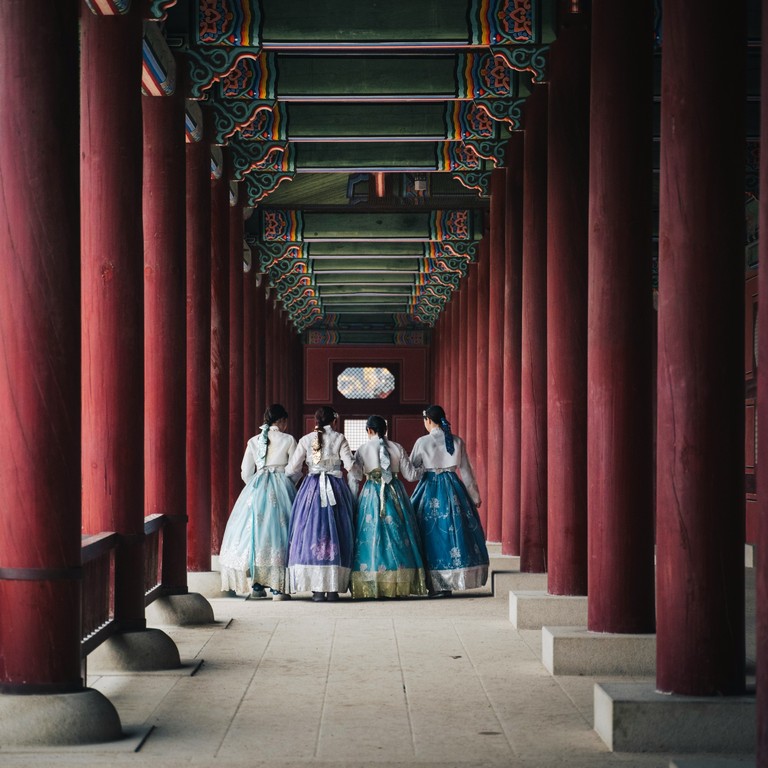} &
    \includegraphics[width=0.09\textwidth]{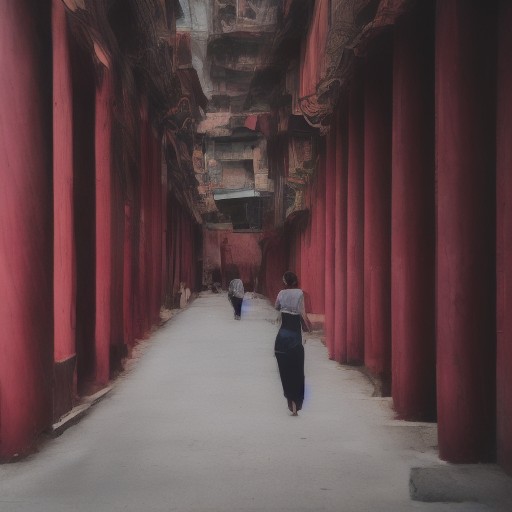} &
    \includegraphics[width=0.09\textwidth]{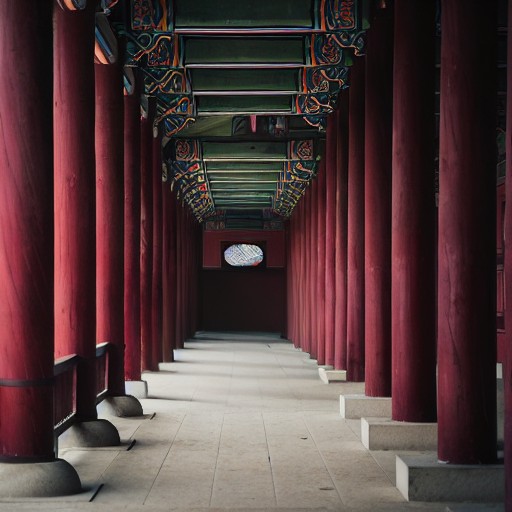} &
    \includegraphics[width=0.09\textwidth]{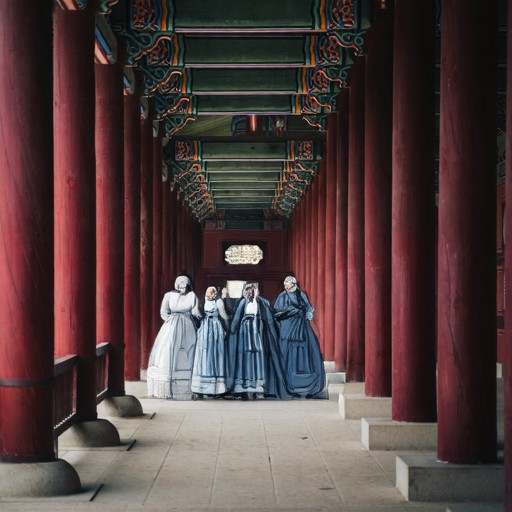} &
    \includegraphics[width=0.09\textwidth]{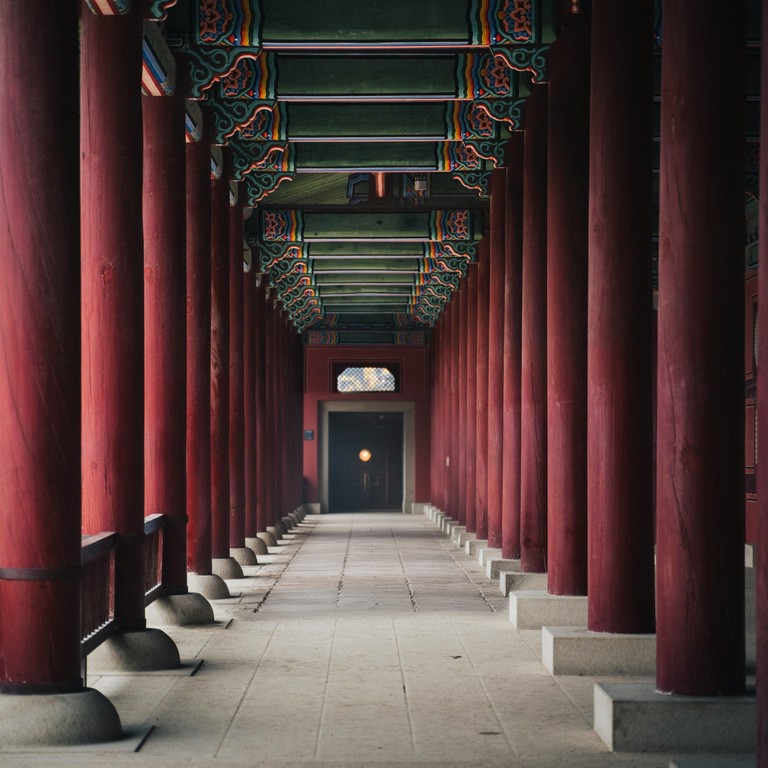} &
    \includegraphics[width=0.09\textwidth]{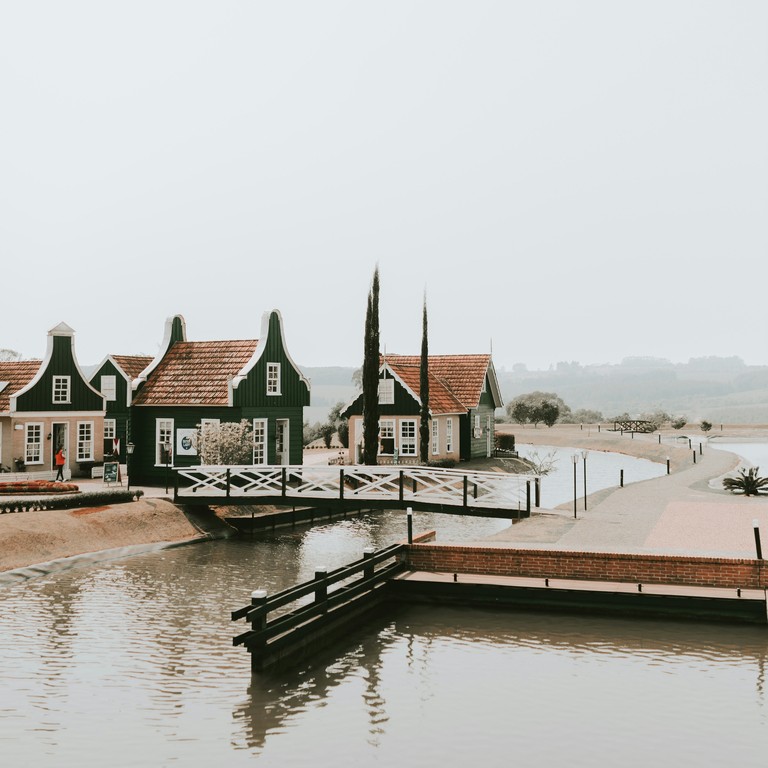} &
    \includegraphics[width=0.09\textwidth]{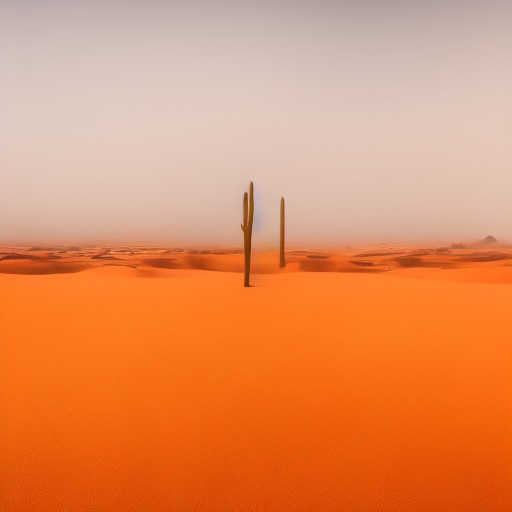} &
    \includegraphics[width=0.09\textwidth]{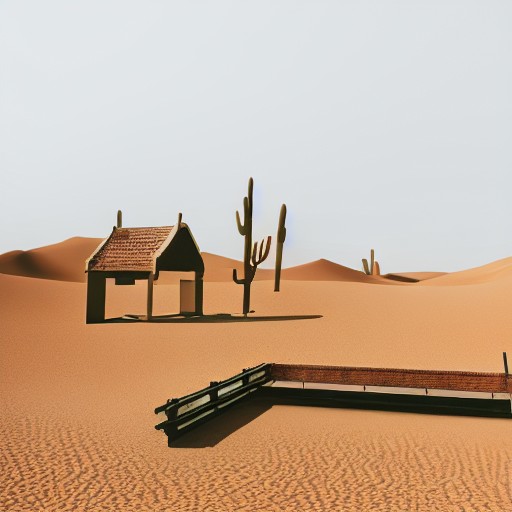} &
    \includegraphics[width=0.09\textwidth]{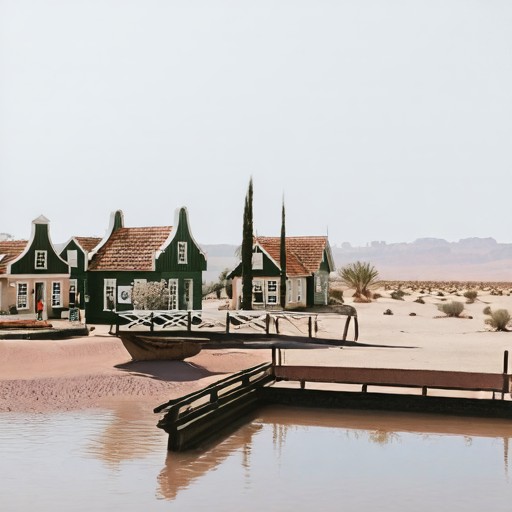} &
    \includegraphics[width=0.09\textwidth]{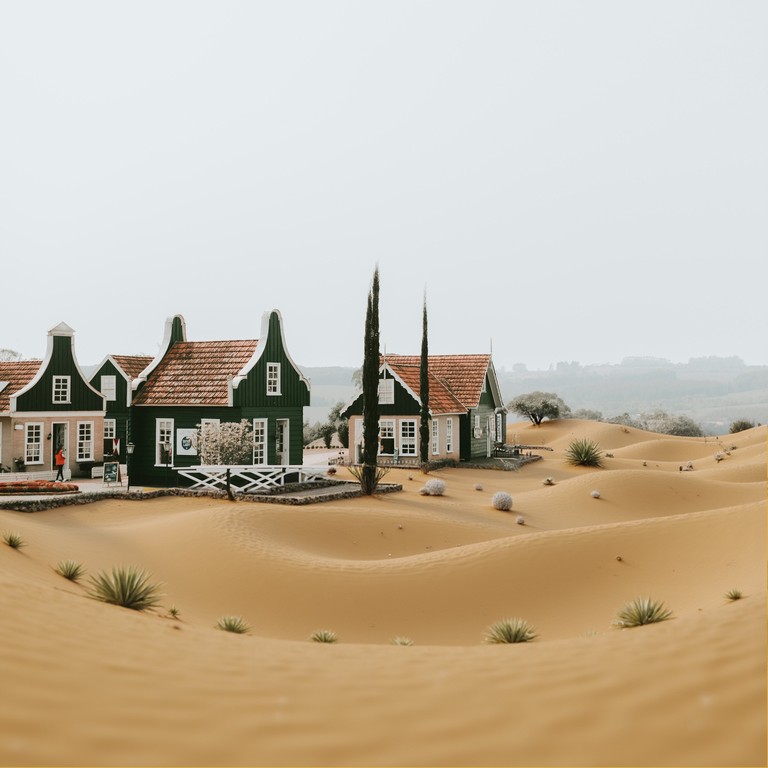} \\
    \multicolumn{5}{c}{\scriptsize\textbf{Object Removal:} Remove the four women.} & \multicolumn{5}{c}{\scriptsize\textbf{Background Change:} Change the riverside to a desert.} \\
    [0.5em]
    \includegraphics[width=0.09\textwidth]{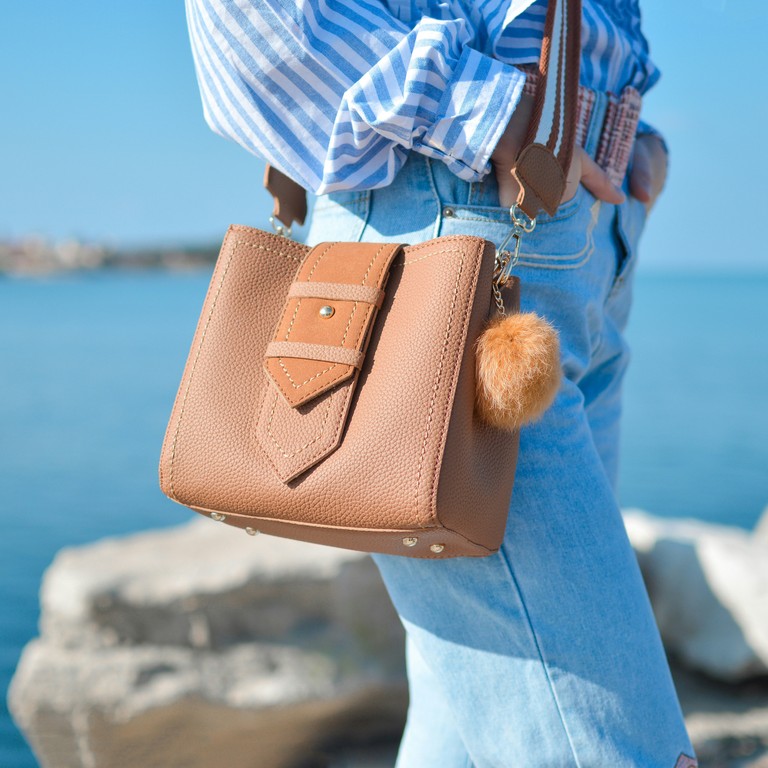} &
    \includegraphics[width=0.09\textwidth]{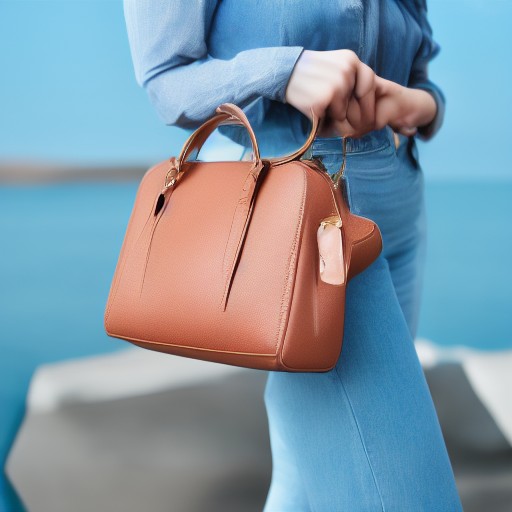} &
    \includegraphics[width=0.09\textwidth]{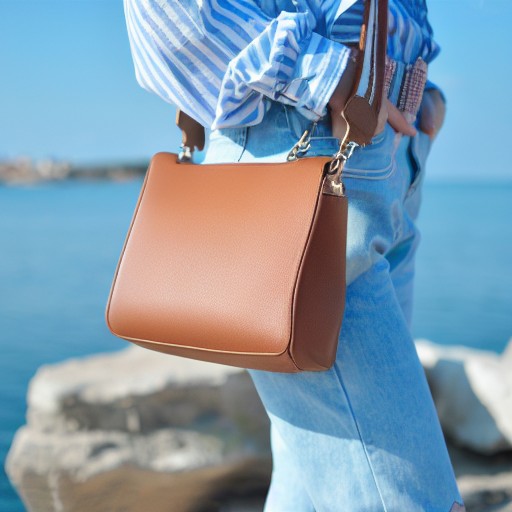} &
    \includegraphics[width=0.09\textwidth]{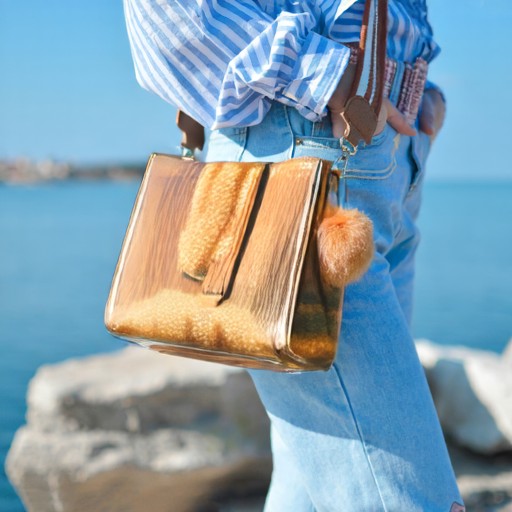} &
    \includegraphics[width=0.09\textwidth]{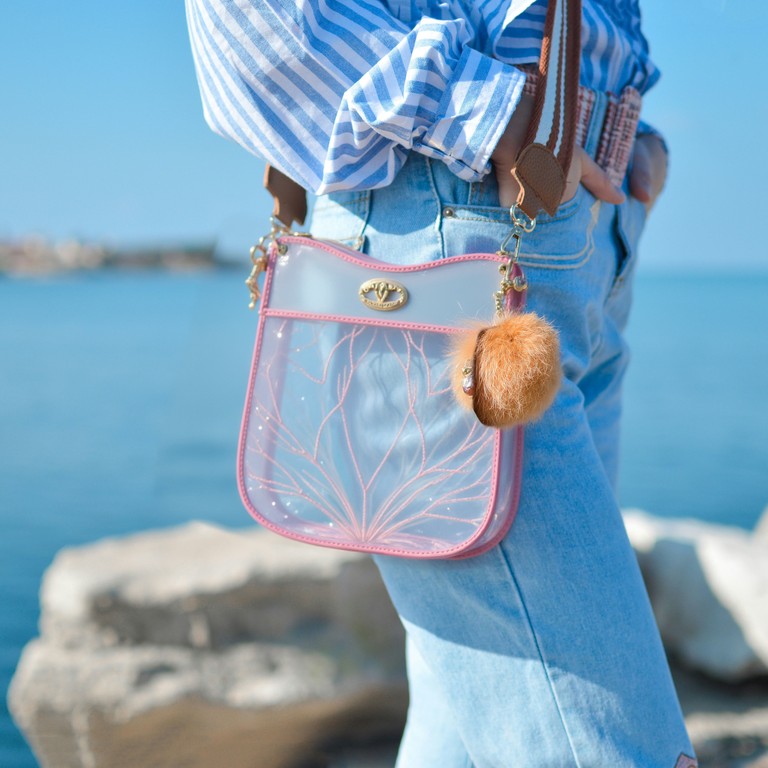} &
    \includegraphics[width=0.09\textwidth]{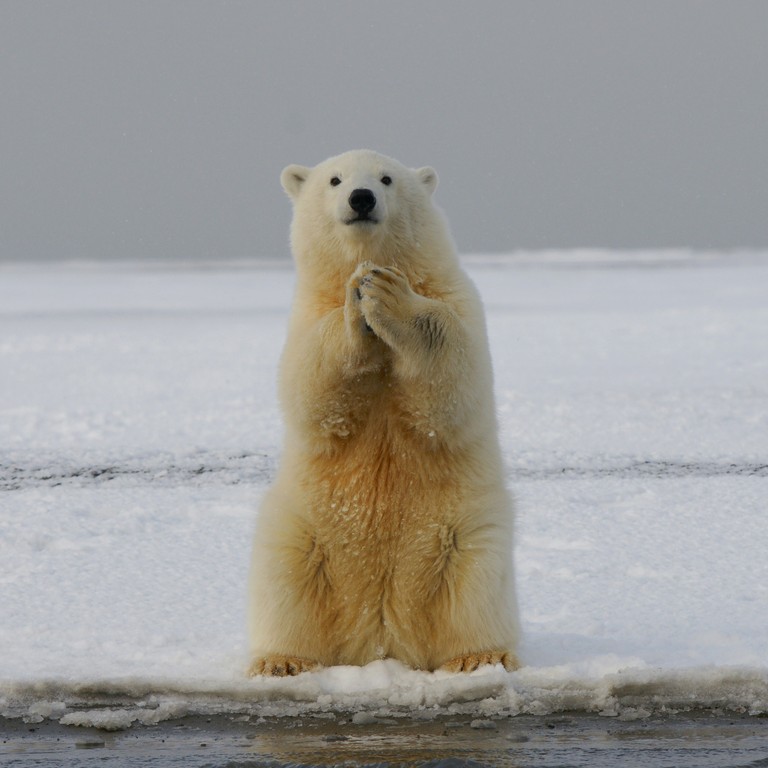} &
    \includegraphics[width=0.09\textwidth]{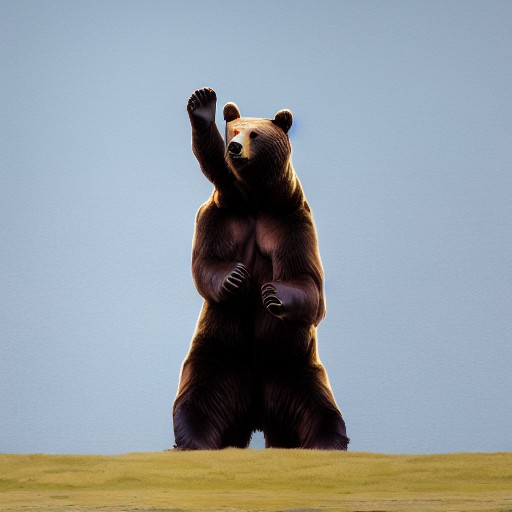} &
    \includegraphics[width=0.09\textwidth]{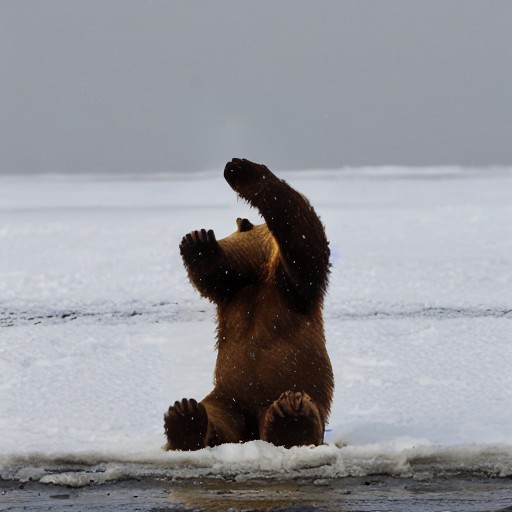} &
    \includegraphics[width=0.09\textwidth]{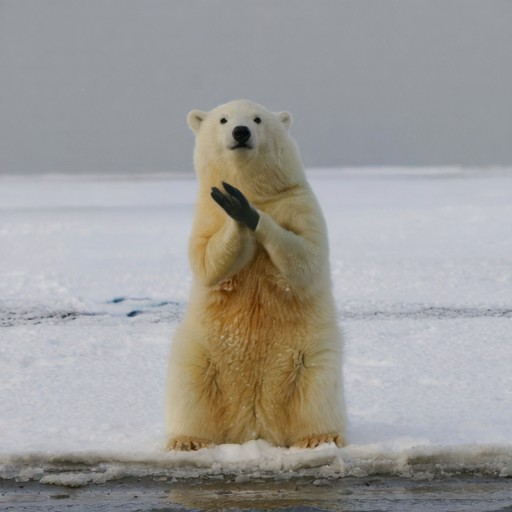} &
    \includegraphics[width=0.09\textwidth]{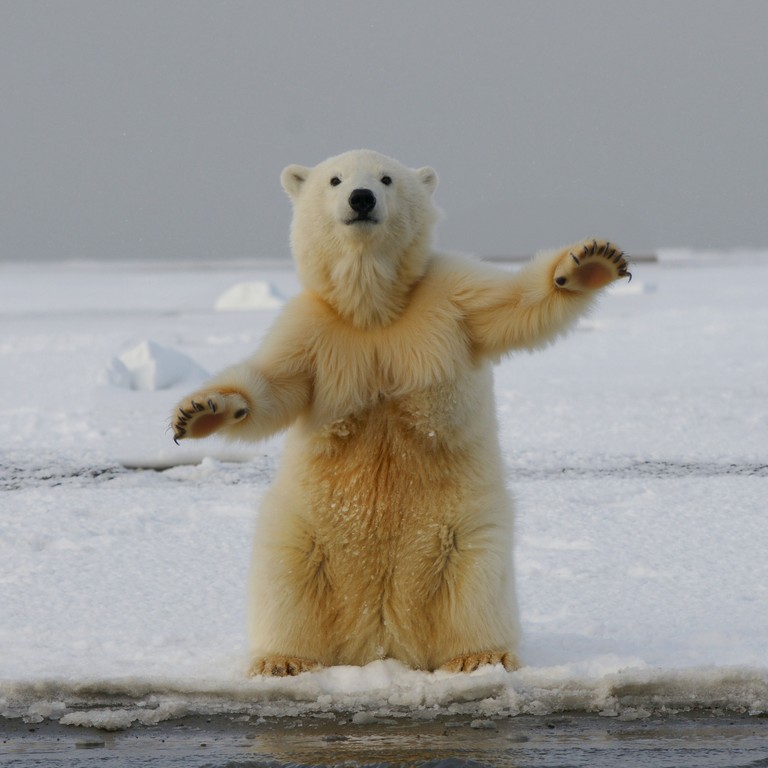} \\
    \multicolumn{5}{c}{\scriptsize\textbf{Texture Change:} Turn the handbag into the glass.} & \multicolumn{5}{c}{\scriptsize\textbf{Action Change:} Turn the bear raising its hand.} \\
\end{tabular}
\caption{\textbf{Our workflow outperforms baseline methods in 6 editing categories from EditEval.} Each group of five images shows an example from an editing task. Each group starts from original image, followed by IP2P, MagicBrush (MB), UltraEdit (UE), and our results.}
\label{fig:benchmark-results-main-text}
\end{center}
\end{figure*}

We evaluate \spice~against baselines of representative open and proprietary models. The results show that \spice~qualitatively outperforms both open (\cref{fig:benchmark-results-main-text}) and proprietary models (\cref{app:challenging-examples}), with a clear advantage in its prompt-following capability. 

\subsection{Precise, Iterative, and Customizable Editing}

Prompt-based image generation or editing models frequently fail to fulfill their promise that users can create content according to their own will. Users are frustrated for two main reasons.

\begin{figure}[t]

\begin{center}
\subfigure[DALL·E 3]{
\scriptsize
\begin{tabular}{c@{\hspace{0.8em}}c}
    \textsc{Round 1} & \textsc{Round 1}\\
    \includegraphics[width=0.09\textwidth]{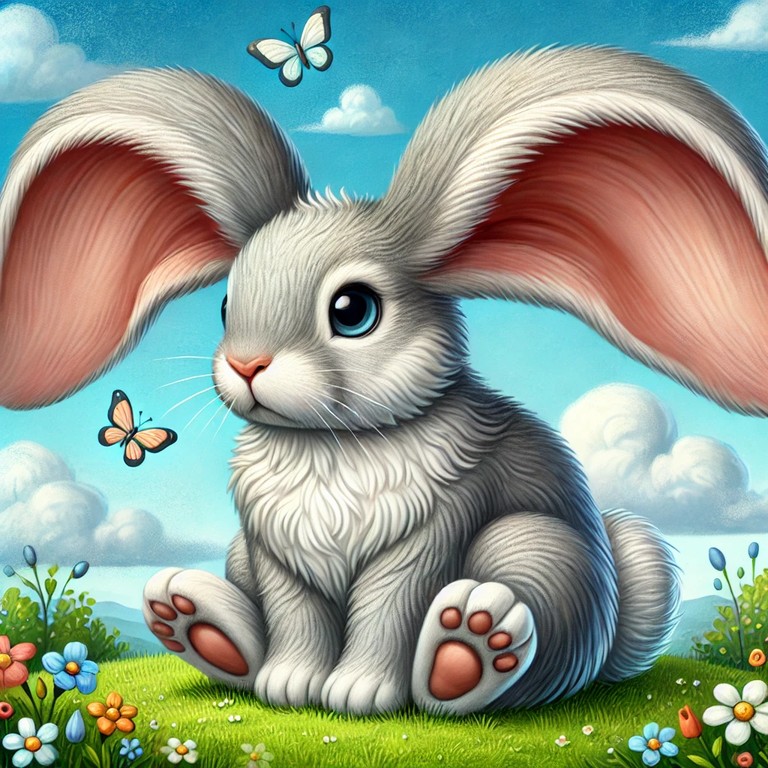} &
    \includegraphics[width=0.09\textwidth]{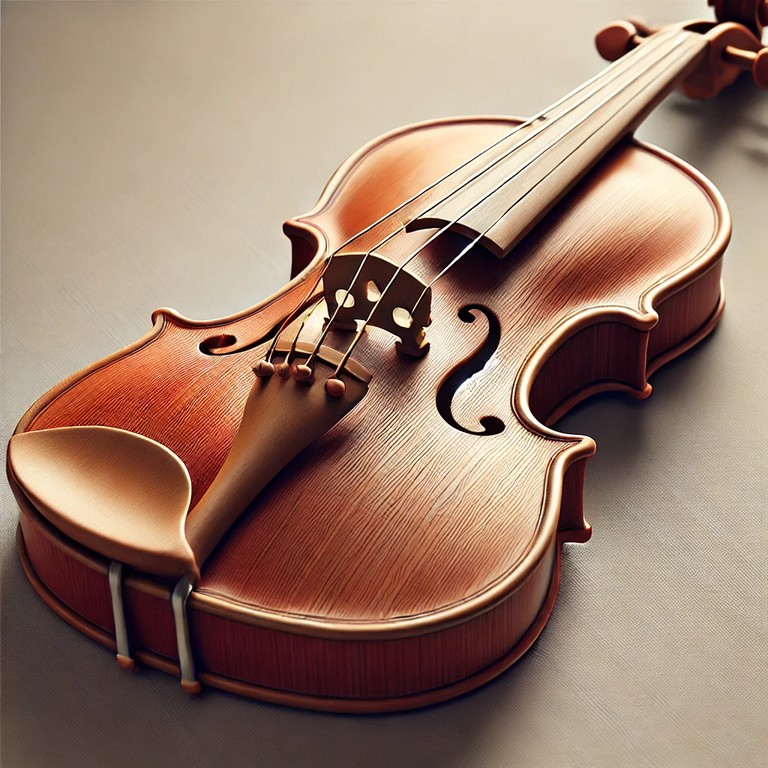} \\
    \textsc{Round 2} & \textsc{Round 2}\\
    \includegraphics[width=0.09\textwidth]{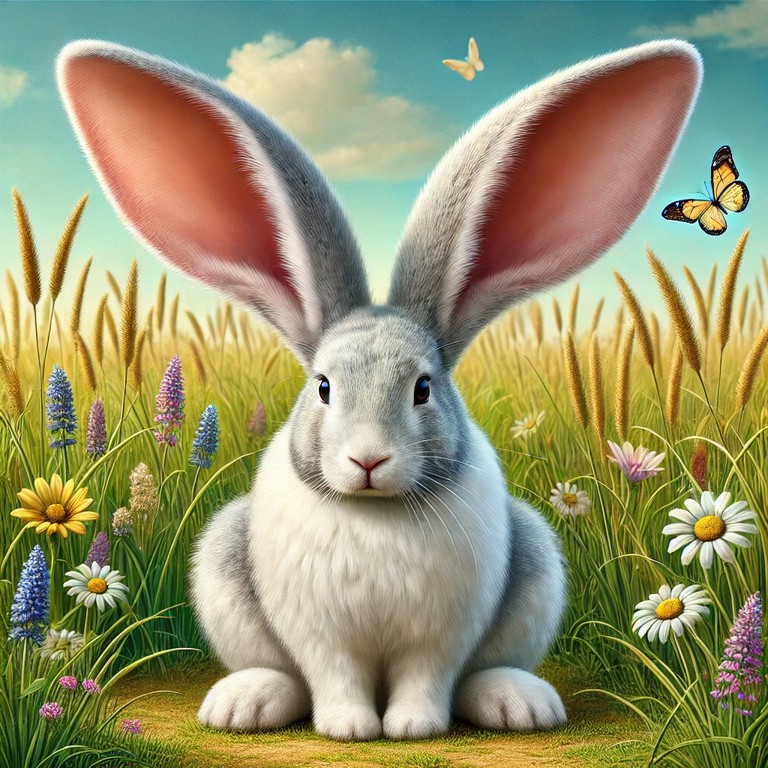} &
    \includegraphics[width=0.09\textwidth]{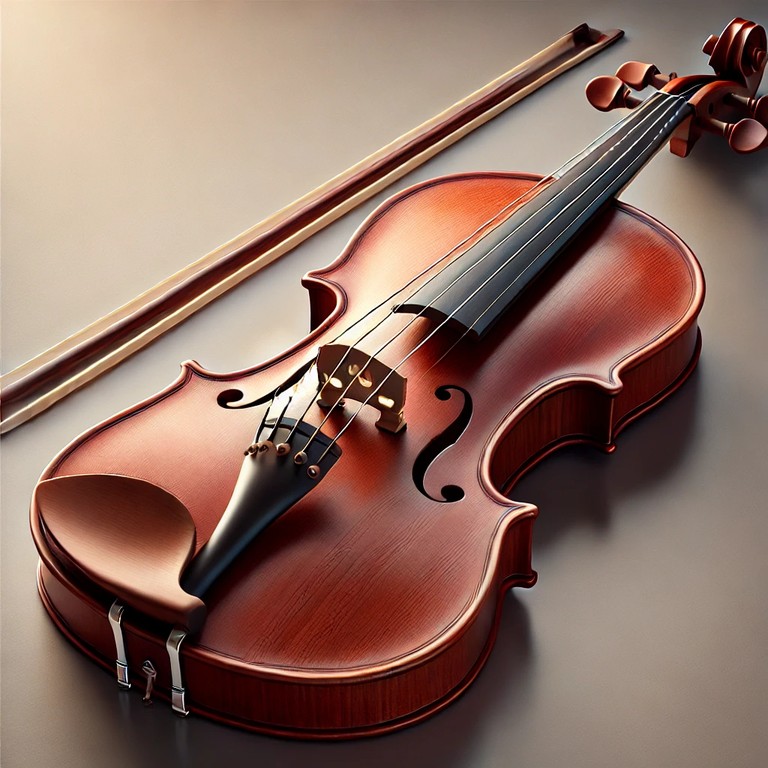} \\
    \textsc{Round 3} & \textsc{Round 3}\\
    \includegraphics[width=0.09\textwidth]{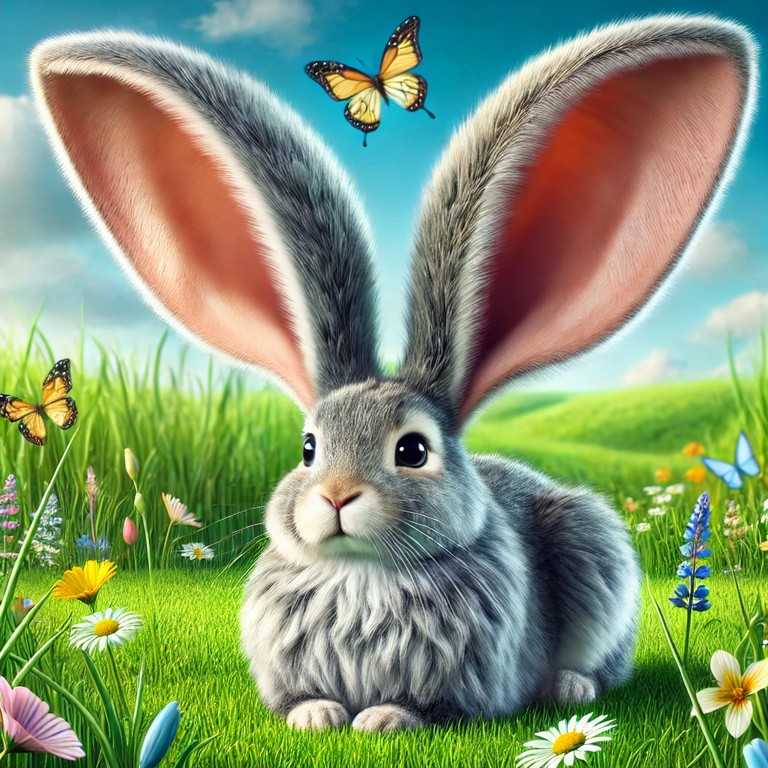} &
    \includegraphics[width=0.09\textwidth]{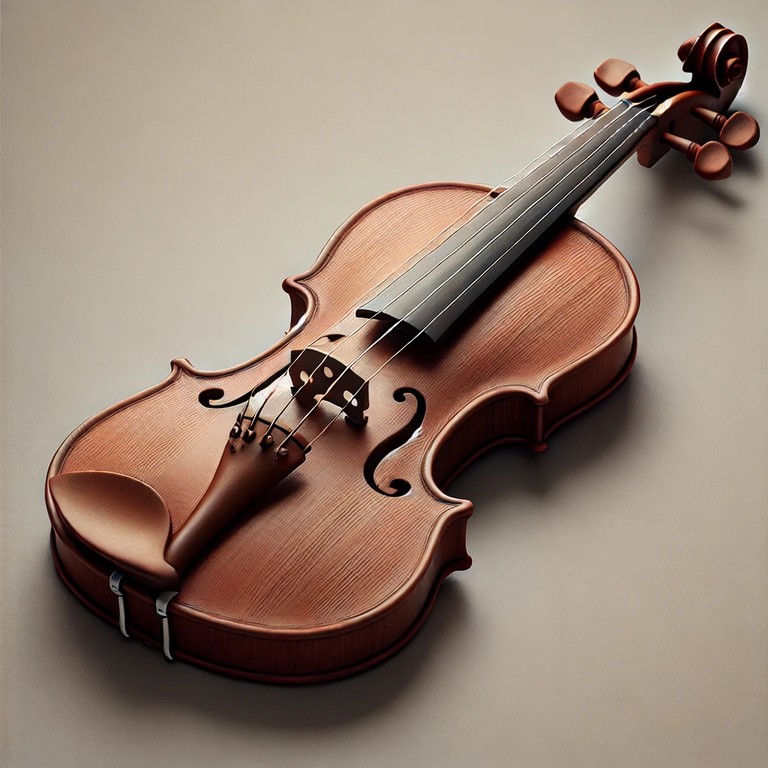} \\
\end{tabular}\label{subfig:dalle-main-text}
}
\subfigure[GPT-4o]{
\begin{tabular}{c}
    \textsc{  } \\
    \includegraphics[width=0.215\columnwidth]{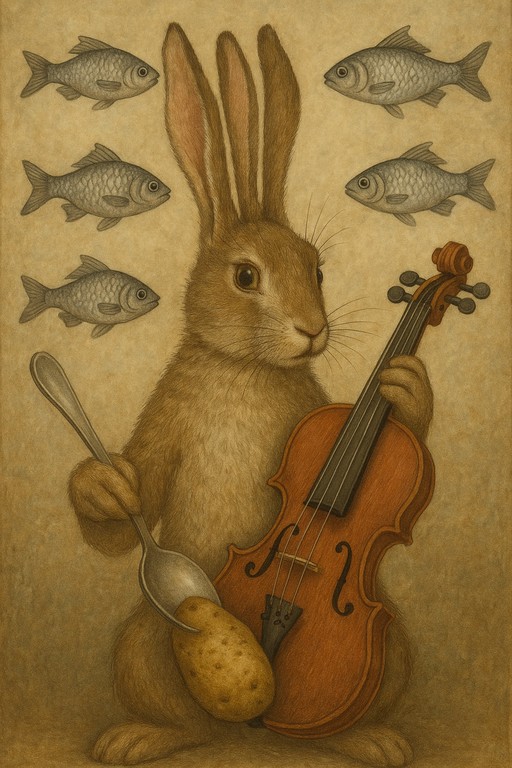}
\end{tabular}
    \label{subfig:gpt4o-rabbit-main-text}
}
\subfigure[Ours]{
\begin{tabular}{c}
    \textsc{  } \\
    \includegraphics[width=0.22\columnwidth]{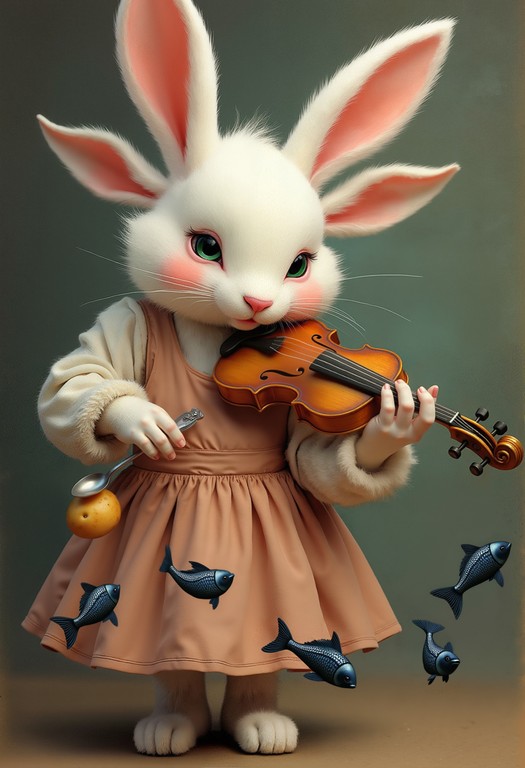}
\end{tabular}
    \label{subfig:customizable-main-text}
}

\caption{\textbf{\spice~can generate content that DALL·E 3 and GPT-4o cannot.} Two examples are a rabbit with 4 ears and a violin without a bridge. For individual objects, DALL·E 3 fails even after the user asks the model to edit the errors multiple times. For combined objects, GPT-4o cannot generate all objects correctly at once. However, \spice~can reliably generate these challenging objects.}
\label{fig:rabbit-main-text}
\end{center}
\end{figure}

First, prompts are hard to design and cannot precisely instruct the model to produce complex and unusual outputs (\cref{subfig:dalle-main-text}). A user assumes that the model can follow the prompt, but the model always fails to interpret the prompt as humans do, even after the user repeatedly corrects the model.

Second, the model’s randomness makes complex images costly—and even impossible—to generate. For example, if an image must include 10 objects and each is generated correctly only 50\% of the time, the chance of getting all 10 right drops below 0.1\%. In practice, object-level correctness rate is even lower and scene complexity higher, so achieving perfect results becomes impossible (\cref{subfig:gpt4o-rabbit-main-text}).

With the high customizability of \spice, instead of desperately engineering prompts, users can effortlessly generate compositions and objects that are impossible for DALL·E 3 and GPT-4o (\cref{fig:rabbit-main-text}, see full prompt in \cref{sec:customizable-editing}).\footnote{\cref{subfig:customizable-main-text} has been accepted to CVPR AI Art Gallery 2025. \url{https://thecvf-art.com/project/compositionality-and-parts/}} Hence, \spice~creates a new paradigm for human-model collaboration, where users can fully realize their creative vision, without ever needing to compromise to the models' lack of prompt compliance. More instructions and examples of iterative and customizable image generation and editing can be found in \cref{app:hints-and-prompts-fridge}, \cref{app:iterative-construction}, and \cref{app:hyperparameter-recommendations}.

\section{Related Work}

Diffusion model-based image editing techniques can be categorized based on the type of inputs used to guide image generation~\cite{croitoru2023diffusion, yang2023diffusion, huang2024diffusion}. Most existing methods rely primarily on textual prompts \cite{brooks2023instructpix2pix, fu2024guiding}, while many also incorporate masks to specify the editing region~\cite{wang2023imagen, zhao2024ultraedit}. Due to the inherent imprecision of both prompts and masks, some methods integrate additional inputs to provide finer control over the generation process~\cite{yang2024imagebrush, shi2024dragdiffusion, liu2024drag}. Among these, MagicQuill~\cite{liu2024magicquill} is most closely related to our workflow. MagicQuill similarly enables user-guided editing using color and edge information. However, its color guidance is constrained to low-resolution 32$\times$32 blocks, regardless of image size, prohibiting fine-grained edits. \spice~overcomes this limitation of MagicQuill by accepting full-resolution color and edge hints, effectively interpreting the provided guidance and leading to precise edits. More discussion to compare \spice~and MagicQuill can be found in \cref{app:magicquill}.

\section{Conclusion}

Existing prompt-based image editing models fail in performing local edits, working under different resolutions, following user instructions, and maintaining image quality during multiple editing steps. We propose \spice, a training-free workflow that addresses all these challenges. We release the workflow to facilitate future research and artistic exploration.

\section*{Acknowledgments}

We thank Anthony Yang for his assistance in preparing the materials presented in \cref{app:iterative-construction}.

\bibliography{references}
\bibliographystyle{plainnat}


\appendix

\clearpage

\startcontents[appendix]
\printcontents[appendix]{ }{0}{\section*{Table of Contents}}

\clearpage

\section{Hints and Prompts of the Fridge Example}
\label{app:hints-and-prompts-fridge}

In \cref{tab:hints-and-prompts-fridge}, we list the simple hints and prompts that we use to produce results in \cref{fig:advantages}. We also explain how each hint is quickly and easily added to the image by using Photoshop. The explanation describes the difference between the hinted image at Step $i$ and the result image at Step $i-1$. The results demonstrate two advantages of \spice. On the one hand, \spice~is highly robust to different types of hints. On the other hand, \spice~allows user to effortlessly achieve sophisticated editing results, without the need to optimize complicated hints or prompts. Sometimes, prompts do not even need to be changed over different steps, as \spice~can infer the user's intent from the hints.

\begin{table}[htbp]
    \centering
    \caption{\textbf{To generate realistic edits, a user only needs to provide simple hints and prompts.} For each step in \cref{fig:advantages}, the image with hints (``Hinted'') and the description prompt are shown below. We also explain the hint for each step. The explanation is not part of the input to the model.}
    \begingroup
    \setlength{\extrarowheight}{3pt}
    \begin{tabular}{ccc p{4.2cm}p{4.2cm}}
        \toprule
        \textbf{Step} & \textbf{Hinted} & \textbf{Result} & \textbf{Description Prompt} & \textbf{Hint Explanation} \\
        \midrule
        1 & \includegraphics[width=0.1\textwidth,valign=t]{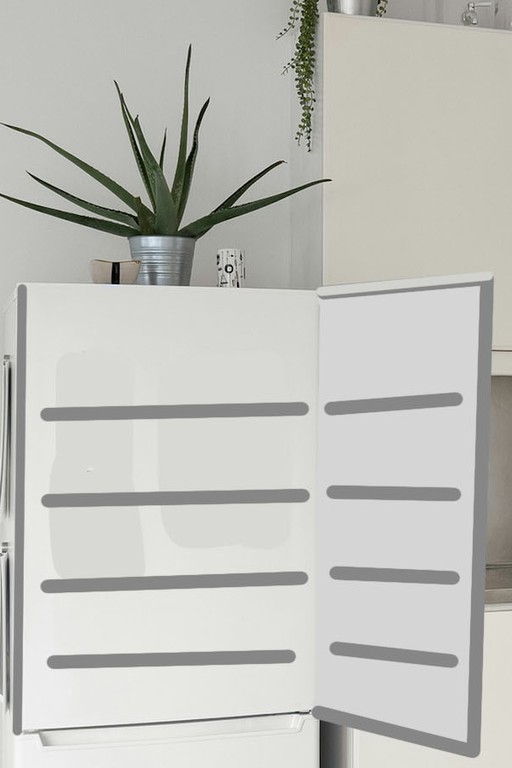} & \includegraphics[width=0.1\textwidth,valign=t]{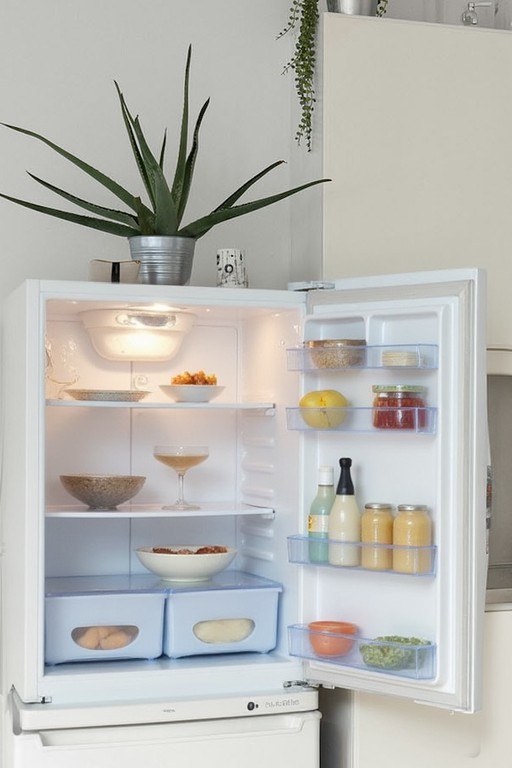} & An open fridge with food in it. & The contours of the fridge door and fridge shelves are roughly sketched using the hard round brush.\\
        2 & \includegraphics[width=0.1\textwidth,valign=t]{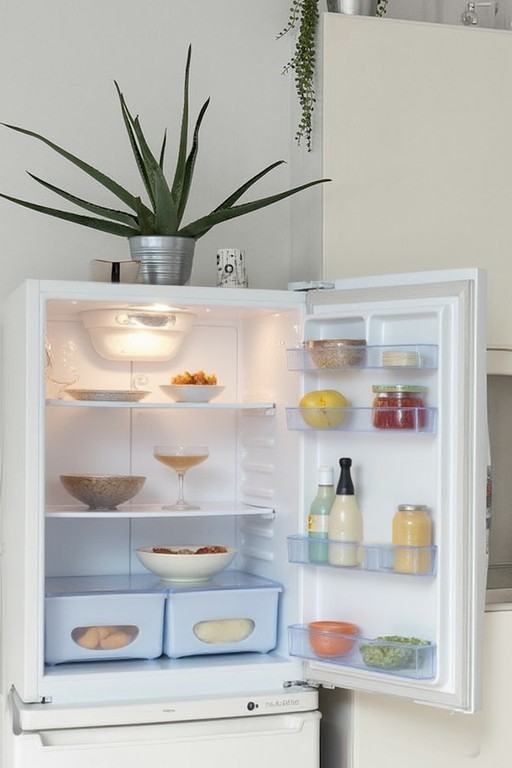} & \includegraphics[width=0.1\textwidth,valign=t]{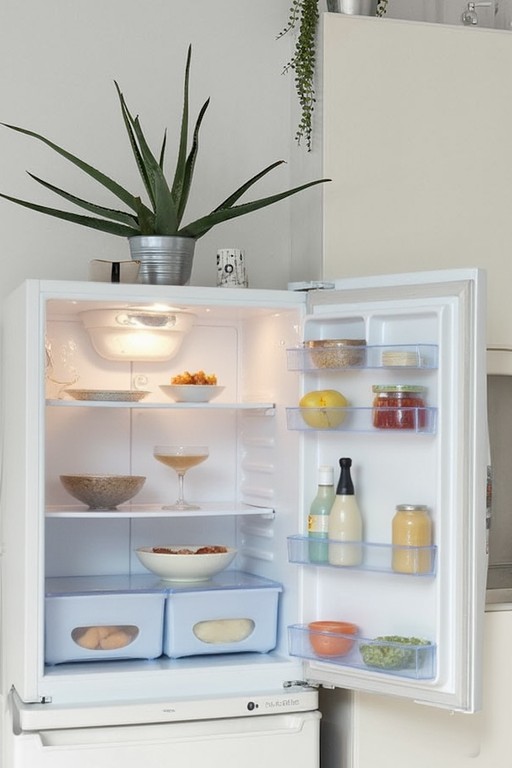} & An open fridge with food in it. & The region occupied by the removed bottle is recolored using the hard round brush and the Clone Stamp tool. \\ 
        3 & \includegraphics[width=0.1\textwidth,valign=t]{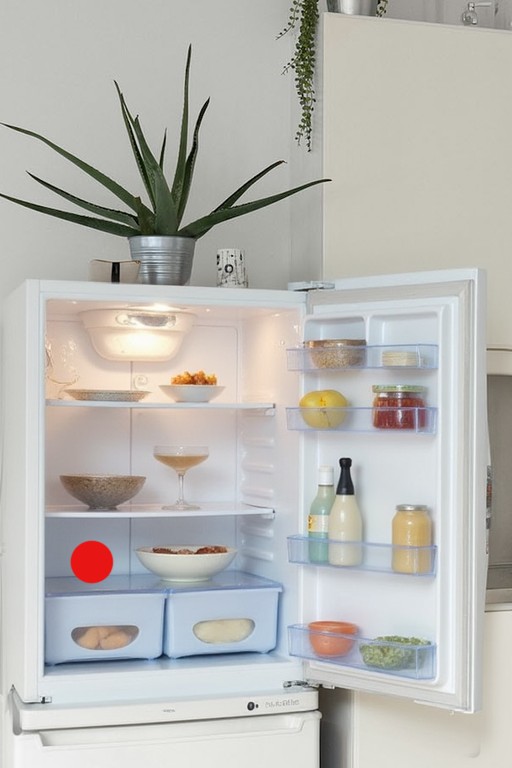} & \includegraphics[width=0.1\textwidth,valign=t]{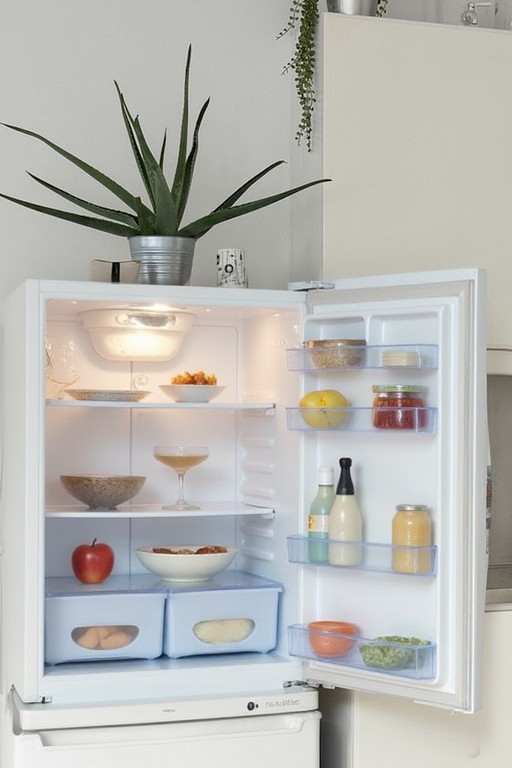} & An open fridge with food in it. & A single red circle is added using the hard round brush. \\ 
        4 & \includegraphics[width=0.1\textwidth,valign=t]{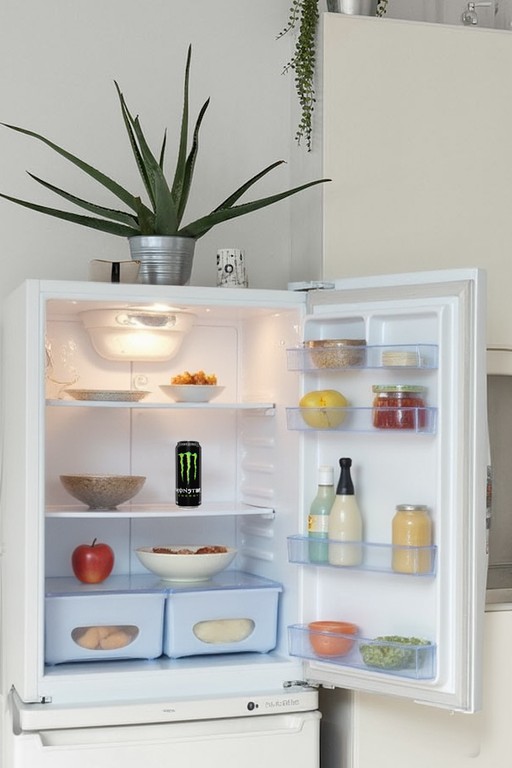} & \includegraphics[width=0.1\textwidth,valign=t]{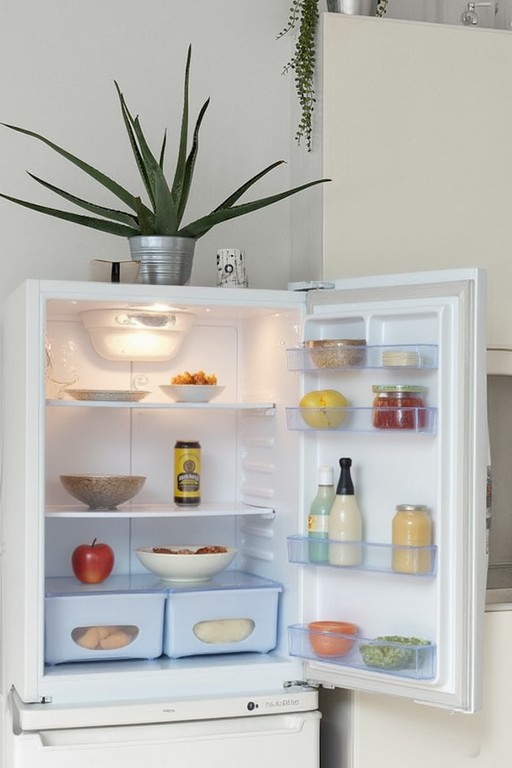} & An open fridge with food in it. A beer. & First, the coupe glass is removed using the Clone Stamp tool. Then, an image of Monster Energy is pasted at the same place.\\ 
        5 & \includegraphics[width=0.1\textwidth,valign=t]{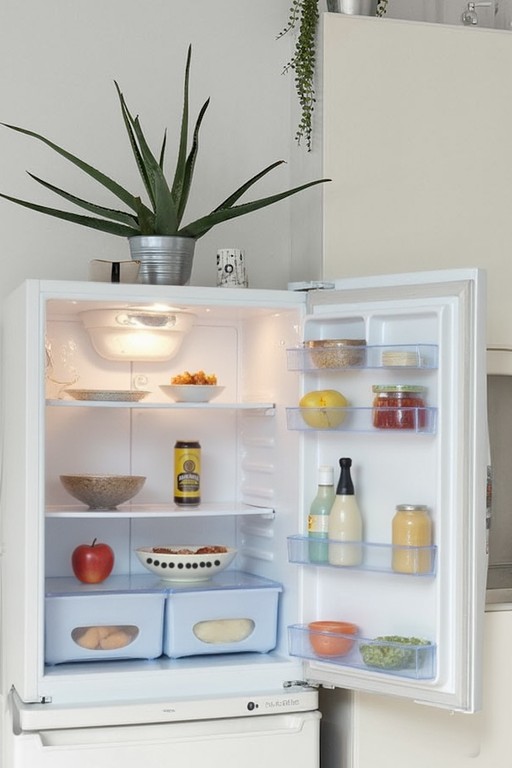} & \includegraphics[width=0.1\textwidth,valign=t]{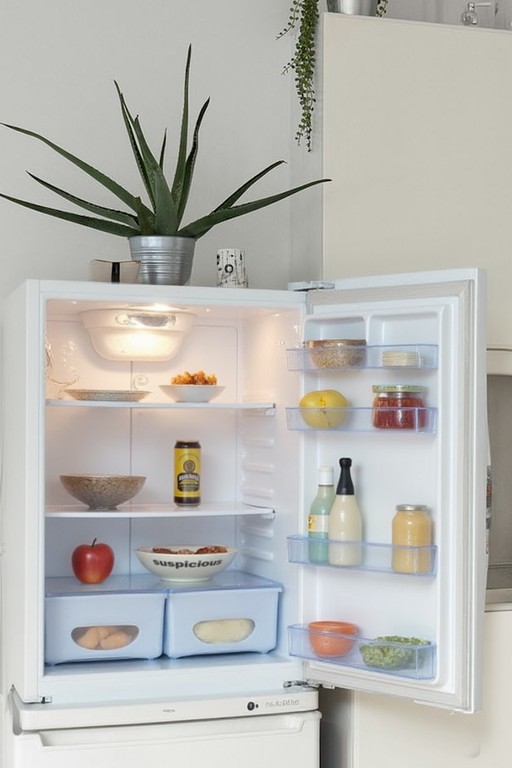} & An open fridge with food in it. A bowl with a word ``suspicious'' on it. & Ten black dots corresponding to ten letters of ``suspicious'' are added on the bowl using the hard round brush. \\ 
        6 & \includegraphics[width=0.1\textwidth,valign=t]{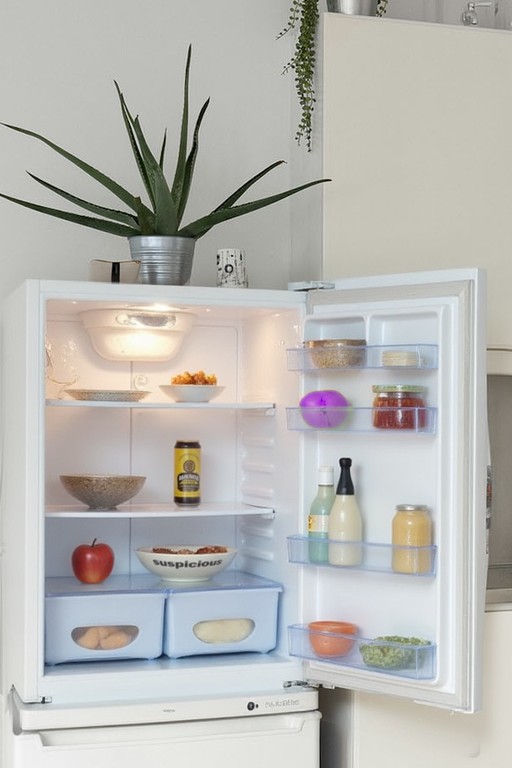} & \includegraphics[width=0.1\textwidth,valign=t]{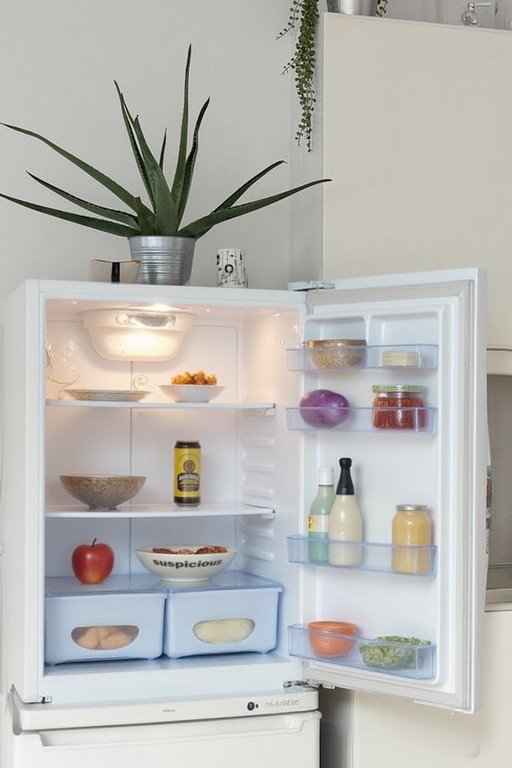} & An open fridge with food in it. A purple potato. & The potato is selected by the Quick Selection tool. The hue of the selected region is changed by the Hue/Saturation tool.\\ 
        7 & \includegraphics[width=0.1\textwidth,valign=t]{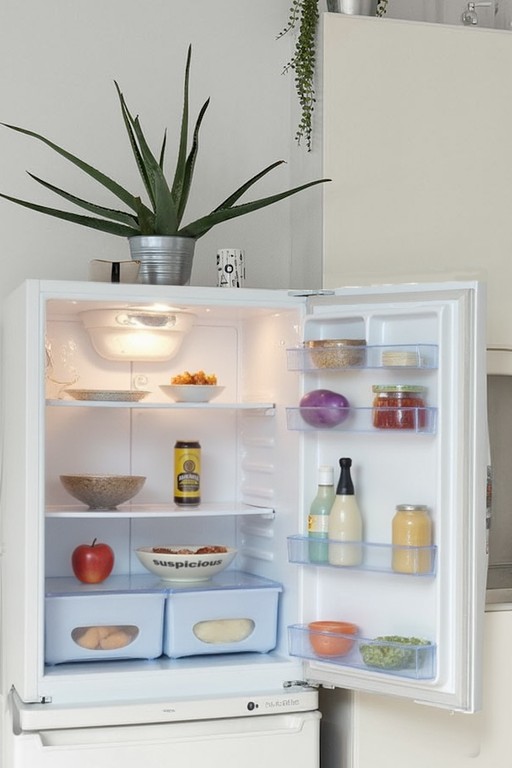} & \includegraphics[width=0.1\textwidth,valign=t]{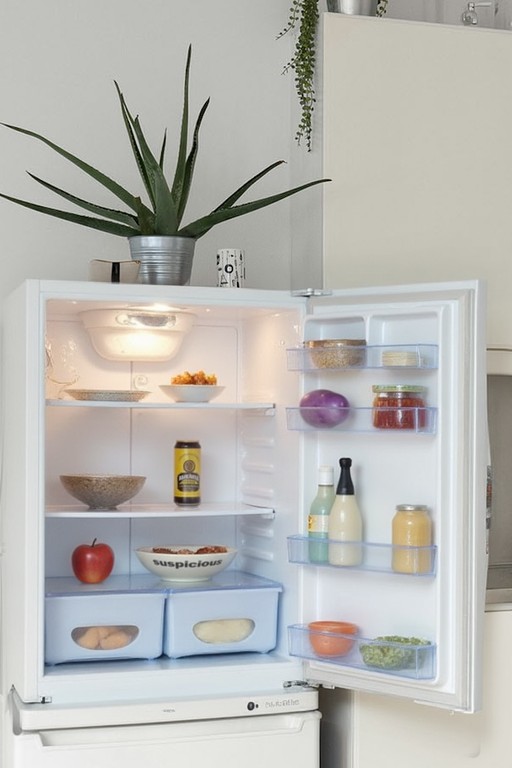} & An open fridge with food in it. & From the image in Step 0, a thin region above the fridge door is selected using the Quick Selection tool and copied to the image. \\ 
        8 & \includegraphics[width=0.1\textwidth,valign=t]{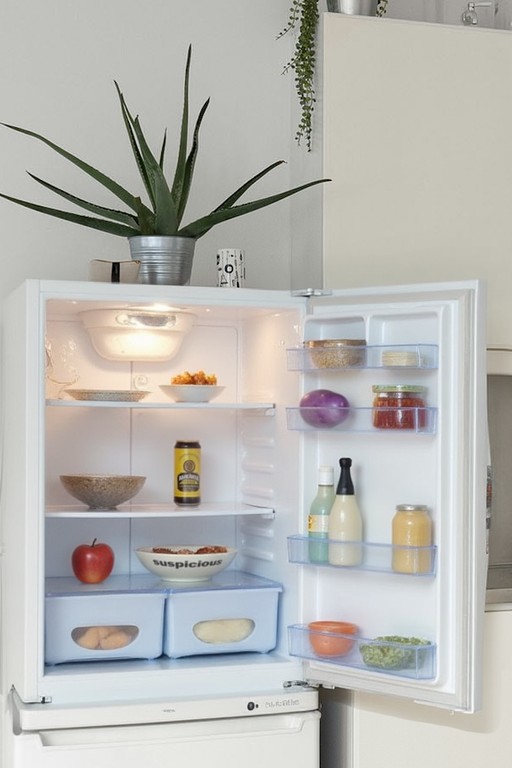} & \includegraphics[width=0.1\textwidth,valign=t]{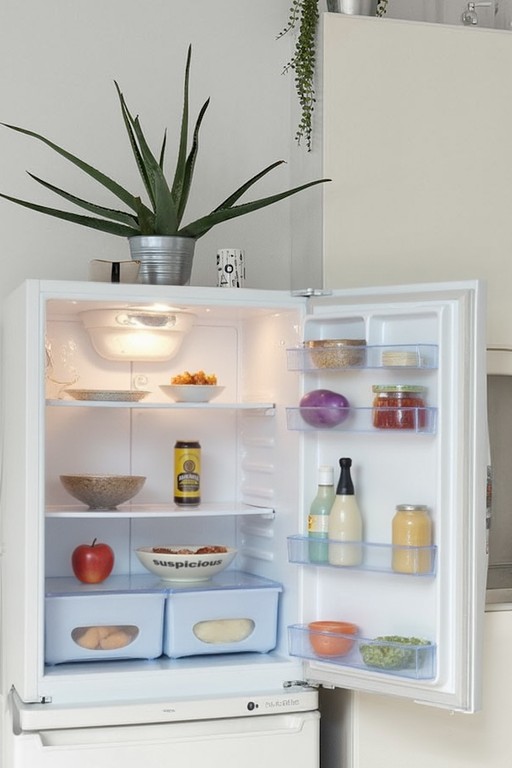} & An open fridge with food in it. & From the image in Step 0, a thin region below the fridge door is selected using the Quick Selection tool and copied to the image. \\ 
        9 & \includegraphics[width=0.1\textwidth,valign=t]{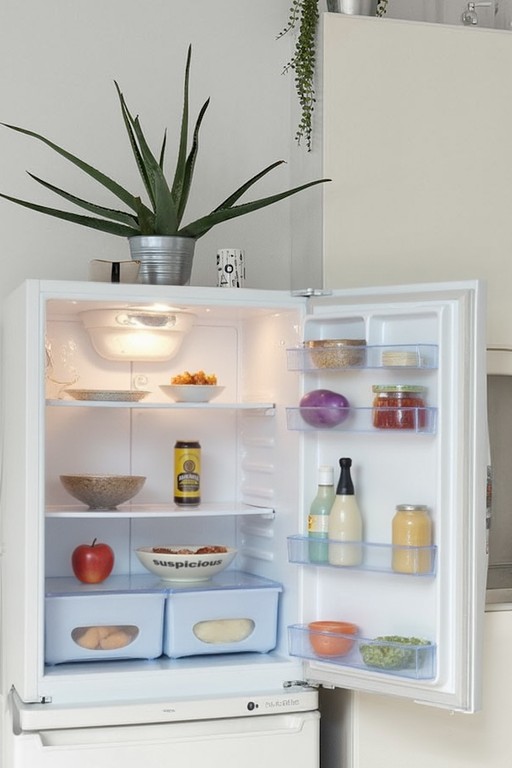} & \includegraphics[width=0.1\textwidth,valign=t]{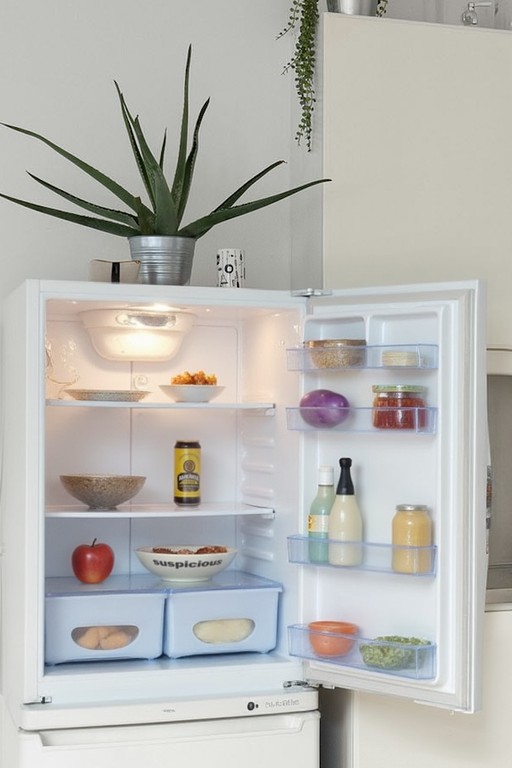} & An open fridge with food in it. & No change is made. \\
        \bottomrule
    \end{tabular}
    \endgroup
    \label{tab:hints-and-prompts-fridge}
\end{table}

\section{Overview of Results}
\label{app:overview-of-results}

\begin{figure*}[htbp]
\begin{center}
\small
\begin{tabular}{c@{\hspace{0.3em}}c@{\hspace{0.3em}}c@{\hspace{0.3em}}c@{\hspace{0.3em}}c@{\hspace{1em}}c@{\hspace{0.3em}}c@{\hspace{0.3em}}c@{\hspace{0.3em}}c@{\hspace{0.3em}}c}
    \scriptsize\textsc{Original} & \scriptsize\textsc{IP2P} & \scriptsize\textsc{MB} & \scriptsize\textsc{UE} & \scriptsize\textsc{\textbf{Ours}} & \scriptsize\textsc{Original} & \scriptsize\textsc{IP2P} & \scriptsize\textsc{MB} & \scriptsize\textsc{UE} & \scriptsize\textsc{\textbf{Ours}} \\
    
    \includegraphics[width=0.09\textwidth]{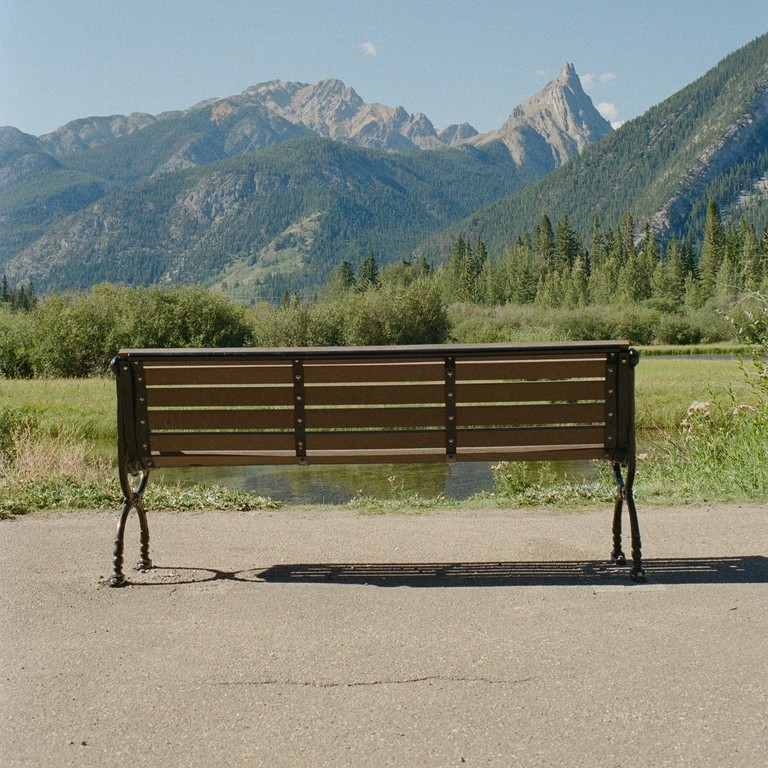} &
    \includegraphics[width=0.09\textwidth]{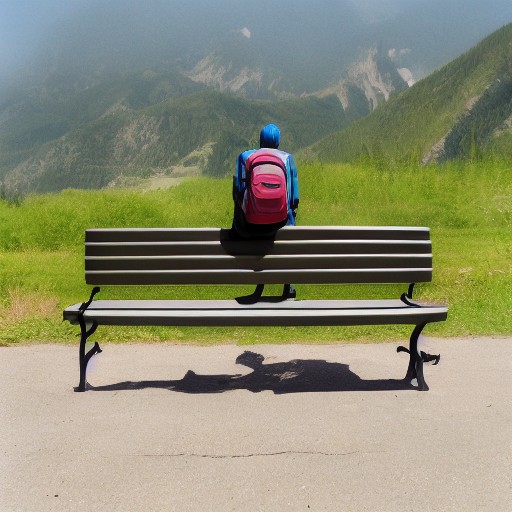} &
    \includegraphics[width=0.09\textwidth]{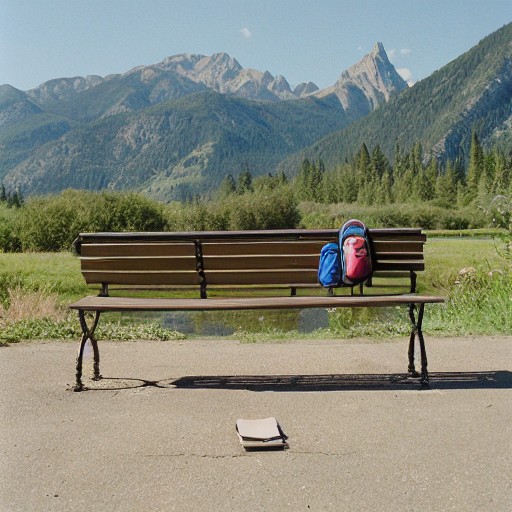} &
    \includegraphics[width=0.09\textwidth]{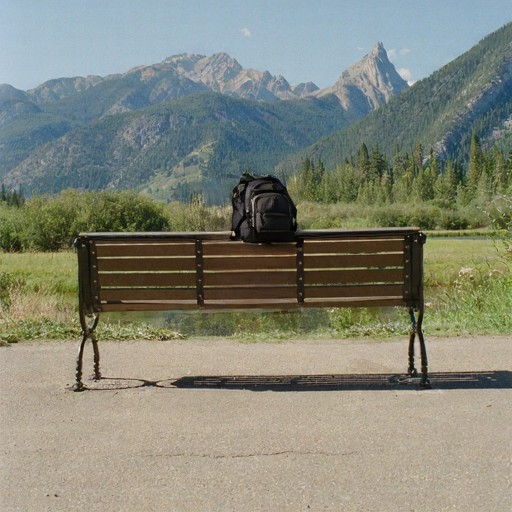} &
    \includegraphics[width=0.09\textwidth]{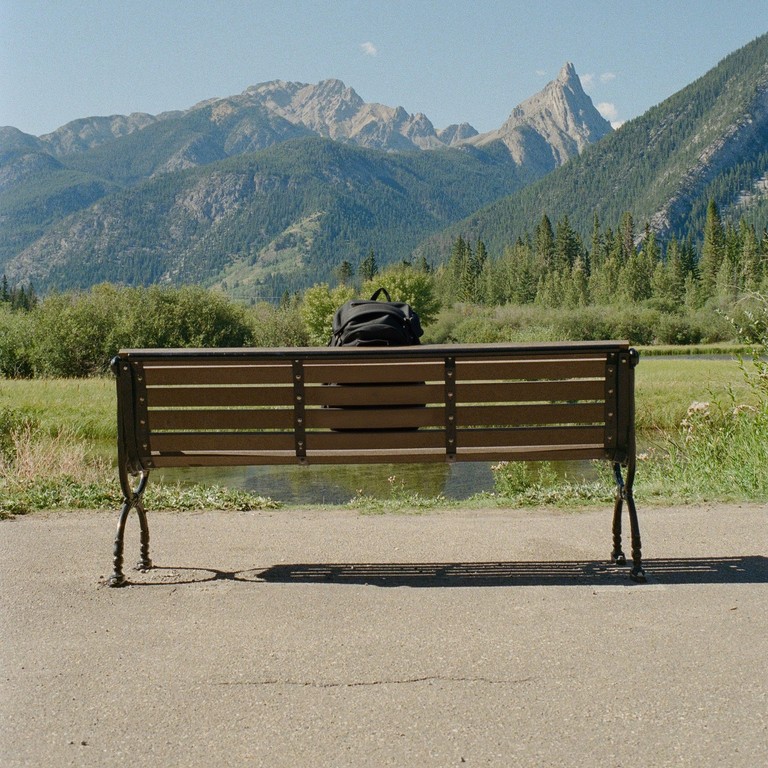} &
    \includegraphics[width=0.09\textwidth]{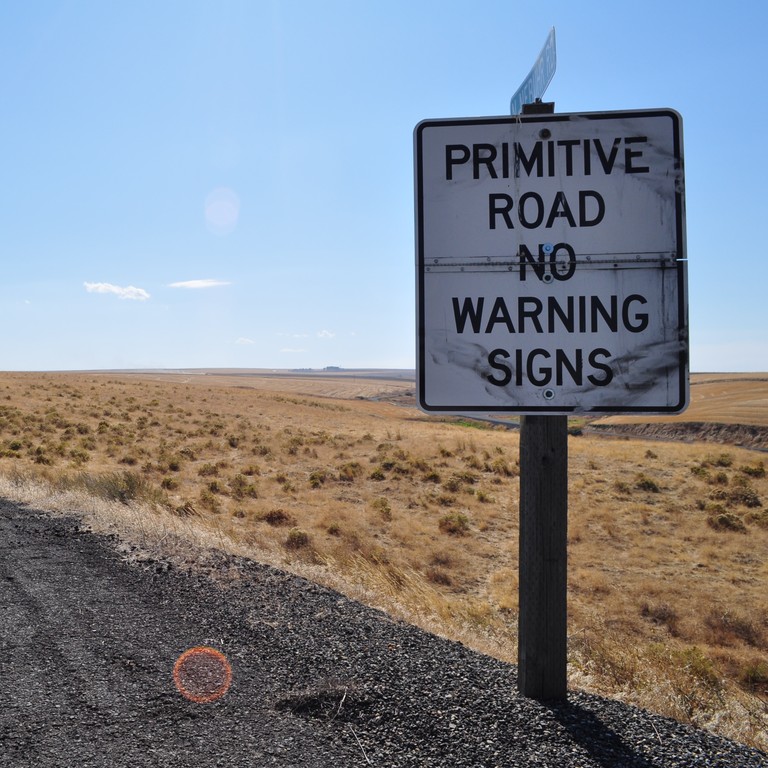} &
    \includegraphics[width=0.09\textwidth]{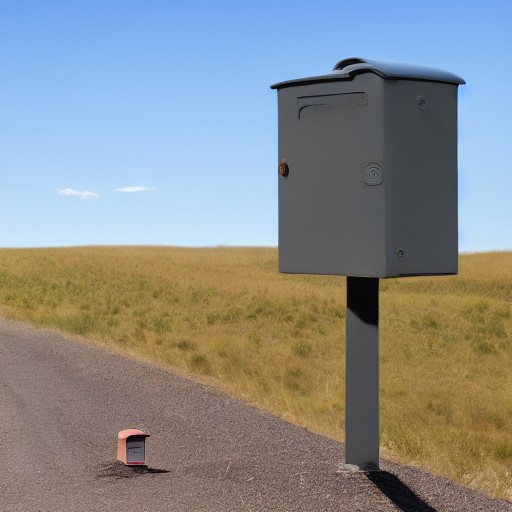} &
    \includegraphics[width=0.09\textwidth]{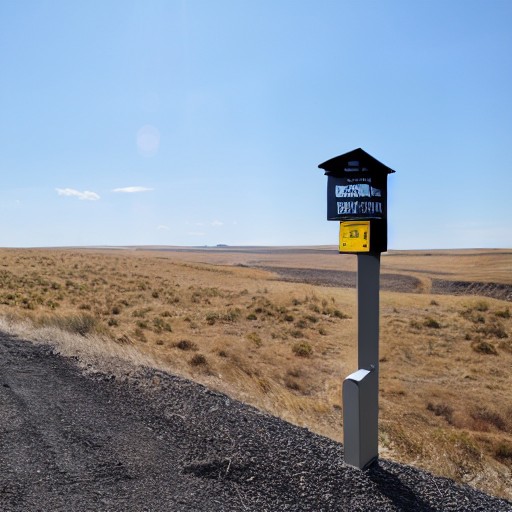} &
    \includegraphics[width=0.09\textwidth]{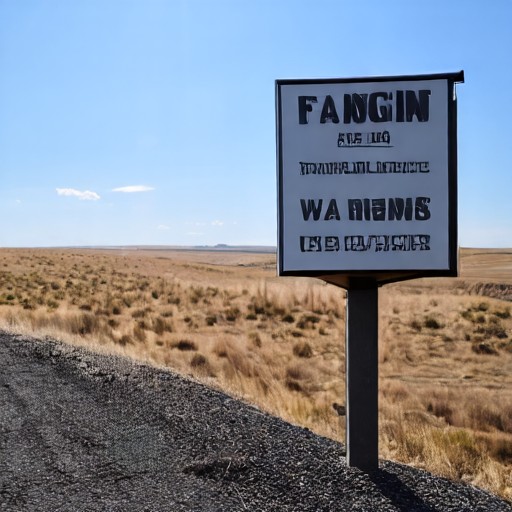} &
    \includegraphics[width=0.09\textwidth]{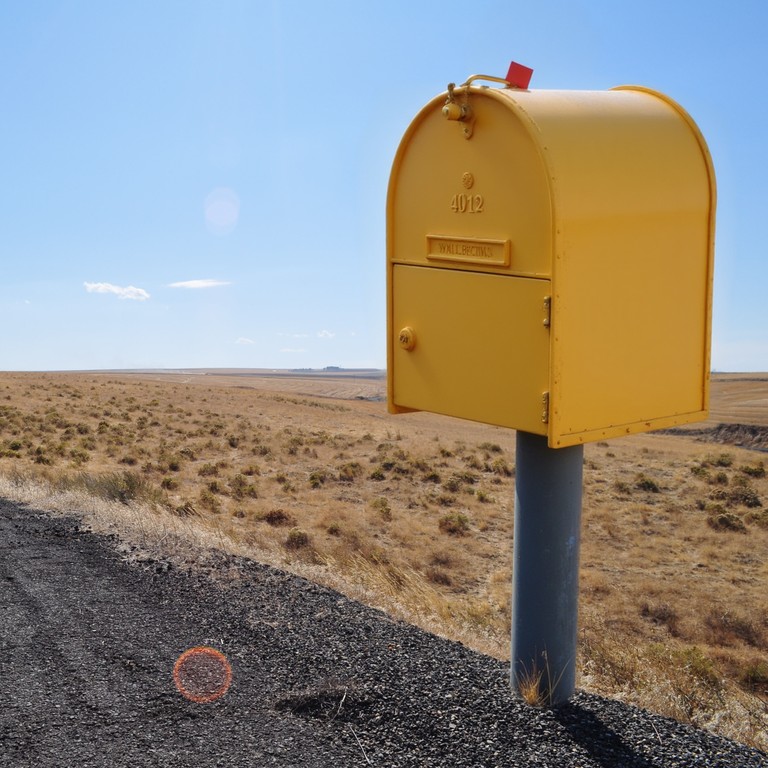} \\
    \multicolumn{5}{c}{\scriptsize \textbf{Object Addition:} Add a backpack placed on the bench.} & \multicolumn{5}{c}{ \scriptsize\textbf{Object Replacement:} Replace the road sign with a mailbox.} \\
    [0.5em]
    \includegraphics[width=0.09\textwidth]{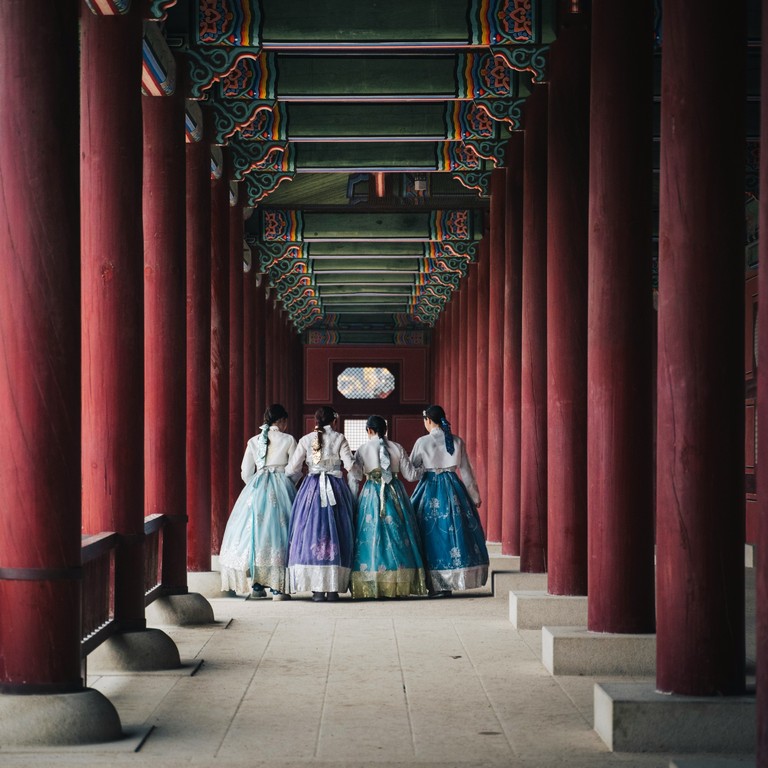} &
    \includegraphics[width=0.09\textwidth]{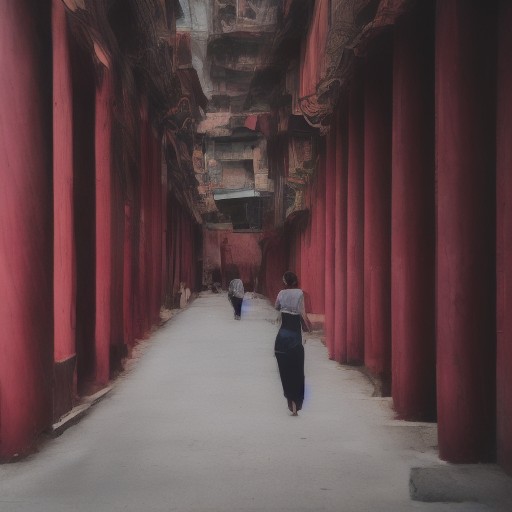} &
    \includegraphics[width=0.09\textwidth]{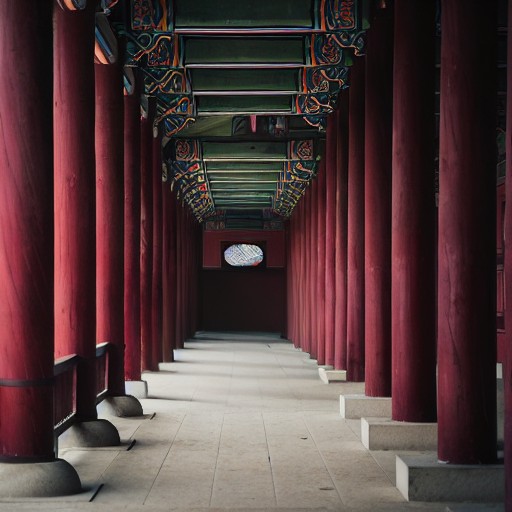} &
    \includegraphics[width=0.09\textwidth]{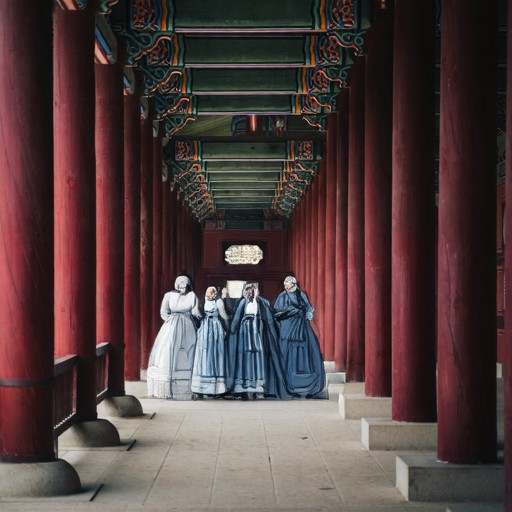} &
    \includegraphics[width=0.09\textwidth]{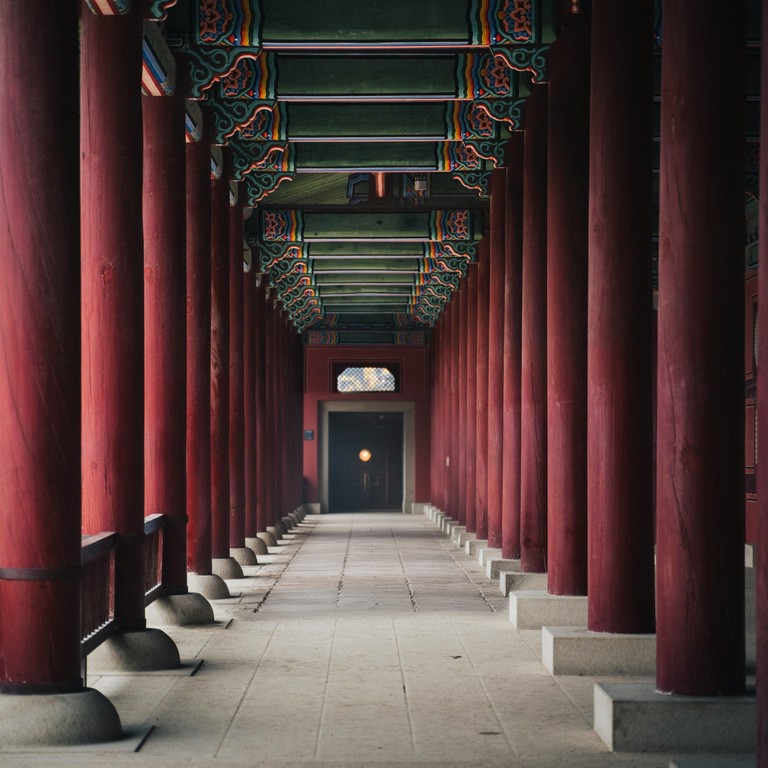} &
    \includegraphics[width=0.09\textwidth]{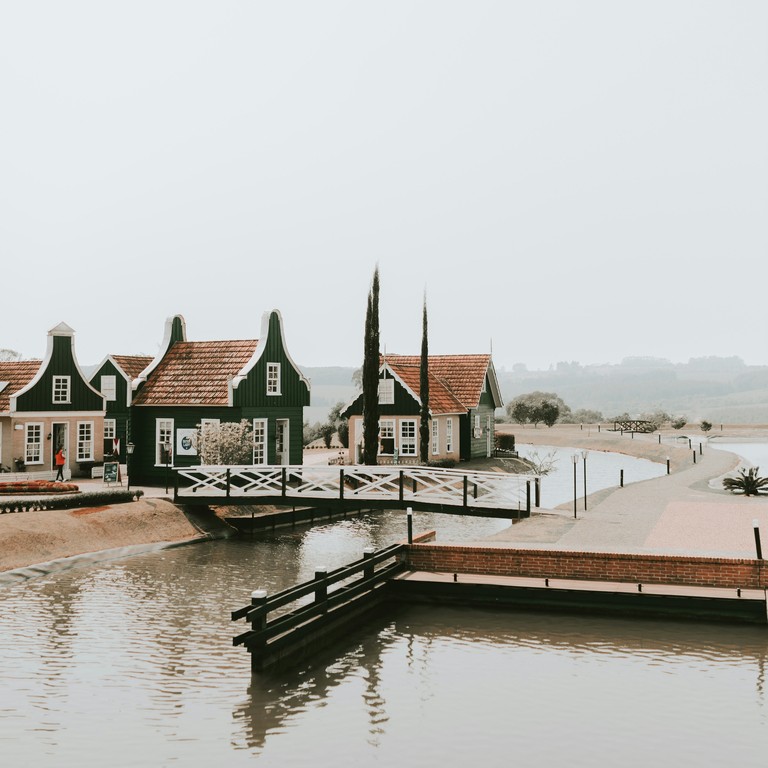} &
    \includegraphics[width=0.09\textwidth]{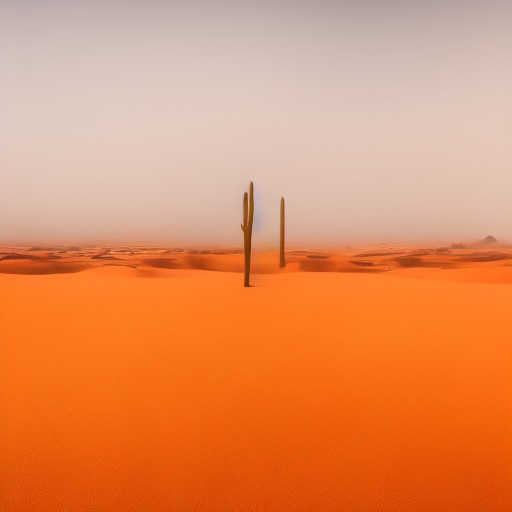} &
    \includegraphics[width=0.09\textwidth]{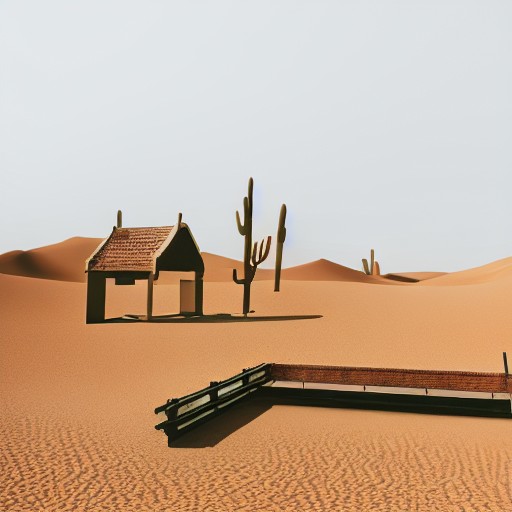} &
    \includegraphics[width=0.09\textwidth]{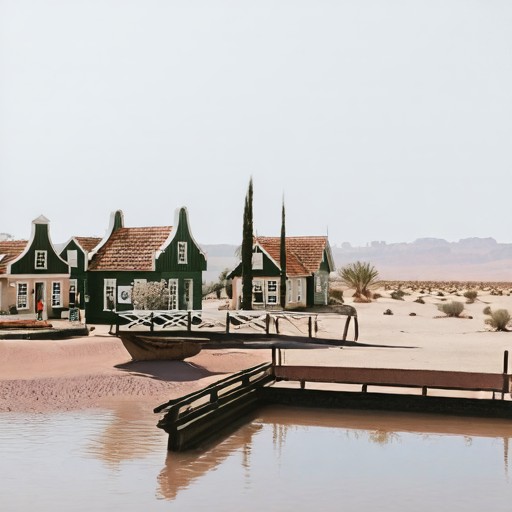} &
    \includegraphics[width=0.09\textwidth]{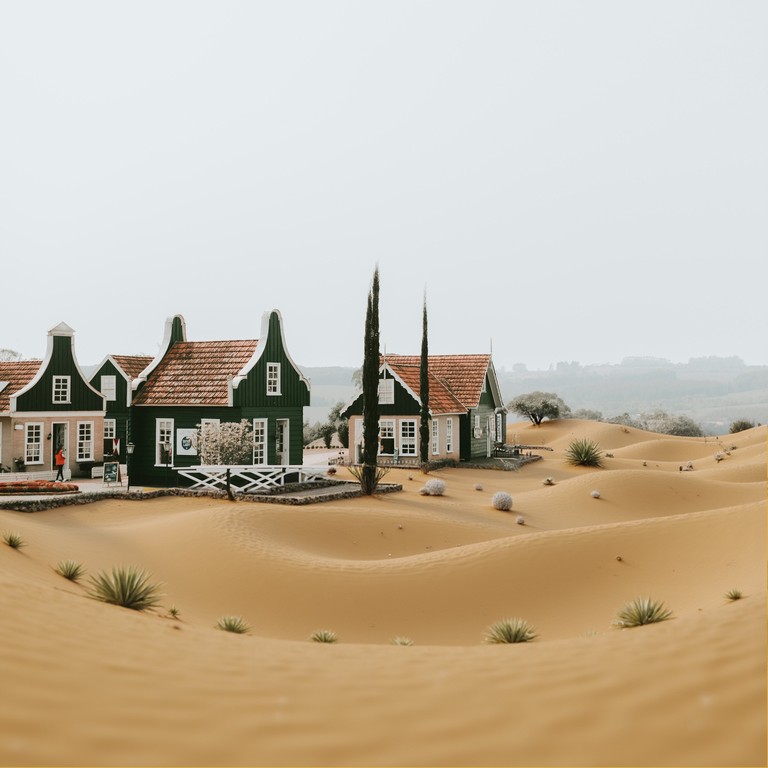} \\
    \multicolumn{5}{c}{\scriptsize\textbf{Object Removal:} Remove the four women.} & \multicolumn{5}{c}{\scriptsize\textbf{Background Change:} Change the riverside to a desert.} \\
    [0.5em]
    \includegraphics[width=0.09\textwidth]{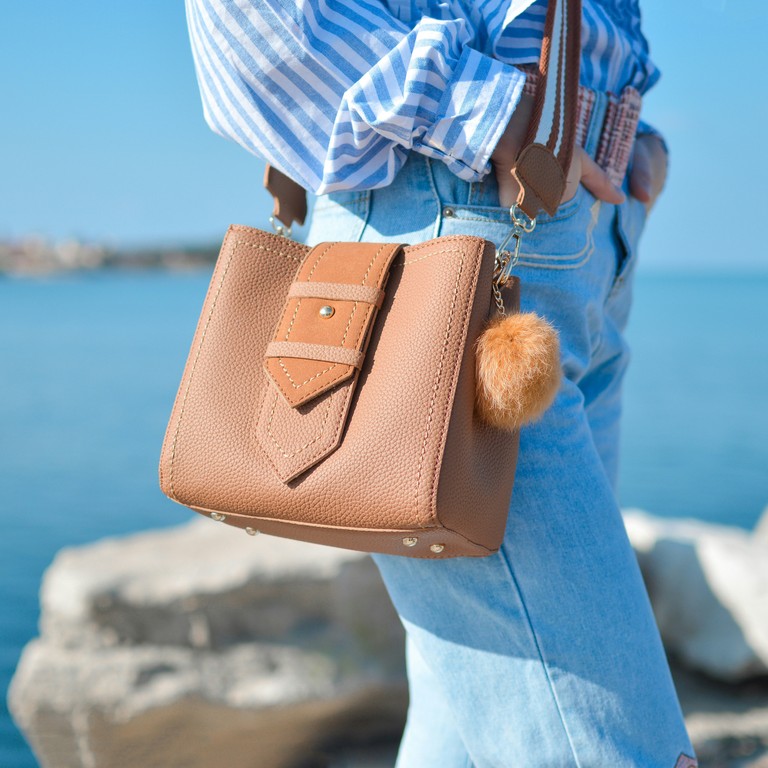} &
    \includegraphics[width=0.09\textwidth]{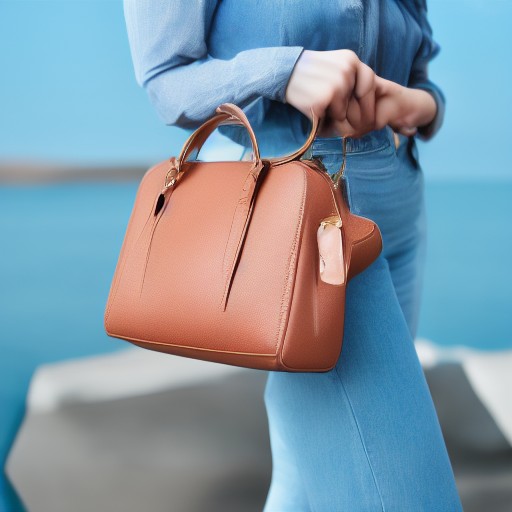} &
    \includegraphics[width=0.09\textwidth]{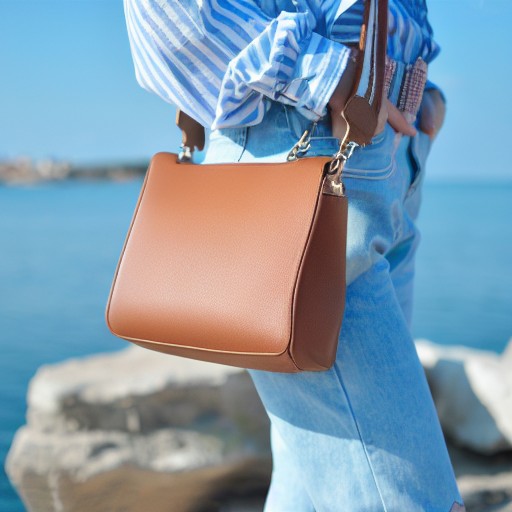} &
    \includegraphics[width=0.09\textwidth]{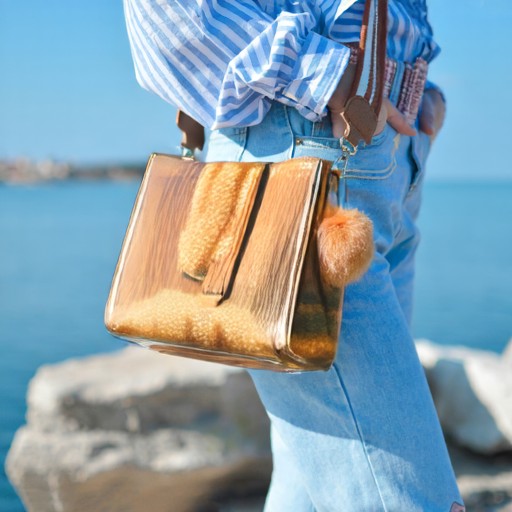} &
    \includegraphics[width=0.09\textwidth]{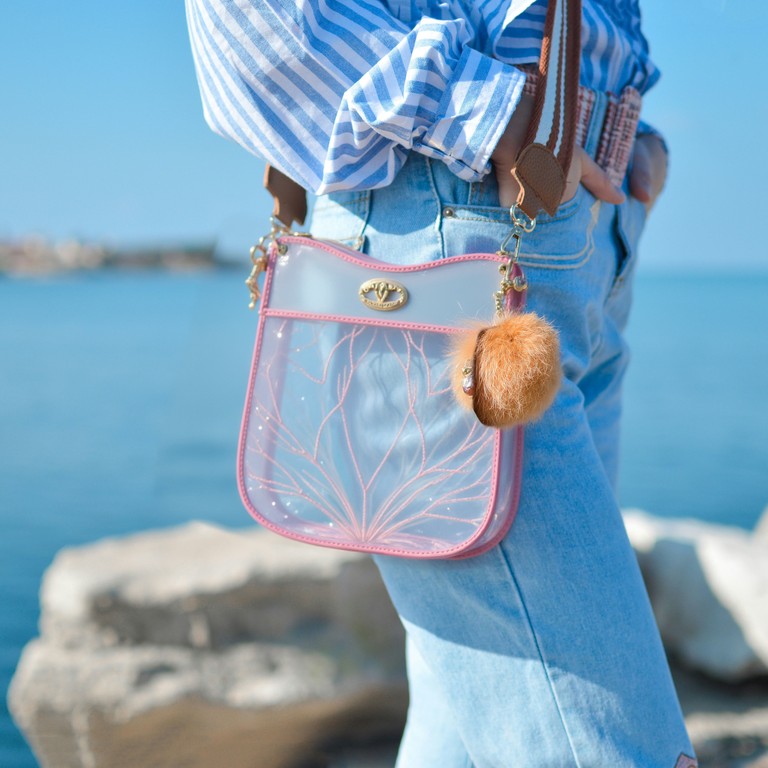} &
    \includegraphics[width=0.09\textwidth]{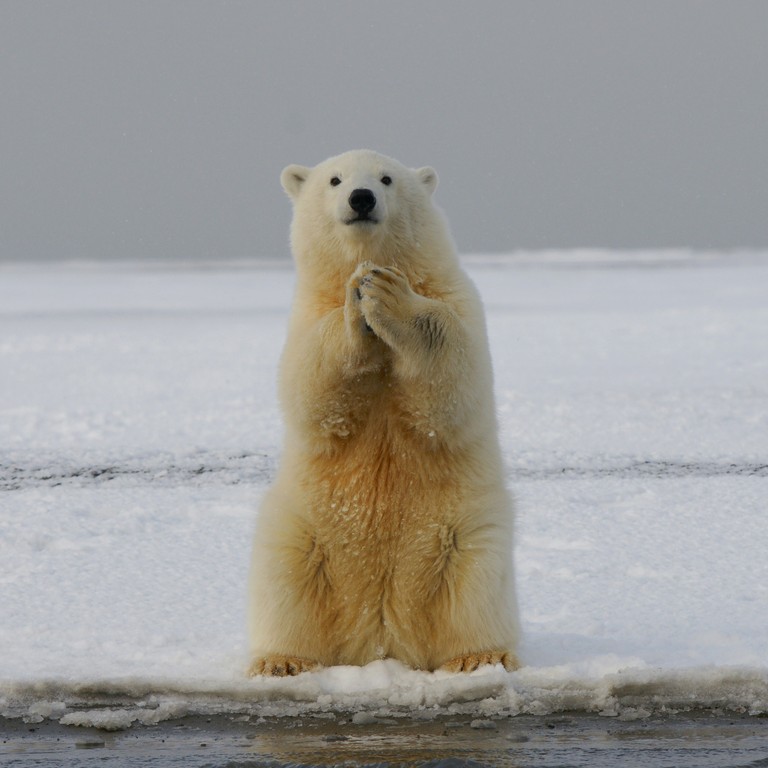} &
    \includegraphics[width=0.09\textwidth]{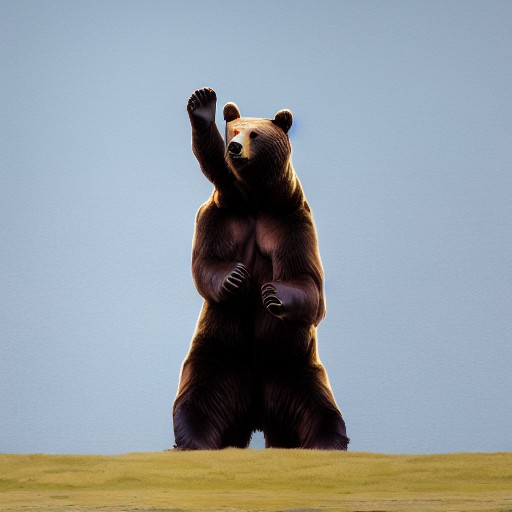} &
    \includegraphics[width=0.09\textwidth]{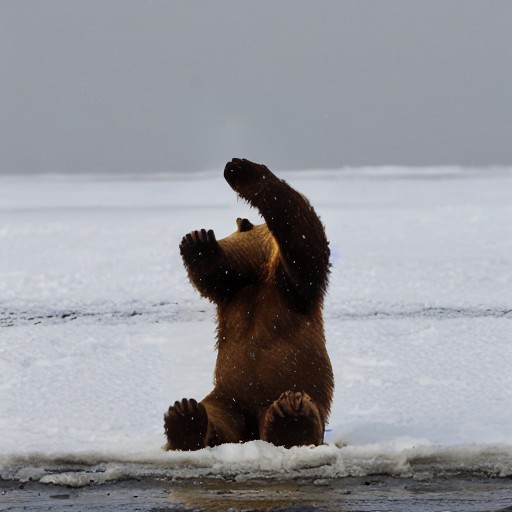} &
    \includegraphics[width=0.09\textwidth]{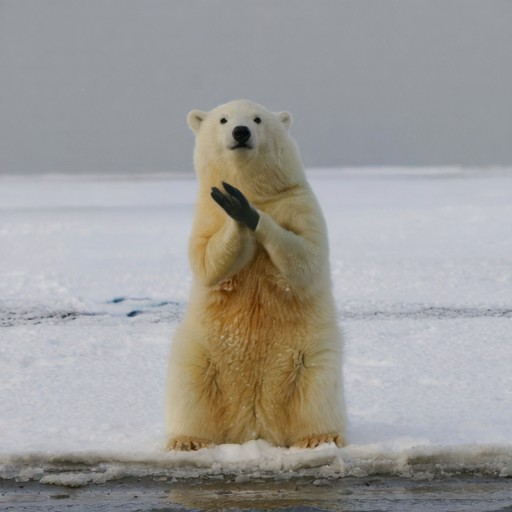} &
    \includegraphics[width=0.09\textwidth]{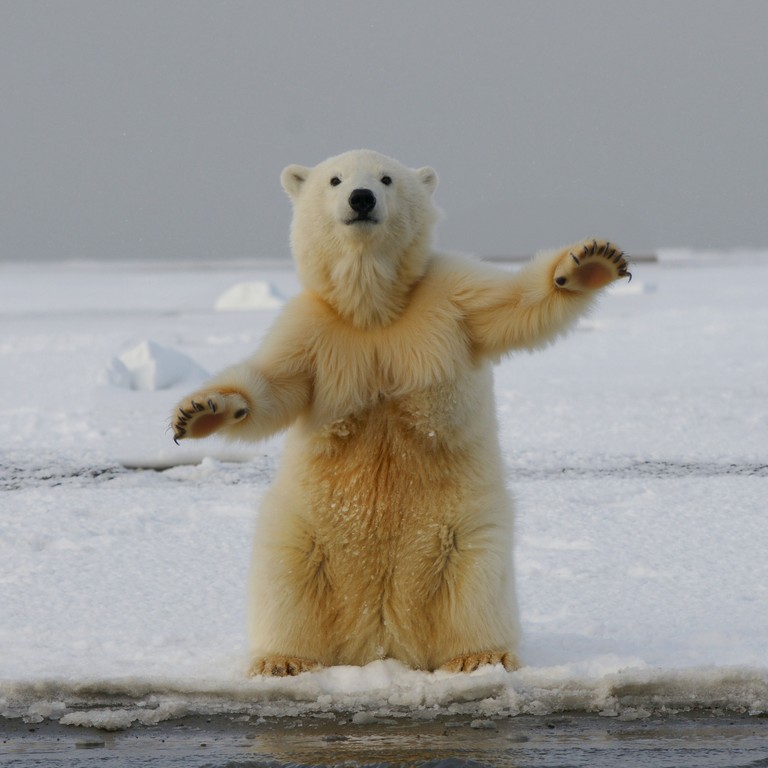} \\
    \multicolumn{5}{c}{\scriptsize\textbf{Texture Change:} Turn the handbag into the glass.} & \multicolumn{5}{c}{\scriptsize\textbf{Action Change:} Turn the bear raising its hand.} \\
\end{tabular}
\caption{\textbf{Our workflow outperforms baseline methods in 6 editing categories from EditEval.} Each group of five images shows an example from an editing task. Each group starts from original image, followed by IP2P, MagicBrush (MB), UltraEdit (UE), and our results.}
\label{fig:benchmark-results}
\end{center}

\end{figure*}

\subsection{Benchmark Results}
\label{sec:benchmark-results}

In this section, we evaluate \spice~under various image-editing scenarios. First, we evaluate our workflow on a standard benchmark of single-step editing. Then, we systematically verify the effectiveness of the three features of \spice, namely precise, iterative, and customizable editing. 

\paragraph{Implementation Details.} We use InstructPix2Pix (IP2P) \cite{brooks2023instructpix2pix}, IP2P trained on MagicBrush \cite{zhang2024magicbrush}, and UltraEdit \cite{zhao2024ultraedit} as baseline methods. We evaluate all models on the EditEval \cite{huang2024diffusion} benchmark, using the standard CLIP \cite{radford2021learning} text-image direction similarity (CLIP$_{dir}$) and CLIP output similarity (CLIP$_{out}$) metrics from the Emu Edit benchmark \cite{sheynin2024emu}. More details can be found in \cref{app:evaluation-details}.

\paragraph{Quantitative and Qualitative Evaluation.} \cref{tab:editeval-quantitative} shows that our workflow achieves the highest scores on the EditEval benchmark. In addition, we compare the edited images by baseline methods and our workflow on all six different editing categories for visual comparison. \cref{fig:benchmark-results} shows the failure patterns of baseline methods including undesired global color shift of the grass (Object Addition), removing the lens flare outside the masked region (Object Replacement), inability to remove multiple humans from the image (Object Removal), and wrong anatomy of the polar bear (Action Change). In contrast, our workflow maintains global color, keeps details outside the mask untouched, removes objects as requested, and generates animals with correct anatomy.

\begin{table}[h]

\caption{\textbf{Our workflow achieves top performance in two quantitative metrics on the EditEval benchmark.} We show mean $\pm$ standard deviation calculated across all images ($n=126$).}
\label{tab:editeval-quantitative}
\begin{center}
\begin{small}
\begin{sc}
\begin{tabular}{lcc}
\toprule
Method & CLIP$_{dir}$ & CLIP$_{out}$\\
\midrule
IP2P       & 0.193 $\pm$ 0.120 & 0.282 $\pm$ 0.042 \\
MagicBrush & 0.210 $\pm$ 0.147 & 0.302 $\pm$ 0.038 \\
UltraEdit  & 0.193 $\pm$ 0.148 & 0.301 $\pm$ 0.039 \\
Ours       & \textbf{0.221 $\pm$ 0.150} & \textbf{0.305 $\pm$ 0.033} \\
\bottomrule
\end{tabular}
\end{sc}
\end{small}
\end{center}

\end{table}

\begin{figure}[ht]

\begin{center}
\includegraphics[width=0.6\columnwidth]{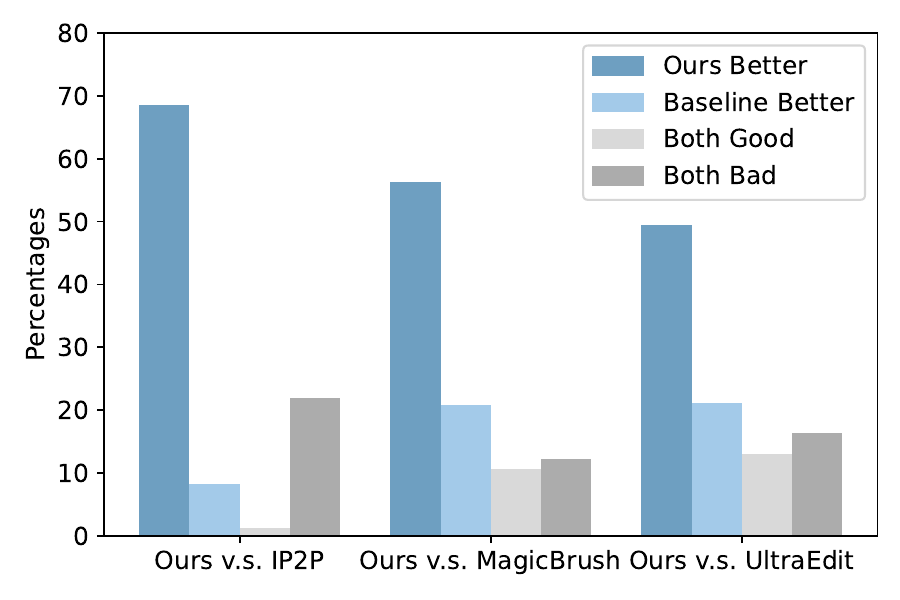}

\caption{\textbf{Our workflow is preferred by the annotators over any baseline method.} In this evaluation, we use only a single editing step and fixed hyperparameters for our workflow, and thus our workflow can still fail (``both bad'' cases). In \cref{sec:iterative-editing}, we show how relaxing these constraints can dramatically improve results. }
\label{fig:human-eval}
\end{center}

\end{figure}

\paragraph{Human Study.} To provide a comprehensive qualitative evaluation, we further conduct a human evaluation to compare results from different models (more details in \cref{app:human-evaluation-details}). 
\cref{fig:human-eval} shows that our workflow is predominantly preferred. Although at a much lower chance than baseline methods, our workflow does sometimes fail to follow instructions, as indicated by the ``both bad'' cases. In \cref{sec:iterative-editing}, we show how the flexibility of our workflow allows us to overcome this limitation easily.

\paragraph{Handling Challenging Edits.} In \cref{app:challenging-examples}, we show that our workflow outperforms baselines on challenging examples from two other popular benchmarks (Emu Edit and MagicBrush). On the challenging examples, our workflow also outperforms GPT-4o, Gemini 2.0 Flash, and SeedEdit \cite{shi2024seededitalignimageregeneration} (a recent mask-based commercial baseline from Doubao AI).

\paragraph{Ablation Study.} While FLUX.1 [dev] is a strong image generation backbone, we demonstrate that our workflow outperforms the backbone model FLUX.1 [dev] by an ablation study. Also, we designed our workflow such that it mitigates many issues of the specialized inpainting model FLUX.1 [dev] Fill.\footnote{\url{https://huggingface.co/black-forest-labs/FLUX.1-Fill-dev}} Hence, the better performance of our workflow than baseline methods does not result from simply selecting a stronger backbone. Results can be found in \cref{app:ablation-studies}.

\paragraph{Minimal Burden on Users.} Despite the superior performance, our workflow imposes minimal burden on the users. While the users need to provide both the masks and the hints, the two inputs can be casually sketched, because our workflow is robust to missing details and inaccurate shapes. Examples of excellent editing results from simple user inputs are shown in \cref{app:simple-inputs}.

\paragraph{Image Editing Benchmarks.}  
Several datasets have been introduced for training and benchmarking image editing models, including EditBench~\cite{wang2023imagen}, MagicBrush~\cite{zhang2024magicbrush}, HQ-Edit~\cite{hui2024hqedithighqualitydatasetinstructionbased}, InstructPix2Pix~\cite{brooks2023instructpix2pix}, UltraEdit~\cite{zhao2024ultraedit}, Seed-Data-Edit~\cite{ge2024seeddataedittechnicalreporthybrid}, and EditEval~\cite{huang2024diffusion}. While dataset sizes have grown over time, the number of challenging test cases remains limited. Future benchmarks should prioritize difficult structural editing tasks, such as modifying object actions or layouts, rather than focusing primarily on simpler semantic edits like color or texture changes.

\begin{figure*}[t!]
    \scriptsize
    \begin{tabular}{@{\hspace{3.4em}}c@{\hspace{0.3em}}c@{\hspace{1.4em}}c@{\hspace{0.3em}}c@{\hspace{1.4em}}c@{\hspace{0.3em}}c@{\hspace{1.4em}}c@{\hspace{0.3em}}c@{\hspace{1.4em}}c@{\hspace{0.3em}}c}
        \scriptsize\textsc{Hinted} & \scriptsize\textsc{Edited} & \scriptsize\textsc{Hinted} & \scriptsize\textsc{Edited} & \scriptsize\textsc{Hinted} & \scriptsize\textsc{Edited} & \scriptsize\textsc{Hinted} & \scriptsize\textsc{Edited} & \scriptsize\textsc{Hinted} & \scriptsize\textsc{Edited} \\
        
        \includegraphics[width=0.08\textwidth]{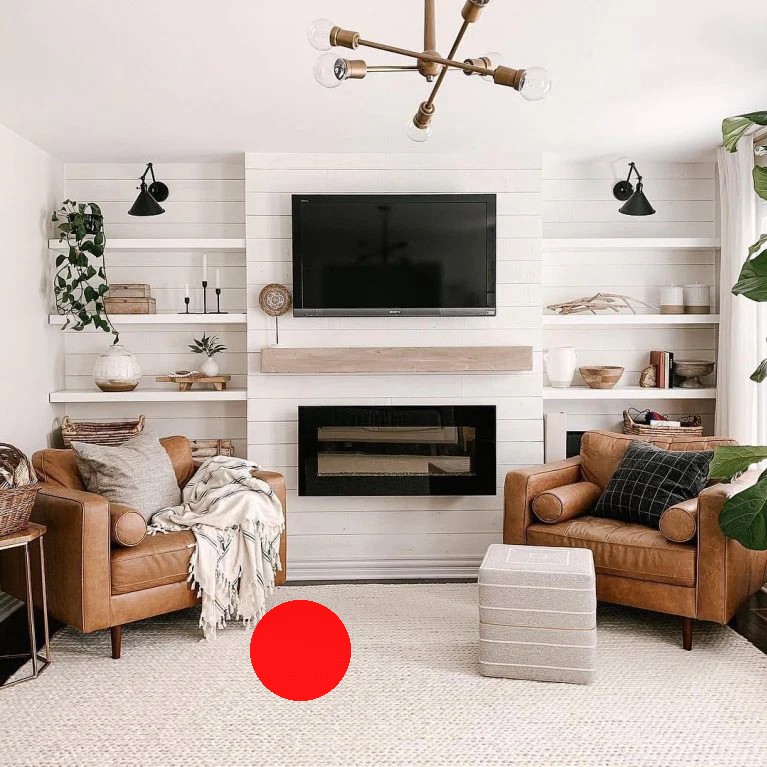} &
        \includegraphics[width=0.08\textwidth]{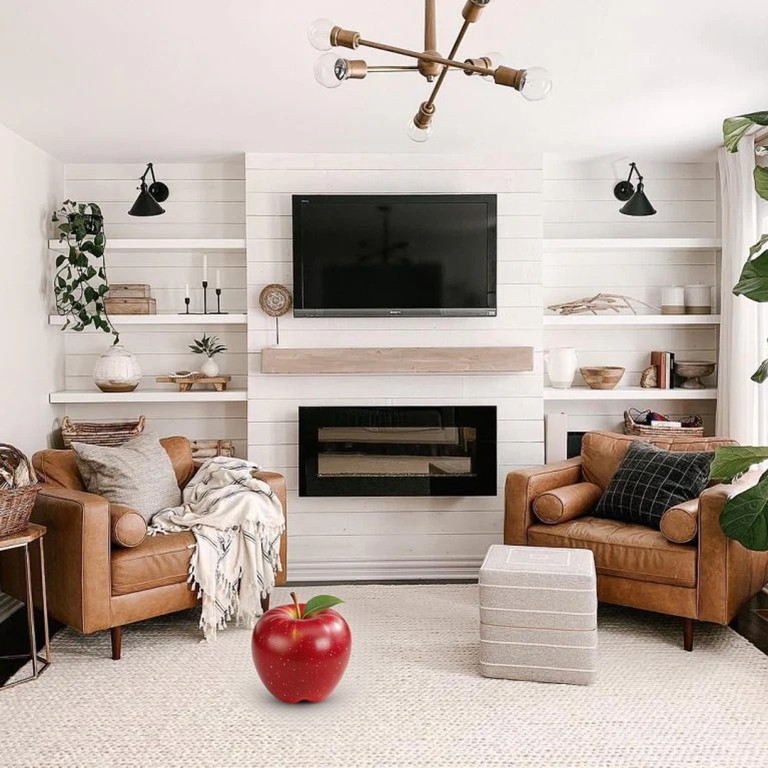} &
        \includegraphics[width=0.08\textwidth]{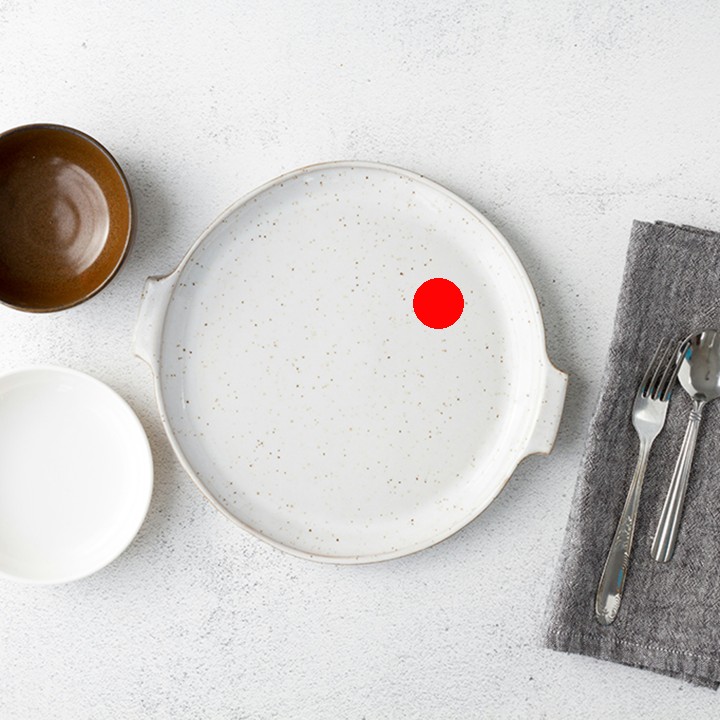} &
        \includegraphics[width=0.08\textwidth]{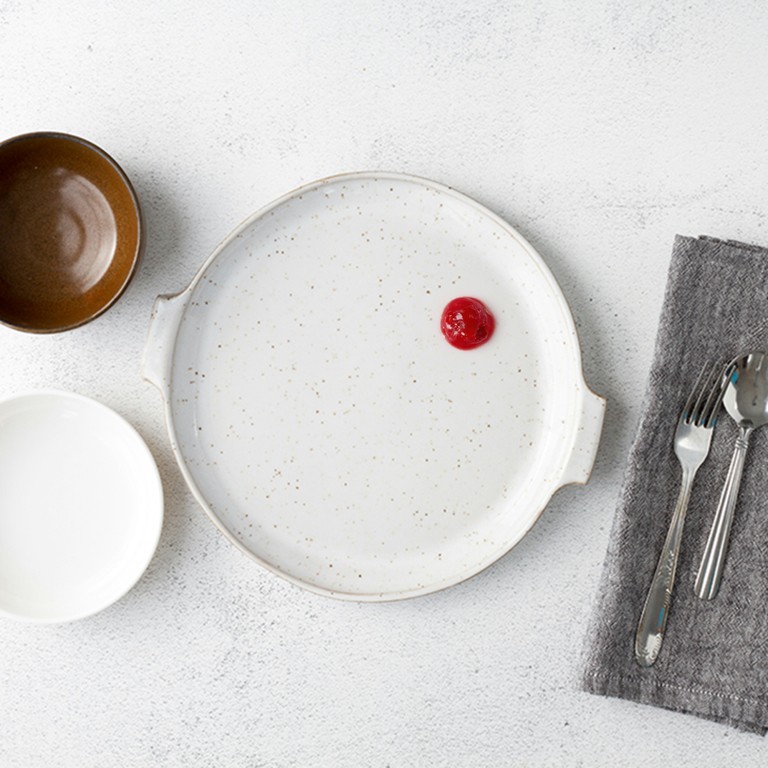} &
        \includegraphics[width=0.08\textwidth]{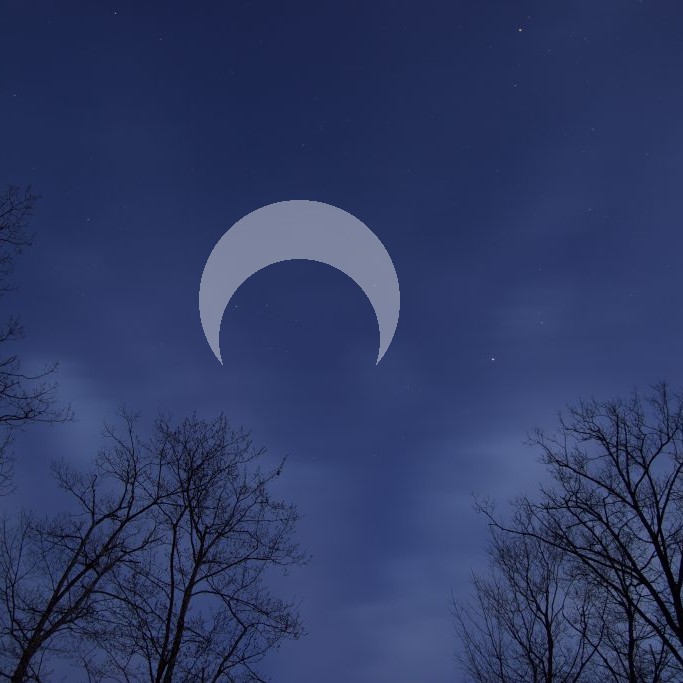} &
        \includegraphics[width=0.08\textwidth]{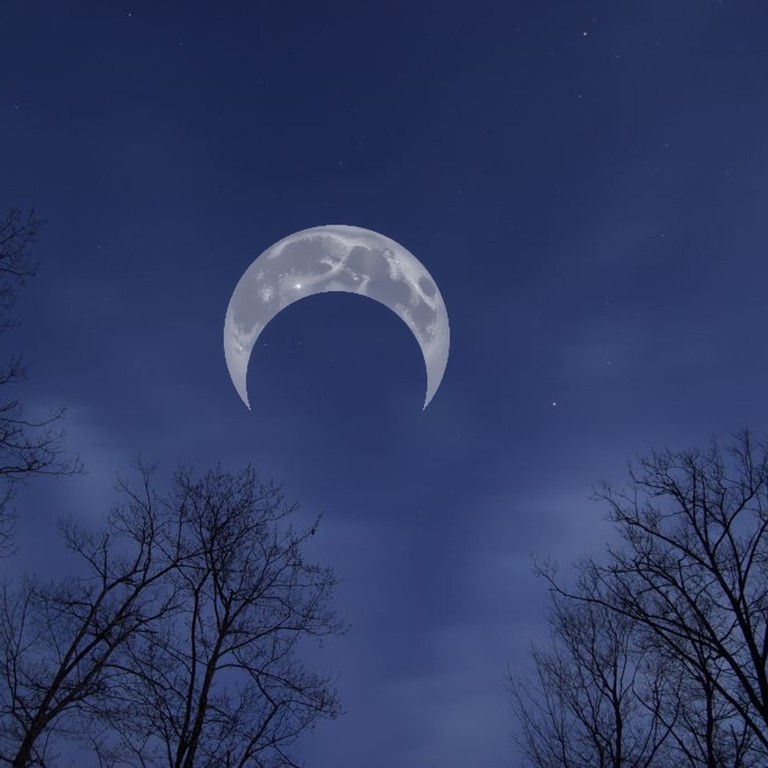} &
        \includegraphics[width=0.08\textwidth]{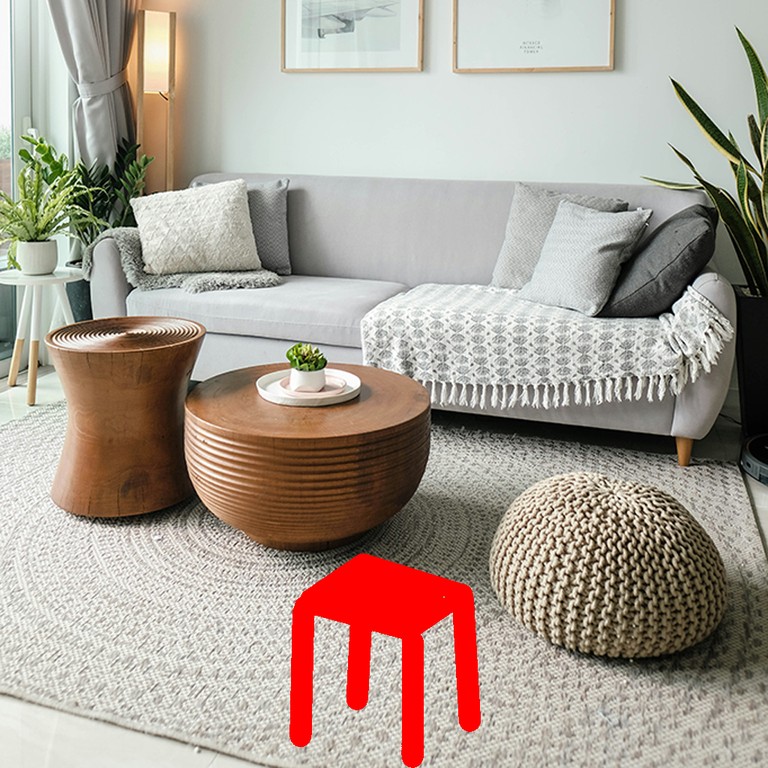} &
        \includegraphics[width=0.08\textwidth]{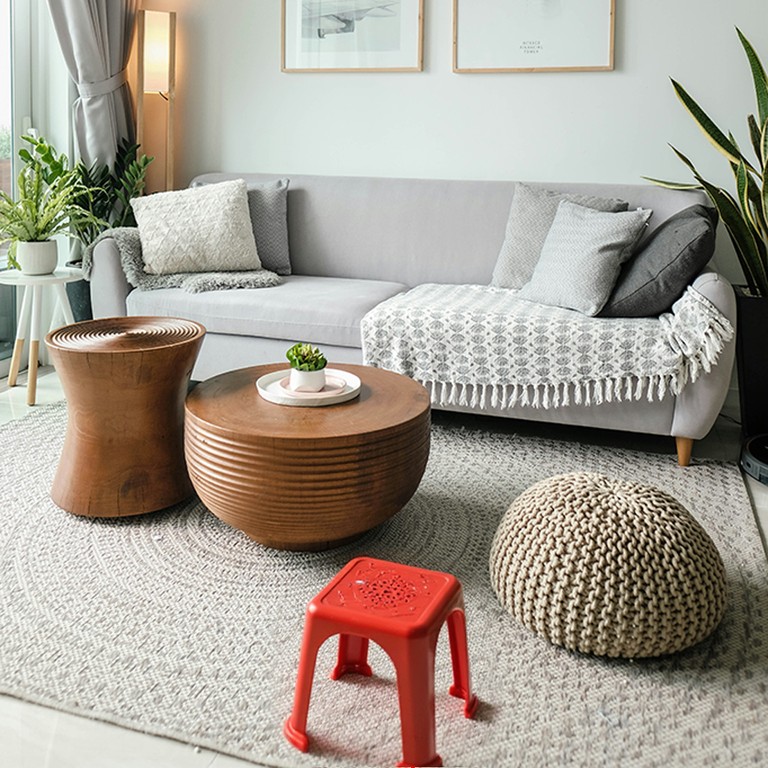} &
        \includegraphics[width=0.08\textwidth]{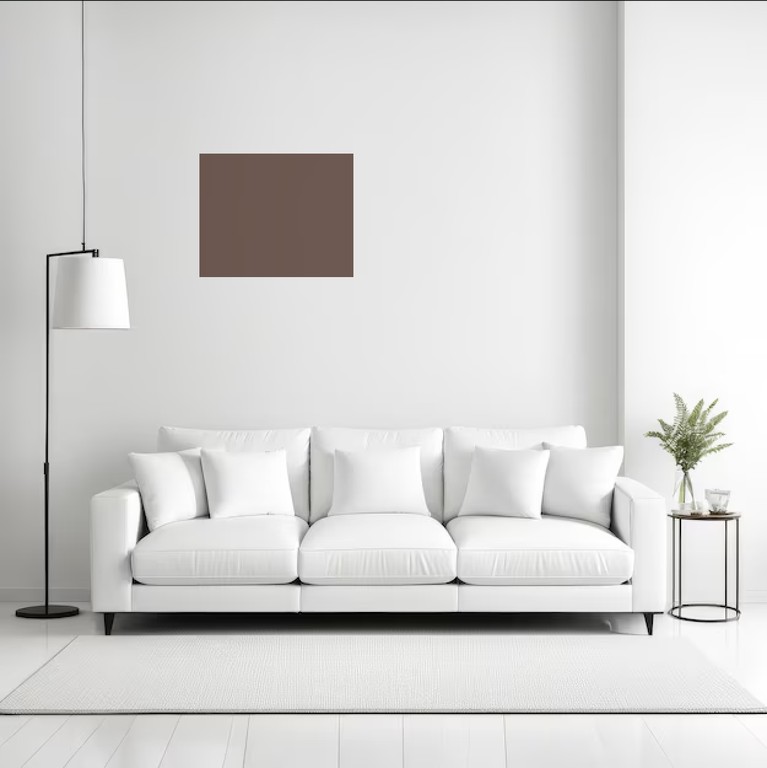} &
        \includegraphics[width=0.08\textwidth]{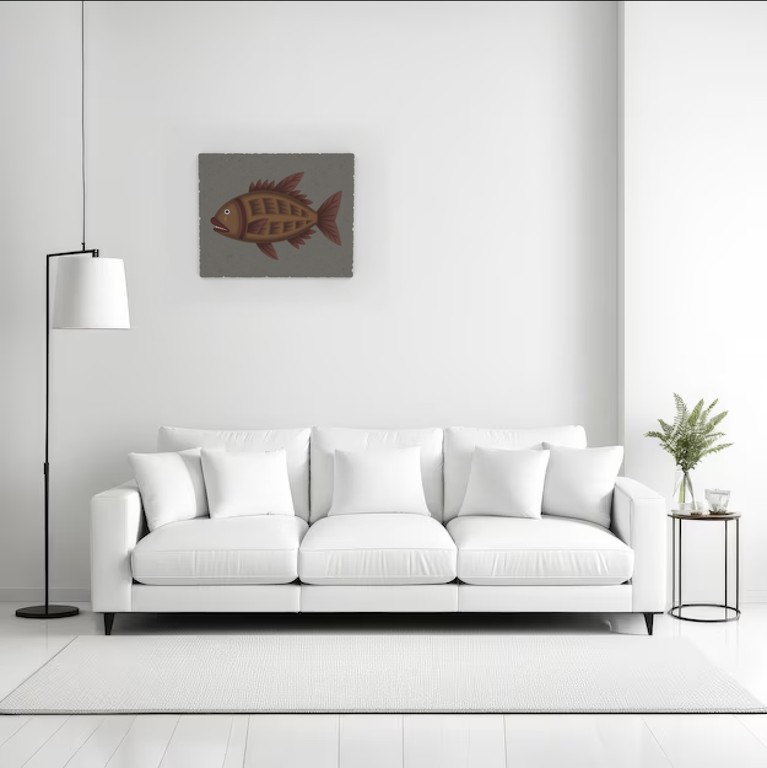} \\
        \includegraphics[width=0.08\textwidth]{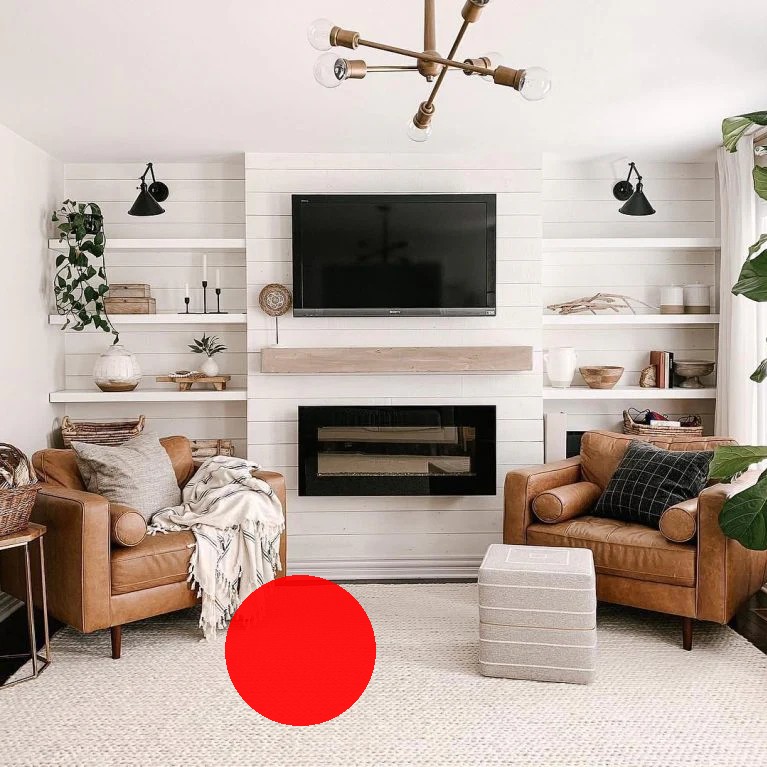} &
        \includegraphics[width=0.08\textwidth]{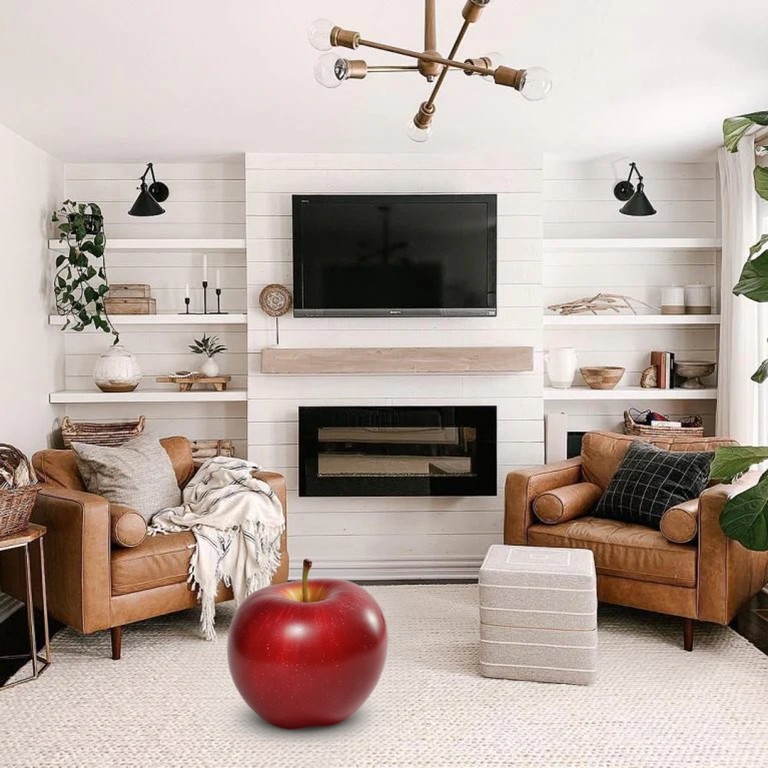} &
        \includegraphics[width=0.08\textwidth]{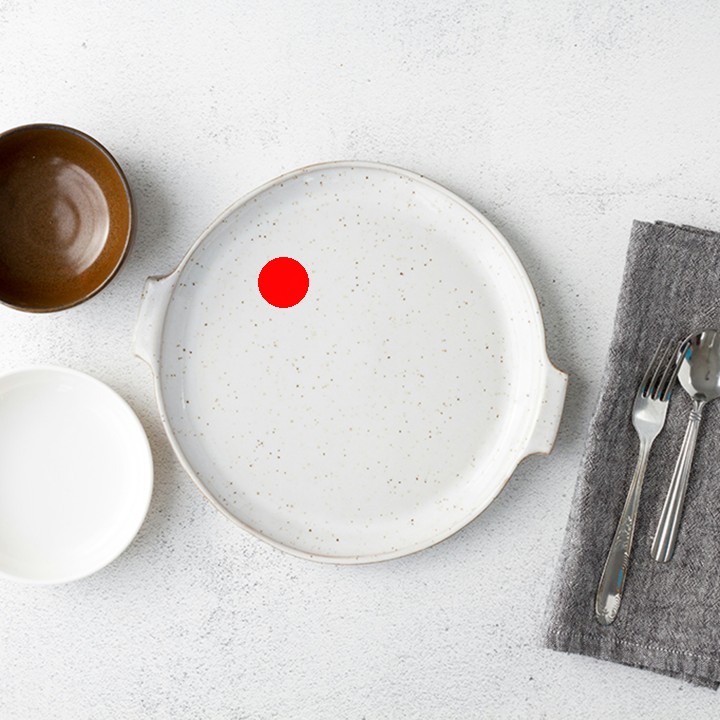} &
        \includegraphics[width=0.08\textwidth]{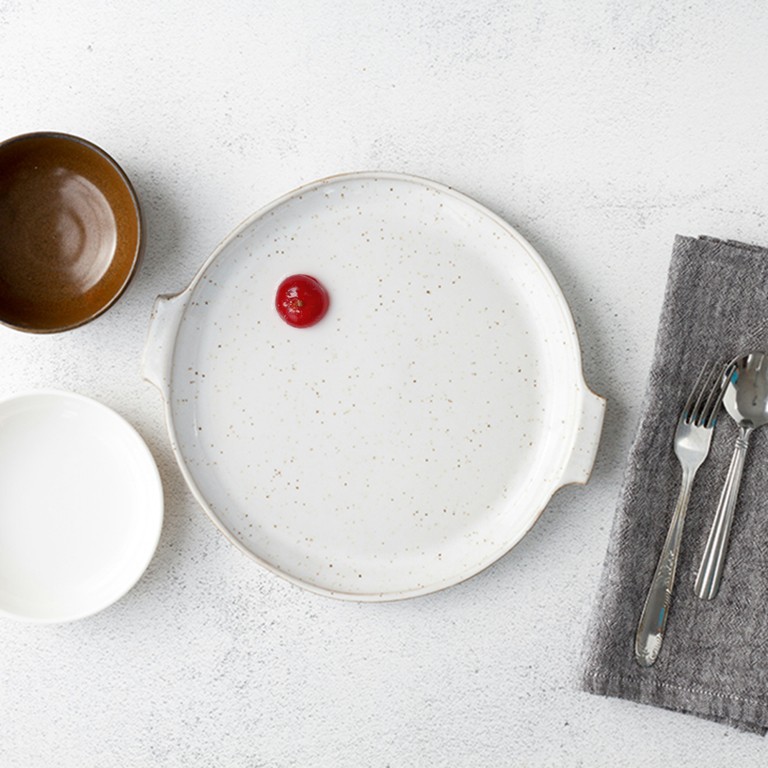} &
        \includegraphics[width=0.08\textwidth]{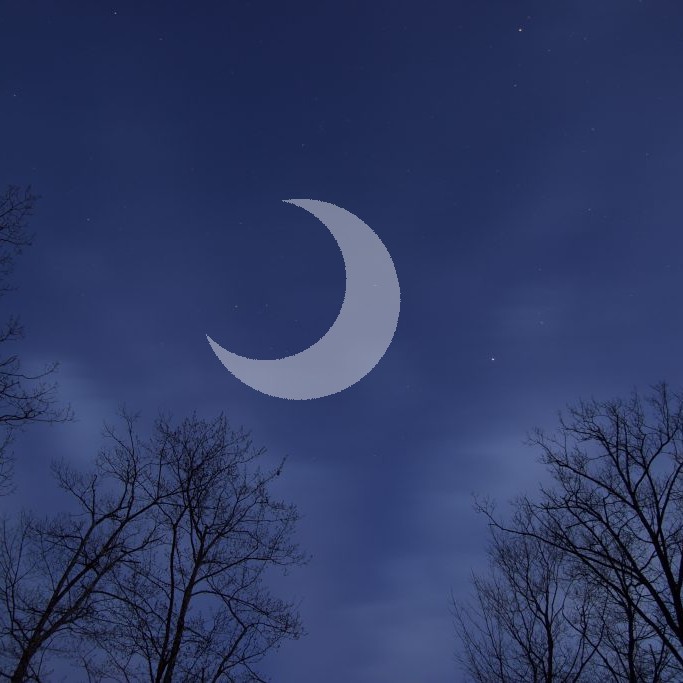} &
        \includegraphics[width=0.08\textwidth]{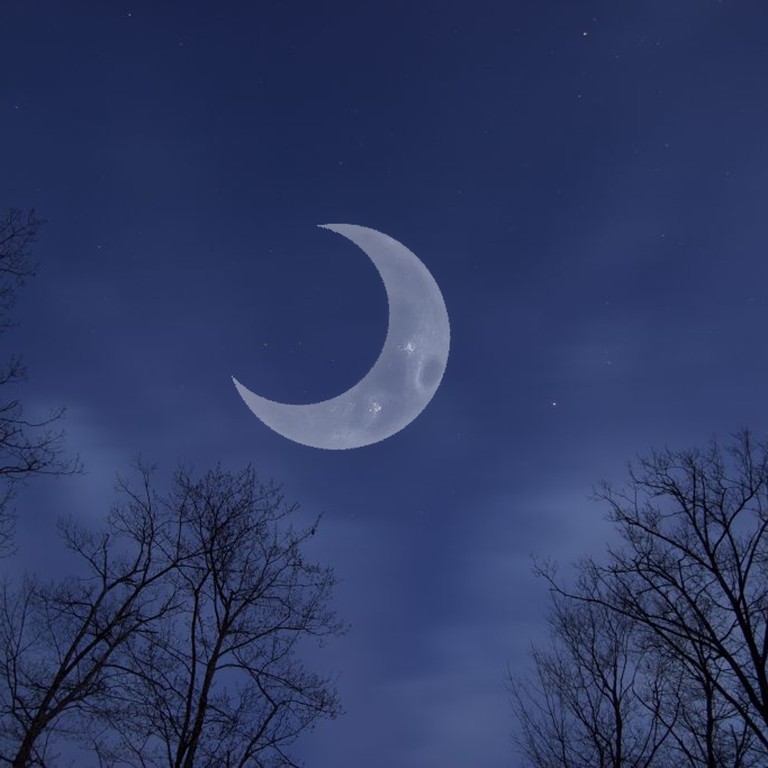} &
        \includegraphics[width=0.08\textwidth]{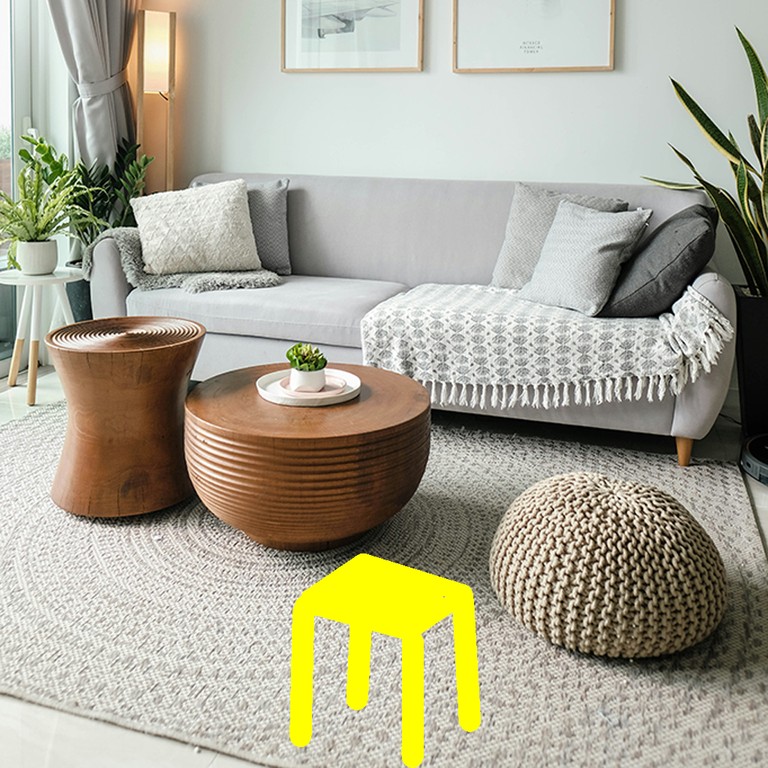} &
        \includegraphics[width=0.08\textwidth]{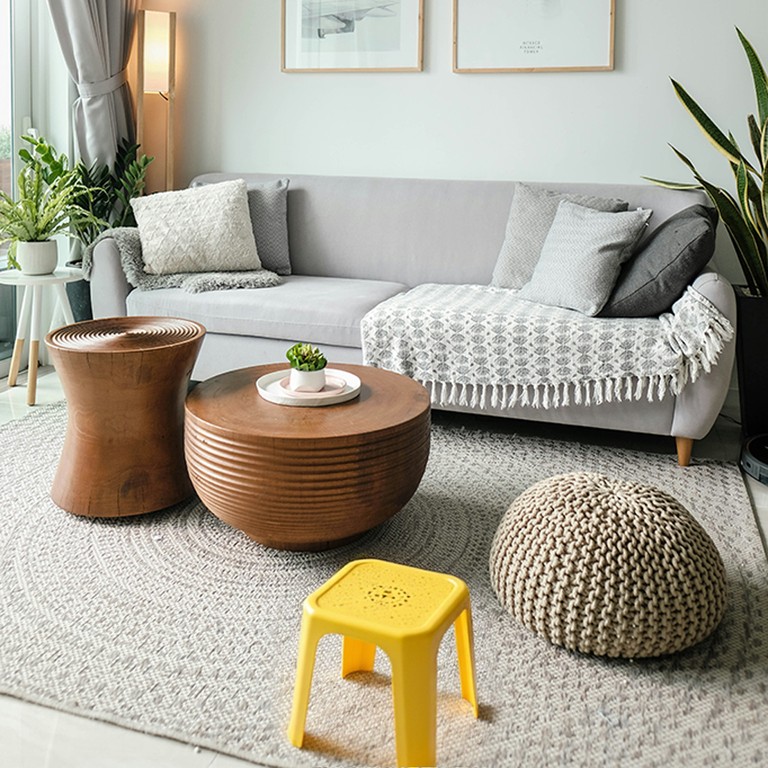} &
        \includegraphics[width=0.08\textwidth]{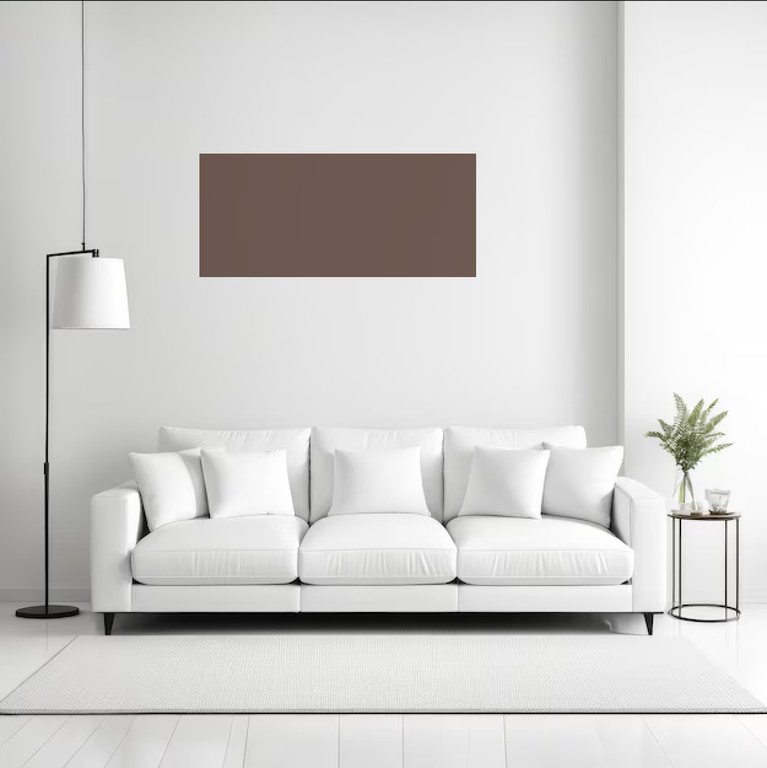} &
        \includegraphics[width=0.08\textwidth]{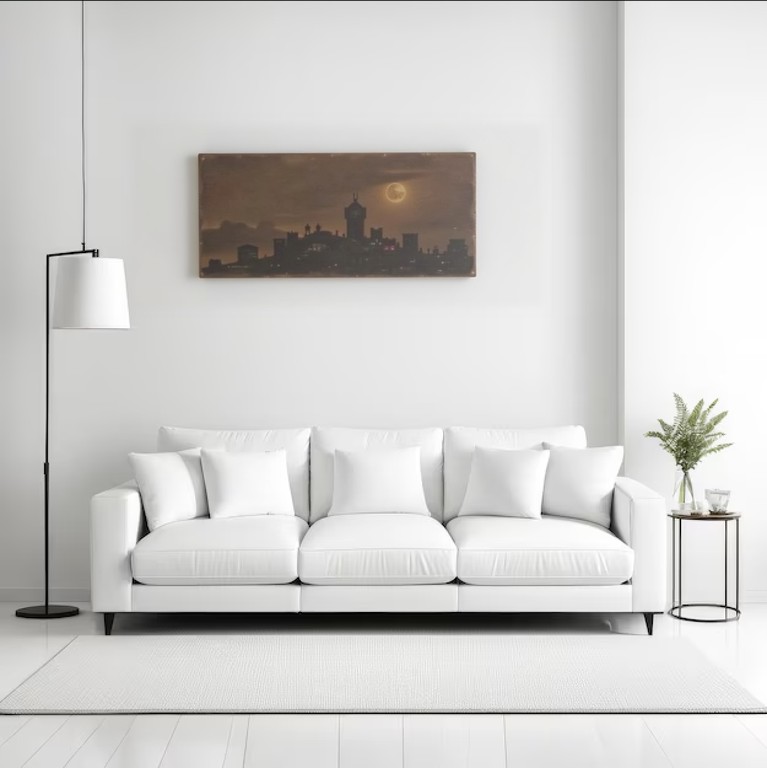} \\
        \includegraphics[width=0.08\textwidth]{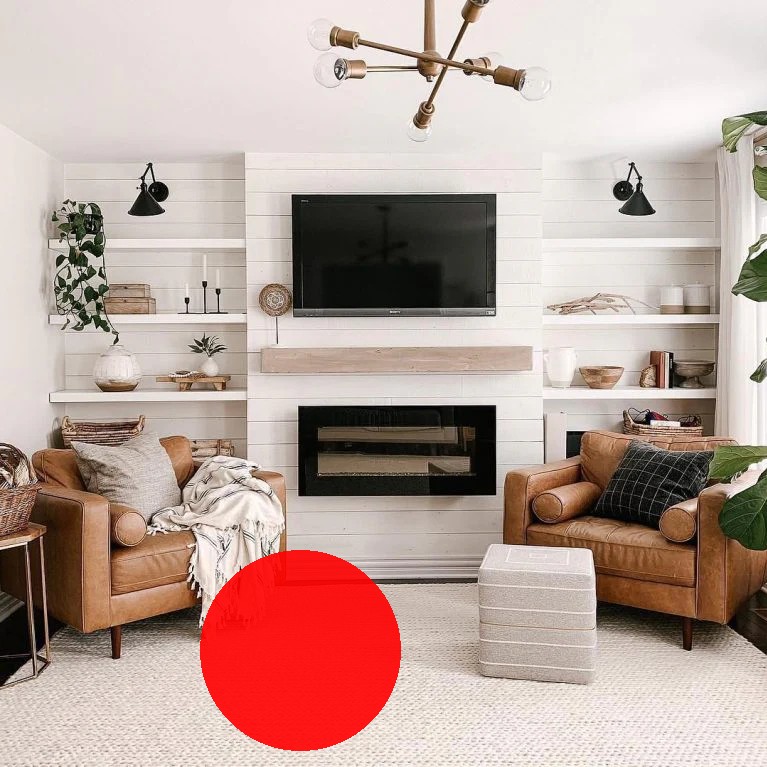} &
        \includegraphics[width=0.08\textwidth]{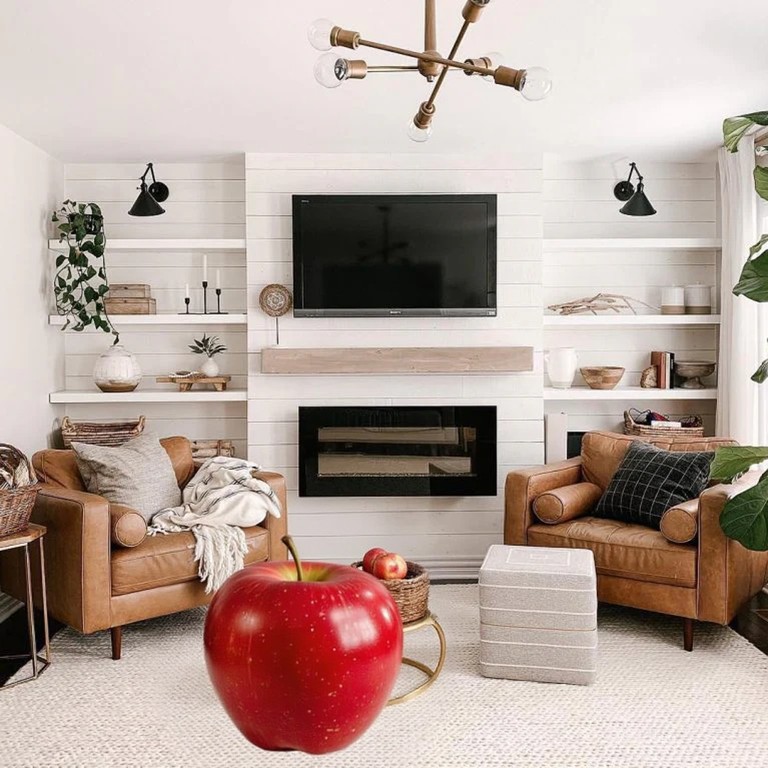} &
        \includegraphics[width=0.08\textwidth]{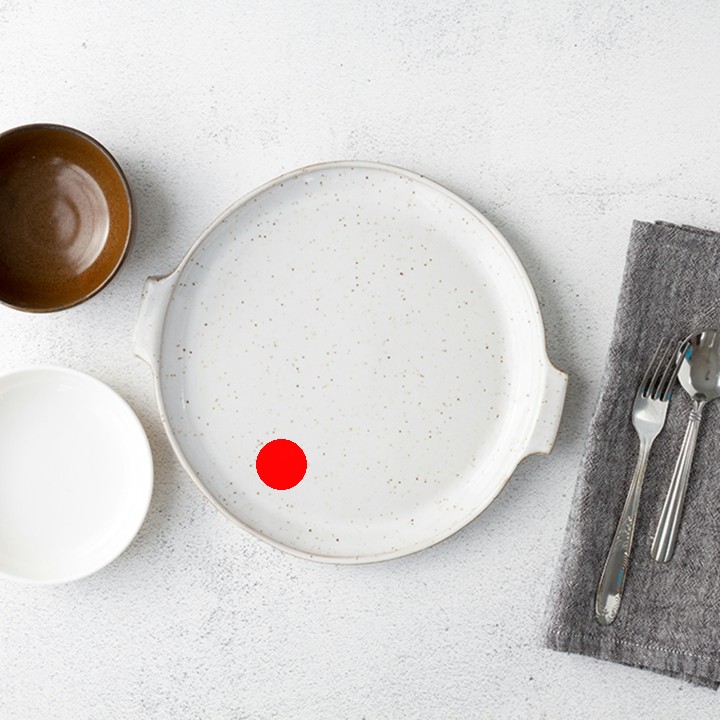} &
        \includegraphics[width=0.08\textwidth]{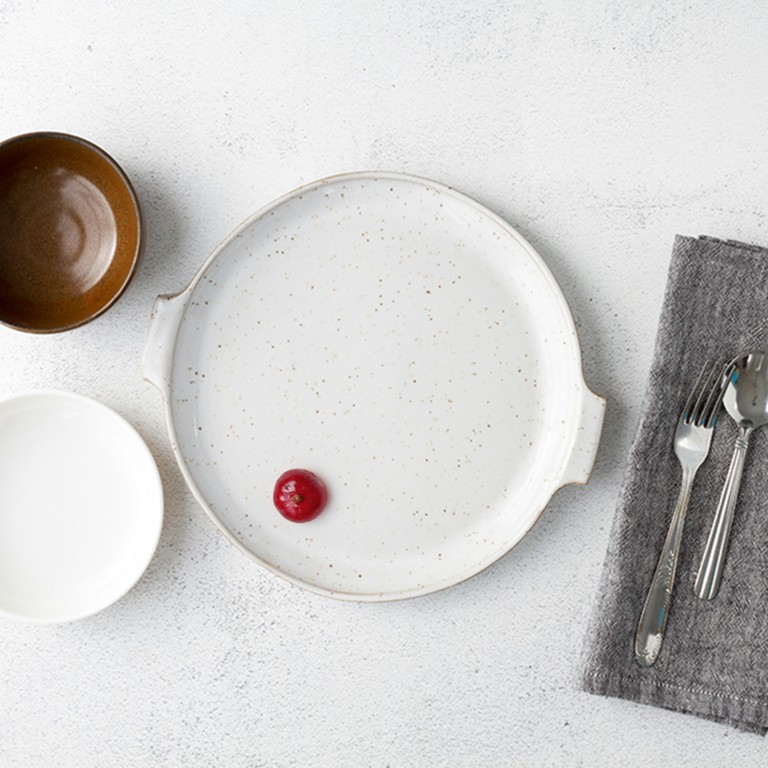} &
        \includegraphics[width=0.08\textwidth]{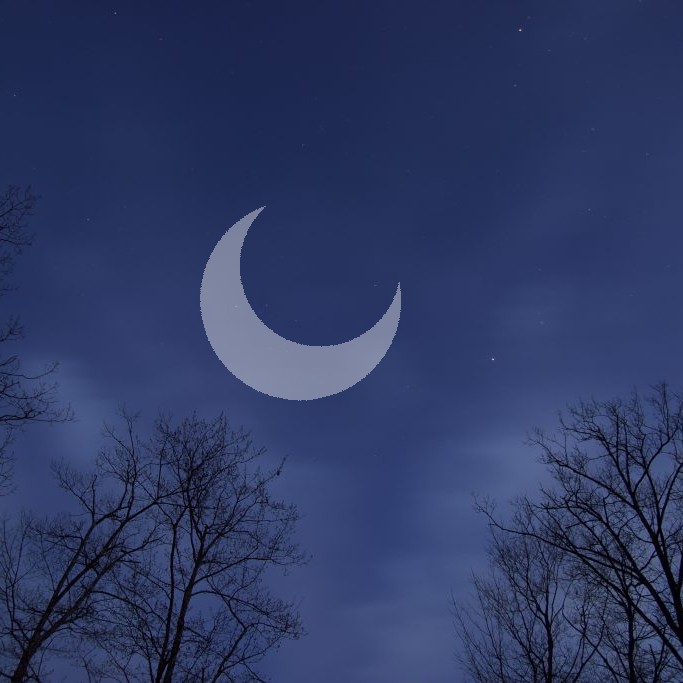} &
        \includegraphics[width=0.08\textwidth]{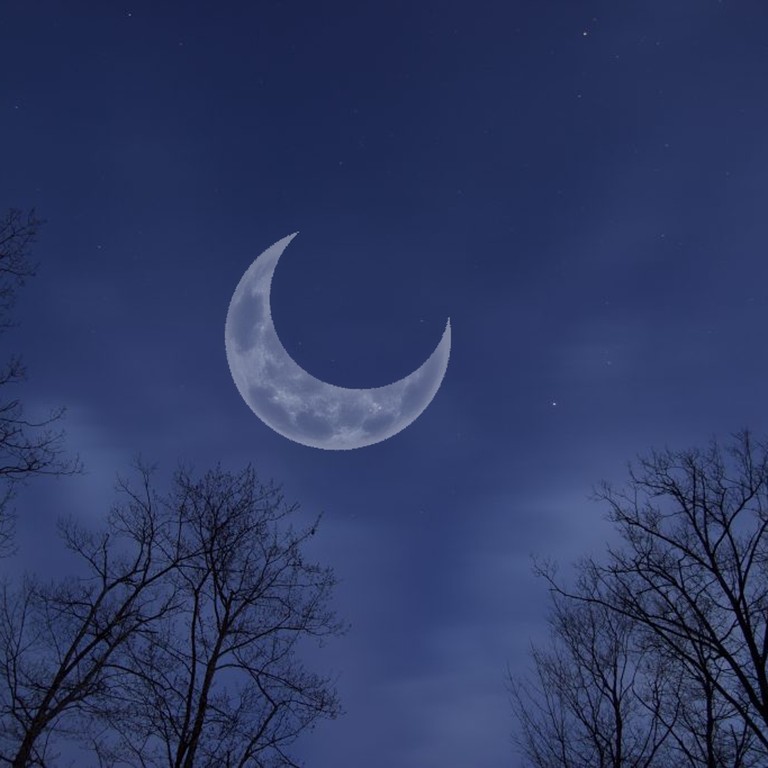} &
        \includegraphics[width=0.08\textwidth]{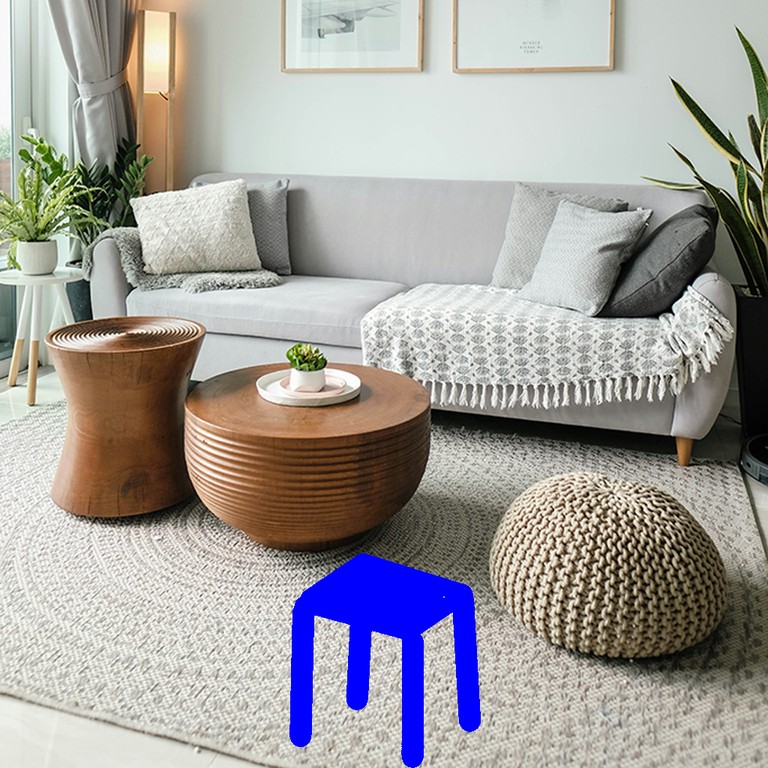} &
        \includegraphics[width=0.08\textwidth]{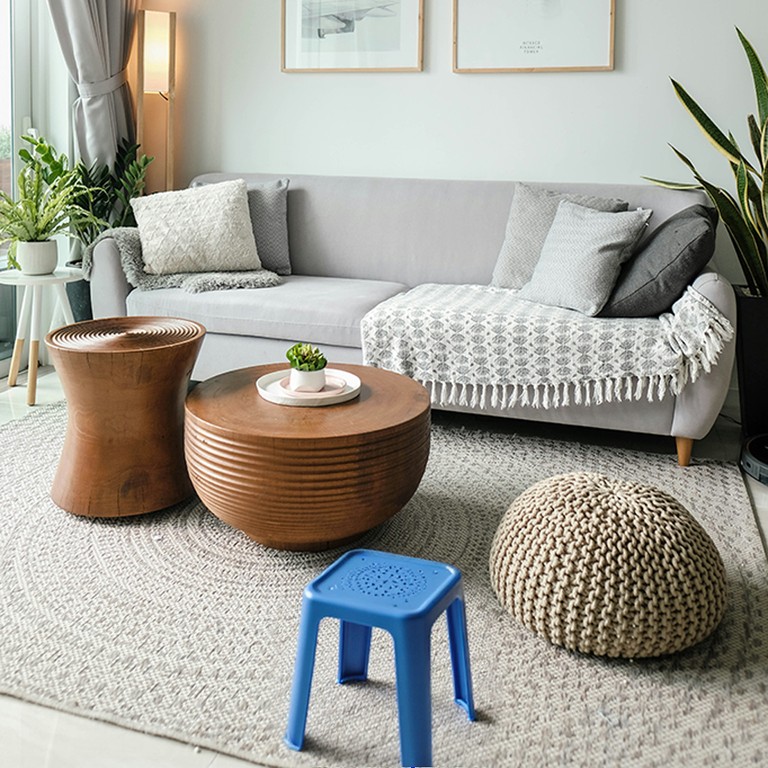} &
        \includegraphics[width=0.08\textwidth]{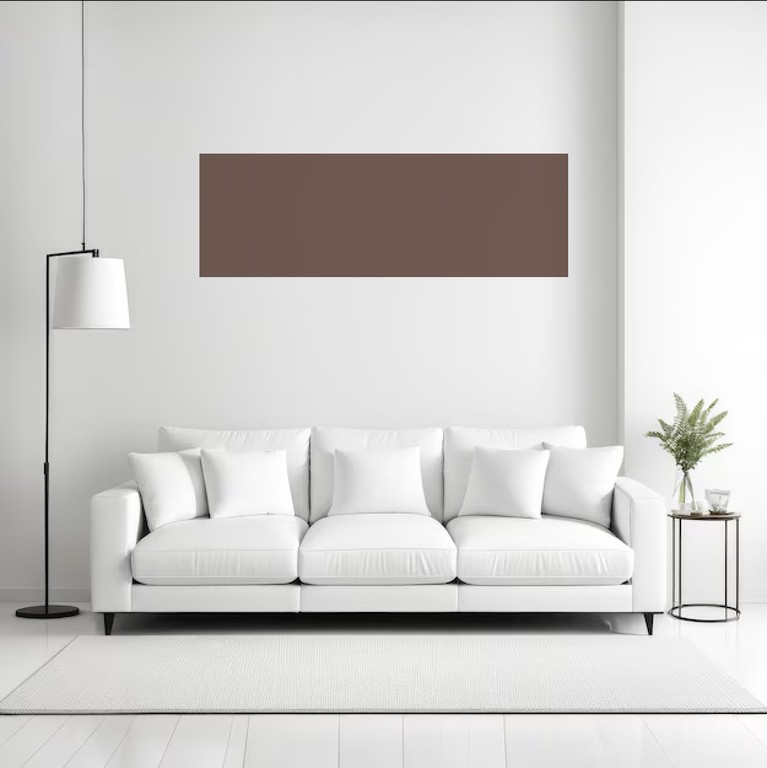} &
        \includegraphics[width=0.08\textwidth]{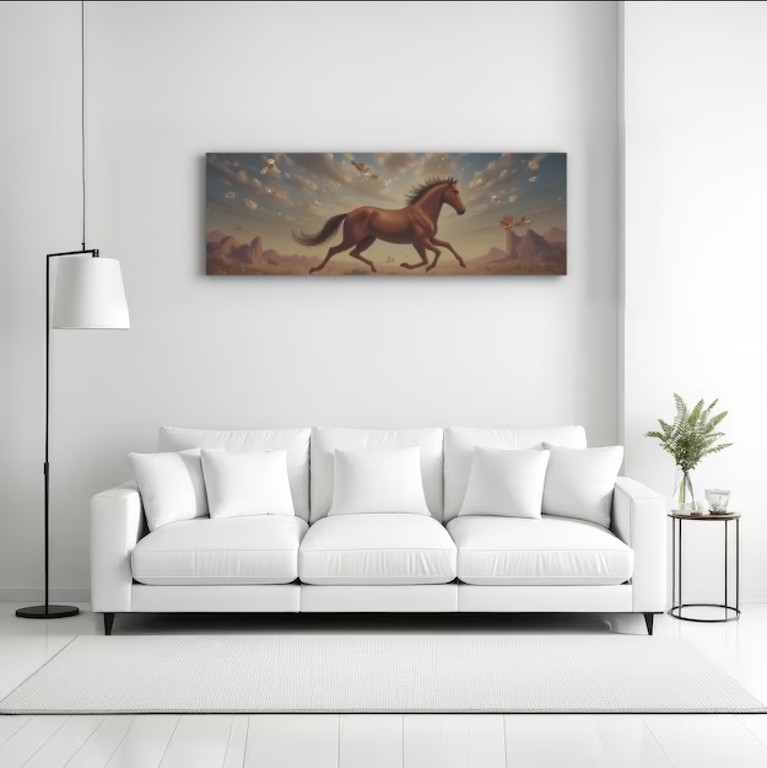} \\
    \end{tabular}
    \centering
    \includegraphics[width=1\linewidth]{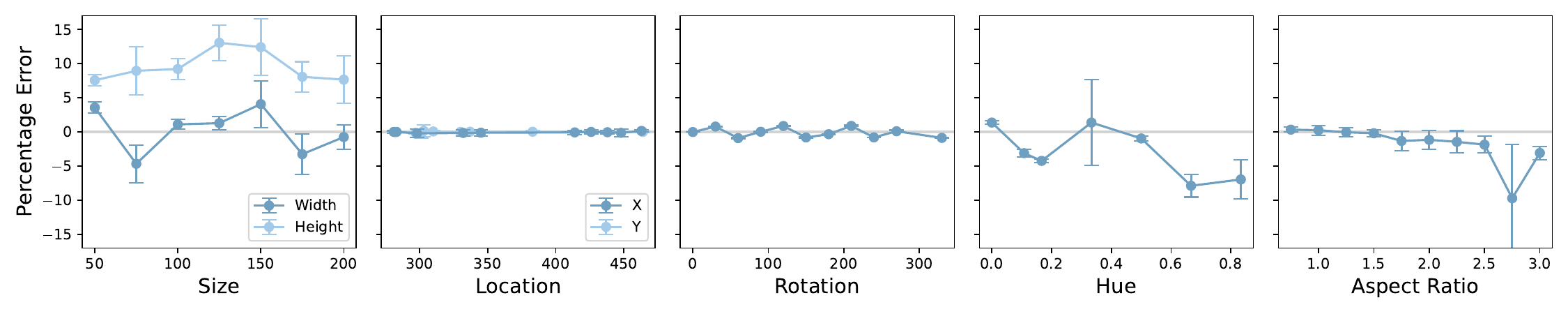}

    \caption{\textbf{Our workflow delivers precise and repeatable edits, where the generated objects align with user specifications.} The top images compare the user-specified color hints with the generated objects, and the bottom plots show the percentage errors. On each plot, we show the mean and standard deviation of percentage errors across 10 random seeds. All mean errors are close to 0.}
    \label{fig:precise_editing}

\end{figure*}

\subsection{Precise Editing}
\label{sec:precise-editing}

To quantify the precision of our workflow, we add objects while specifying 5 properties, including size, location, rotation, color, and aspect ratio (\cref{fig:precise_editing}). To specify the value of each property, we use a hint in the form of a color patch. In the edited image, we measure the property value of the added object and calculate the percentage error against the user-specified property value (\cref{app:measuring-properties-of-generated-objects}). We use 10 random seeds to account for generative variability.

\cref{fig:precise_editing} shows that the mean of percentage errors for each specified property value is close to 0, with a small standard deviation across random seeds. There are still systematic errors, such as the height of the apple being consistently 10\% larger than specified. However, for the apple, the height is larger because the apple's stem is unspecified by the hint. Overall, the results confirm that our workflow delivers precise, repeatable edits critical for real-world scenarios.

\subsection{Iterative Editing}
\label{sec:iterative-editing}

In this subsection, we consider two iterative editing regimes, one using fewer than 5 editing steps, another more than 100. Multiple editing steps are often necessary, as one single editing or generation step usually fails to align with complicated user requirements (further discussed in \cref{sec:customizable-editing}).

\begin{figure}[t!]
\begin{center}
\begin{tabular}{c@{\hspace{0.3em}}c@{\hspace{0.3em}}c@{\hspace{0.3em}}c@{\hspace{0.3em}}c}
    \scriptsize\textsc{Original} & \scriptsize\textsc{Step 1} & \scriptsize\textsc{Step 2} & \scriptsize\textsc{Step 3} & \scriptsize\textsc{Step 4} \\
    \includegraphics[width=0.18\textwidth]{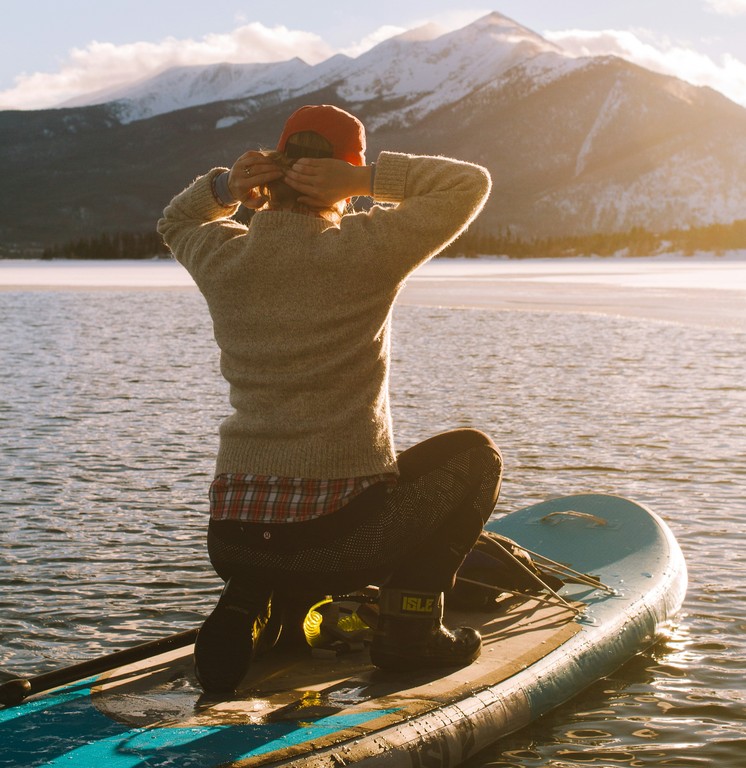} &
    \includegraphics[width=0.18\textwidth]{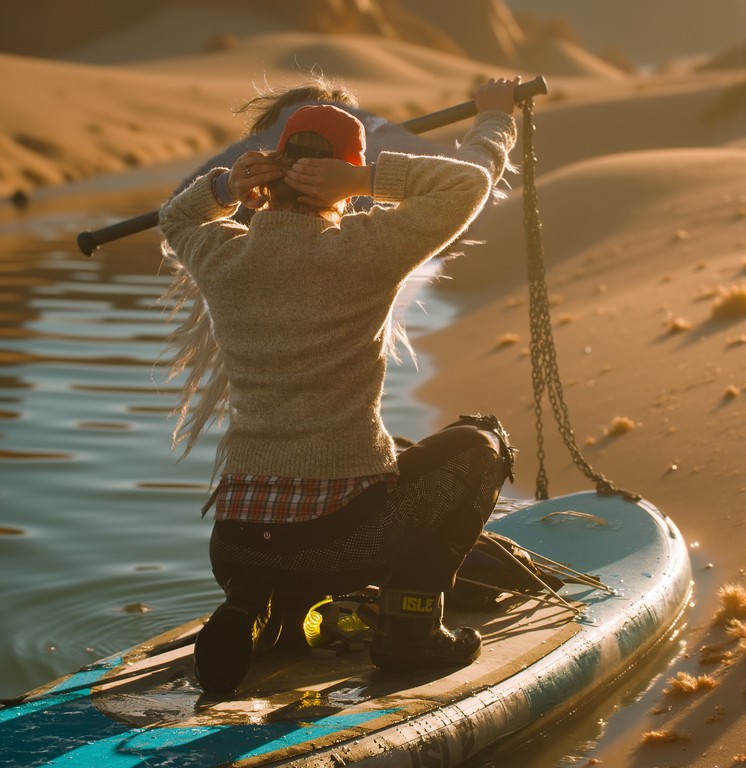} &
    \includegraphics[width=0.18\textwidth]{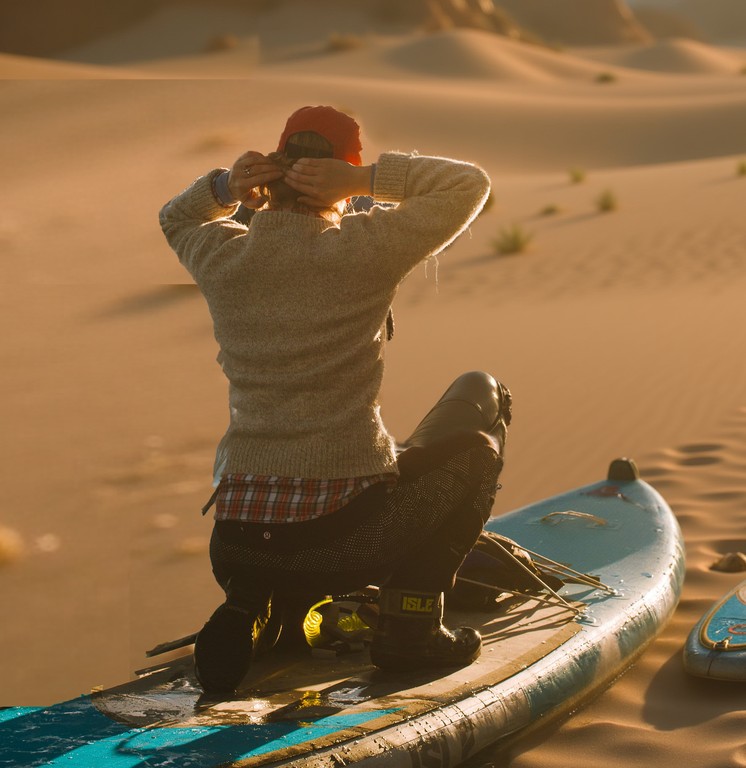} &
    \includegraphics[width=0.18\textwidth]{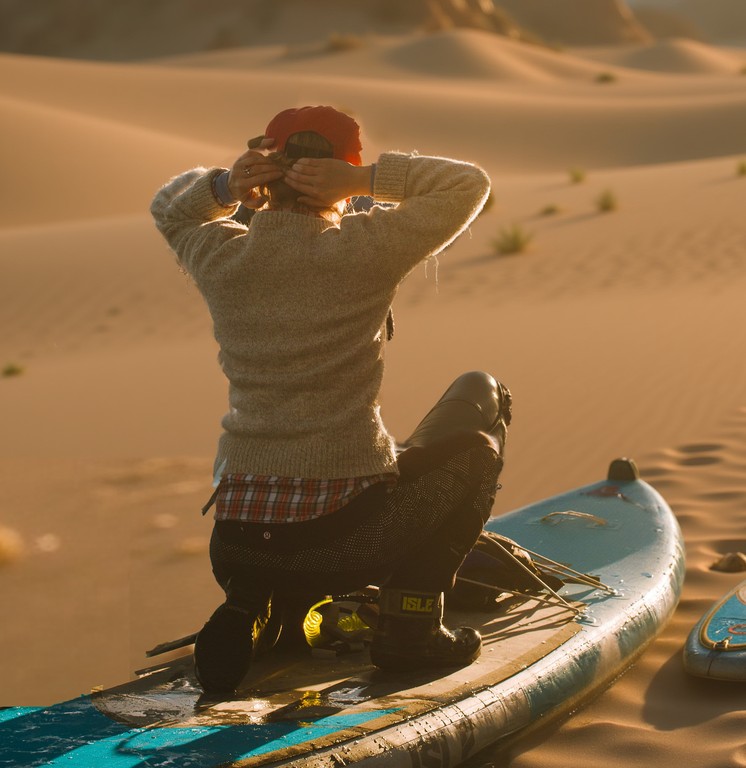} &
    \includegraphics[width=0.18\textwidth]{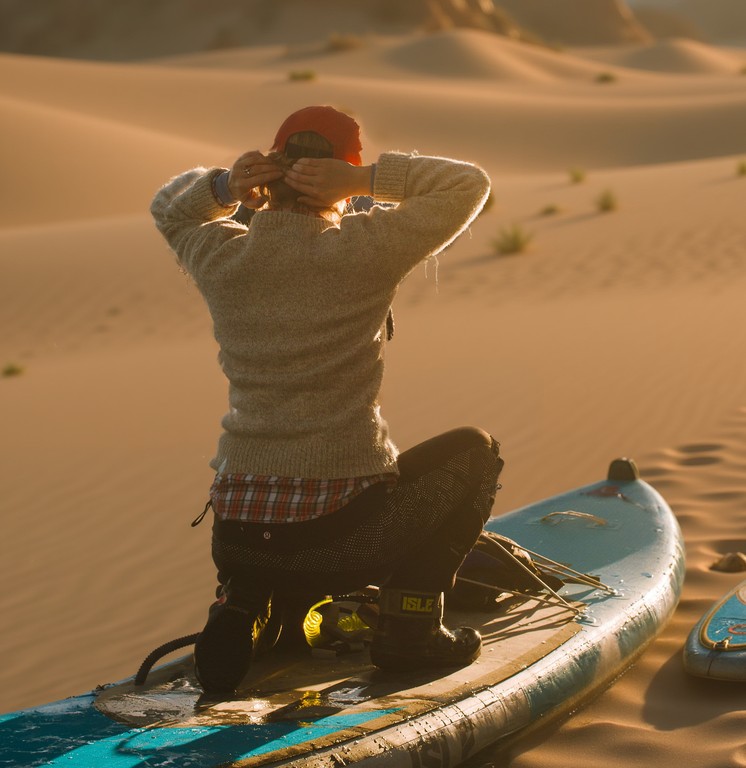} \\
\end{tabular}

\caption{\textbf{We use \spice~to recycle failures.} In this example, the editing instruction is ``change the lake to a desert''. After Step 1, \spice~fails to remove the water from the background. However, 3 more editing steps greatly improve the outcome by removing obvious artifacts. In each step, the mask and hint are redrawn to emphasize different parts of the image for editing, but the description prompt is fixed, imposing low burden on the user.}
\label{fig:recycling}
\end{center}

\end{figure}

In the low-steps regime, if unsatisfied with results from the first editing step, we can improve them using another editing step with our workflow. In \cref{sec:benchmark-results}, we identify that our workflow does not work well for a few cases in single-step editing. In these cases, a large proportion of the edited region is satisfactory, leaving only a small proportion to be fixed. For baseline methods, as the image quality degrades after the first editing step, we cannot run a second editing step but can only rerun the model with the same input and different random seeds, and the results are unsatisfactory (\cref{app:changing-the-random-seed}). However, with our workflow, we can improve the result by recycling the first output (\cref{fig:recycling}).

\begin{figure}[t!]

\begin{center}
\subfigure[Iteratively constructing an image]{
\begin{tabular}{c@{\hspace{0.3em}}c}
    \scriptsize\textsc{Step 1} & \scriptsize\textsc{Step 40} \\
    \includegraphics[width=0.45\textwidth]{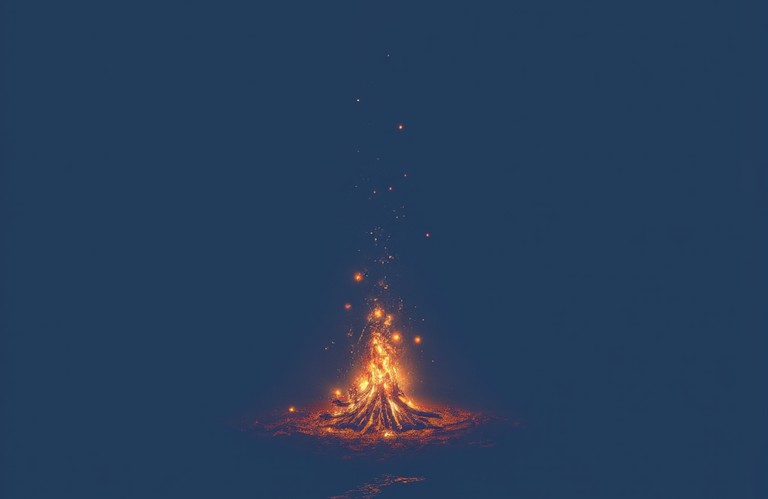} &
    \includegraphics[width=0.45\textwidth]{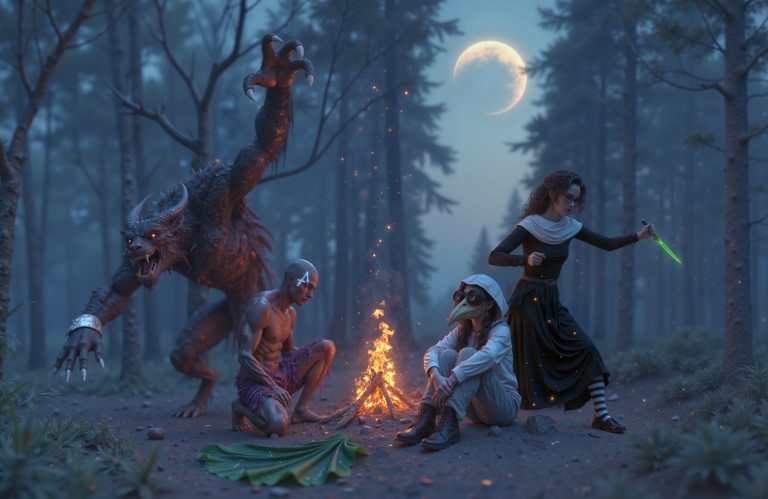} \\
\end{tabular}
\label{subfig:iterative-heart}
}
\subfigure[Iteratively refining an image]{
\begin{tabular}{c@{\hspace{0.3em}}c}
    \scriptsize\textsc{Step 1} & \scriptsize\textsc{Step 100+} \\
    \includegraphics[width=0.45\textwidth]{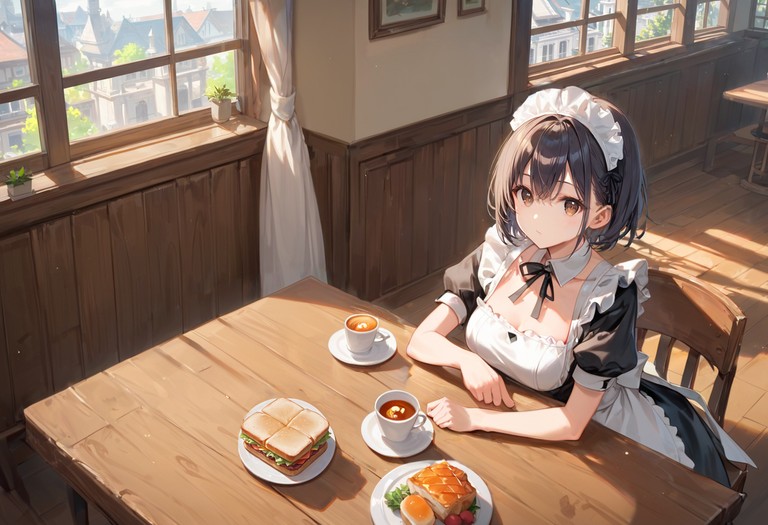} &
    \includegraphics[width=0.45\textwidth]{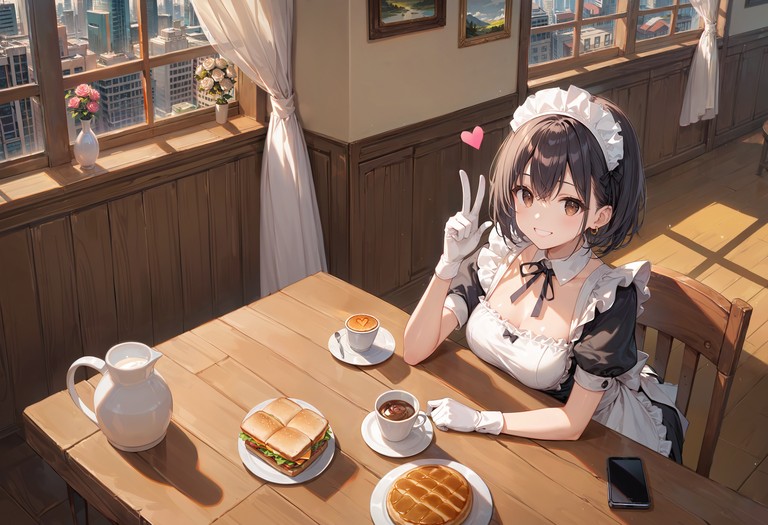} \\
\end{tabular}
\label{subfig:iterative-maid}
}

\caption{\textbf{We use \spice~to consistently improve image quality during a large number of editing steps.} The editing steps can either change the structure or refine the details. The top images are generated using Flux.1 [dev]. The bottom images are generated using BoleroMix (Pony) v1.41.}
\end{center}

\end{figure}

In the high-steps regime, while many previous methods support multi-step editing, heavy artifacts are generated and propagated in each step, disqualifying these methods for long editing tasks. For example, when GPT-4o is prompted to return the same image to the user,\footnote{\url{https://replicateimage.com/examples}} the image quality quickly deteriorates within even a few steps, disqualifying GPT-4o from iterative editing. In contrast, our workflow steadily improves the image after multiple steps. In \cref{subfig:iterative-heart}, we show a result after 40 diverse editing steps, including outpainting, structural edits, upscaling, relighting, and composition adjustment (forming a heart shape using the sky, the crescent, the claw, and the branches). Full steps are shown in \cref{app:iterative-construction}. Interestingly, to relight characters according to the environment, we do not need to provide any color hints. We also do not need a specialized relighting model, such as IC-Light \cite{zhang2025scaling}. In \cref{subfig:iterative-maid}, we start with a model-generated image with various errors in shadows, objects, and anatomy. Using the same model as the backbone in our workflow, we fix these errors one by one across more than 100 editing steps. This image is produced by BoleroMix (Pony) v1.41, a checkpoint derived from SDXL. More examples of final edited results in different styles can be found in \cref{app:art-styles}, where around 200 editing steps are performed to achieve the desired level of detail. Each image is created from scratch within 4 hours in one sitting.


\subsection{Customizable Editing}
\label{sec:customizable-editing}

Prompt-based image generation or editing models frequently fail to fulfill their promise that users can create content according to their own will. Users are frustrated for two main reasons.

\begin{figure}[t]

\begin{center}
\subfigure[DALL·E 3]{
\scriptsize
\begin{tabular}{c@{\hspace{0.8em}}c}
    \textsc{Round 1} & \textsc{Round 1}\\
    \includegraphics[width=0.09\textwidth]{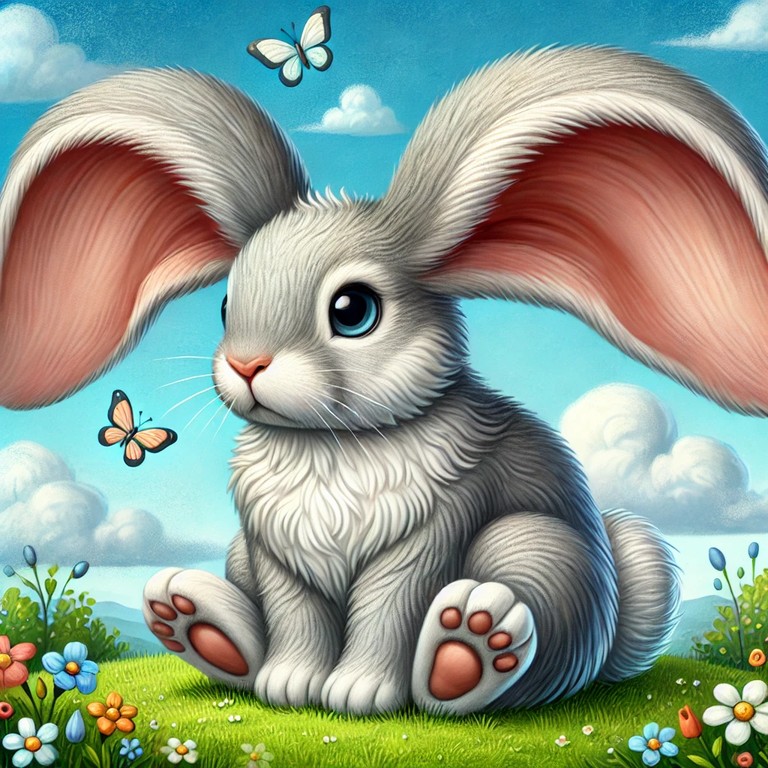} &
    \includegraphics[width=0.09\textwidth]{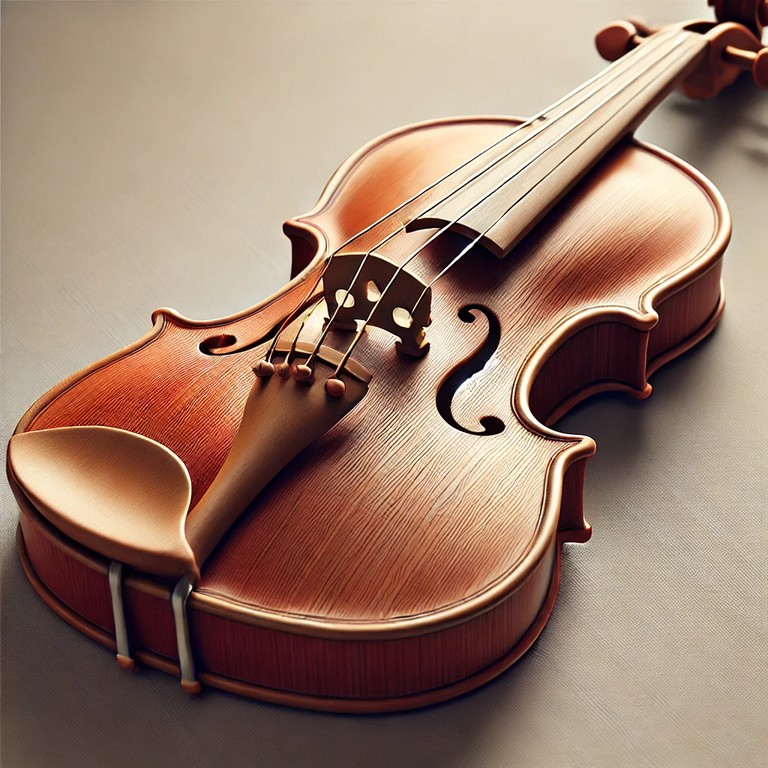} \\
    \textsc{Round 2} & \textsc{Round 2}\\
    \includegraphics[width=0.09\textwidth]{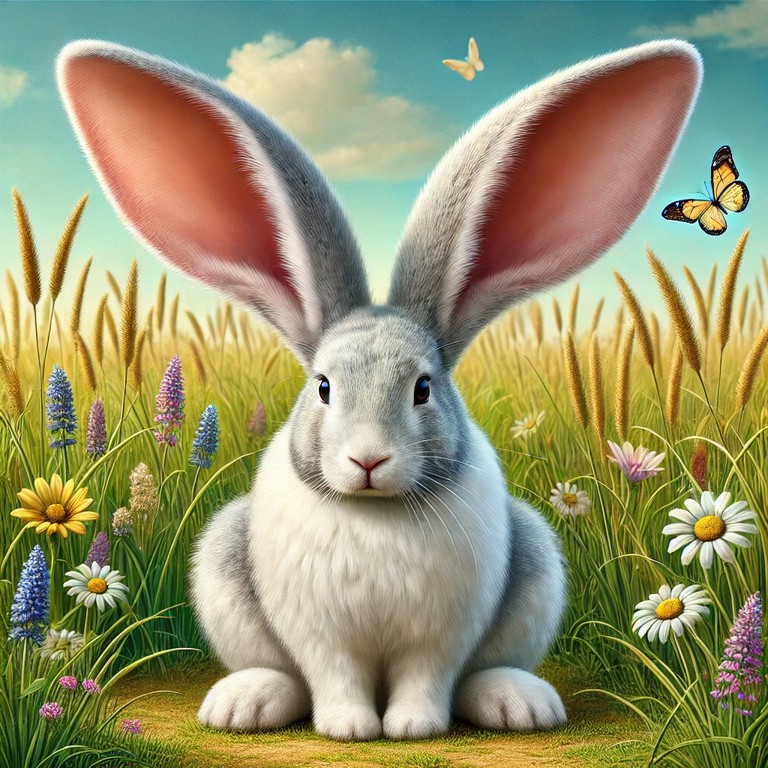} &
    \includegraphics[width=0.09\textwidth]{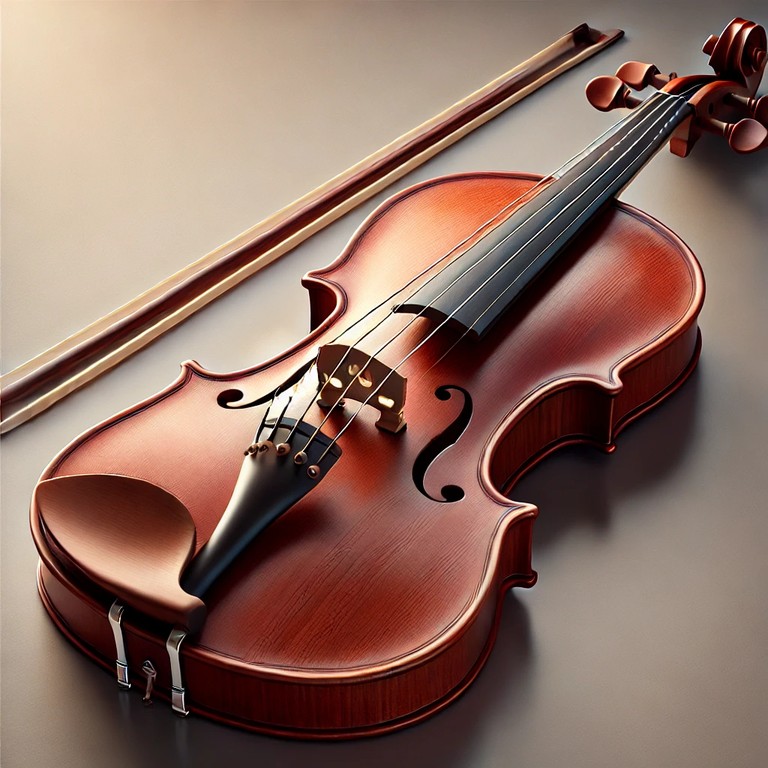} \\
    \textsc{Round 3} & \textsc{Round 3}\\
    \includegraphics[width=0.09\textwidth]{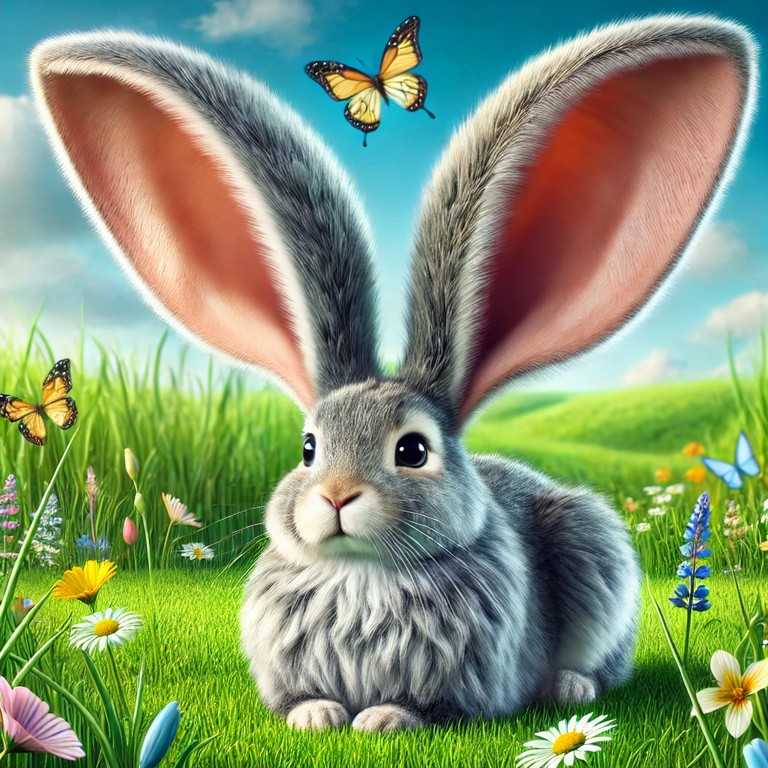} &
    \includegraphics[width=0.09\textwidth]{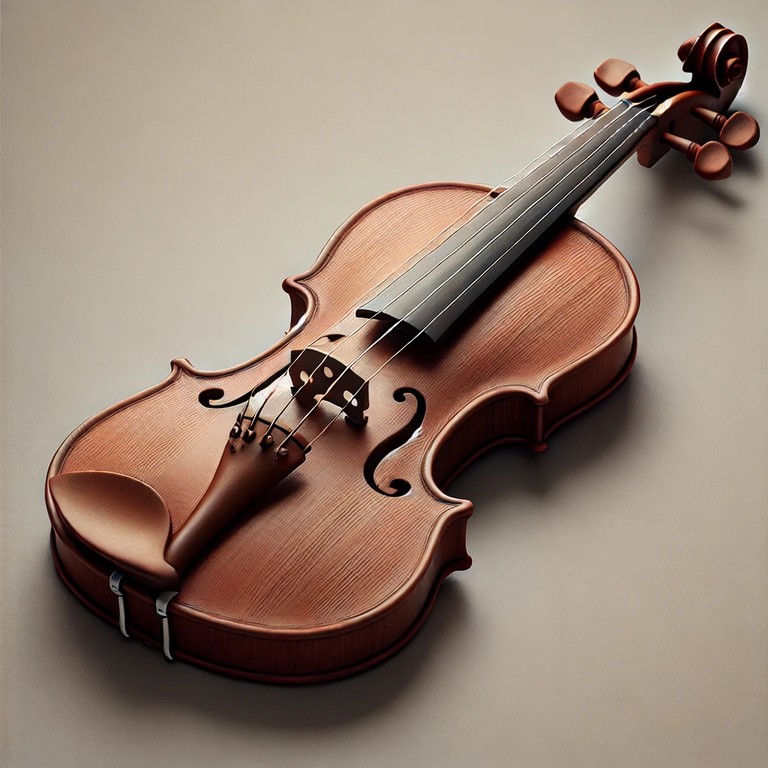} \\
\end{tabular}\label{subfig:dalle}
}
\subfigure[GPT-4o]{
\begin{tabular}{c}
    \textsc{  } \\
    \includegraphics[width=0.215\columnwidth]{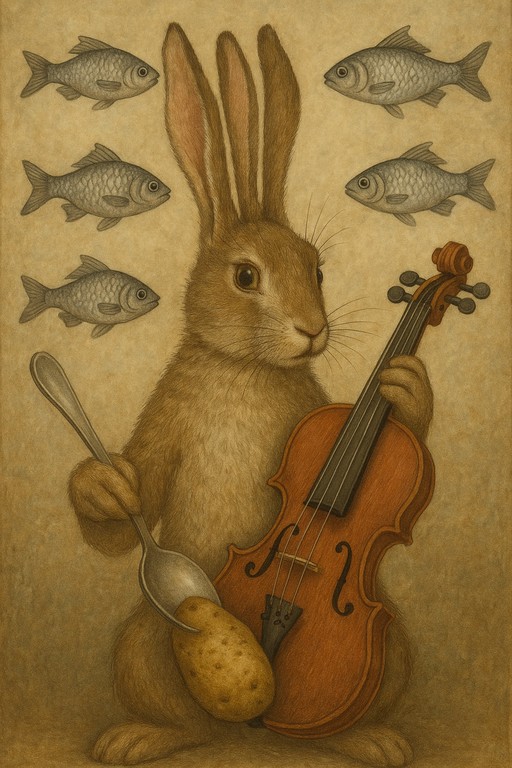}
\end{tabular}
    \label{subfig:gpt4o-rabbit}
}
\subfigure[Ours]{
\begin{tabular}{c}
    \textsc{  } \\
    \includegraphics[width=0.22\columnwidth]{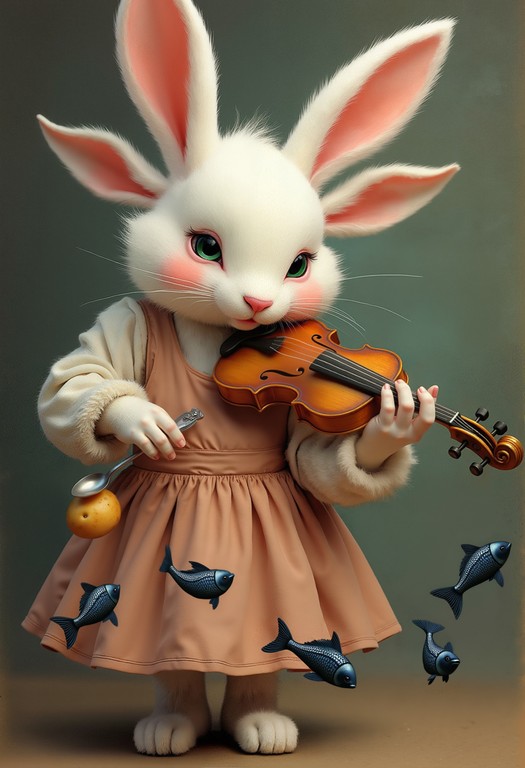}
\end{tabular}
    \label{subfig:customizable}
}

\caption{\textbf{\spice~can generate content that DALL·E 3 and GPT-4o cannot.} Two examples are a rabbit with 4 ears and a violin without a bridge. By January 2025, DALL·E 3 fails even after the user points out the error and asks the model to edit the errors multiple times. By May 2025, GPT-4o also cannot generate all elements correctly at the same time. In contrast, our workflow generates all elements correctly in one image, demonstrating superior customizability.}
\label{fig:rabbit}
\end{center}

\end{figure}

First, prompts are hard to design and often cannot instruct the model to produce complex output. A user assumes that the model follows the prompt, but the model does not interpret the prompt as humans do. One example is that DALL·E 3 \cite{betker2023improving} in November 2024 still fails to generate a rabbit with four ears or a violin without a bridge,\footnote{\url{https://cs.nyu.edu/~davise/papers/DALL-E-Parts/PartsNovember24/DALL-E-Parts-November24.html}} as shown in \cref{subfig:dalle}. Another example is that DALL·E 3 fails to interpret relations or numbers, such as a potato under a spoon or five fish \cite{conwell2024relations}.

Second, the innate randomness of the model dramatically increases the generation cost or even prohibits the generation of complicated images. Suppose a user wants to generate an image with 10 different objects at various locations on this image. If each object has a 50\% chance to be generated perfectly, the success rate of all objects being perfect is below 0.1\%. In real scenarios, the perfection rate for each object will only be lower, and the number of objects larger. Then, generating such an image is practically impossible.

With \spice, instead of desperately engineering prompts or varying the random seed, users handle these difficulties by tuning 3 hyperparameters, namely context size, denoising strength, and Canny model steps. By slightly tuning the 3 hyperparameters (\cref{app:hyperparameter-recommendations}), we are able to generate \cref{subfig:customizable}, an image with a rabbit of four ears, holding a violin without a bridge with the left hand and a spoon with the right hand. There is a potato under the spoon and five fish in the air.\footnote{\cref{subfig:customizable} has been accepted to CVPR AI Art Gallery 2025. \url{https://thecvf-art.com/project/compositionality-and-parts/}} In contrast, DALL·E 3 and GPT-4o both fail to generate the components correctly. The prompt for GPT-4o is ``Generate a rabbit with four ears holding a violin without a bridge. The rabbit holds a spoon in her hand. There is a potato under the spoon. There are five fish floating in the air.''

\section{Evaluation Details}
\label{app:evaluation-details}

\paragraph{Evaluation Benchmark.} We use the second version of the EditEval \cite{huang2024diffusion} benchmark to evaluate our workflow and baseline methods. EditEval is a single-step editing benchmark consisting of original images, source captions, target captions, and editing prompts. The original images have resolutions ranging from 1863$\times$1863 to 8742$\times$8742. EditEval covers semantic editing (object addition, object removal, object replacement, and background change), stylistic editing (style change and texture change), and structural editing tasks (action change). Due to the extremely high resolutions of images, many editing tasks in EditEval require challenging fine-grained edits. An example is removing a tiny insect from a bird's beak, which requires editing fewer than 1\% of all pixels. EditEval allows a fair comparison of all methods, as none of them were trained on a dataset with the same distribution as EditEval.

\paragraph{Baseline Methods.} The baseline methods include InstructPix2Pix (IP2P) \cite{brooks2023instructpix2pix}, IP2P trained on MagicBrush \cite{zhang2024magicbrush}, and UltraEdit \cite{zhao2024ultraedit}. For these methods, we use the original editing prompts from EditEval. The inference hyperparameters are set to recommended values. For UltraEdit, we use the mask-based checkpoint and the same binary masks (with context dots) as in our workflow. Hence, UltraEdit similarly benefits from the extra user input, ensuring a fair comparison between our workflow and this strong baseline.

For InstructPix2Pix, we use the model and hyperparameters according the Hugging Face official page.\footnote{\url{https://huggingface.co/timbrooks/instruct-pix2pix}} The number of inference steps is set to 10, and the image guidance scale is set to 1.

For MagicBrush, we also refer to the Hugging Face page.\footnote{\url{https://huggingface.co/osunlp/InstructPix2Pix-MagicBrush}} Specifically, we use the recommended ``recent best checkpoint'', which is \texttt{MagicBrush-epoch-52-step-4999.ckpt}. We do not provide extra hyperparameters when calling the inference script from the command line, except that we provide a random seed. Hence, all parameters are fixed at their default values, presumably recommended by the model developers.

For UltraEdit, we first use the model version that supports masks.\footnote{\url{https://huggingface.co/BleachNick/SD3_UltraEdit_w_mask}} We use the recommended hyperparameter values. We also use the free-form version,\footnote{\url{https://huggingface.co/BleachNick/SD3_UltraEdit_freeform}} since binary masks are not available in Emu Edit and MagicBrush test sets. Since there is not a recommended set of hyperparameters on the official page for the free-form version, we use the same hyperparameters for the free-form version and the masked version.

For our workflow, we use 0.9 denoising strength, 5 Canny model steps, and 25 base model steps. We use Flux.1 [dev] Canny as the Canny model and Flux.1 [dev] as the base model. We also use a LoRA (Midjourney Dreamlike Fantasy FLUX LoRA\footnote{\url{https://civitai.com/models/679736/midjourney-dreamlike-fantasy-flux-lora}}) with 1.0 strength on the base model to stabilize the output style. We manually draw color patches and masks with a hard round brush, which is a coarse-grained brush that limits the complexity of user input. We also used Photoshop's selection tools whenever necessary, such as the Rectangular Marquee tool and Quick Selection tool. For each original image, we spend less than one minute to do the selection and add the color patch. The color patch opacity is 0.8. To prevent cherry-picking, we fix the random seed at 0. We use target captions from EditEval as description prompts.

\paragraph{Evaluation Metrics.} We use the CLIP \cite{radford2021learning} text-image direction similarity (CLIP$_{dir}$) and CLIP output similarity (CLIP$_{out}$) metrics from the Emu Edit benchmark \cite{sheynin2024emu}. These two metrics are suitable for reference-image-free evaluation. We do not calculate the L1 distance between edited and original images, as the L1 distance does not monotonically increase with the editing quality (higher distance can simultaneously indicate better global editing performance and worse local editing performance). In EditEval, we exclude the style change task, as the task is better handled by style LoRAs or specialized checkpoints (e.g., Flux.1 [dev] Redux).\footnote{\url{https://huggingface.co/black-forest-labs/FLUX.1-Redux-dev}} After excluding 24 images, there are $n=126$ images remaining.

\section{Human Evaluation Details}
\label{app:human-evaluation-details}

For human evaluation, we adopt a preference setting. An annotator is asked to choose a better image out of a pair or consider both images to be ``good'' or ``bad''. We construct an \textit{image pair} by using one image from our workflow and another image from one of the baselines. In total, there are $3\times126=378$ image pairs for 3 baselines. For each of the 3 subsets (126 image pairs), we ask 3 annotators for evaluation, totaling 9 annotators. For each image pair, the annotator is shown 3 images, including the original image, an edited image from one method, and an edited image from another method. The order of our workflow and the baseline is randomly chosen for each image pair. The order is hidden from the annotators. We show the editing category and the editing prompt, but not the source prompt or the target prompt. Before the annotation process begins, each annotator is given the same evaluation instructions, where we do not specify detailed criteria for preference other than the adherence to the editing instruction.

\begin{figure*}[ht]

\begin{center}
\includegraphics[width=\textwidth]{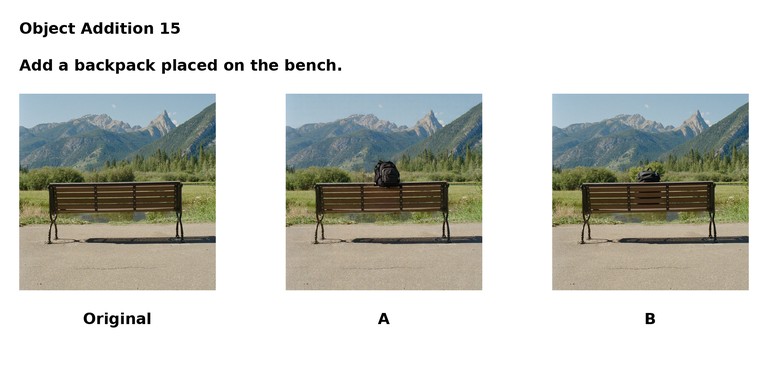}
\caption{\textbf{An example of an image pair that the annotator sees.} The annotator does not know which option comes from which model. While they can see the name of the baseline method they are evaluating, none of the annotators know about the baseline methods before performing annotation.}
\label{fig:image-pair-annotator}
\end{center}

\end{figure*}

An example of an image pair the annotator sees is shown in \cref{fig:image-pair-annotator}. To ensure fairness, we send the same instructions to the annotators. To ensure minimal bias, we do not explain the relative strengths and weaknesses of each method. We do not further communicate with annotators or provide clarifications in written or spoken form, until the annotators finishes the task. The only message we send to annotators is shown below (from ``Hi'' to ``winner'').

Hi [NAME],

We are working on a project about using diffusion models to edit images. We would like you to help evaluate the results. This evaluation requires you to choose the better image from a pair. There are a total of 126 pairs, so the evaluation would not take long. 

The image pairs can be found at

[LINK TO THE FOLDER CONTAINING IMAGE PAIRS]

Please submit your evaluation using this spreadsheet

[LINK TO THE SPREADSHEET]

Put a 1 in the option you choose. Please feel free to discuss anything you observe with us. Thanks for your help!

The evaluation instruction:

Please choose from A and B the image that better follows the editing instruction. If the instruction is followed by both A and B, choose the image that looks better. Only choose “both good” or “both bad” if there is no clear winner.

\section{Challenging Examples from More Benchmarks}
\label{app:challenging-examples}

\begin{figure*}[htbp]

\begin{center}
\small
\begin{tabular}{c@{\hspace{0.3em}}c@{\hspace{0.3em}}c@{\hspace{0.3em}}c@{\hspace{0.3em}}c@{\hspace{0.3em}}c@{\hspace{0.3em}}c@{\hspace{0.3em}}c@{\hspace{0.3em}}c}
    \scriptsize\textsc{Original} & \scriptsize\textsc{Emu Edit} & \scriptsize\textsc{IP2P} & \scriptsize\textsc{MB} & \scriptsize\textsc{UE} & \scriptsize\textsc{Doubao} & \scriptsize\textsc{Gemini} & \scriptsize\textsc{GPT-4o} &\scriptsize\textsc{\textbf{Ours}} \\
    
    \includegraphics[width=0.09\textwidth]{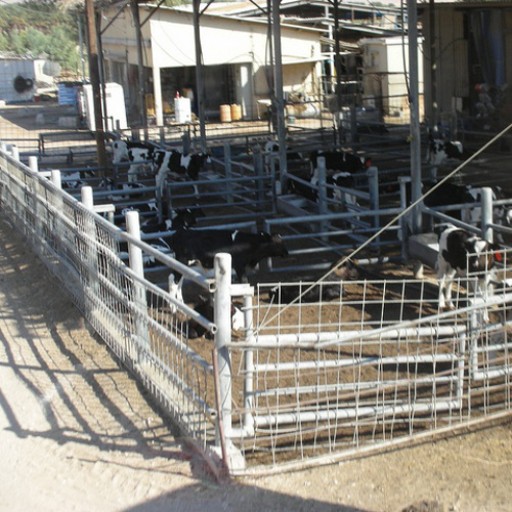} &
    \includegraphics[width=0.09\textwidth]{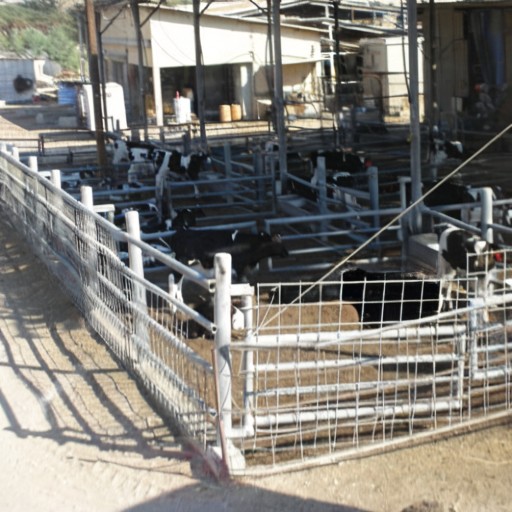} &
    \includegraphics[width=0.09\textwidth]{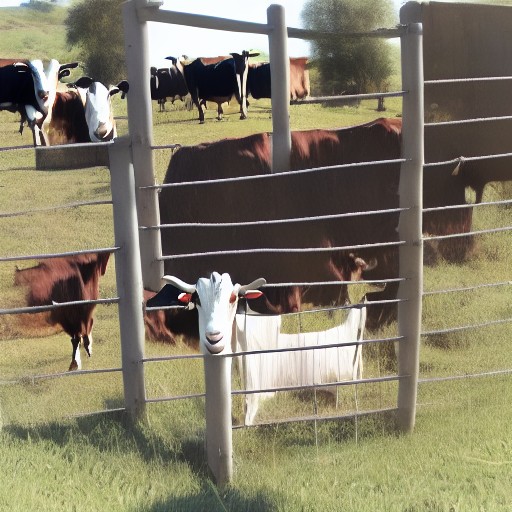} &
    \includegraphics[width=0.09\textwidth]{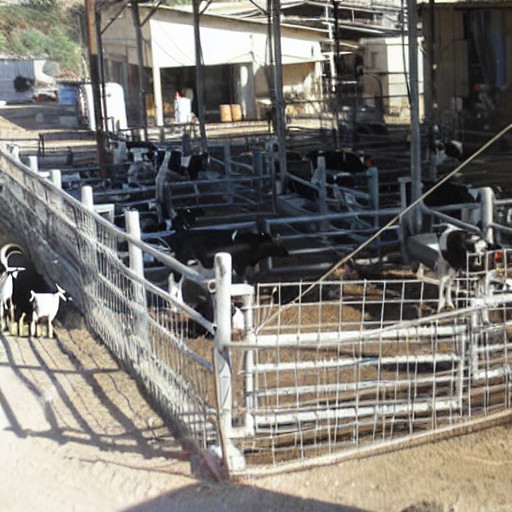} &
    \includegraphics[width=0.09\textwidth]{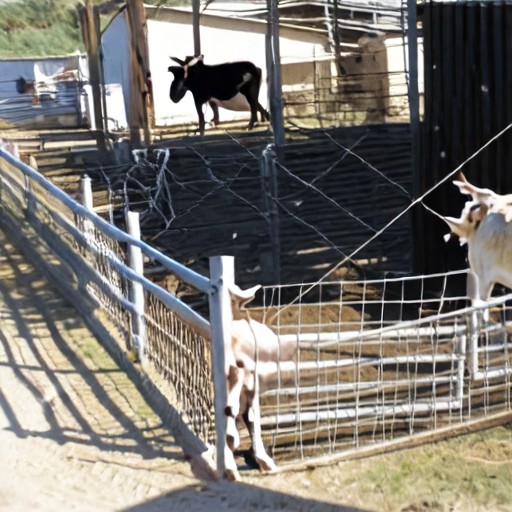} &
    \includegraphics[width=0.09\textwidth]{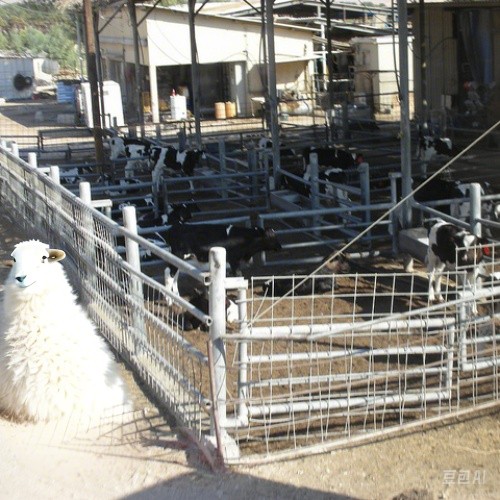} &
    \includegraphics[width=0.09\textwidth]{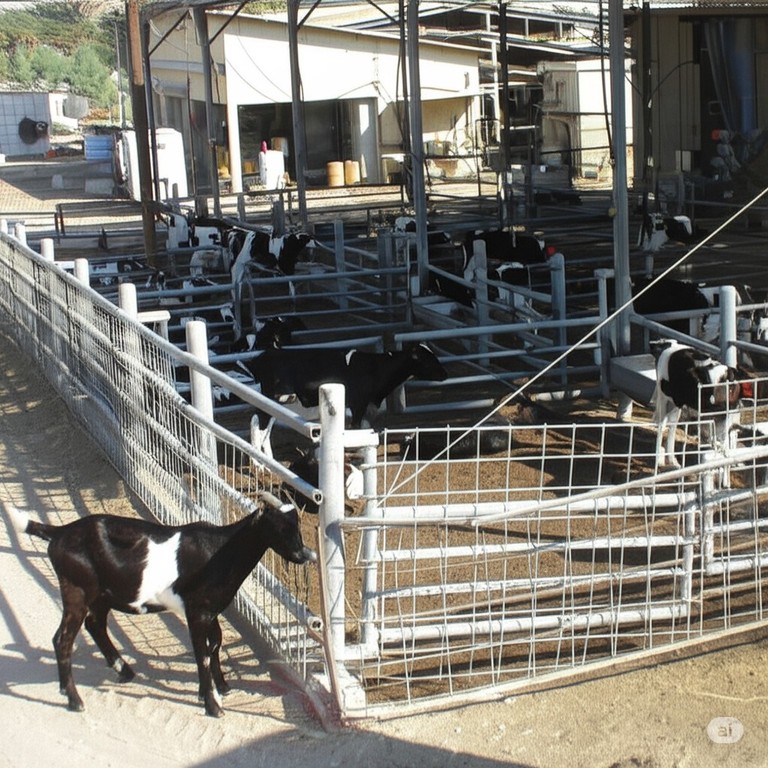} &
    \includegraphics[width=0.09\textwidth]{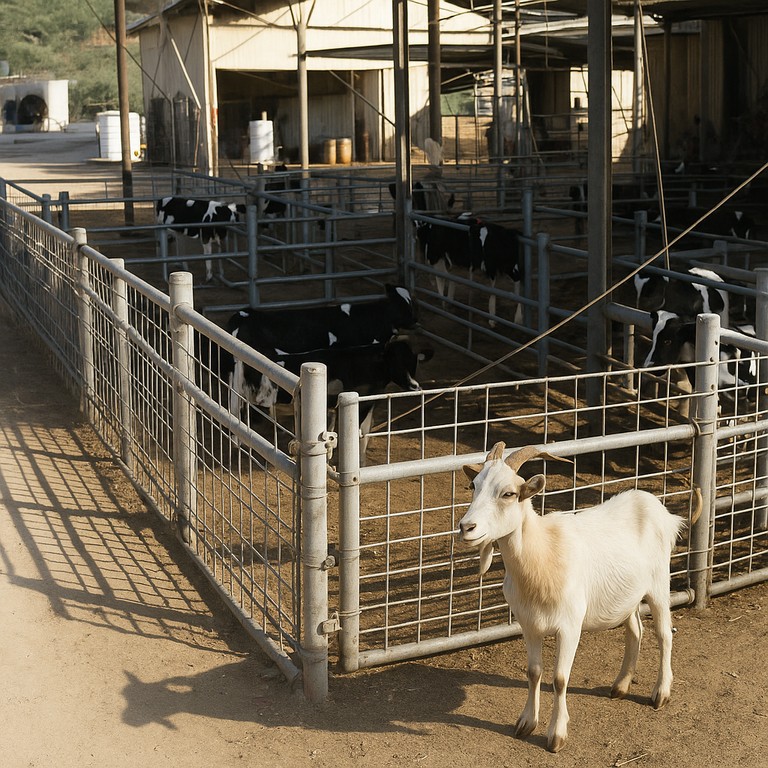} &
    \includegraphics[width=0.09\textwidth]{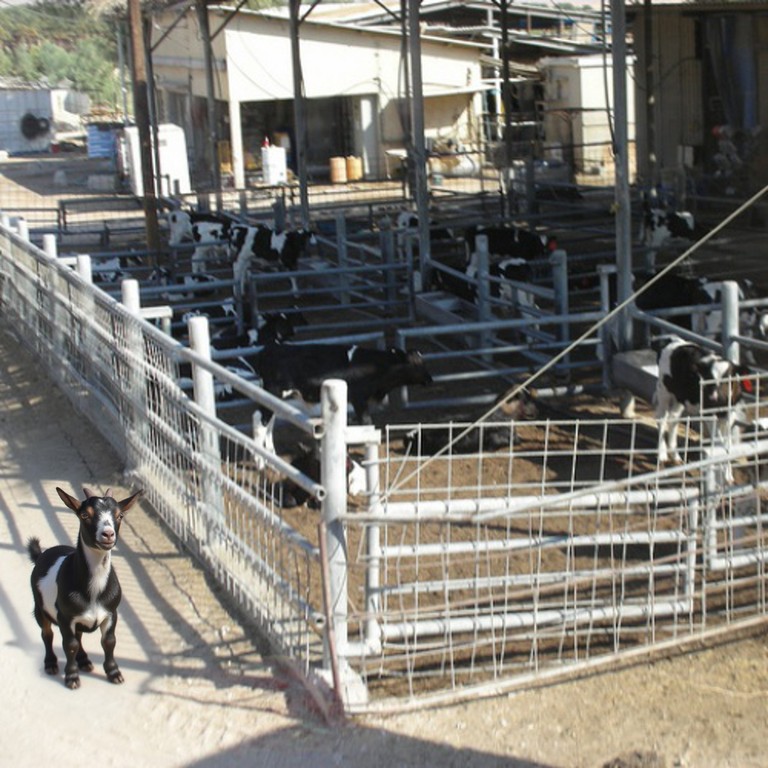} \\
    \multicolumn{9}{c}{\scriptsize\textbf{Add:} Add a goat outside the fence looking at the cows.} \\
    [0.5em]
    \includegraphics[width=0.09\textwidth]{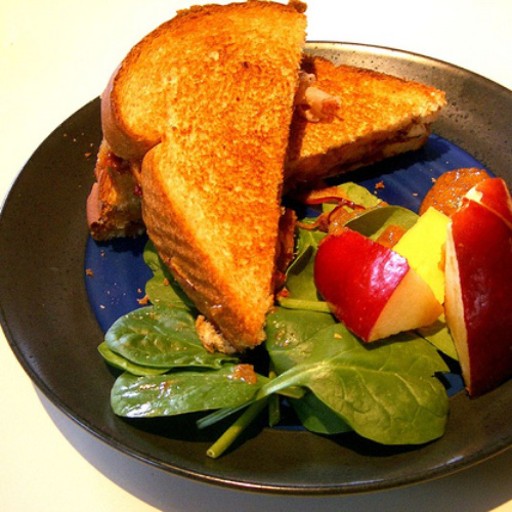} &
    \includegraphics[width=0.09\textwidth]{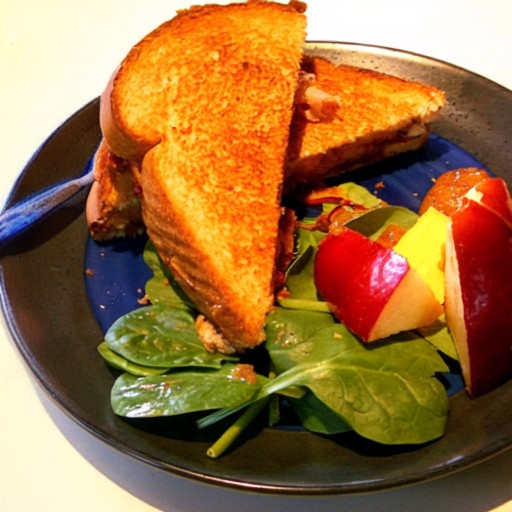} &
    \includegraphics[width=0.09\textwidth]{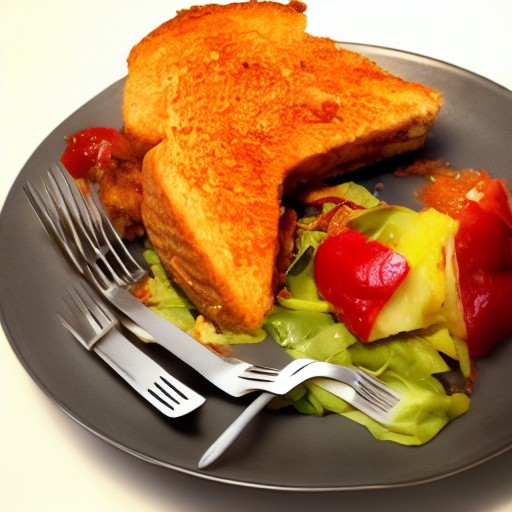} &
    \includegraphics[width=0.09\textwidth]{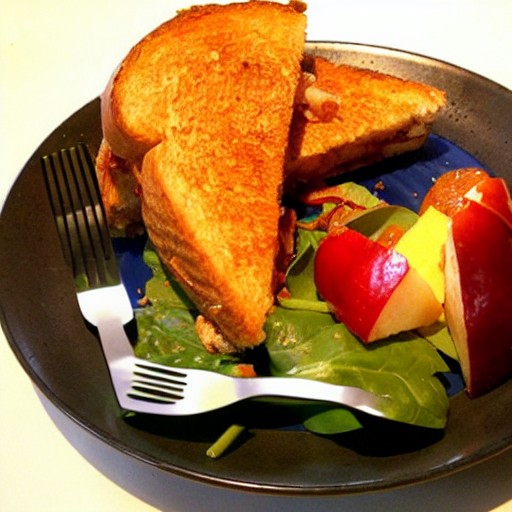} &
    \includegraphics[width=0.09\textwidth]{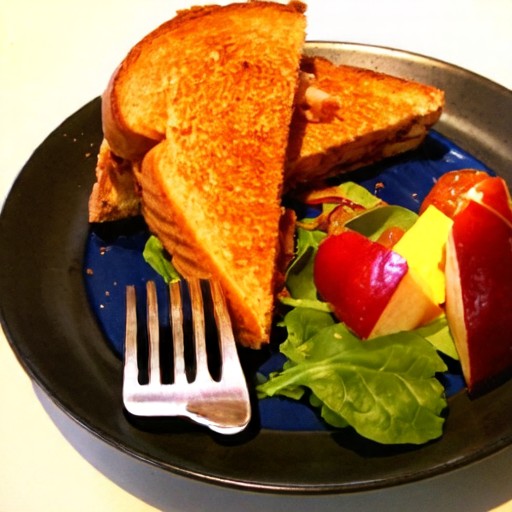} &
    \includegraphics[width=0.09\textwidth]{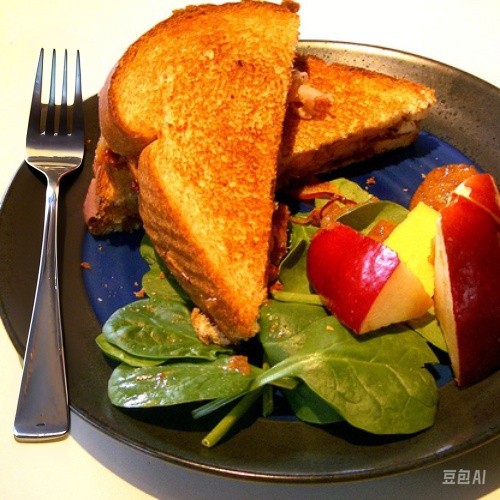} &
    \includegraphics[width=0.09\textwidth]{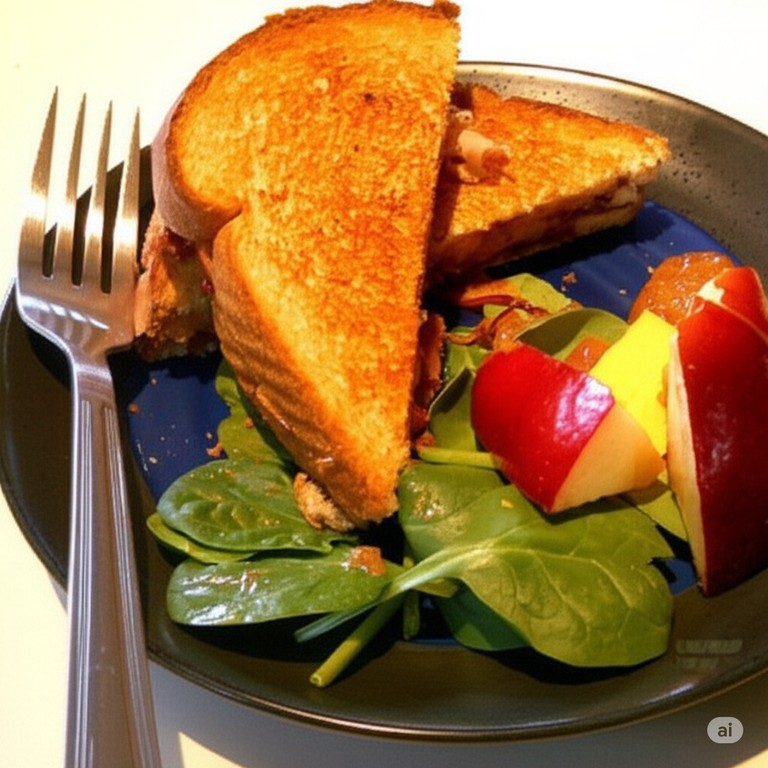} &
    \includegraphics[width=0.09\textwidth]{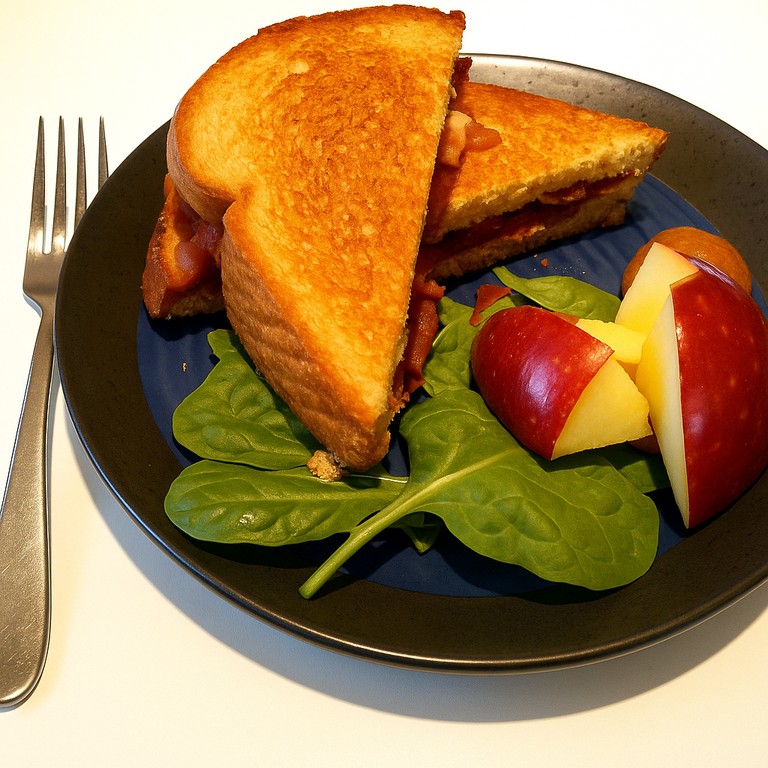} &
    \includegraphics[width=0.09\textwidth]{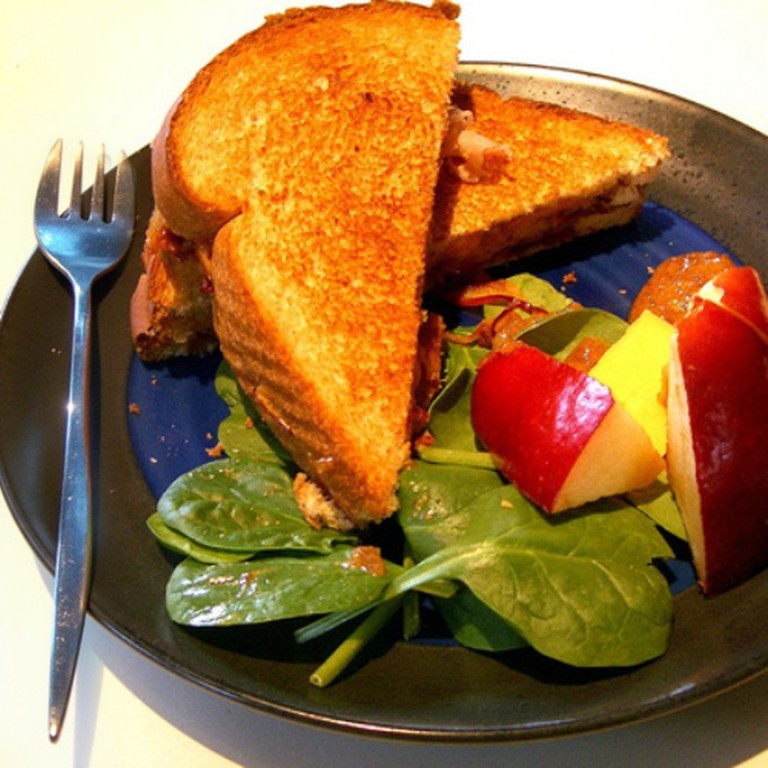} \\
    \multicolumn{9}{c}{\scriptsize\textbf{Add:} Add a fork on the left side of the plate.} \\
    [0.5em]
    \includegraphics[width=0.09\textwidth]{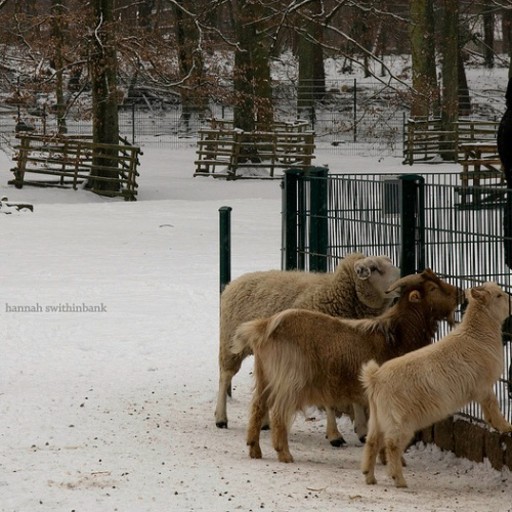} &
    \includegraphics[width=0.09\textwidth]{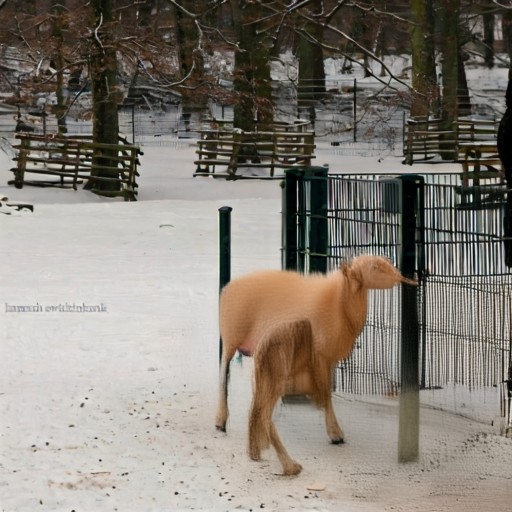} &
    \includegraphics[width=0.09\textwidth]{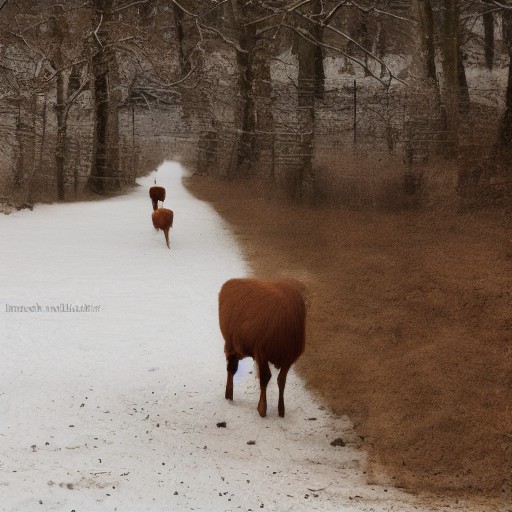} &
    \includegraphics[width=0.09\textwidth]{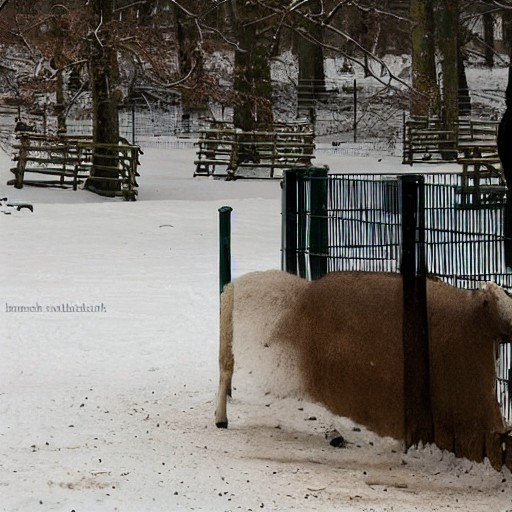} &
    \includegraphics[width=0.09\textwidth]{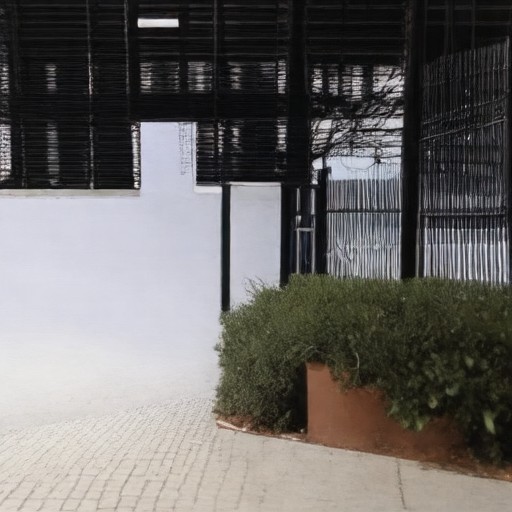} &
    \includegraphics[width=0.09\textwidth]{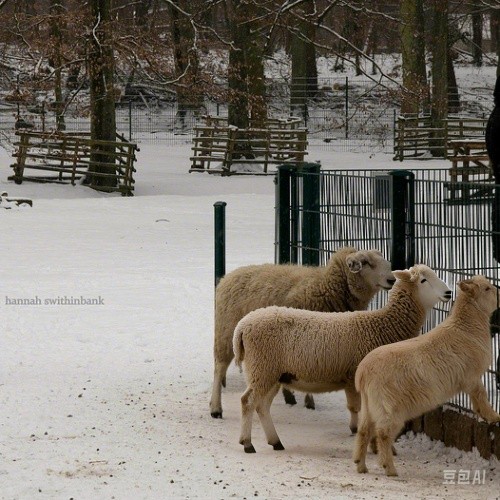} &
    \includegraphics[width=0.09\textwidth]{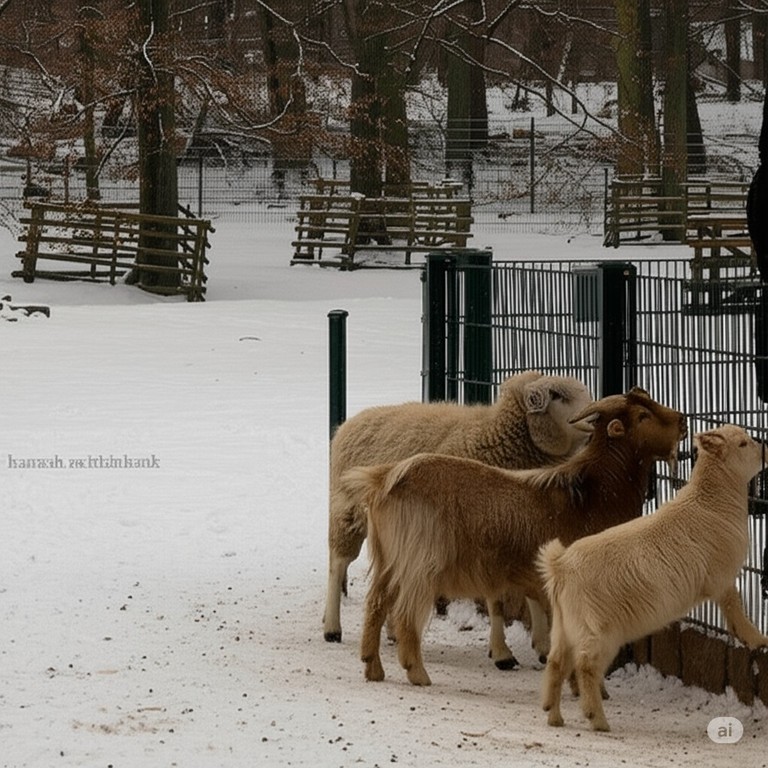} &
    \includegraphics[width=0.09\textwidth]{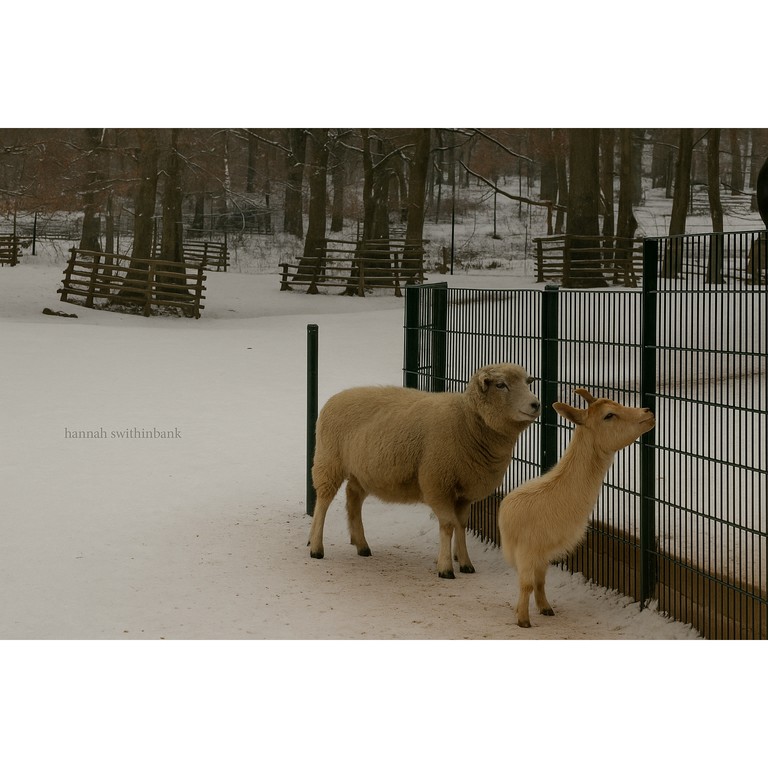} &
    \includegraphics[width=0.09\textwidth]{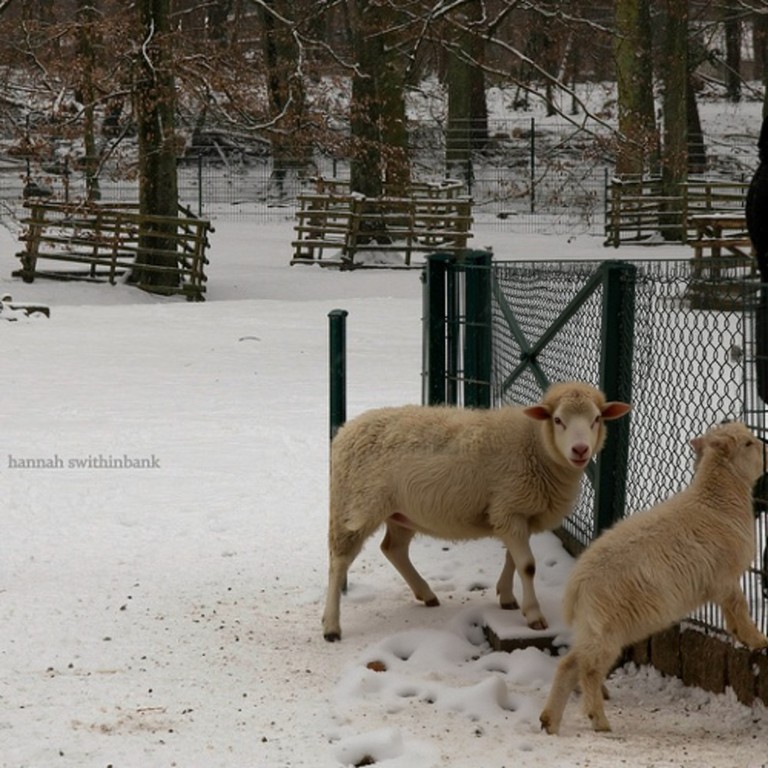} \\
    \multicolumn{9}{c}{\scriptsize\textbf{Remove:} Remove the brown goat from the image.} \\
    [0.5em]
    \includegraphics[width=0.09\textwidth]{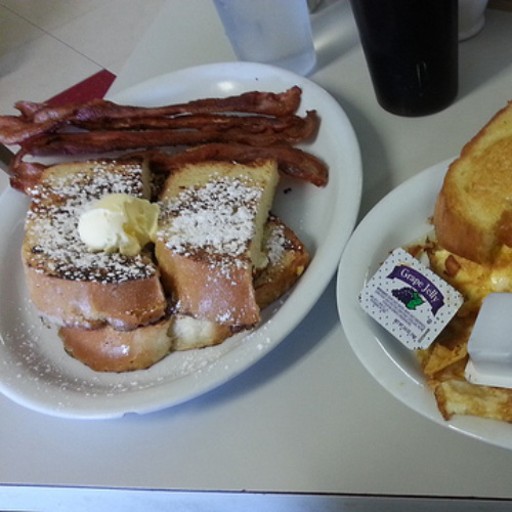} &
    \includegraphics[width=0.09\textwidth]{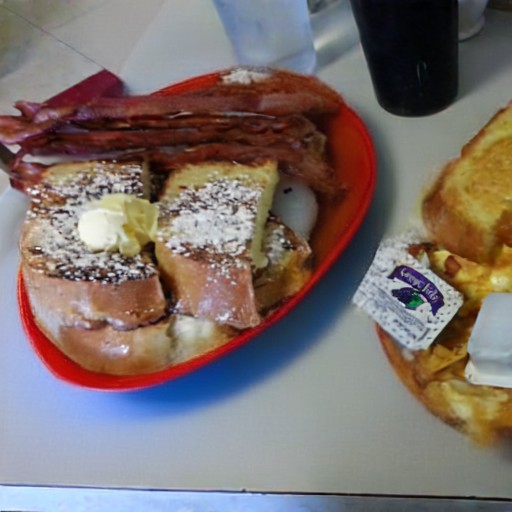} &
    \includegraphics[width=0.09\textwidth]{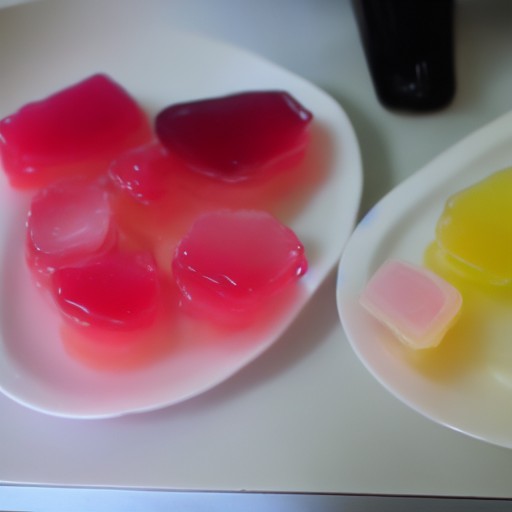} &
    \includegraphics[width=0.09\textwidth]{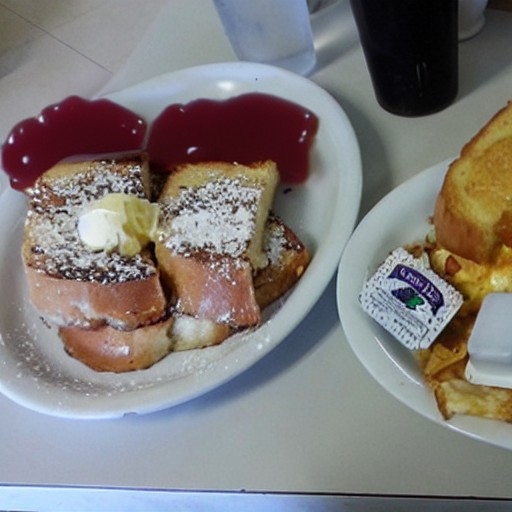} &
    \includegraphics[width=0.09\textwidth]{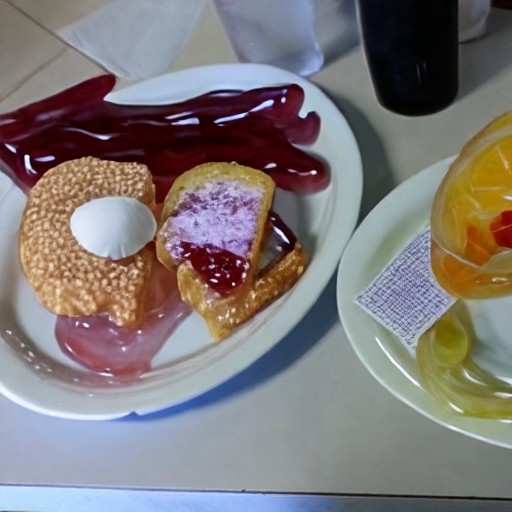} &
    \includegraphics[width=0.09\textwidth]{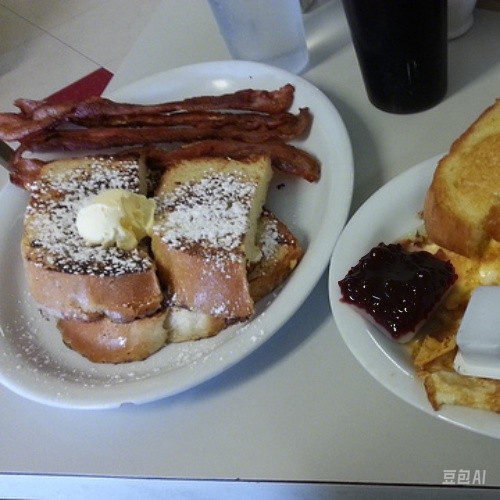} &
    \includegraphics[width=0.09\textwidth]{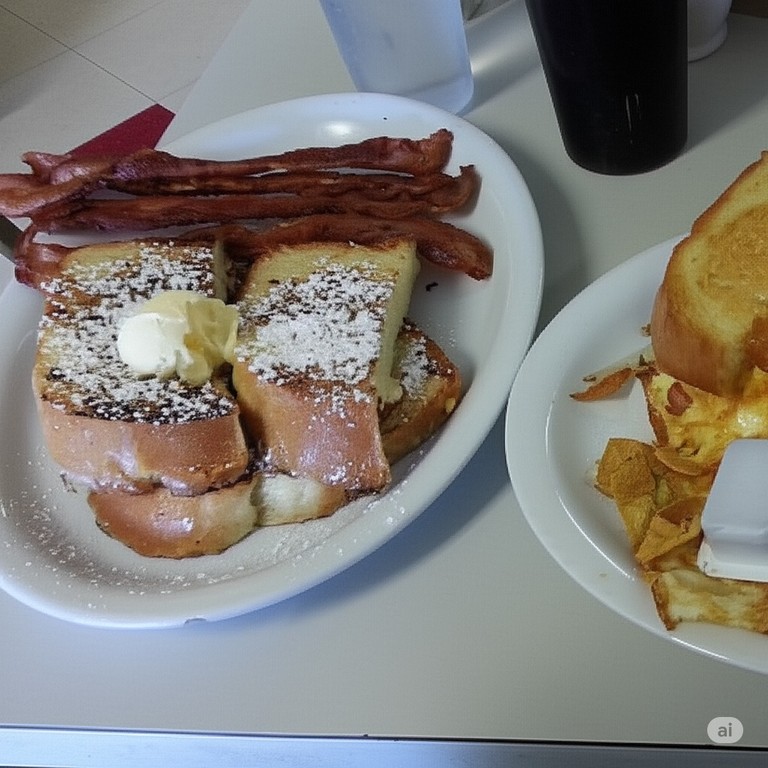} &
    \includegraphics[width=0.09\textwidth]{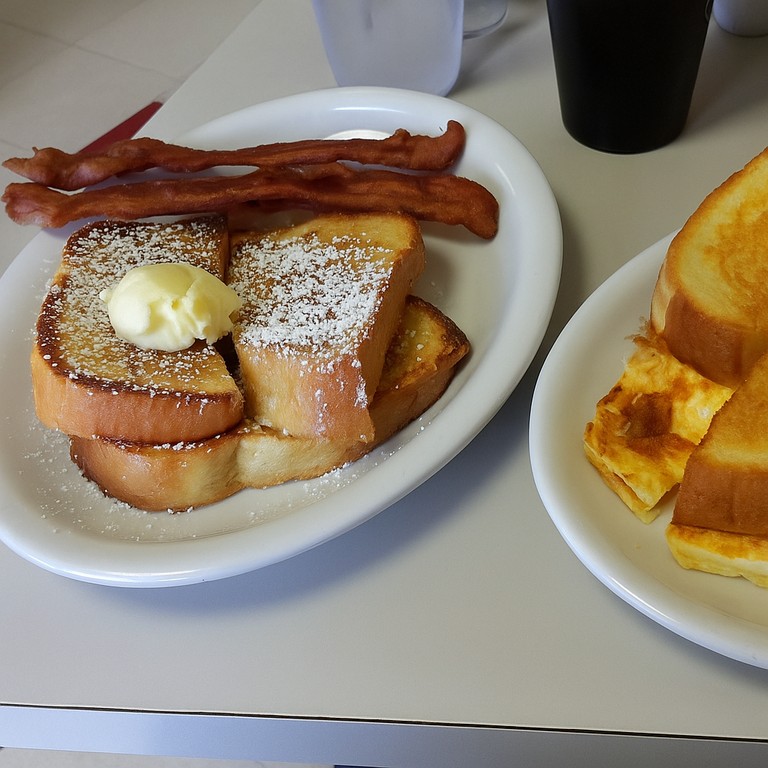} &
    \includegraphics[width=0.09\textwidth]{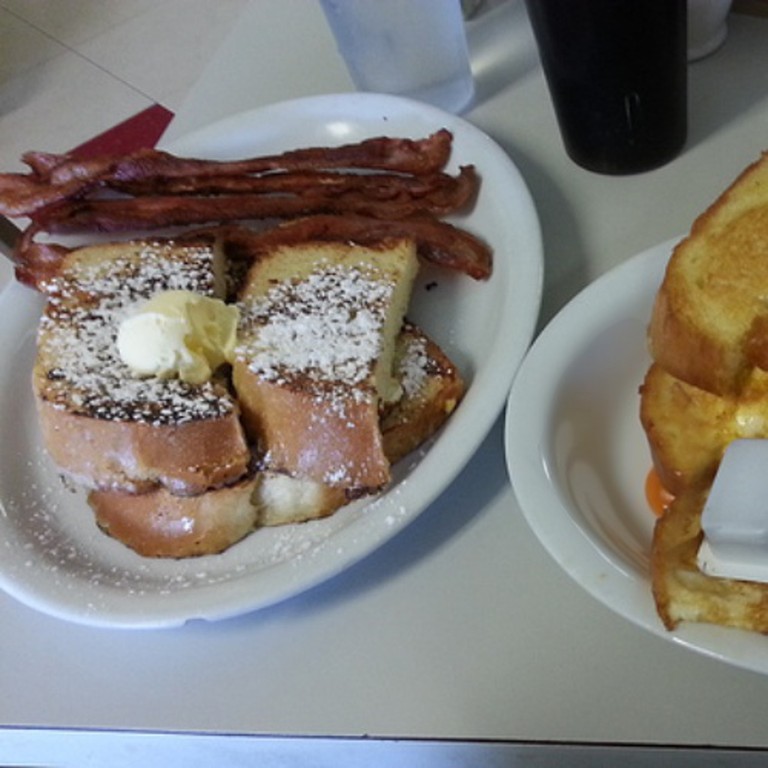} \\
    \multicolumn{9}{c}{\scriptsize\textbf{Remove:} Delete the packet of jelly on the right plate.} \\
    [0.5em]
    \includegraphics[width=0.09\textwidth]{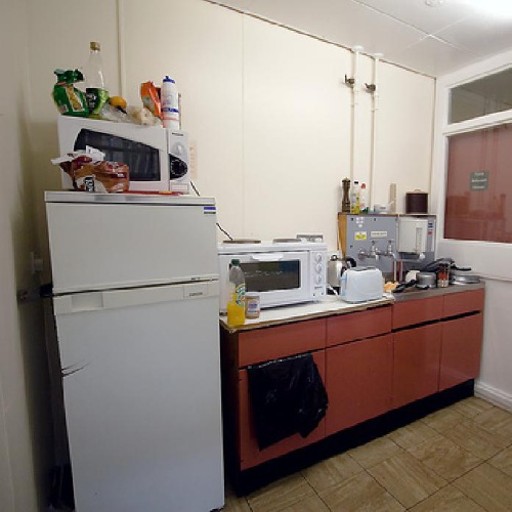} &
    \includegraphics[width=0.09\textwidth]{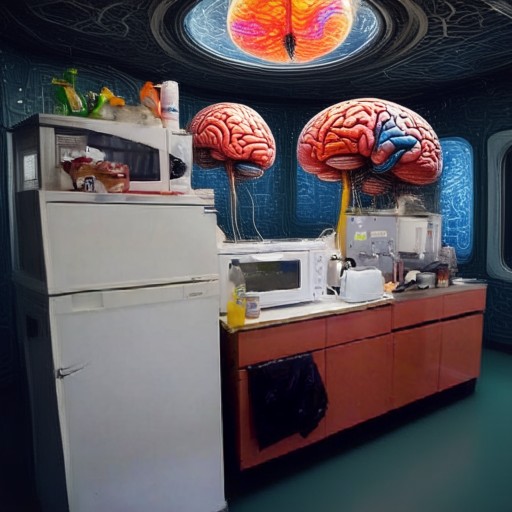} &
    \includegraphics[width=0.09\textwidth]{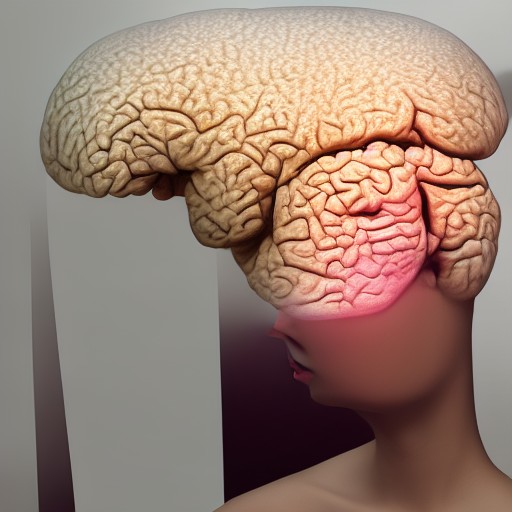} &
    \includegraphics[width=0.09\textwidth]{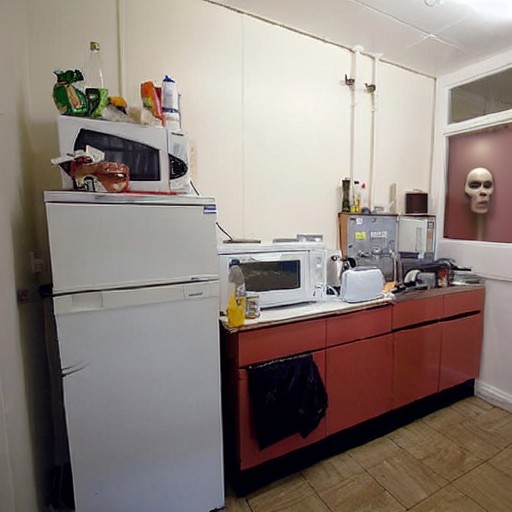} &
    \includegraphics[width=0.09\textwidth]{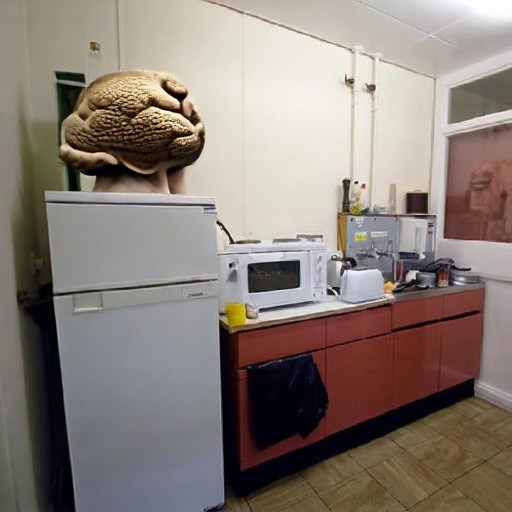} &
    \includegraphics[width=0.09\textwidth]{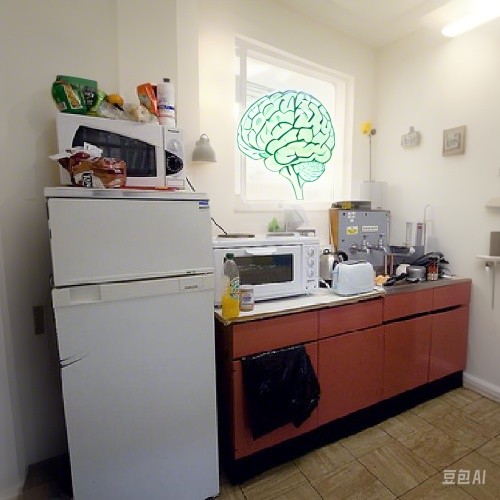} &
    \includegraphics[width=0.09\textwidth]{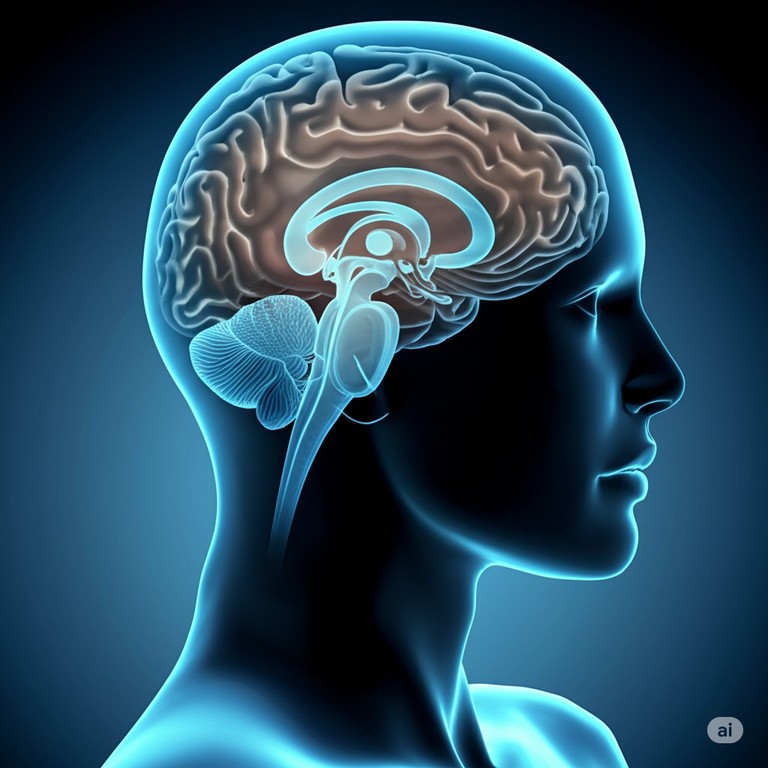} &
    \includegraphics[width=0.09\textwidth]{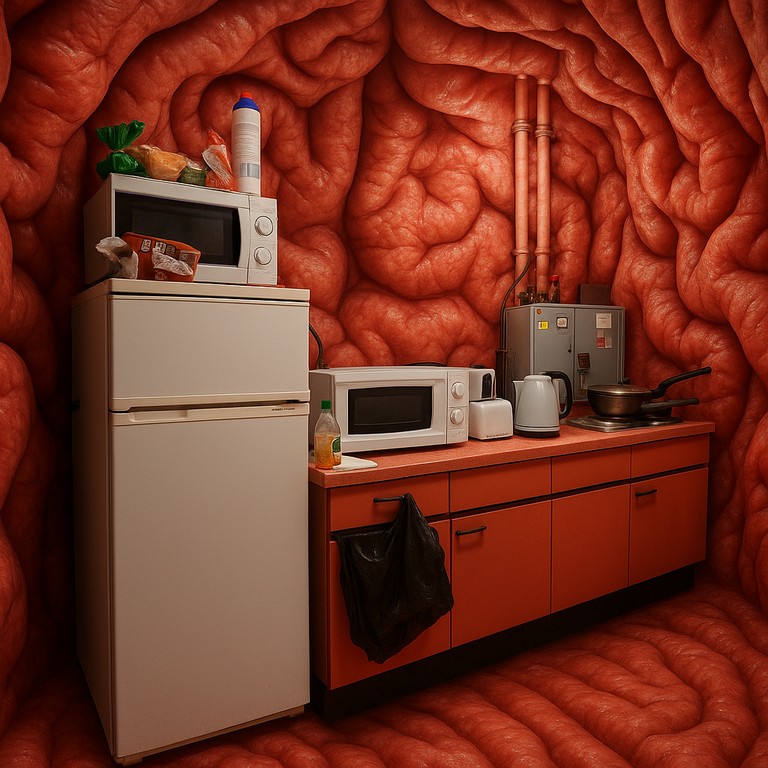} &
    \includegraphics[width=0.09\textwidth]{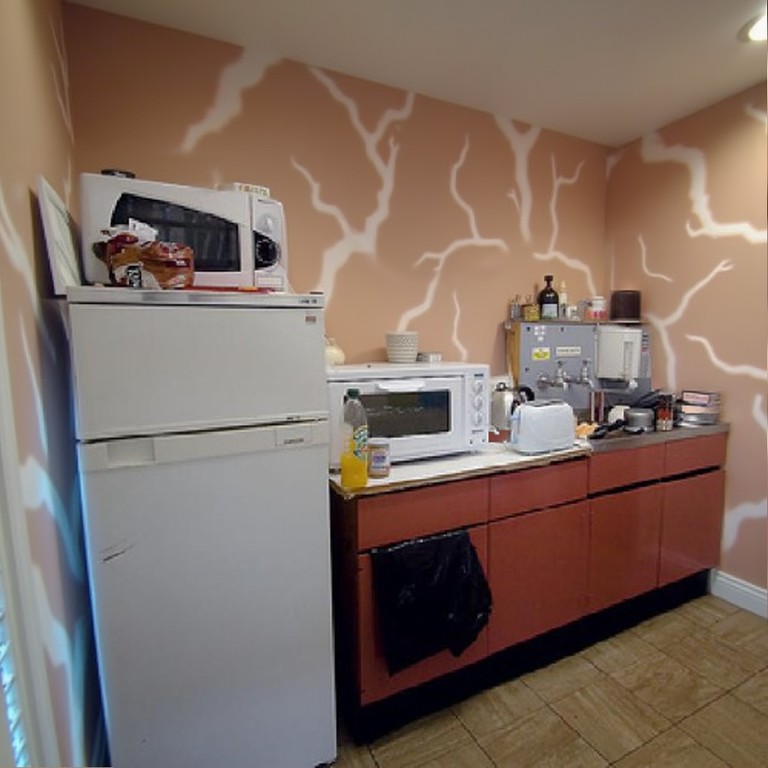} \\
    \multicolumn{9}{c}{\scriptsize\textbf{Background:} Change the background to the inside of a human brain.} \\
    [0.5em]
    \includegraphics[width=0.09\textwidth]{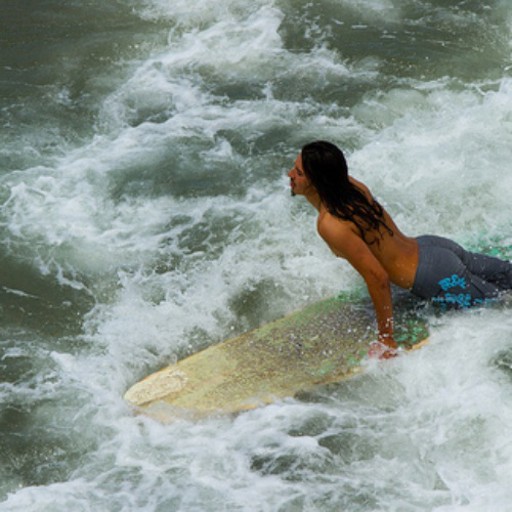} &
    \includegraphics[width=0.09\textwidth]{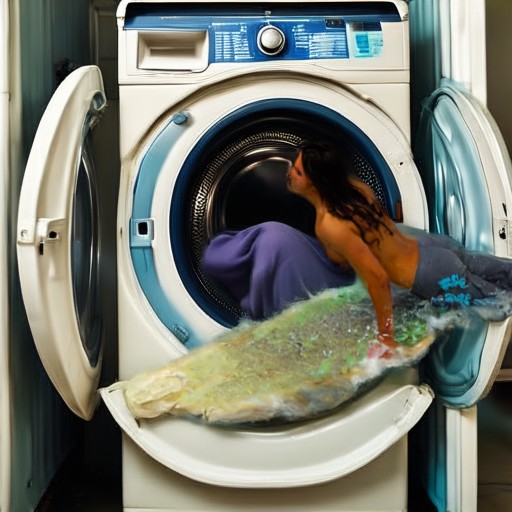} &
    \includegraphics[width=0.09\textwidth]{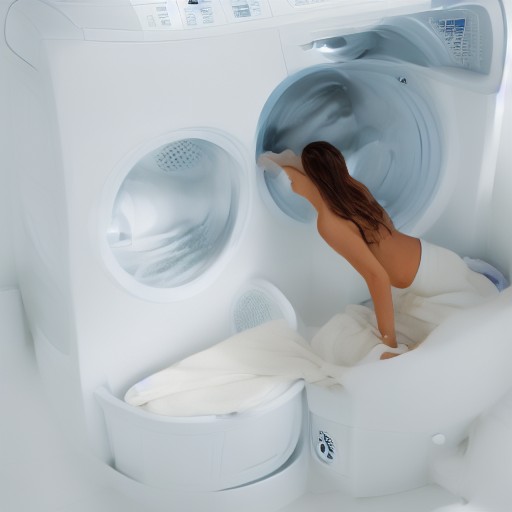} &
    \includegraphics[width=0.09\textwidth]{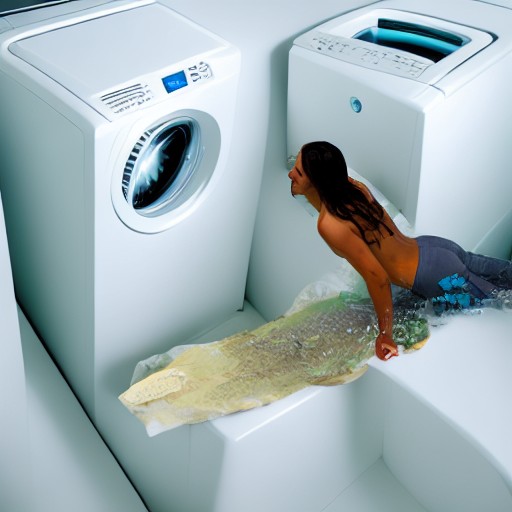} &
    \includegraphics[width=0.09\textwidth]{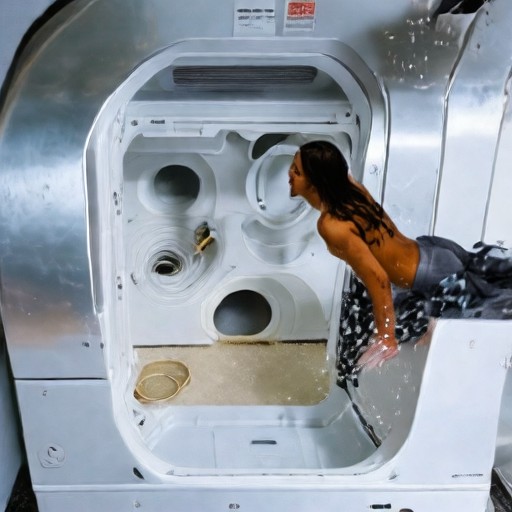} &
    \includegraphics[width=0.09\textwidth]{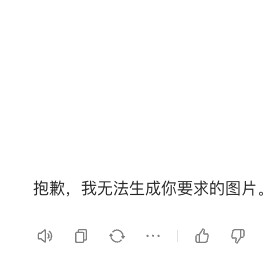} &
    \includegraphics[width=0.09\textwidth]{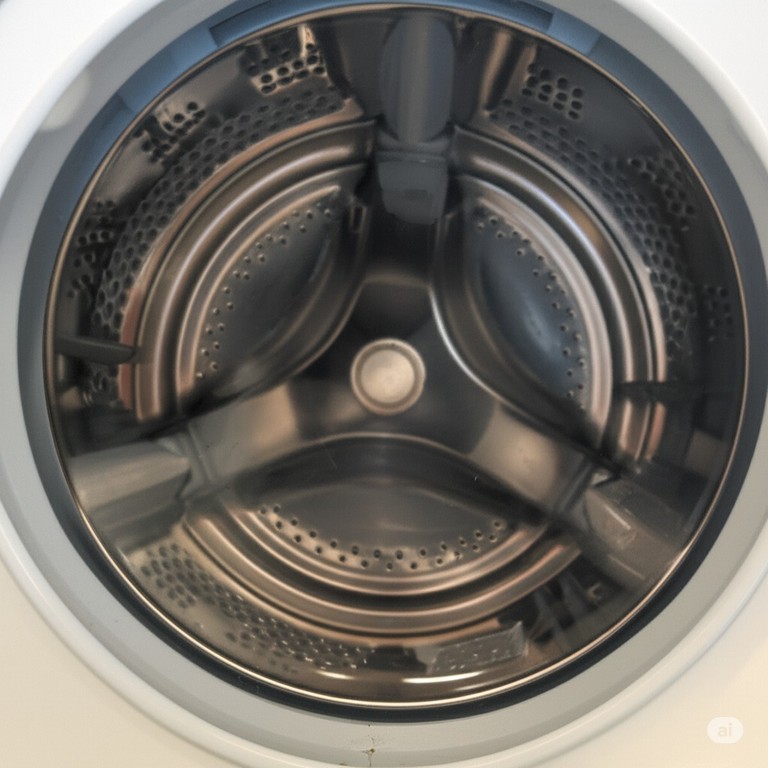} &
    \includegraphics[width=0.09\textwidth]{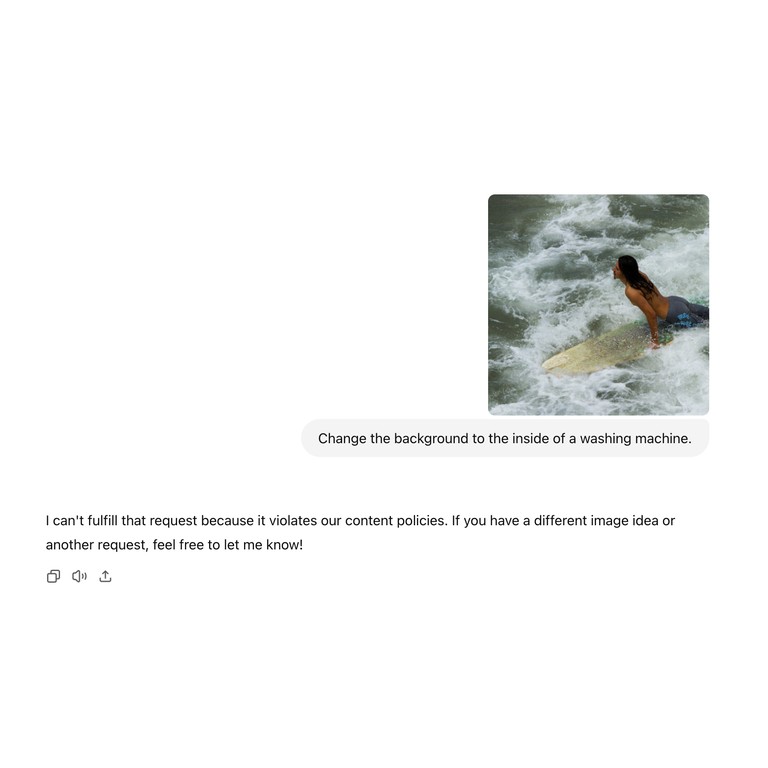} &
    \includegraphics[width=0.09\textwidth]{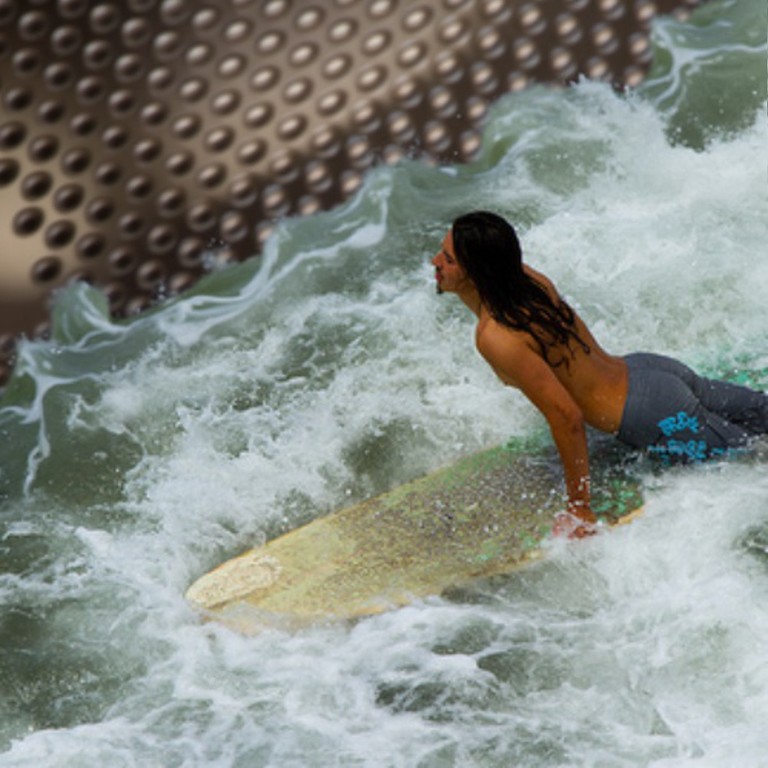} \\
    \multicolumn{9}{c}{\scriptsize\textbf{Background:} Change the background to the inside of a washing machine.} \\
    [0.5em]
    \includegraphics[width=0.09\textwidth]{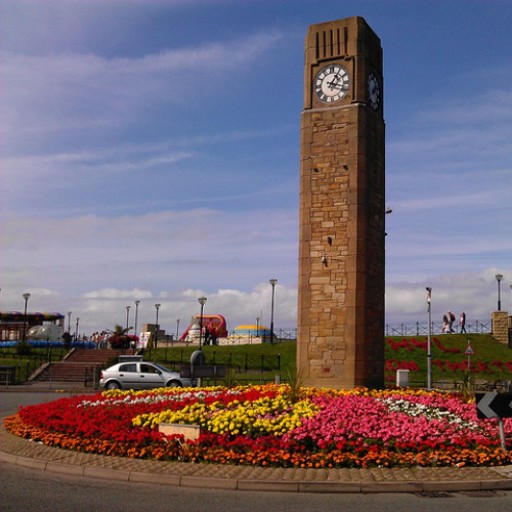} &
    \includegraphics[width=0.09\textwidth]{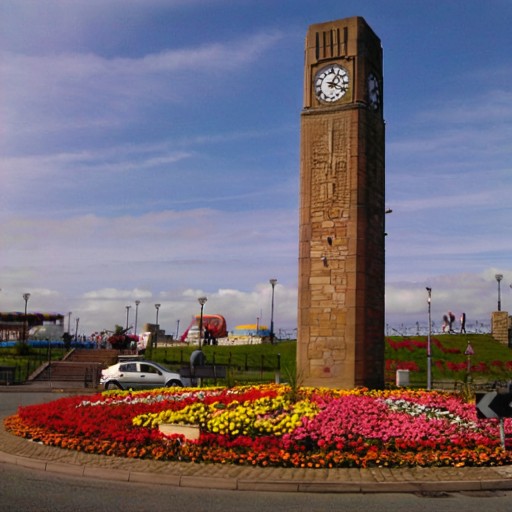} &
    \includegraphics[width=0.09\textwidth]{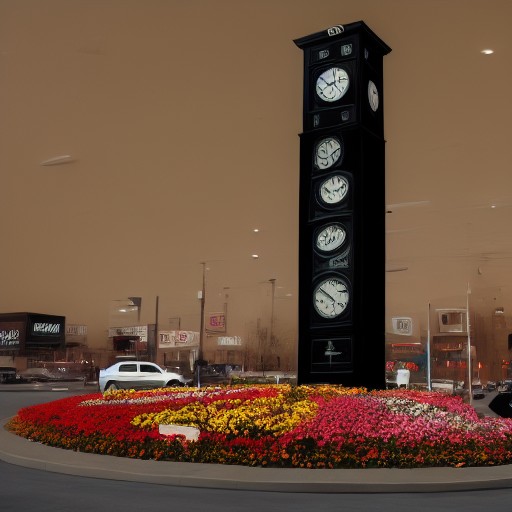} &
    \includegraphics[width=0.09\textwidth]{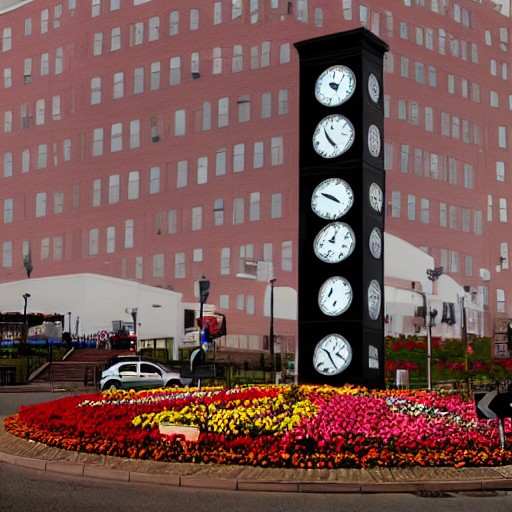} &
    \includegraphics[width=0.09\textwidth]{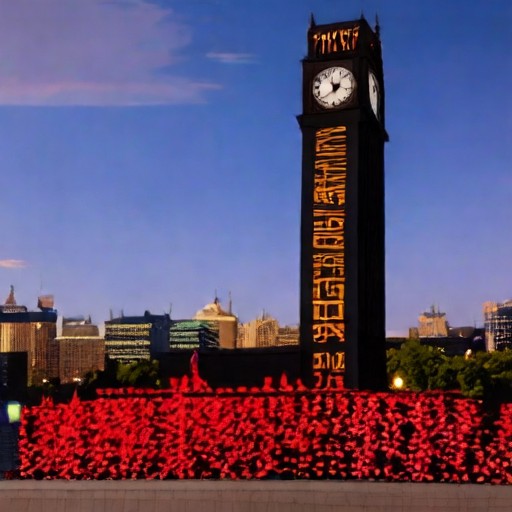} &
    \includegraphics[width=0.09\textwidth]{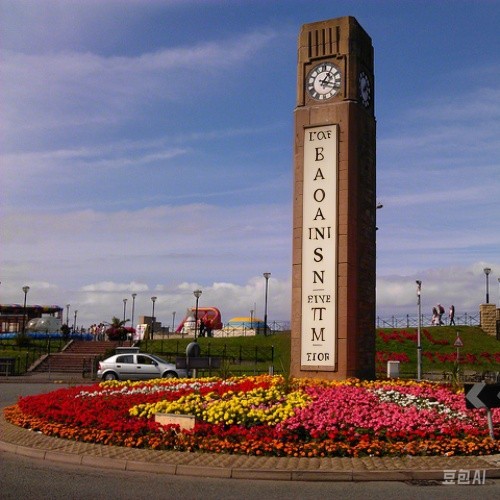} &
    \includegraphics[width=0.09\textwidth]{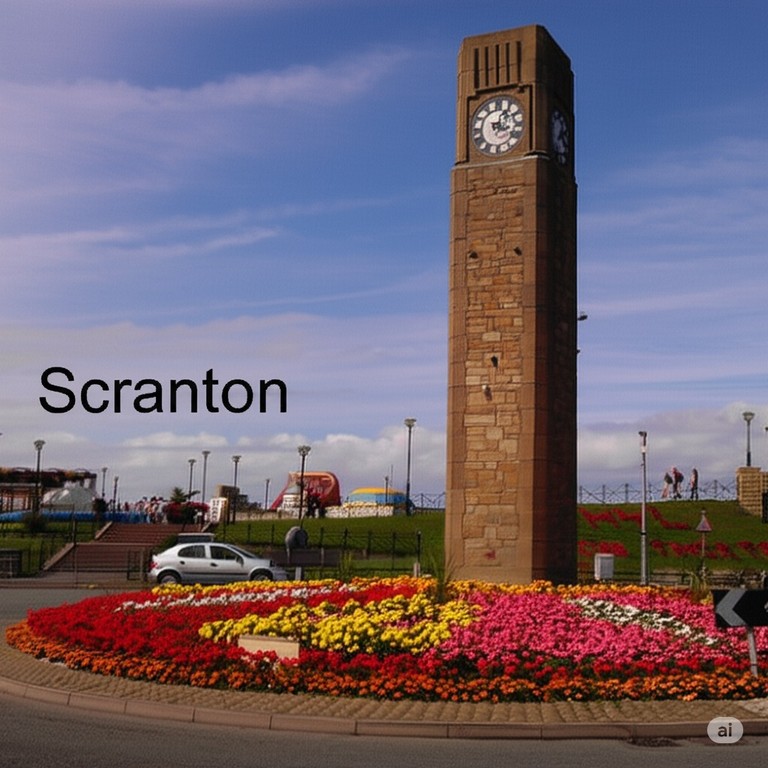} &
    \includegraphics[width=0.09\textwidth]{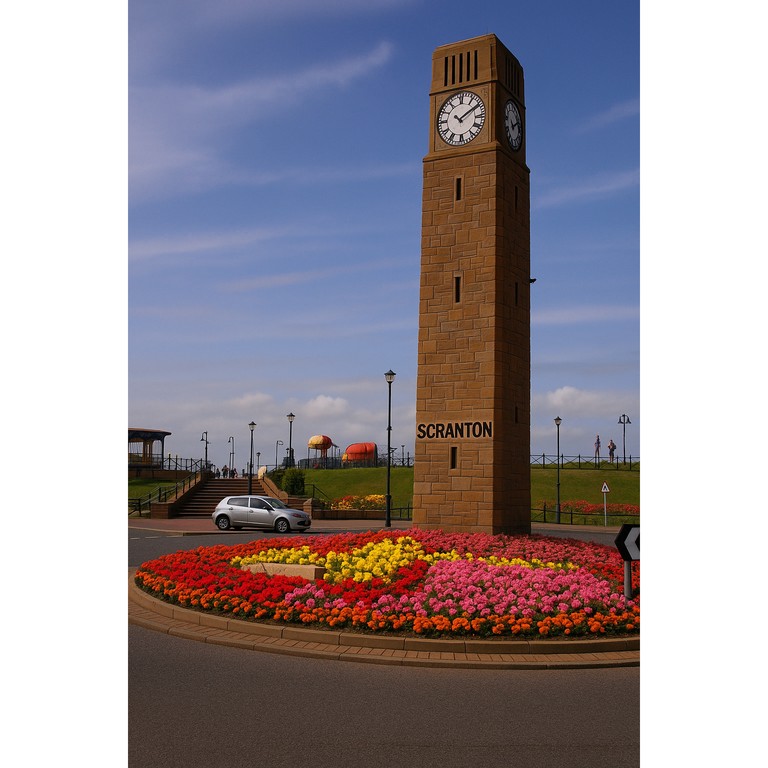} &
    \includegraphics[width=0.09\textwidth]{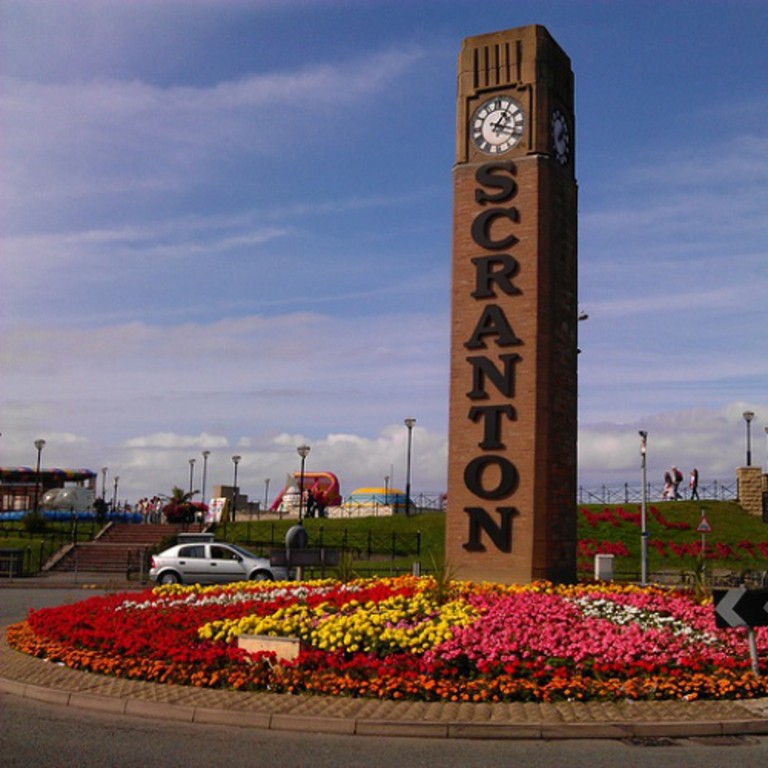} \\
    \multicolumn{9}{c}{\scriptsize\textbf{Text:} Add the word ``Scranton'', in black, to the space below the clock that is front facing.} \\
\end{tabular}
\caption{\textbf{Our workflow performs the best on the first half of 14 challenging examples from Emu Edit.} Both Doubao AI and GPT-4o refuses to generate an image for the washing machine example, showing an error message instead.}
\label{fig:challenging-emuedit-first-half}
\end{center}

\end{figure*}

\begin{figure*}[htbp]

\begin{center}
\small
\begin{tabular}{c@{\hspace{0.3em}}c@{\hspace{0.3em}}c@{\hspace{0.3em}}c@{\hspace{0.3em}}c@{\hspace{0.3em}}c@{\hspace{0.3em}}c@{\hspace{0.3em}}c@{\hspace{0.3em}}c}
    \scriptsize\textsc{Original} & \scriptsize\textsc{Emu Edit} & \scriptsize\textsc{IP2P} & \scriptsize\textsc{MB} & \scriptsize\textsc{UE} & \scriptsize\textsc{Doubao} & \scriptsize\textsc{Gemini} & \scriptsize\textsc{GPT-4o} &\scriptsize\textsc{\textbf{Ours}} \\
    
    \includegraphics[width=0.09\textwidth]{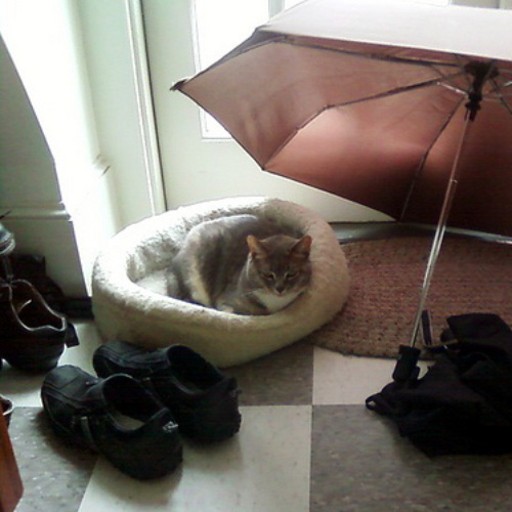} &
    \includegraphics[width=0.09\textwidth]{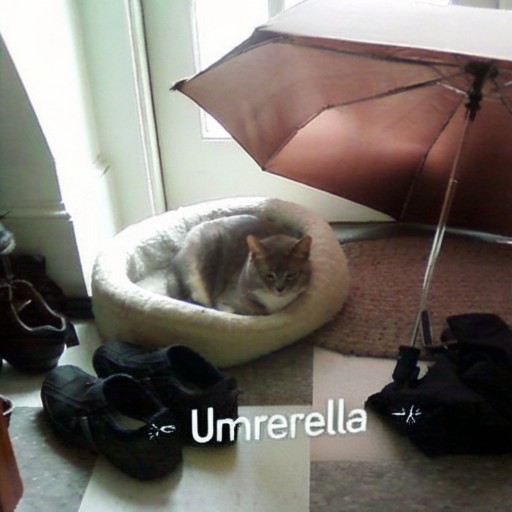} &
    \includegraphics[width=0.09\textwidth]{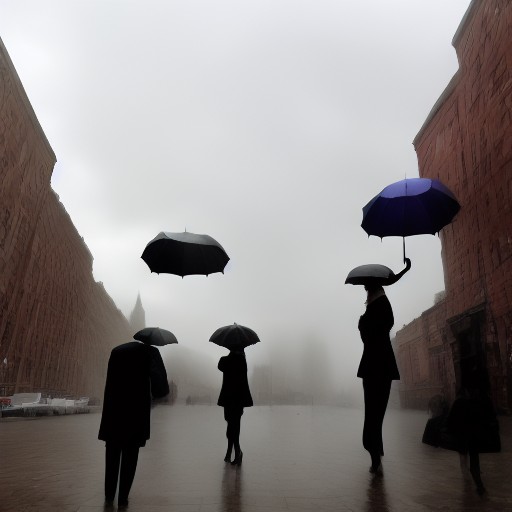} &
    \includegraphics[width=0.09\textwidth]{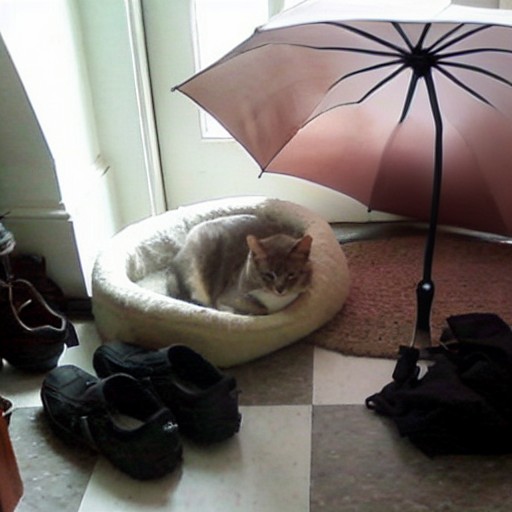} &
    \includegraphics[width=0.09\textwidth]{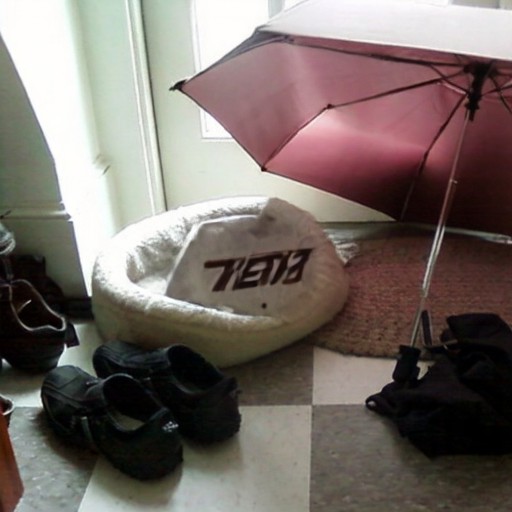} &
    \includegraphics[width=0.09\textwidth]{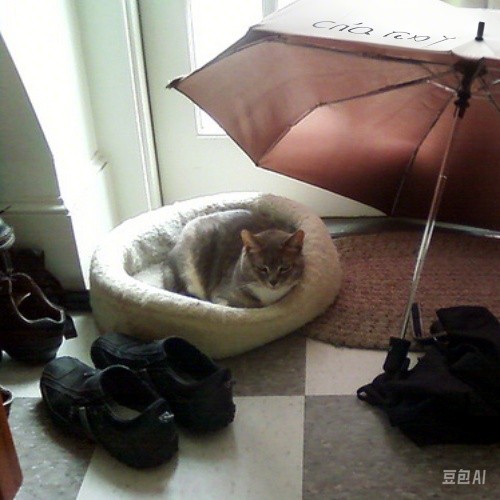} &
    \includegraphics[width=0.09\textwidth]{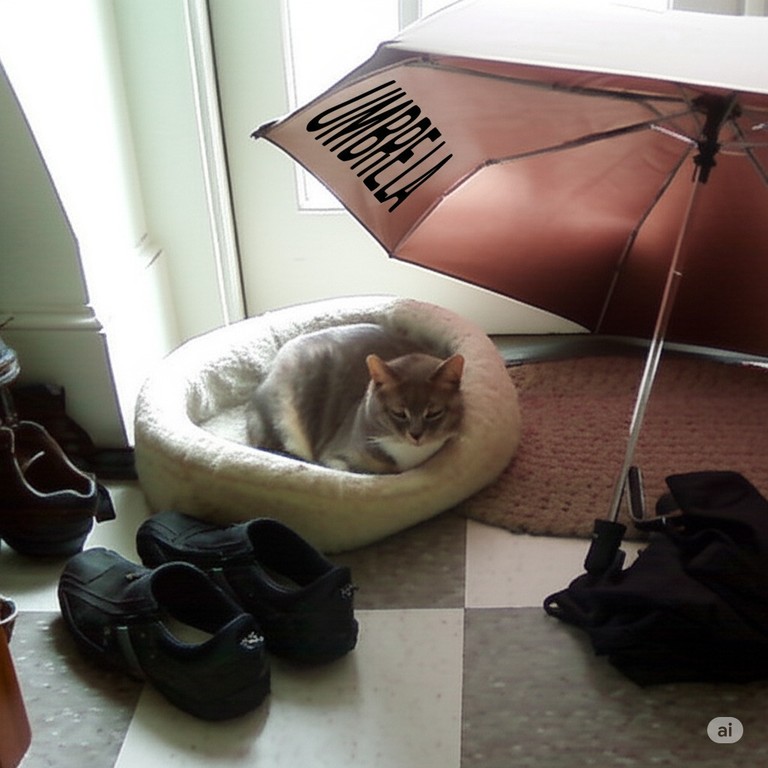} &
    \includegraphics[width=0.09\textwidth]{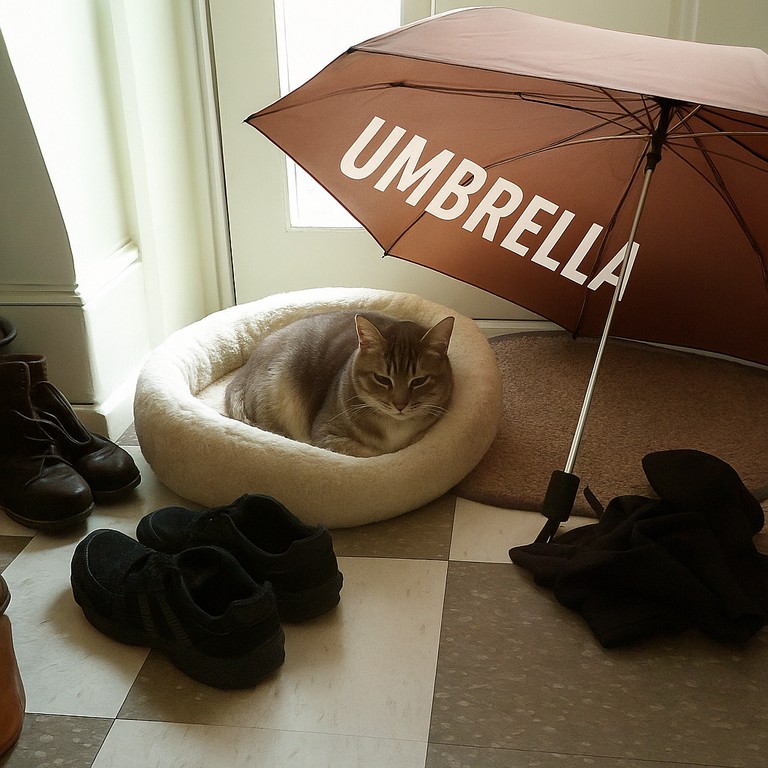} &
    \includegraphics[width=0.09\textwidth]{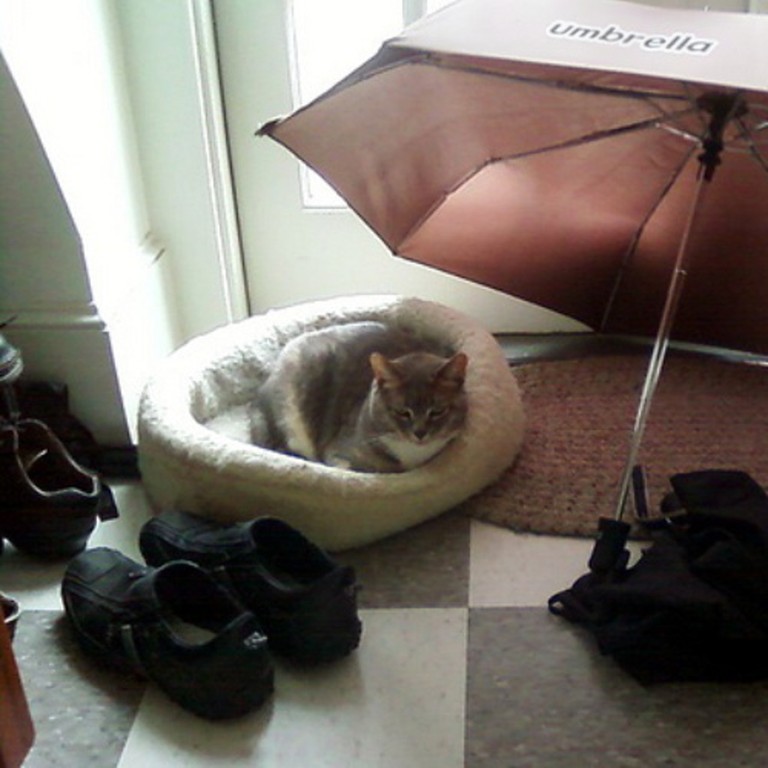} \\
    \multicolumn{9}{c}{\scriptsize\textbf{Text:} Add the word ``umbrella'' onto the umbrella.} \\
    [0.5em]
    \includegraphics[width=0.09\textwidth]{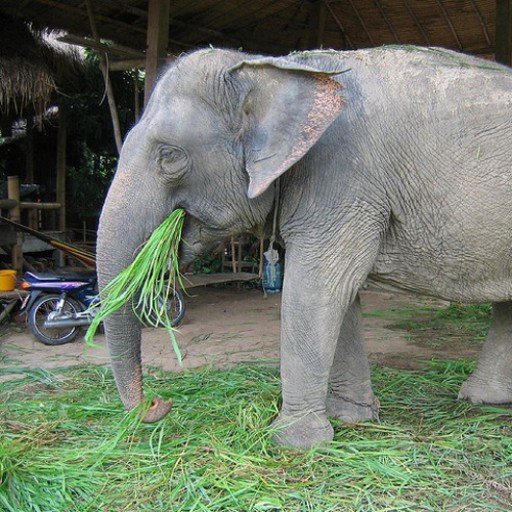} &
    \includegraphics[width=0.09\textwidth]{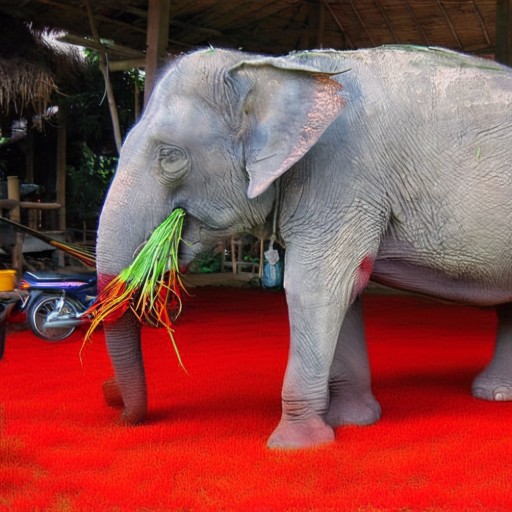} &
    \includegraphics[width=0.09\textwidth]{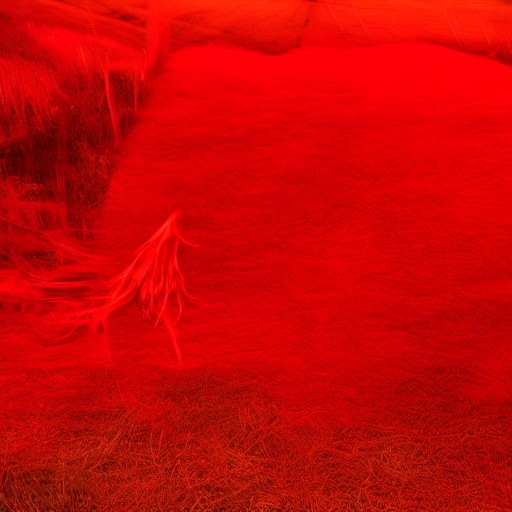} &
    \includegraphics[width=0.09\textwidth]{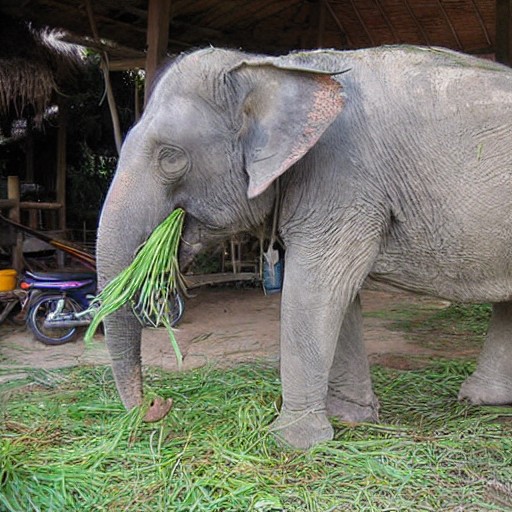} &
    \includegraphics[width=0.09\textwidth]{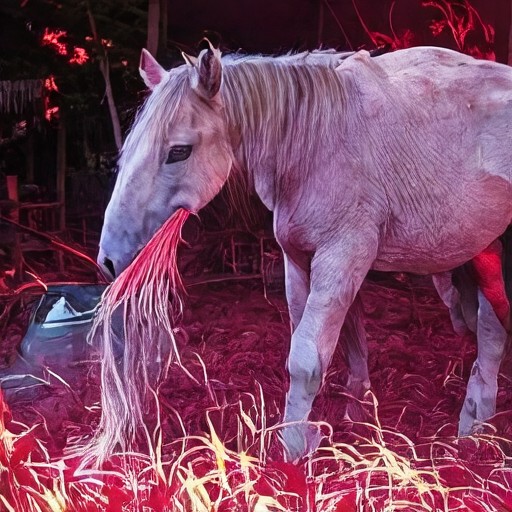} &
    \includegraphics[width=0.09\textwidth]{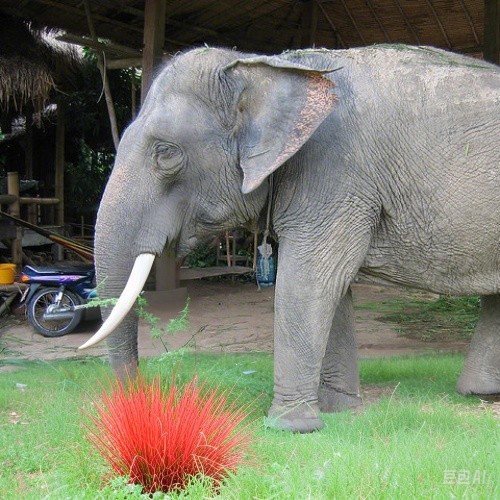} &
    \includegraphics[width=0.09\textwidth]{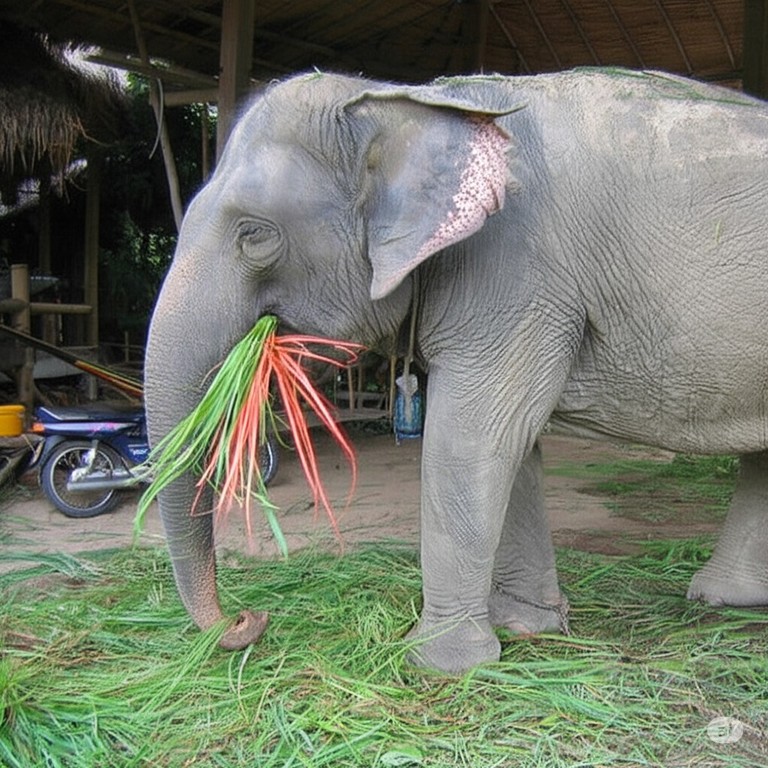} &
    \includegraphics[width=0.09\textwidth]{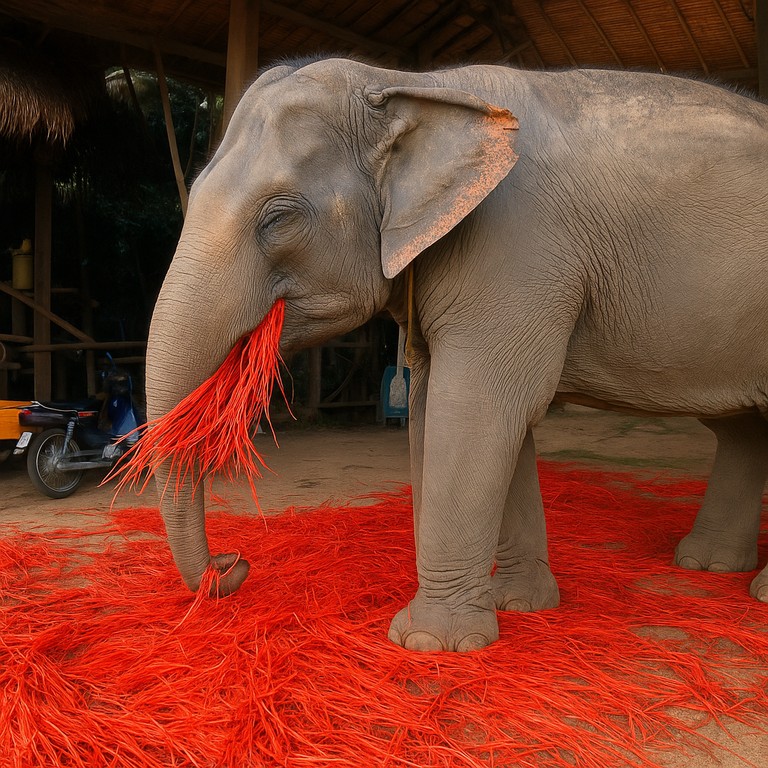} &
    \includegraphics[width=0.09\textwidth]{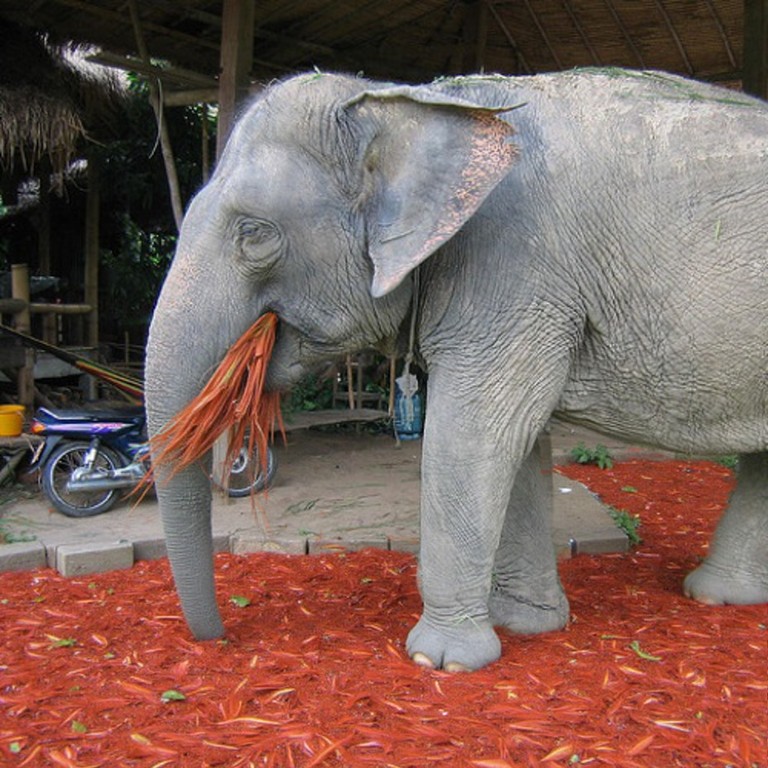} \\
    \multicolumn{9}{c}{\scriptsize\textbf{Color:} Change the color of the grass to fiery red.} \\
    [0.5em]
    \includegraphics[width=0.09\textwidth]{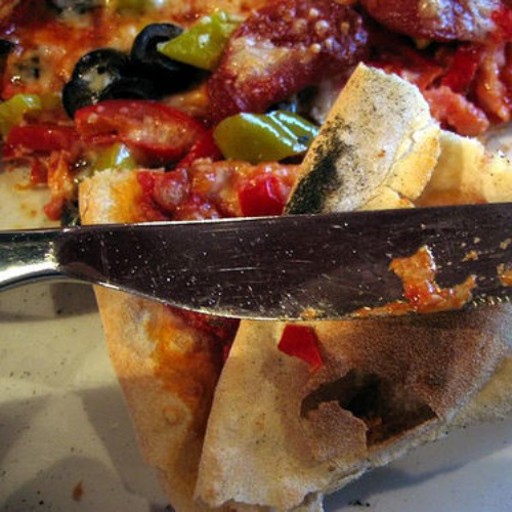} &
    \includegraphics[width=0.09\textwidth]{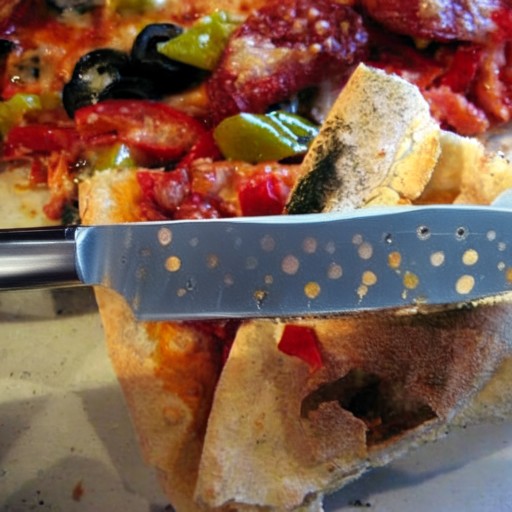} &
    \includegraphics[width=0.09\textwidth]{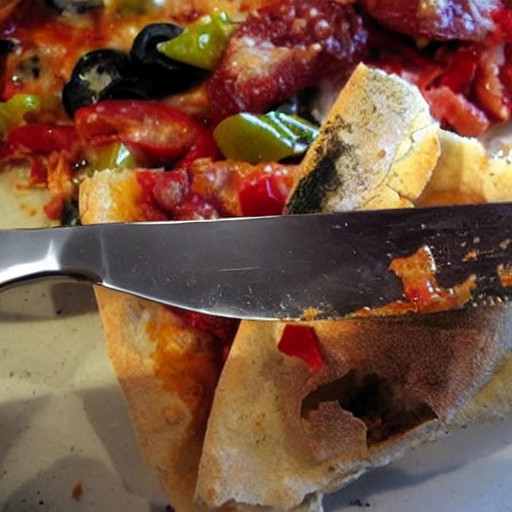} &
    \includegraphics[width=0.09\textwidth]{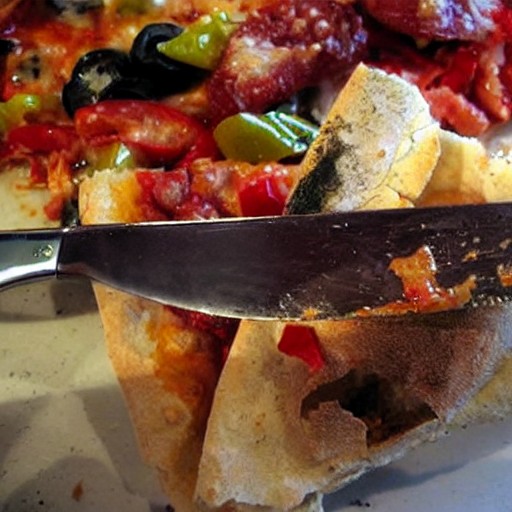} &
    \includegraphics[width=0.09\textwidth]{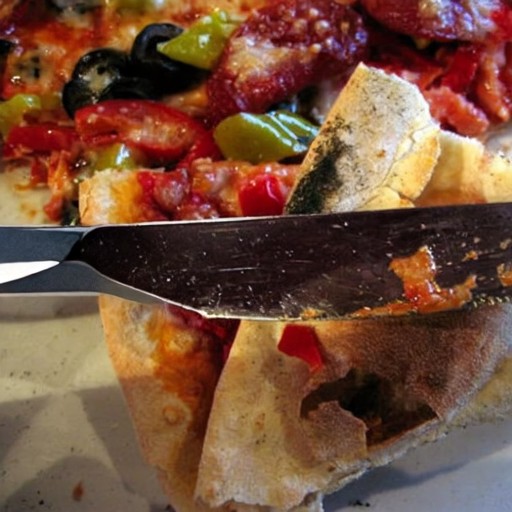} &
    \includegraphics[width=0.09\textwidth]{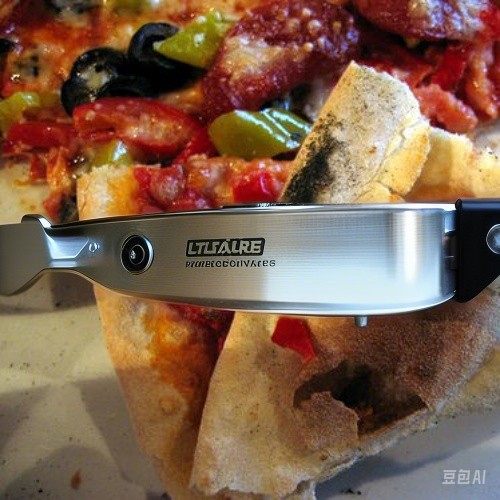} &
    \includegraphics[width=0.09\textwidth]{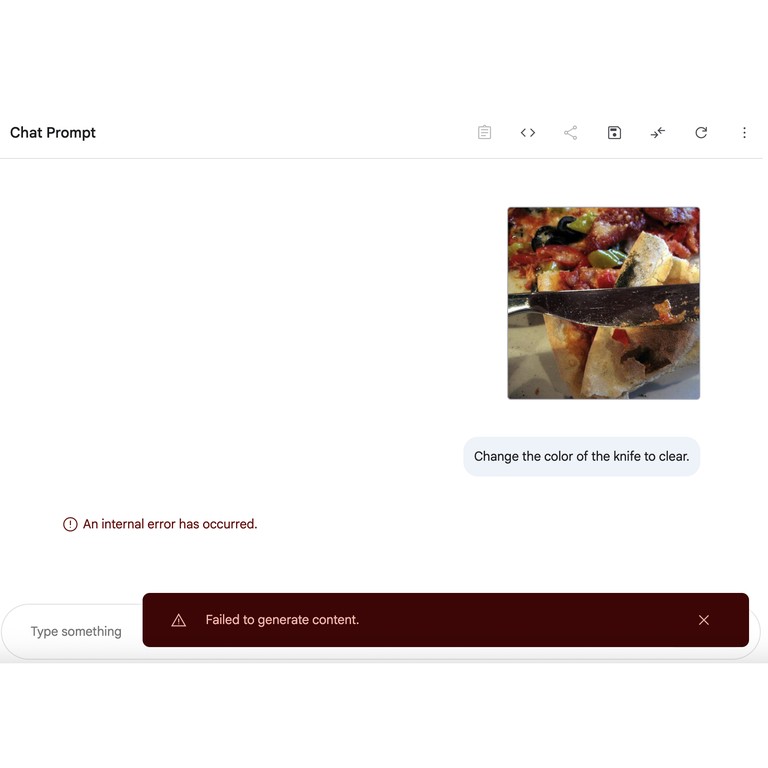} &
    \includegraphics[width=0.09\textwidth]{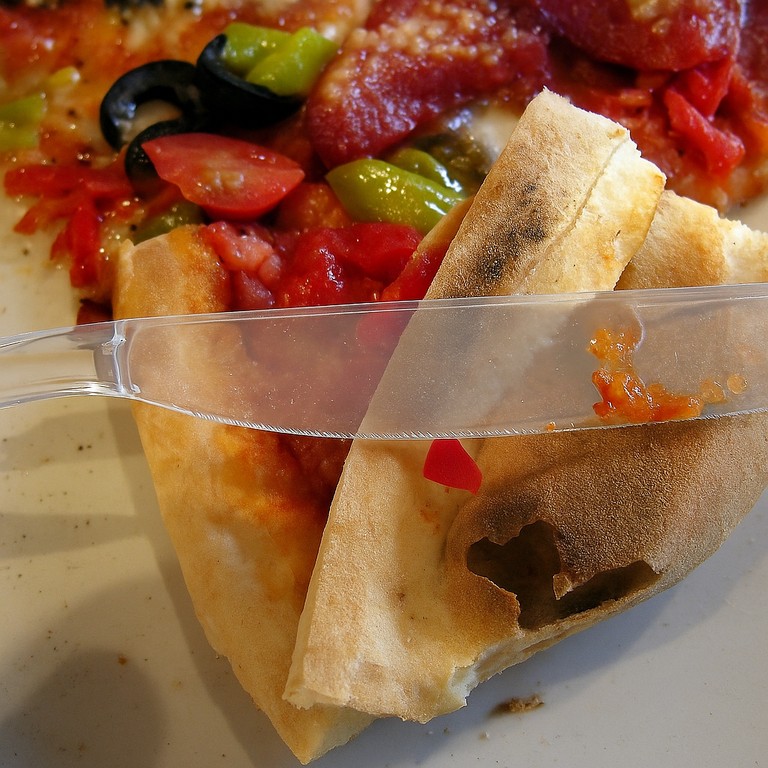} &
    \includegraphics[width=0.09\textwidth]{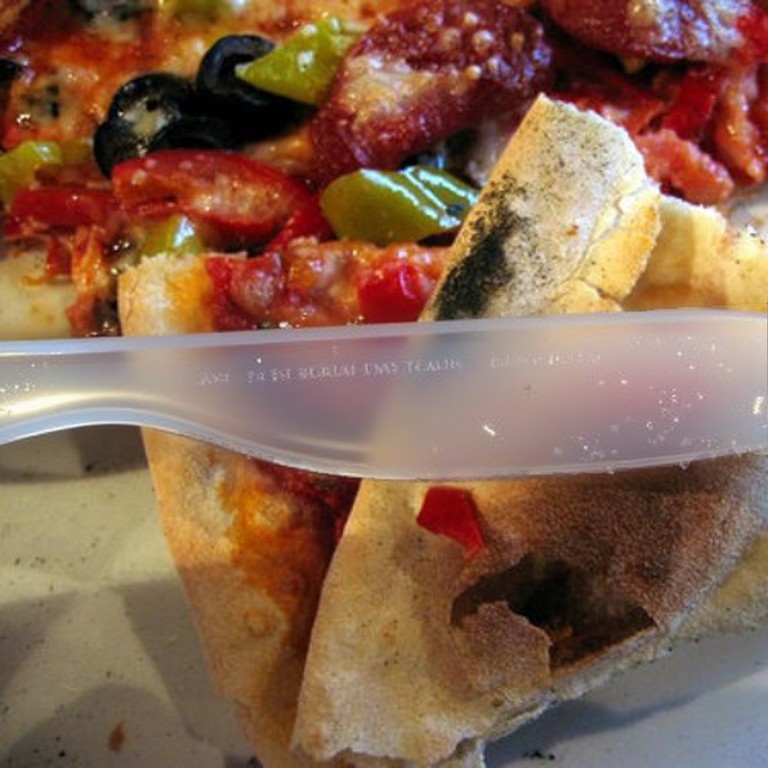} \\
    \multicolumn{9}{c}{\scriptsize\textbf{Color:} Change the color of the knife to clear.} \\
    [0.5em]
    \includegraphics[width=0.09\textwidth]{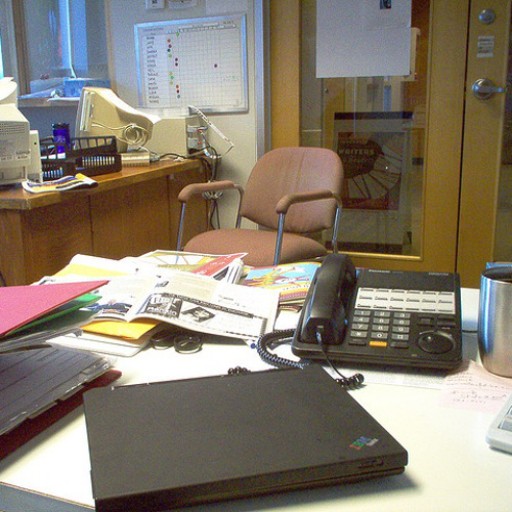} &
    \includegraphics[width=0.09\textwidth]{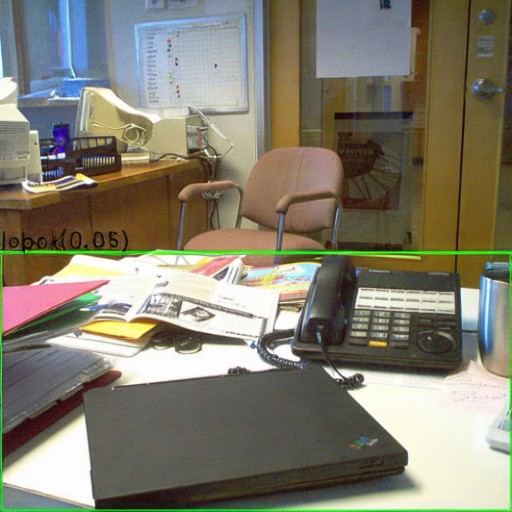} &
    \includegraphics[width=0.09\textwidth]{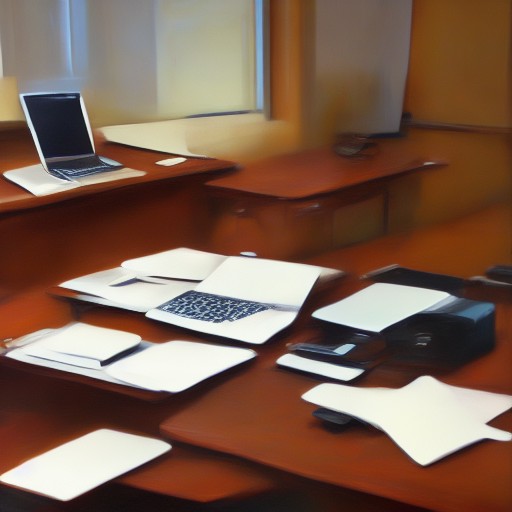} &
    \includegraphics[width=0.09\textwidth]{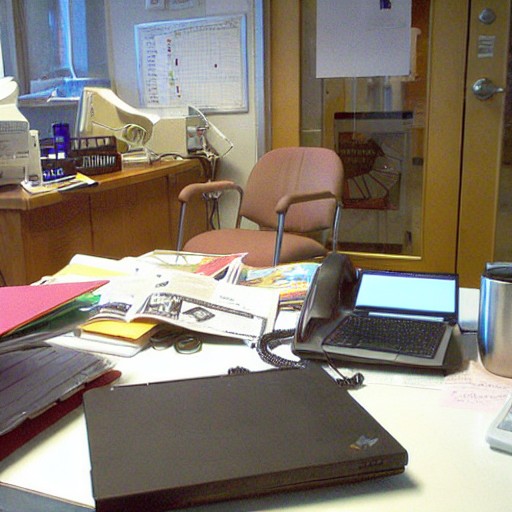} &
    \includegraphics[width=0.09\textwidth]{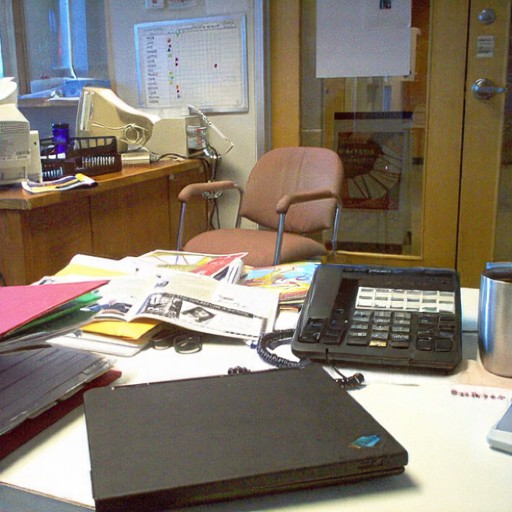} &
    \includegraphics[width=0.09\textwidth]{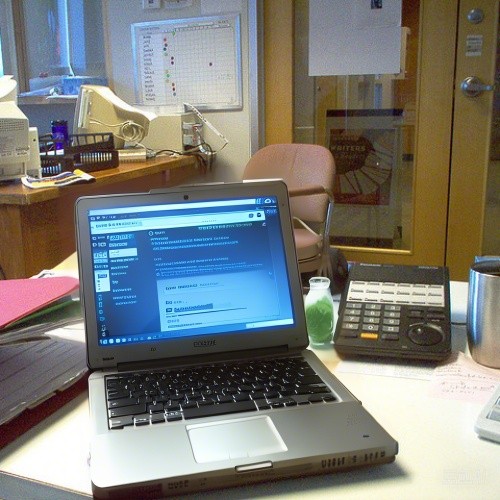} &
    \includegraphics[width=0.09\textwidth]{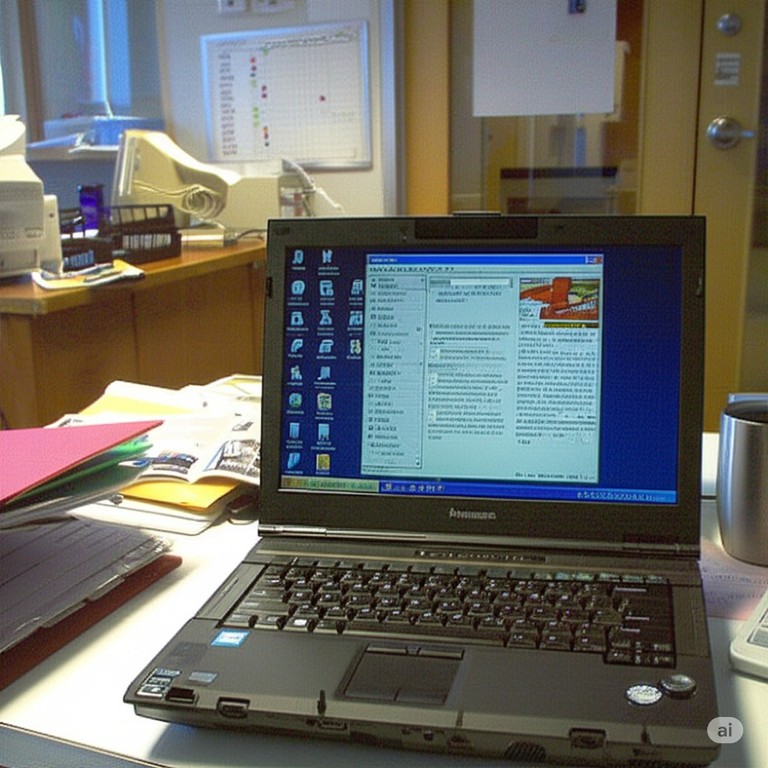} &
    \includegraphics[width=0.09\textwidth]{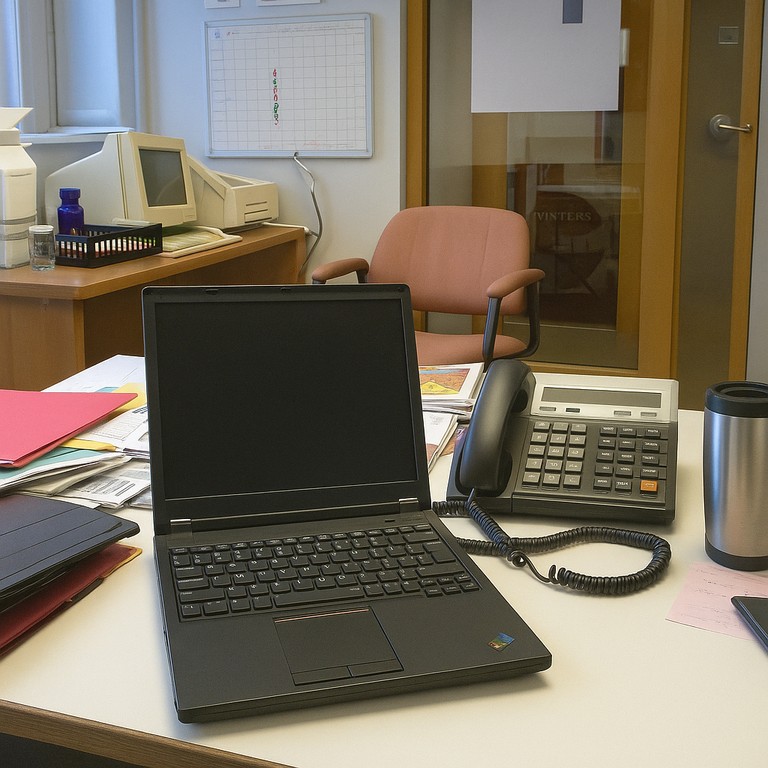} &
    \includegraphics[width=0.09\textwidth]{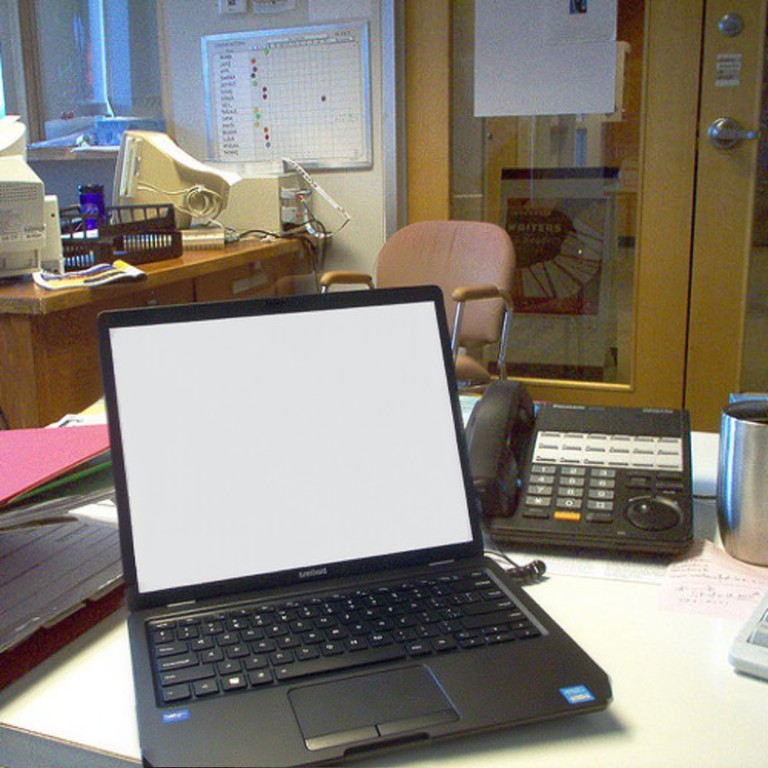} \\
    \multicolumn{9}{c}{\scriptsize\textbf{Local:} Open the laptop that is on the desk.} \\
    [0.5em]
    \includegraphics[width=0.09\textwidth]{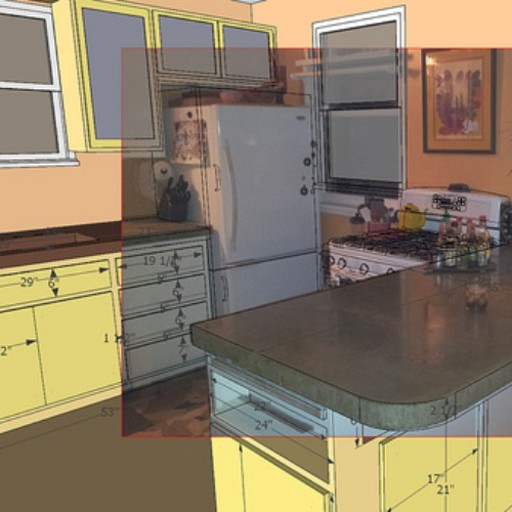} &
    \includegraphics[width=0.09\textwidth]{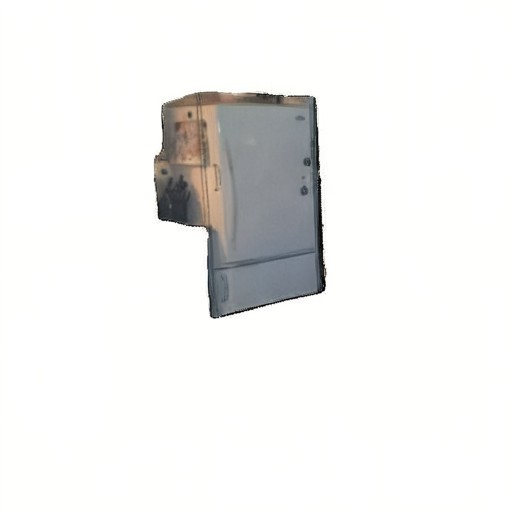} &
    \includegraphics[width=0.09\textwidth]{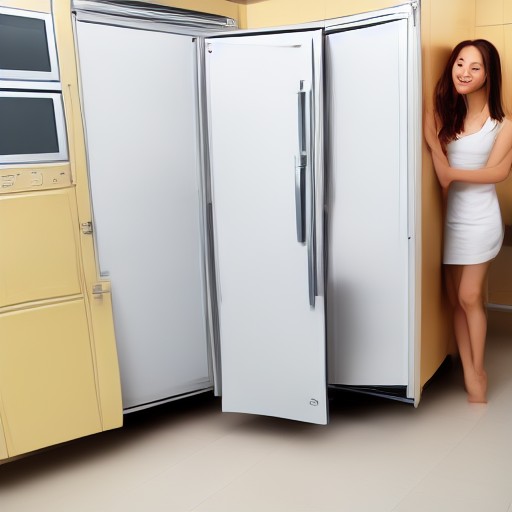} &
    \includegraphics[width=0.09\textwidth]{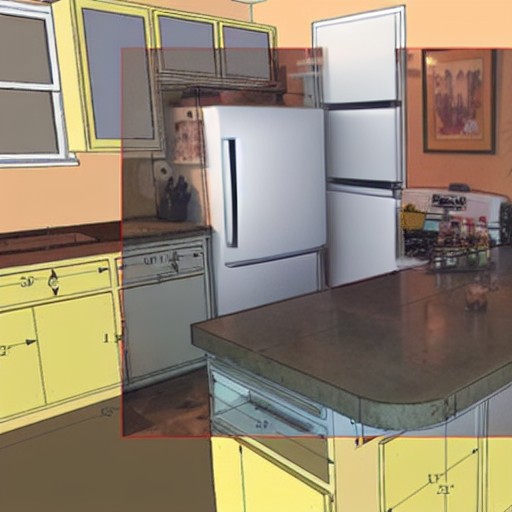} &
    \includegraphics[width=0.09\textwidth]{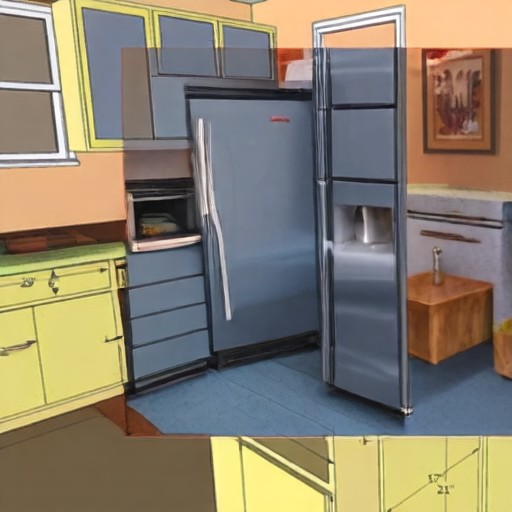} &
    \includegraphics[width=0.09\textwidth]{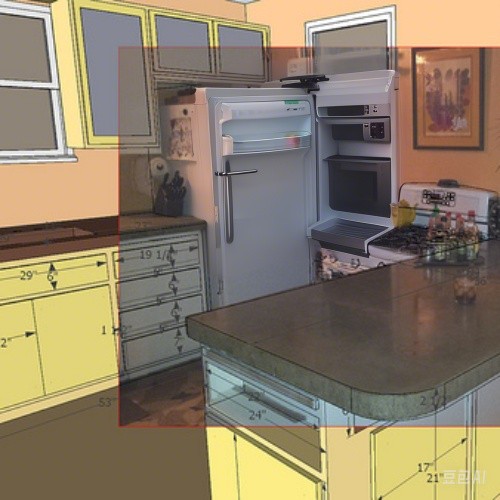} &
    \includegraphics[width=0.09\textwidth]{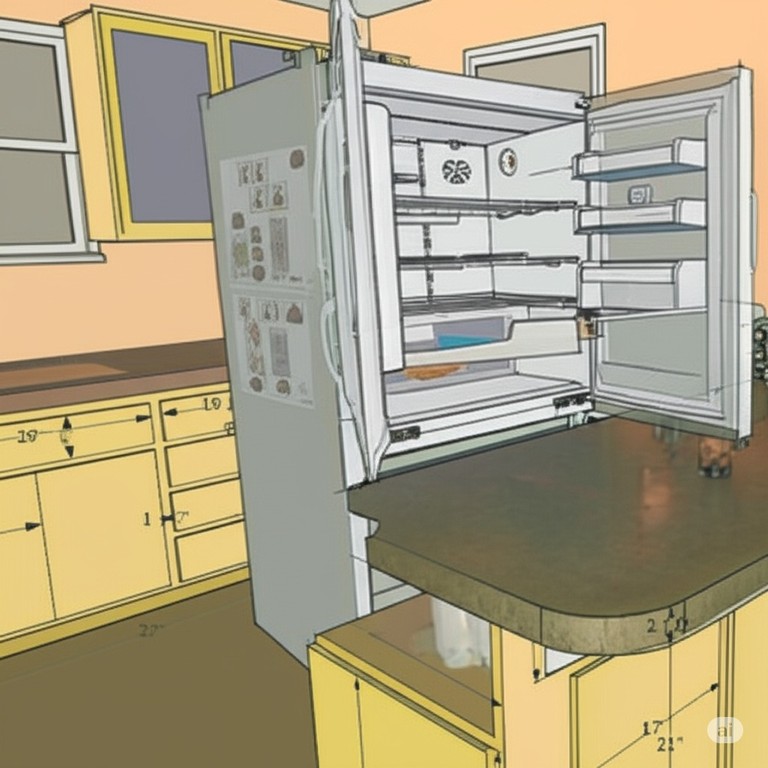} &
    \includegraphics[width=0.09\textwidth]{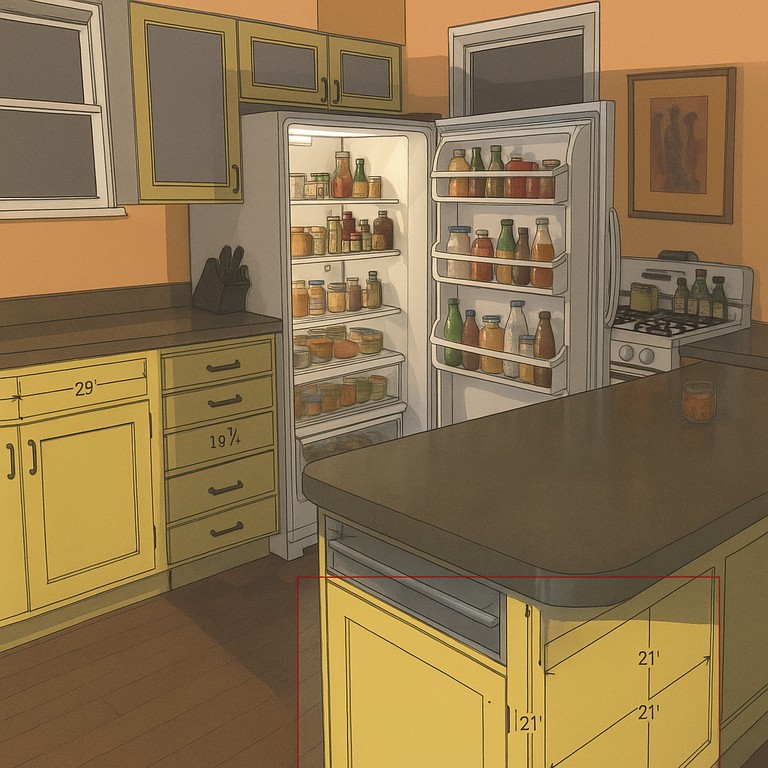} &
    \includegraphics[width=0.09\textwidth]{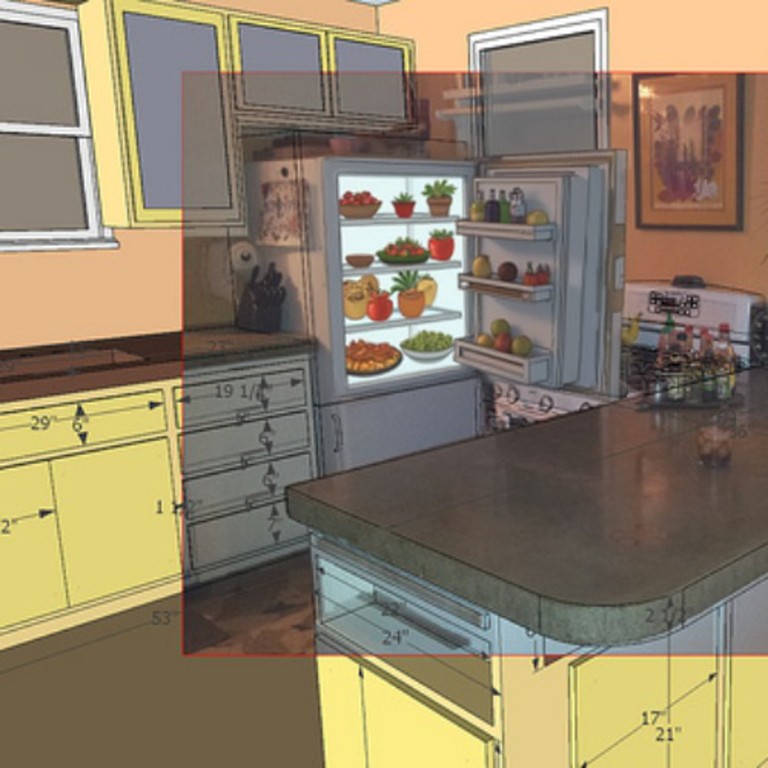} \\
    \multicolumn{9}{c}{\scriptsize\textbf{Local:} Open the refrigerator door in the image.} \\
    [0.5em]
    \includegraphics[width=0.09\textwidth]{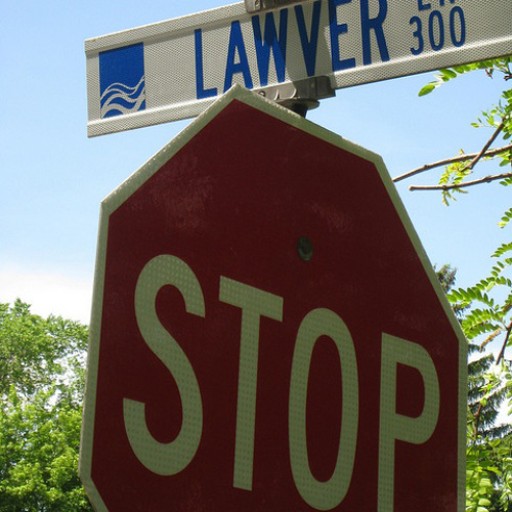} &
    \includegraphics[width=0.09\textwidth]{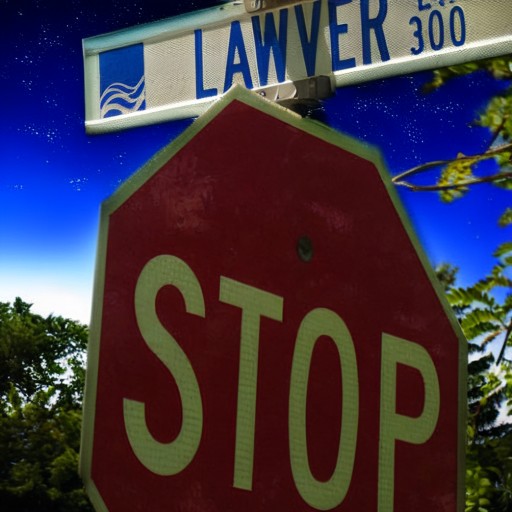} &
    \includegraphics[width=0.09\textwidth]{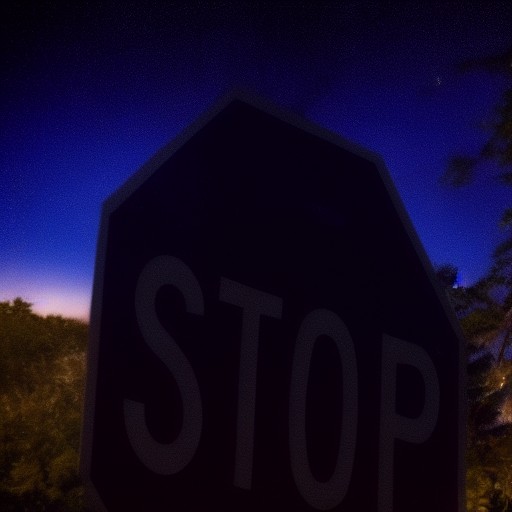} &
    \includegraphics[width=0.09\textwidth]{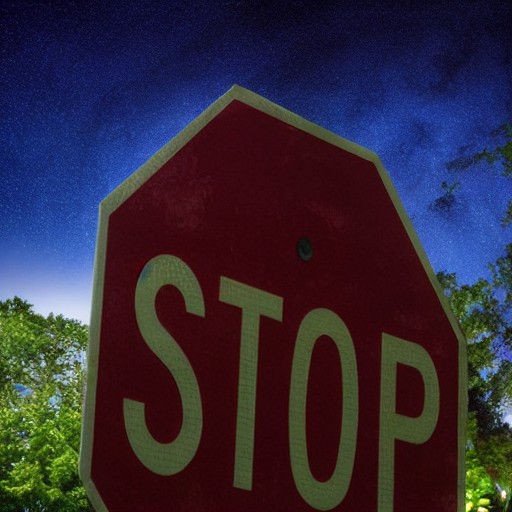} &
    \includegraphics[width=0.09\textwidth]{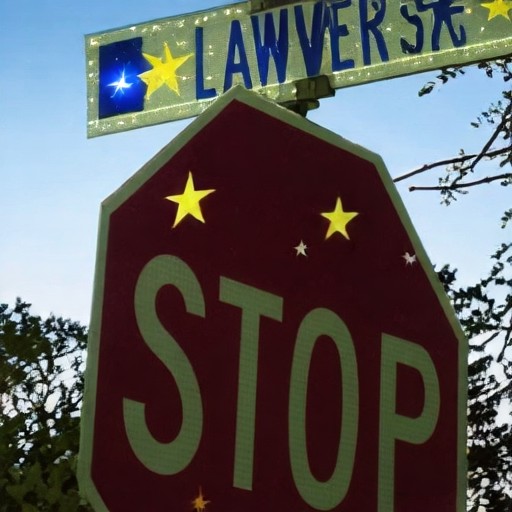} &
    \includegraphics[width=0.09\textwidth]{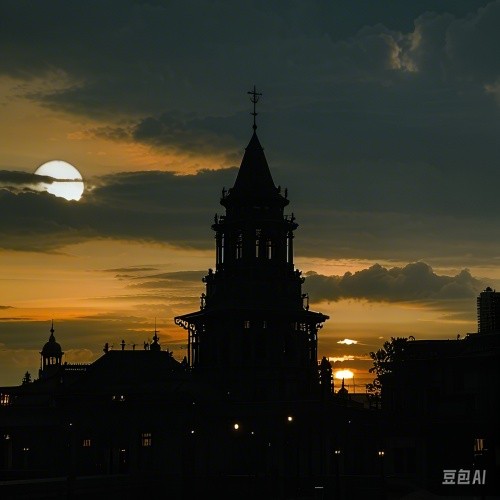} &
    \includegraphics[width=0.09\textwidth]{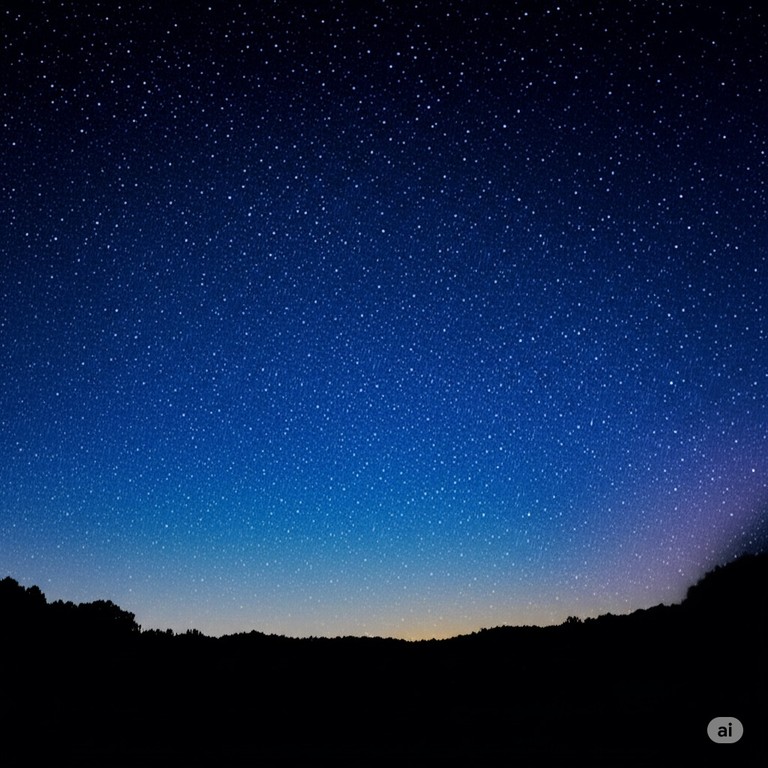} &
    \includegraphics[width=0.09\textwidth]{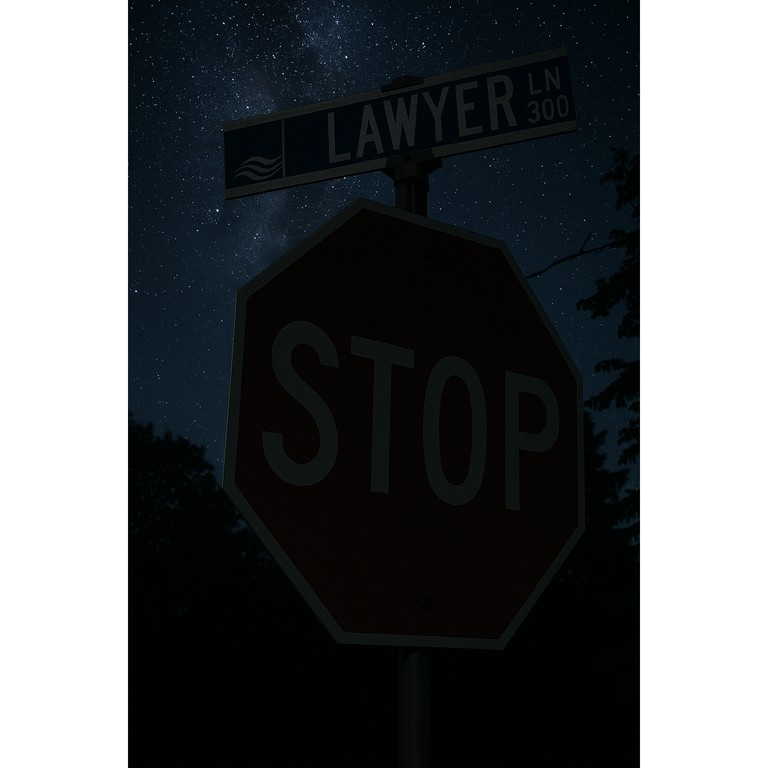} &
    \includegraphics[width=0.09\textwidth]{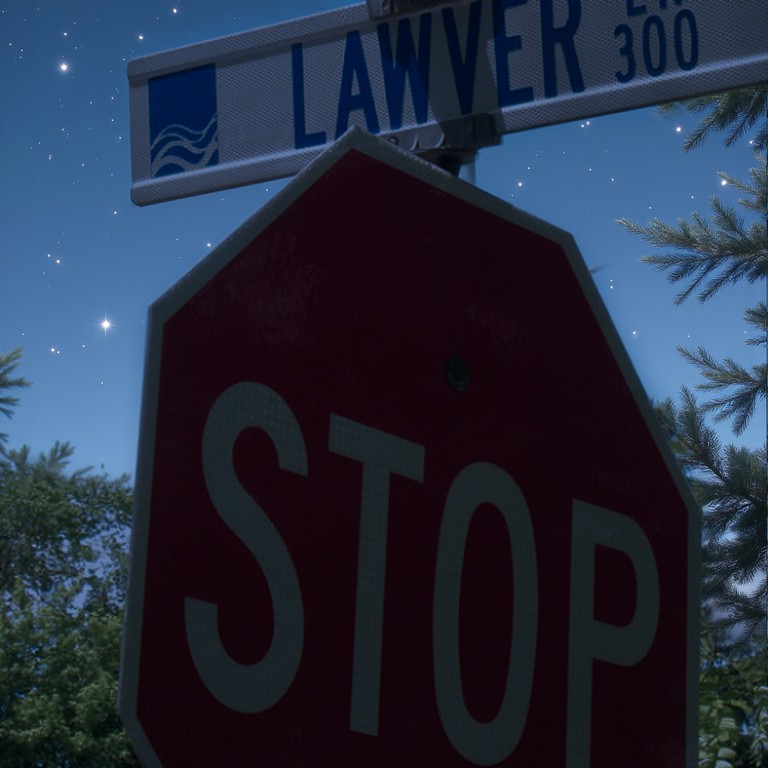} \\
    \multicolumn{9}{c}{\scriptsize\textbf{Global:} Change the image so it appears it is night and there are millions of bright stars.} \\
    [0.5em]
    \includegraphics[width=0.09\textwidth]{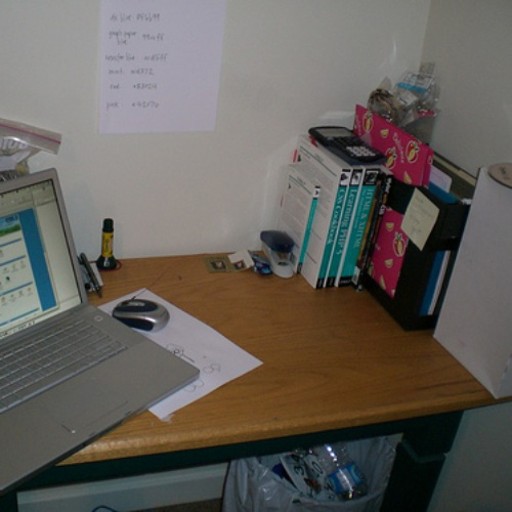} &
    \includegraphics[width=0.09\textwidth]{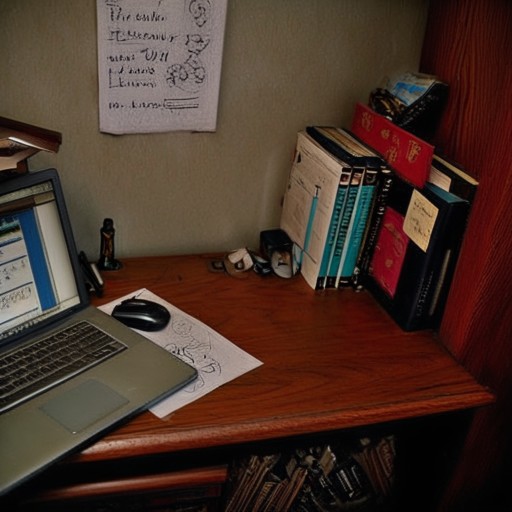} &
    \includegraphics[width=0.09\textwidth]{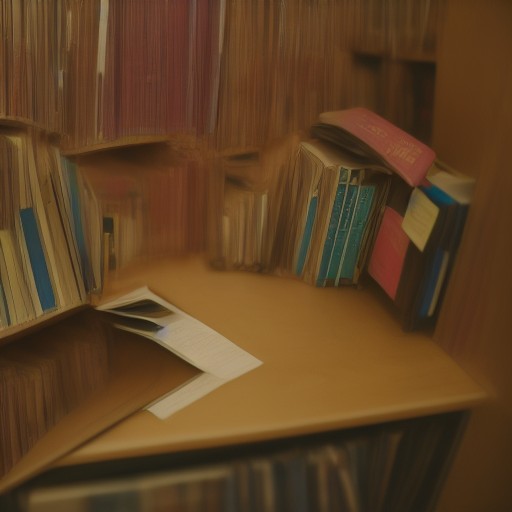} &
    \includegraphics[width=0.09\textwidth]{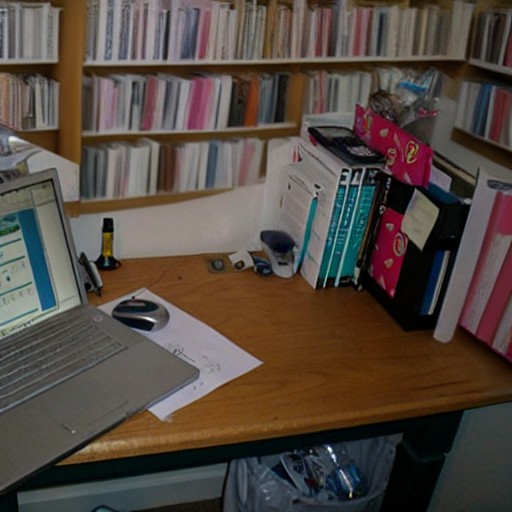} &
    \includegraphics[width=0.09\textwidth]{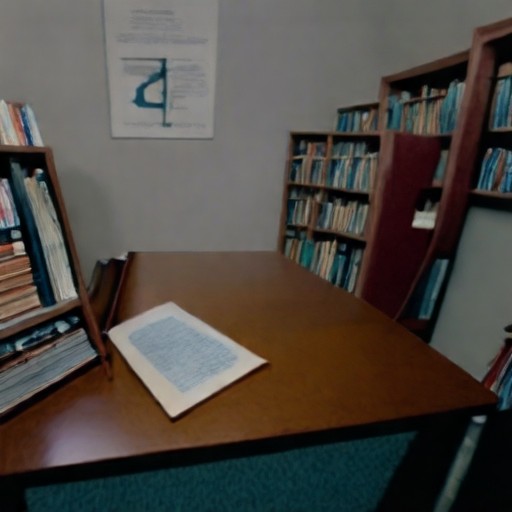} &
    \includegraphics[width=0.09\textwidth]{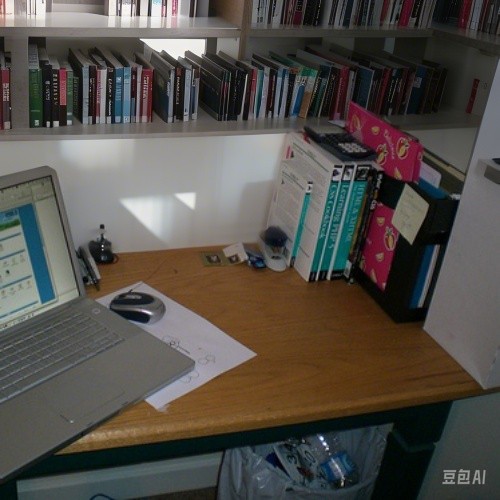} &
    \includegraphics[width=0.09\textwidth]{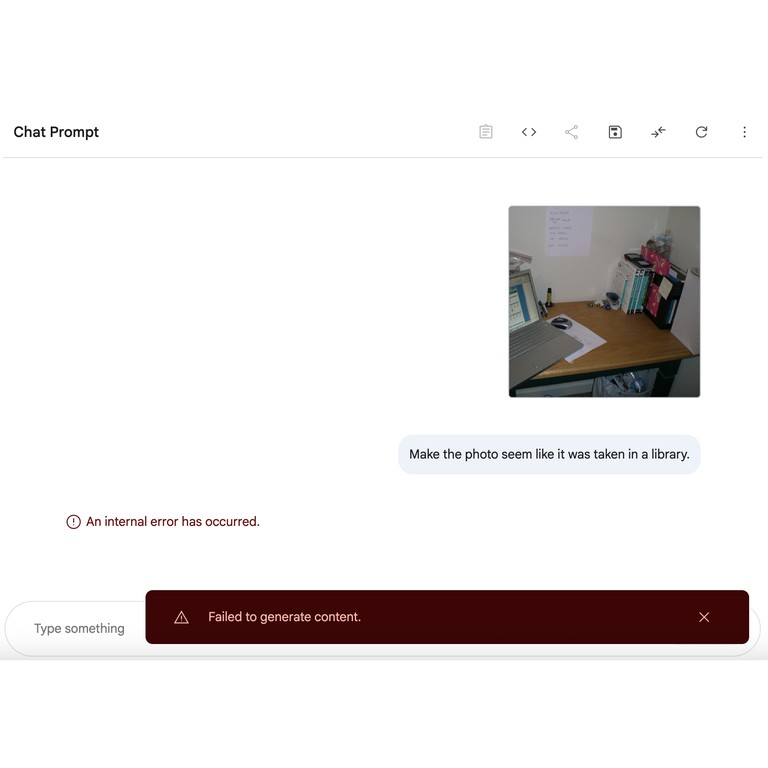} &
    \includegraphics[width=0.09\textwidth]{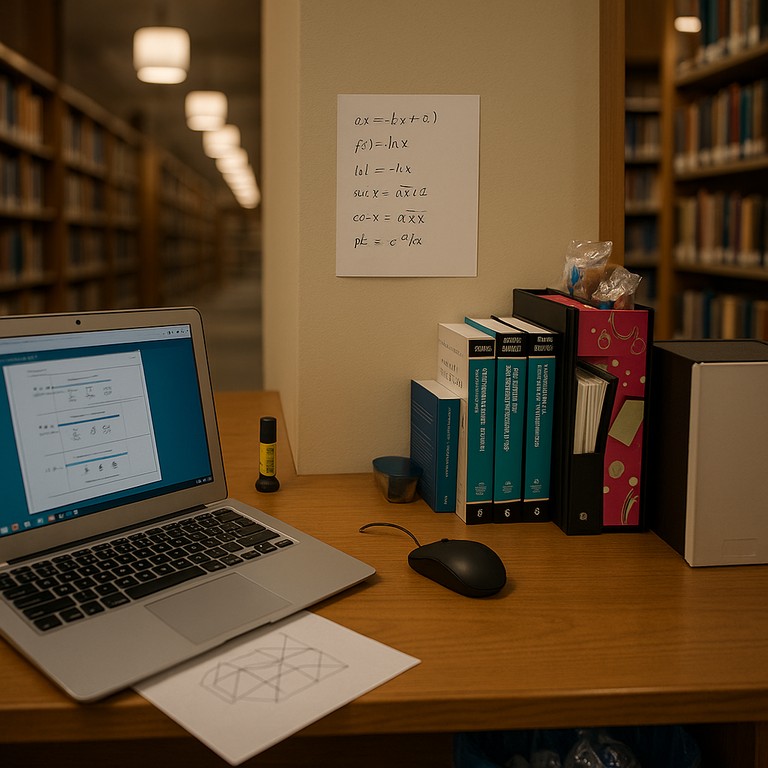} &
    \includegraphics[width=0.09\textwidth]{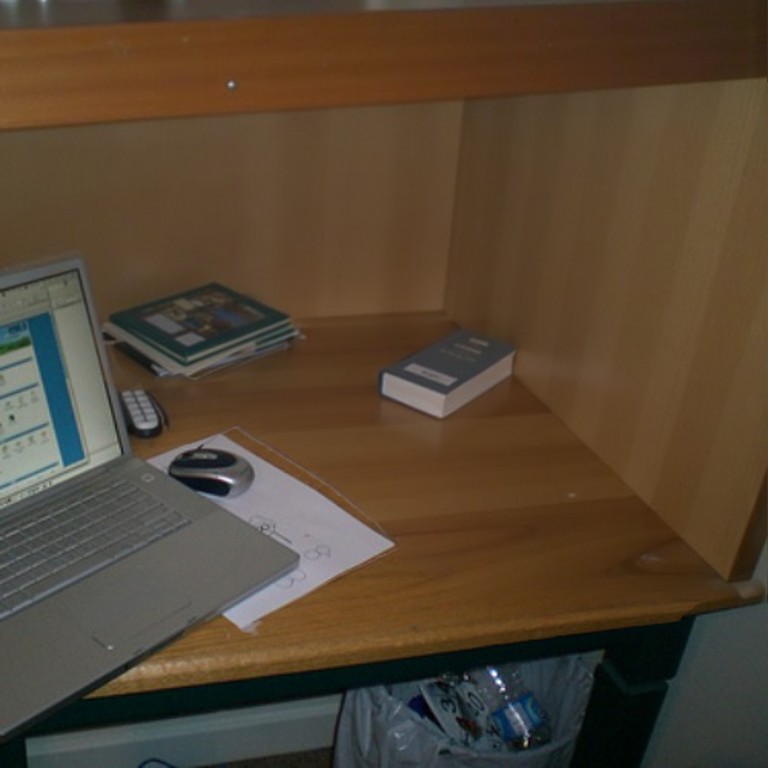} \\
    \multicolumn{9}{c}{\scriptsize\textbf{Global:} Make the photo seem like it was taken in a library.} \\
\end{tabular}
\caption{\textbf{Our workflow performs the best on the second half of 14 challenging examples from Emu Edit.}}
\label{fig:challenging-emuedit-second-half}
\end{center}

\end{figure*}

\begin{figure*}[htbp]

\begin{center}
\small
\begin{tabular}{c@{\hspace{0.3em}}c@{\hspace{0.3em}}c@{\hspace{0.3em}}c@{\hspace{0.3em}}c@{\hspace{0.3em}}c@{\hspace{0.3em}}c@{\hspace{0.3em}}c@{\hspace{0.3em}}c}
    \scriptsize\textsc{Original} & \scriptsize\textsc{Reference} & \scriptsize\textsc{IP2P} & \scriptsize\textsc{MB} & \scriptsize\textsc{UE} & \scriptsize\textsc{Doubao} & \scriptsize\textsc{Gemini} & \scriptsize\textsc{GPT-4o} & \scriptsize\textsc{\textbf{Ours}} \\
    
    \includegraphics[width=0.09\textwidth]{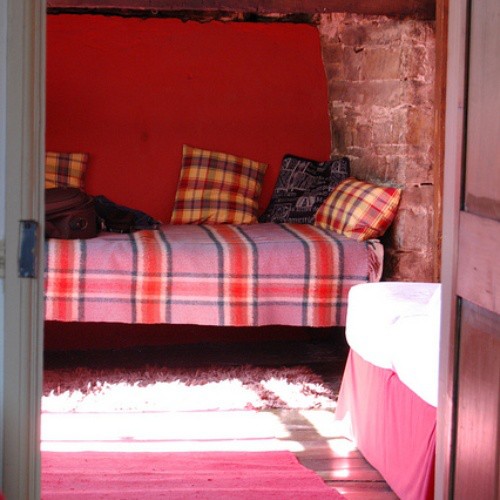} &
    \includegraphics[width=0.09\textwidth]{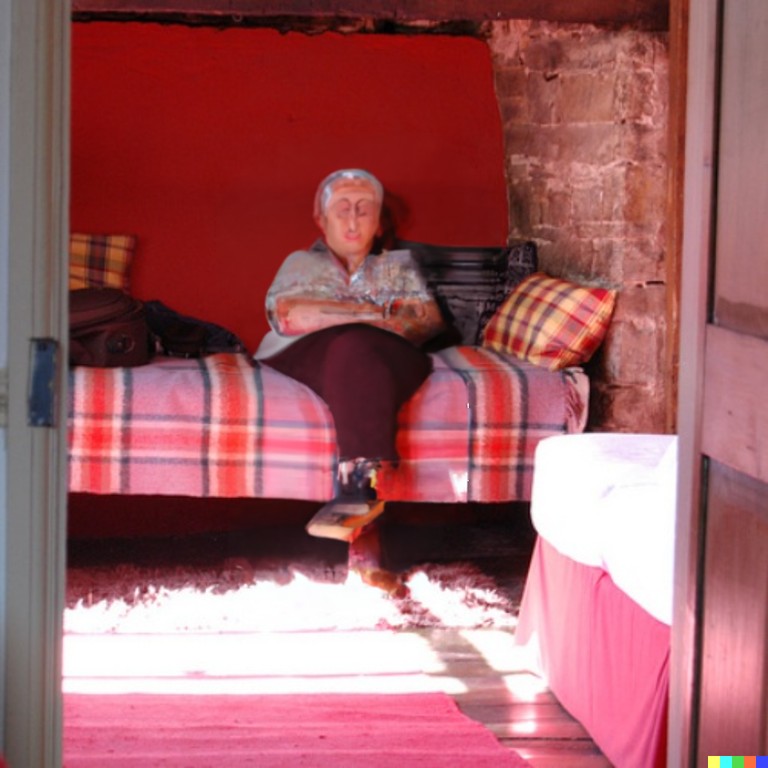} &
    \includegraphics[width=0.09\textwidth]{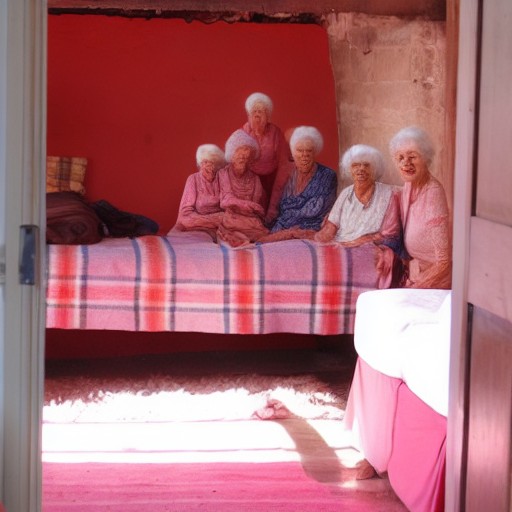} &
    \includegraphics[width=0.09\textwidth]{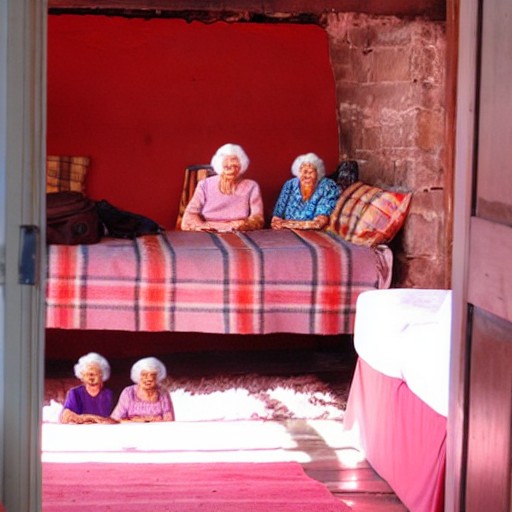} &
    \includegraphics[width=0.09\textwidth]{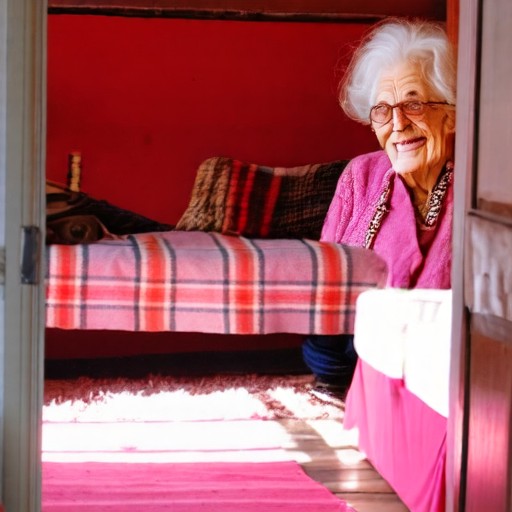} &
    \includegraphics[width=0.09\textwidth]{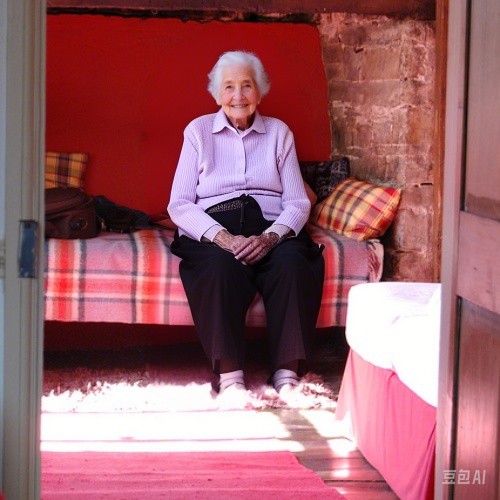} &
    \includegraphics[width=0.09\textwidth]{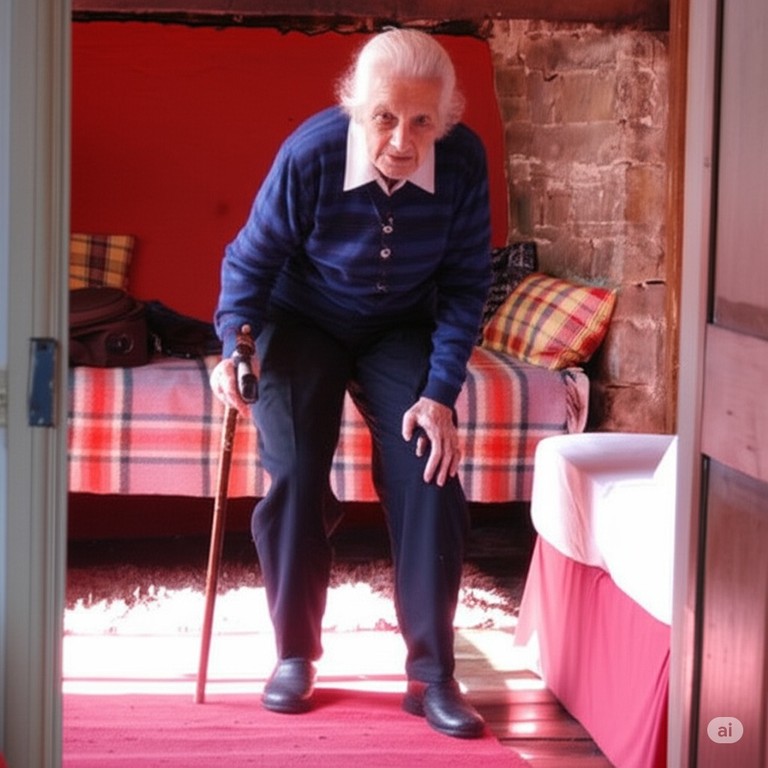} &
    \includegraphics[width=0.09\textwidth]{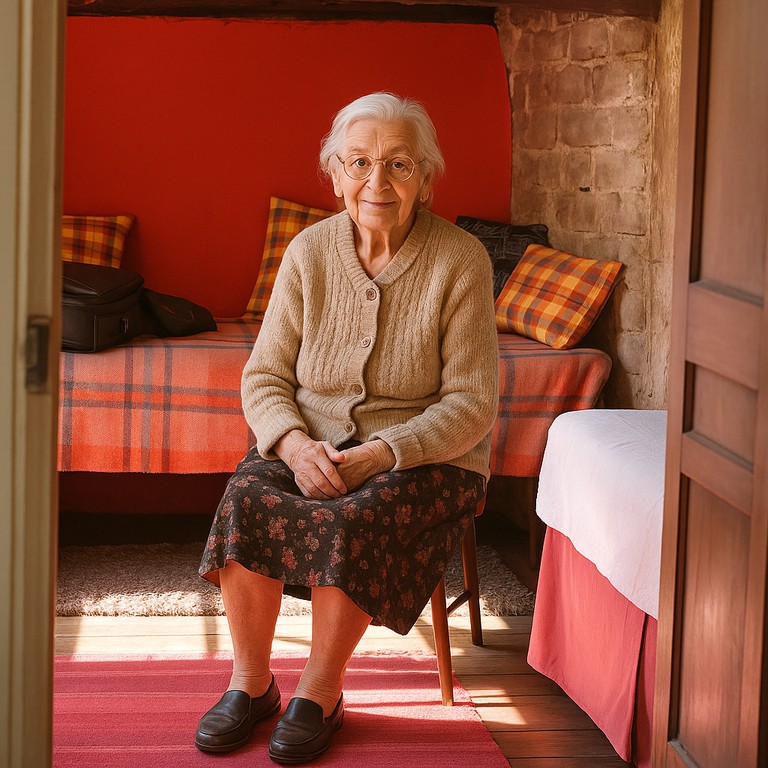} &
    \includegraphics[width=0.09\textwidth]{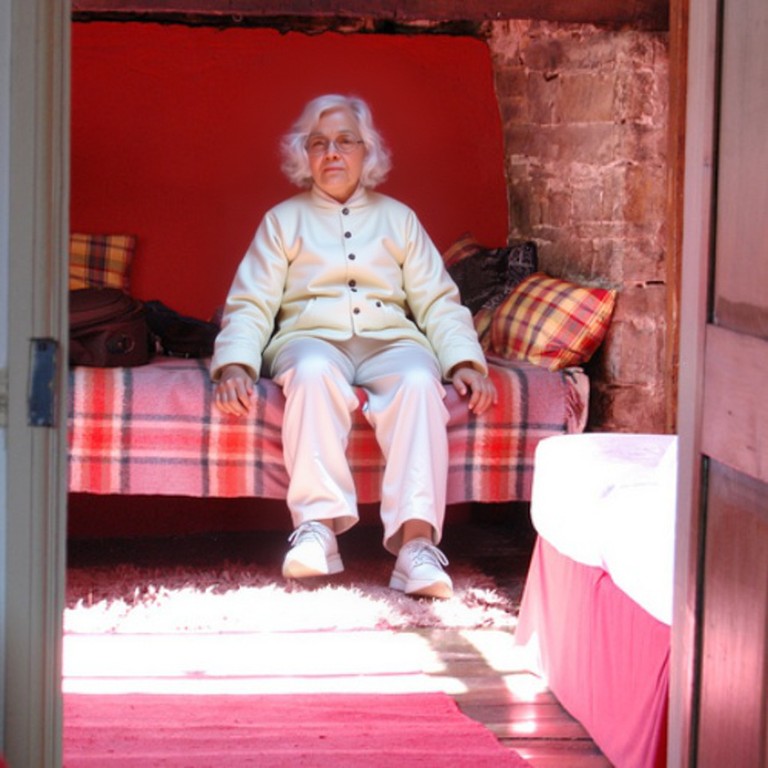} \\
    \multicolumn{9}{c}{\scriptsize\textbf{Add:} Add a grandma.} \\
    [0.5em]
    \includegraphics[width=0.09\textwidth]{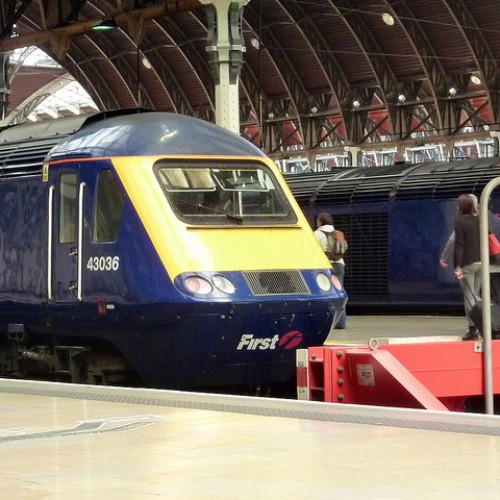} &
    \includegraphics[width=0.09\textwidth]{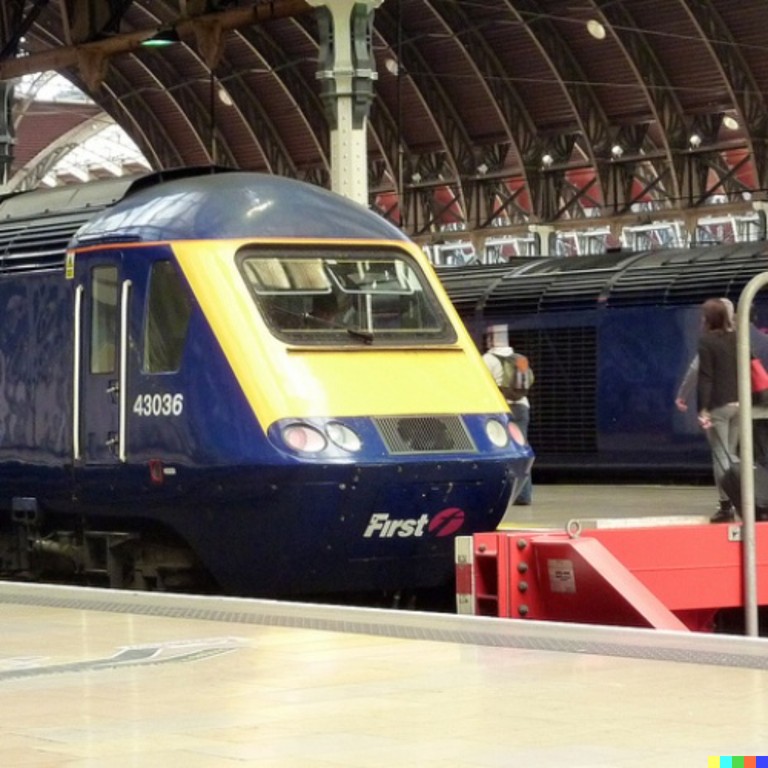} &
    \includegraphics[width=0.09\textwidth]{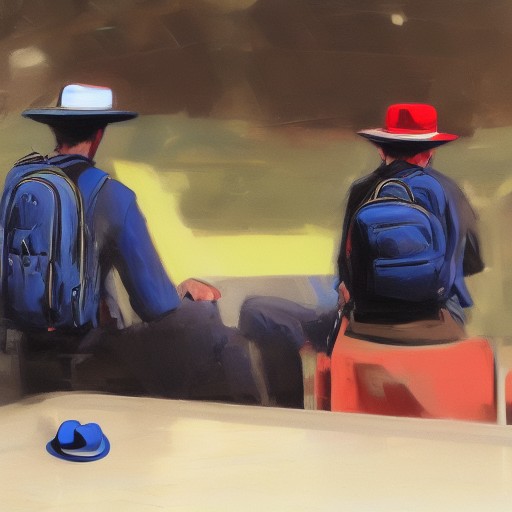} &
    \includegraphics[width=0.09\textwidth]{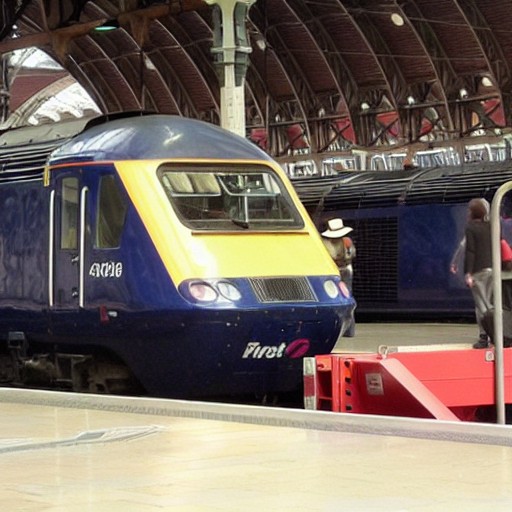} &
    \includegraphics[width=0.09\textwidth]{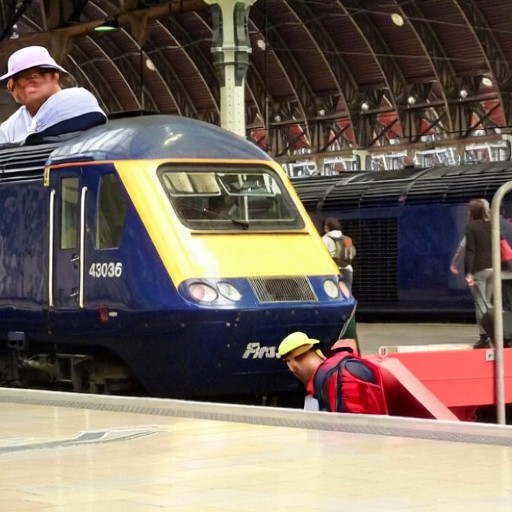} &
    \includegraphics[width=0.09\textwidth]{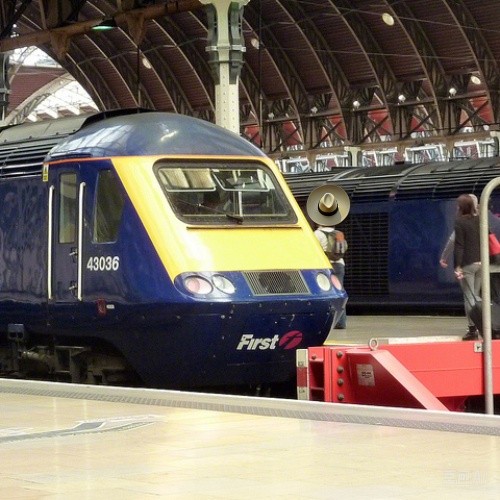} &
    \includegraphics[width=0.09\textwidth]{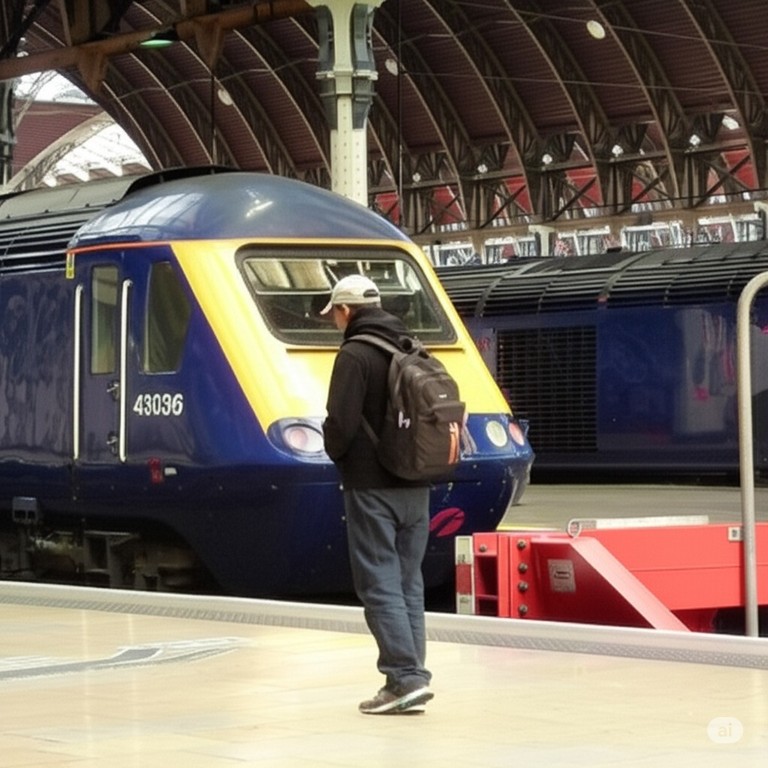} &
    \includegraphics[width=0.09\textwidth]{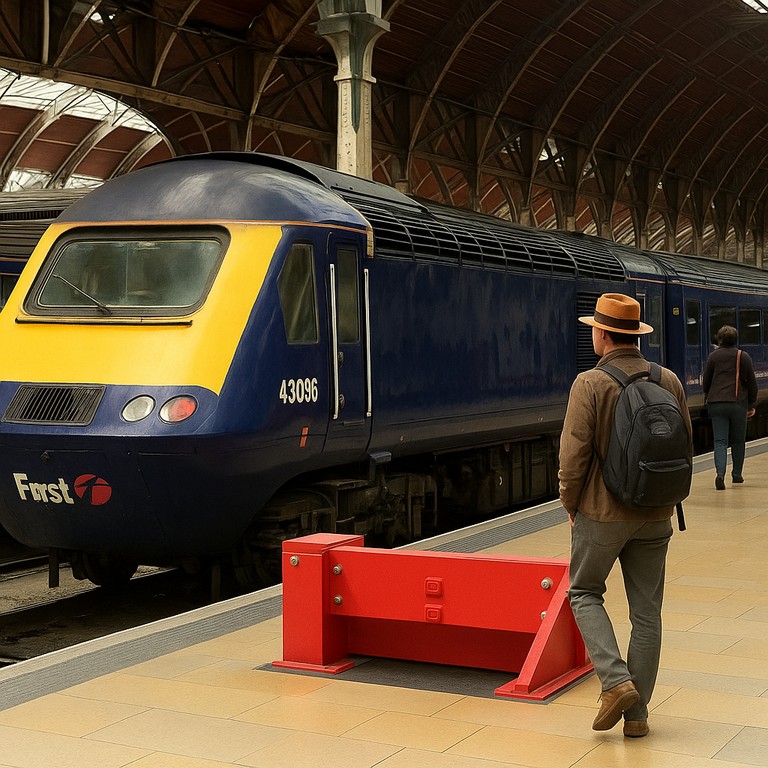} &
    \includegraphics[width=0.09\textwidth]{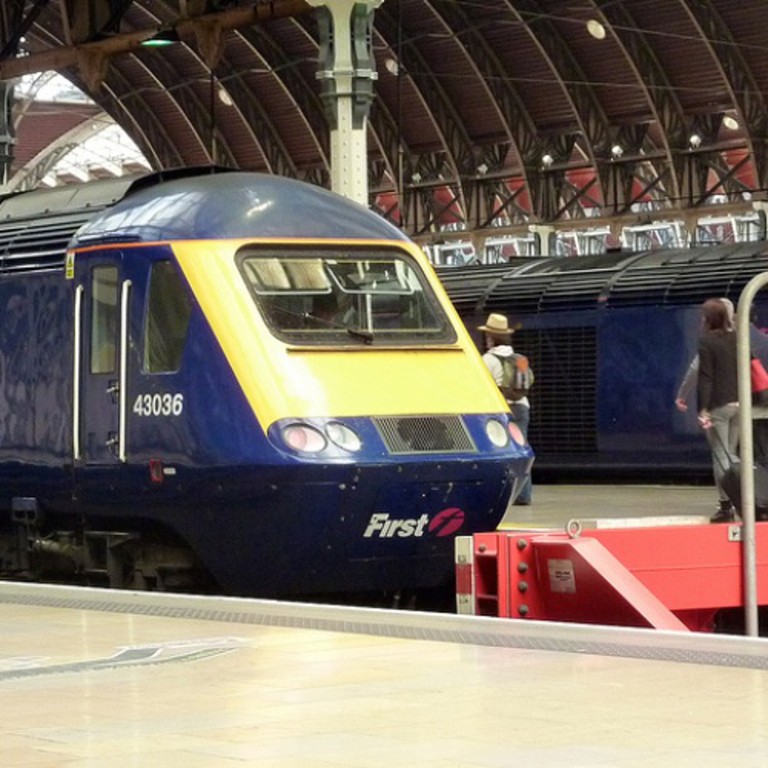} \\
    \multicolumn{9}{c}{\scriptsize\textbf{Add:} Let's add a hat to the man with the backpack.} \\
    [0.5em]
    \includegraphics[width=0.09\textwidth]{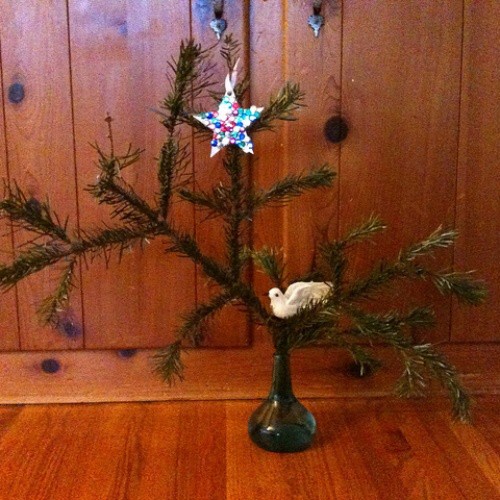} &
    \includegraphics[width=0.09\textwidth]{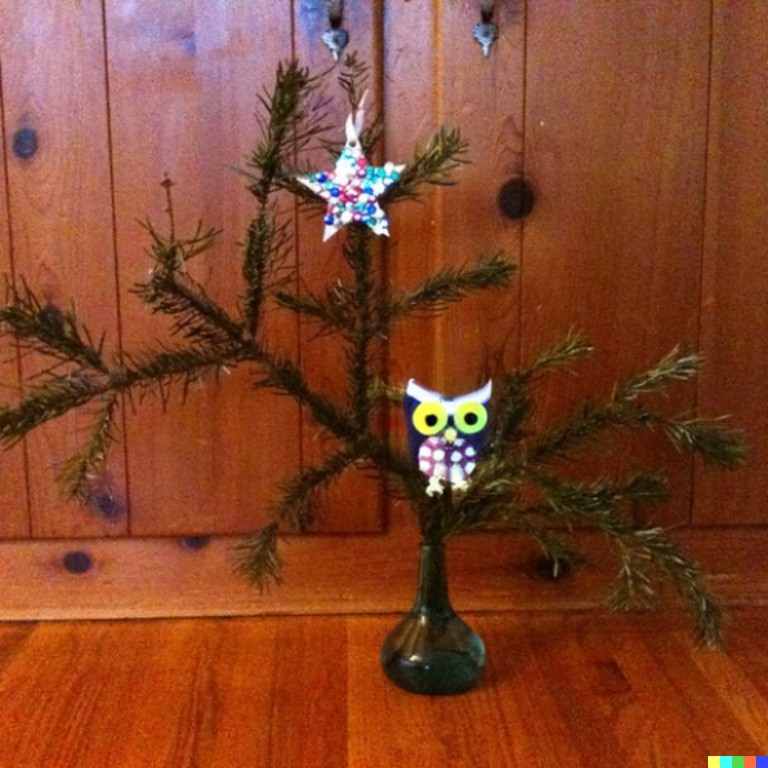} &
    \includegraphics[width=0.09\textwidth]{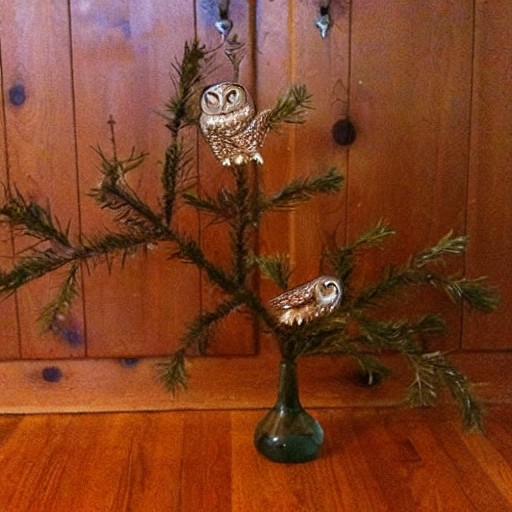} &
    \includegraphics[width=0.09\textwidth]{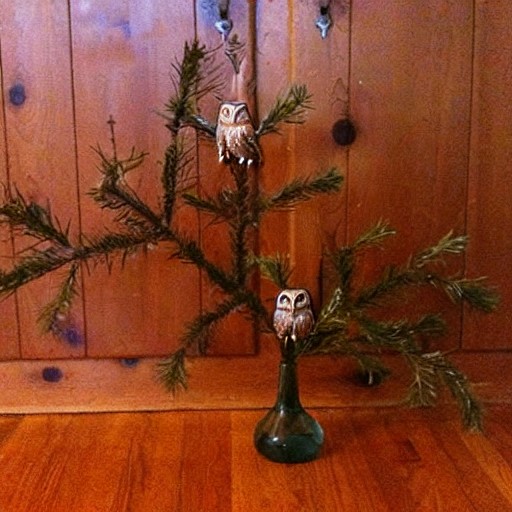} &
    \includegraphics[width=0.09\textwidth]{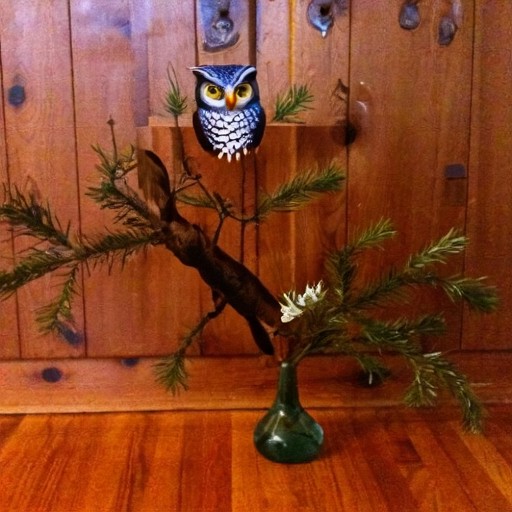} &
    \includegraphics[width=0.09\textwidth]{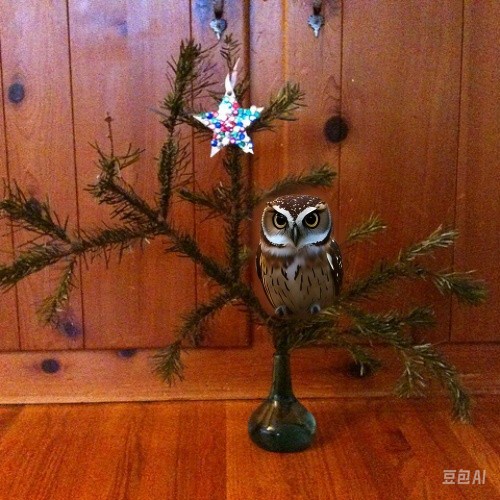} &
    \includegraphics[width=0.09\textwidth]{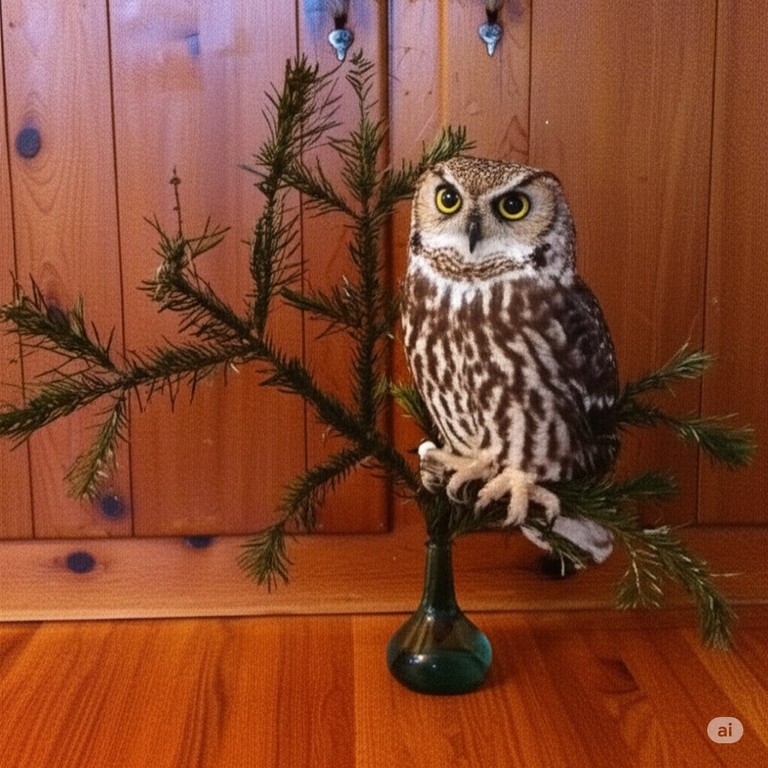} &
    \includegraphics[width=0.09\textwidth]{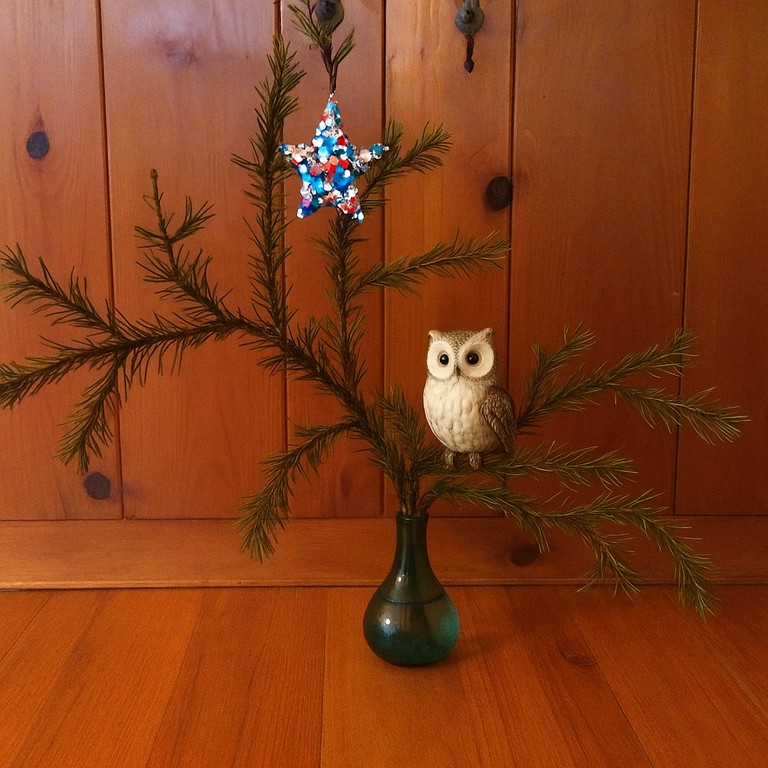} &
    \includegraphics[width=0.09\textwidth]{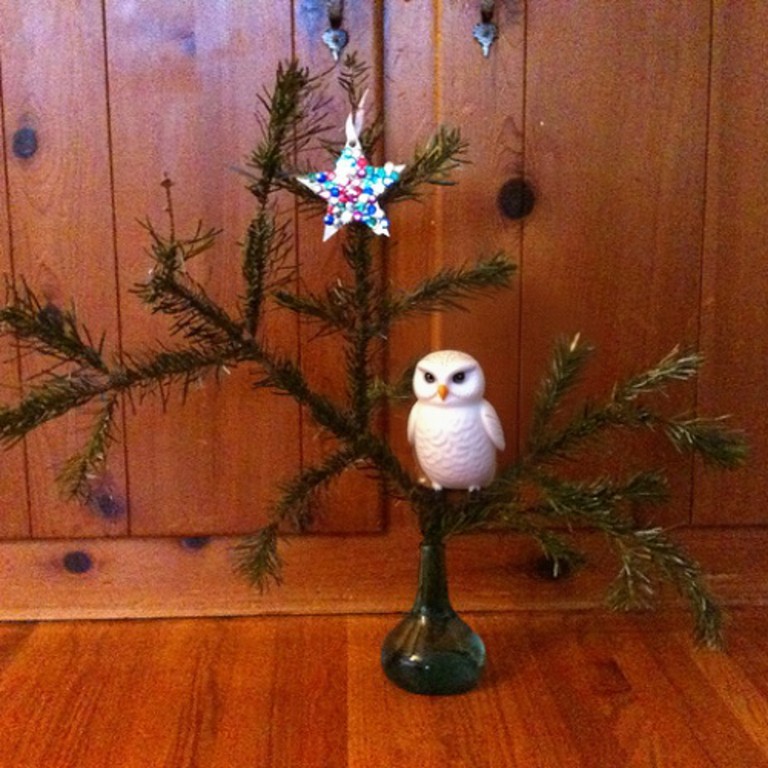} \\
    \multicolumn{9}{c}{\scriptsize\textbf{Replace:} Replace the dove with an owl.} \\
    [0.5em]
    \includegraphics[width=0.09\textwidth]{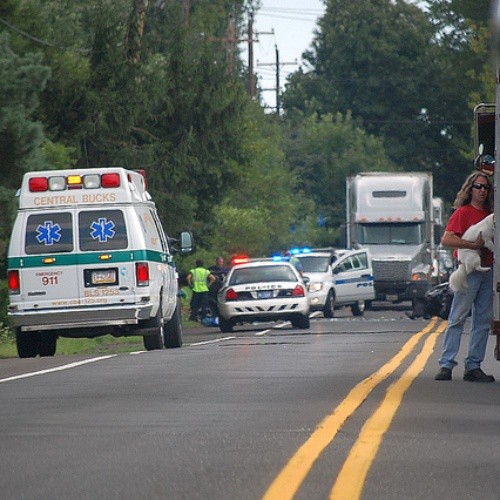} &
    \includegraphics[width=0.09\textwidth]{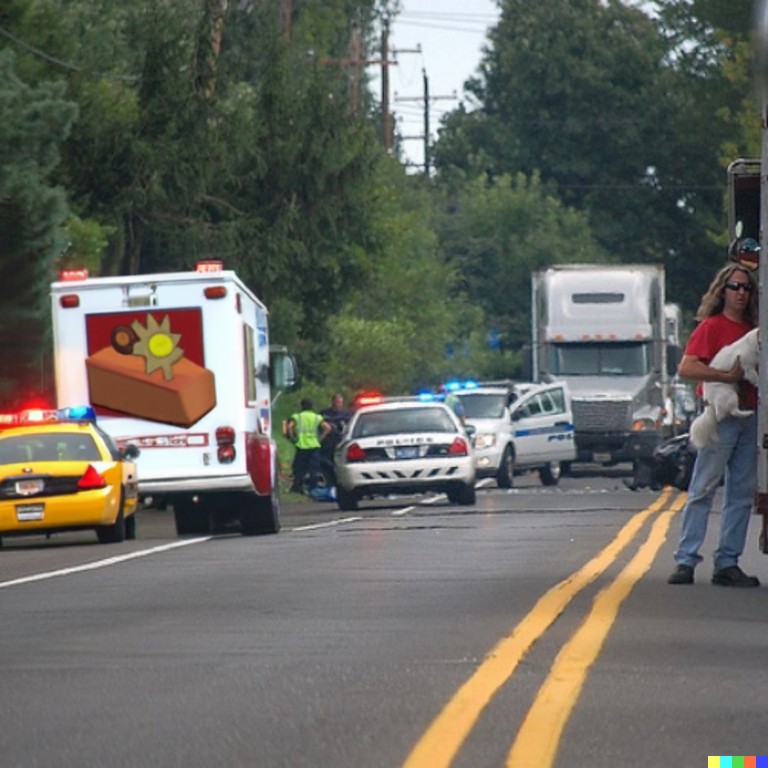} &
    \includegraphics[width=0.09\textwidth]{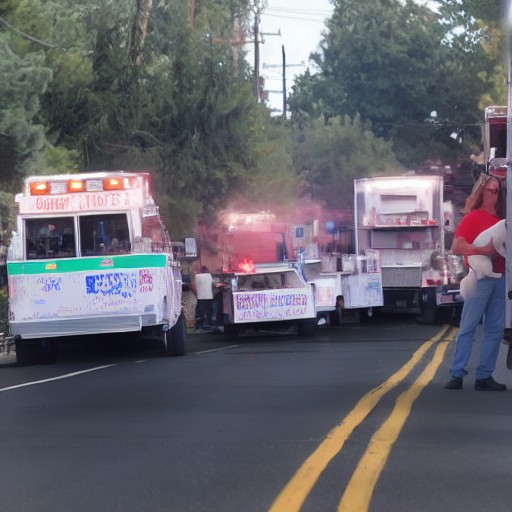} &
    \includegraphics[width=0.09\textwidth]{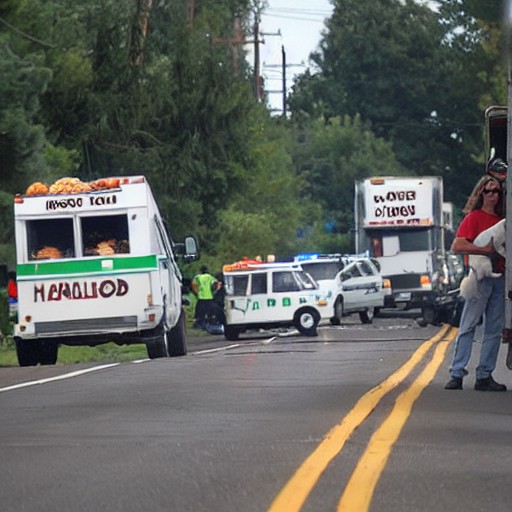} &
    \includegraphics[width=0.09\textwidth]{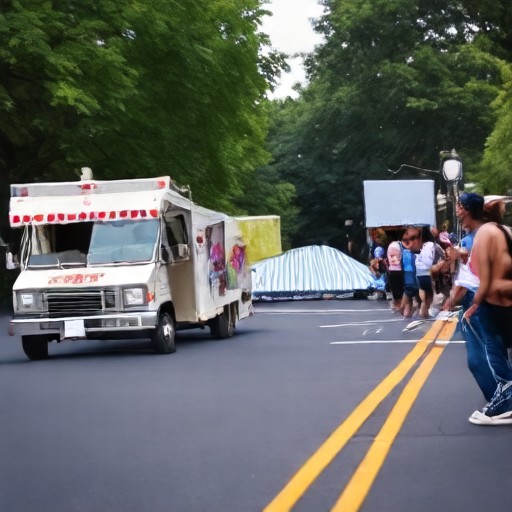} &
    \includegraphics[width=0.09\textwidth]{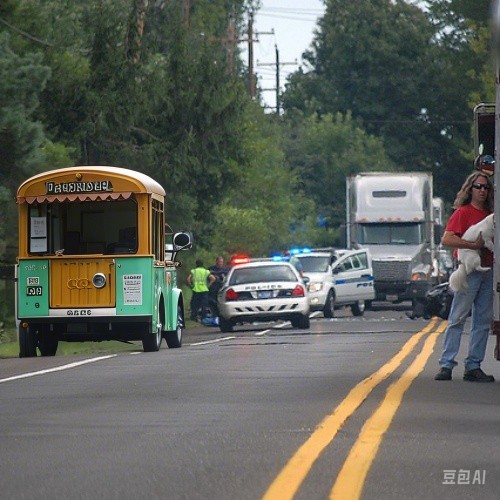} &
    \includegraphics[width=0.09\textwidth]{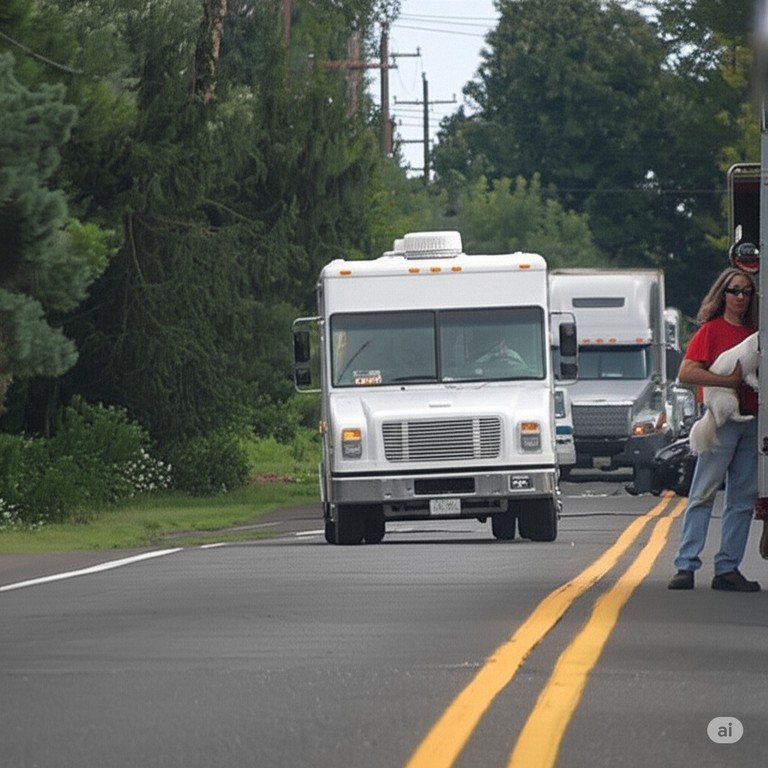} &
    \includegraphics[width=0.09\textwidth]{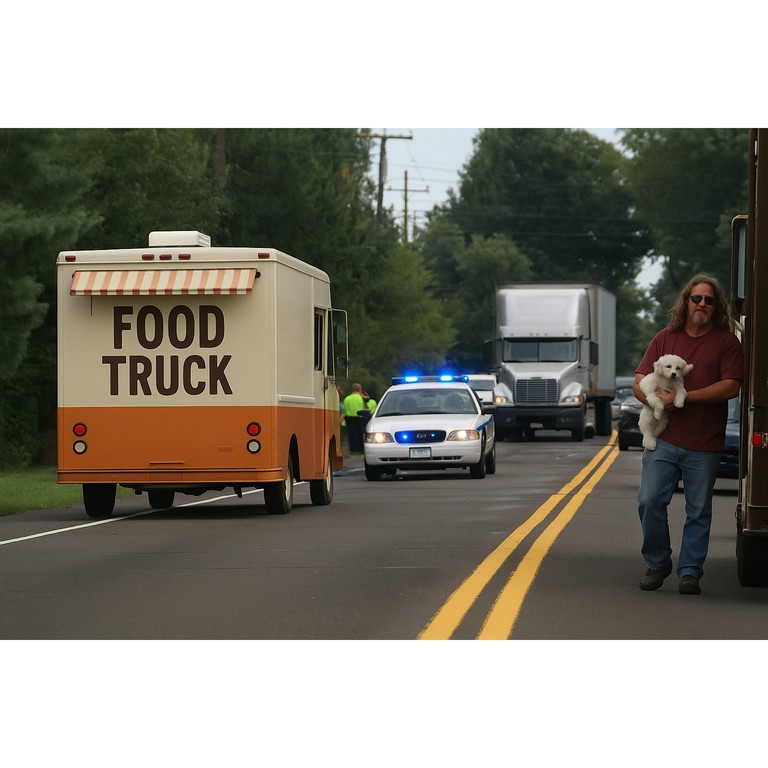} &
    \includegraphics[width=0.09\textwidth]{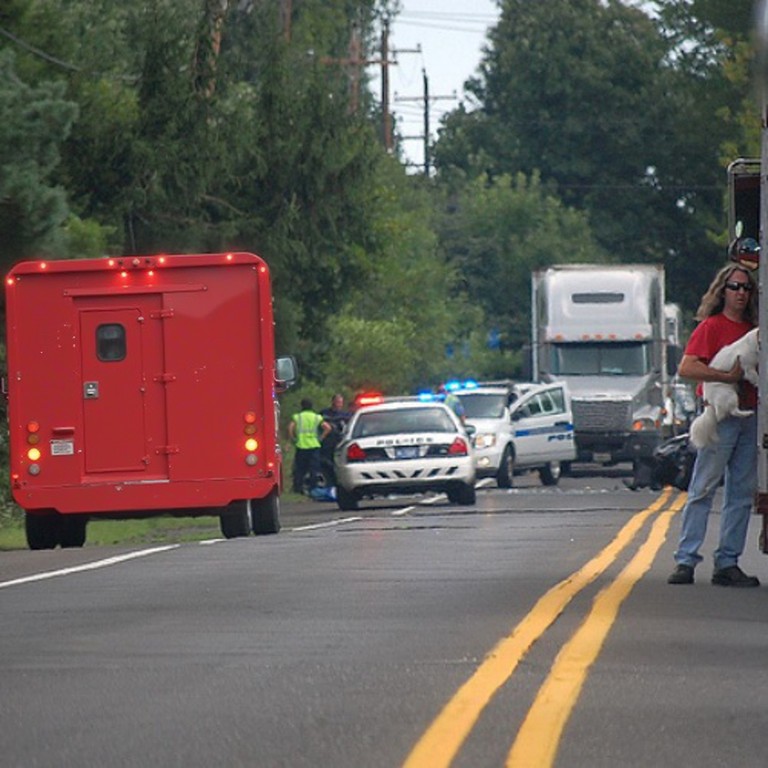} \\
    \multicolumn{9}{c}{\scriptsize\textbf{Replace:} Change the ambulance into a food truck.} \\
    [0.5em]
    \includegraphics[width=0.09\textwidth]{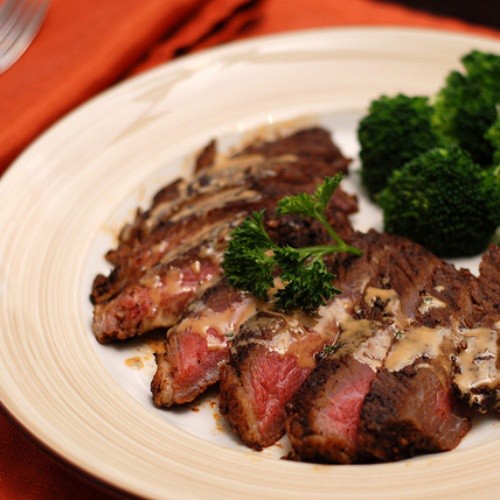} &
    \includegraphics[width=0.09\textwidth]{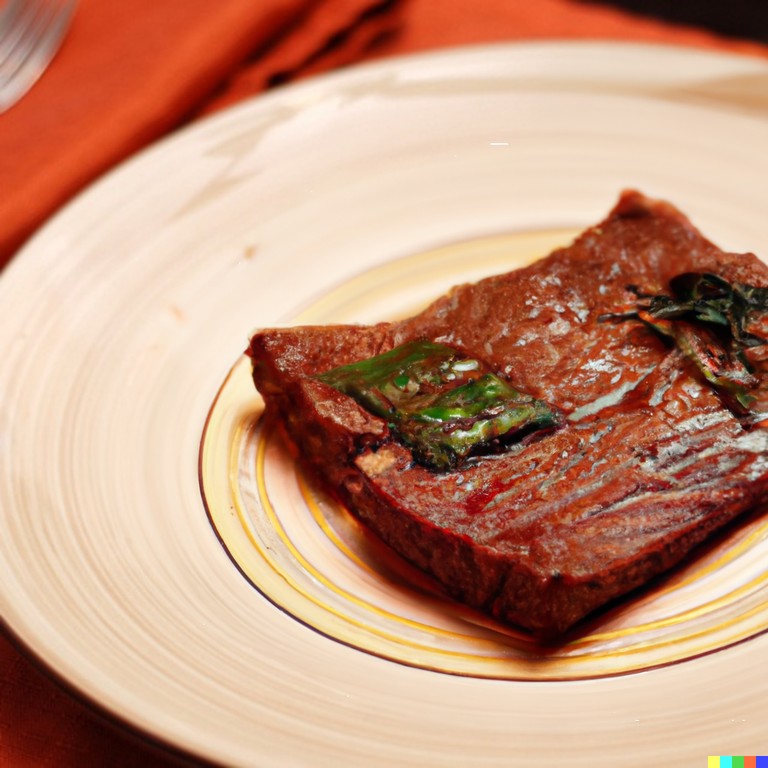} &
    \includegraphics[width=0.09\textwidth]{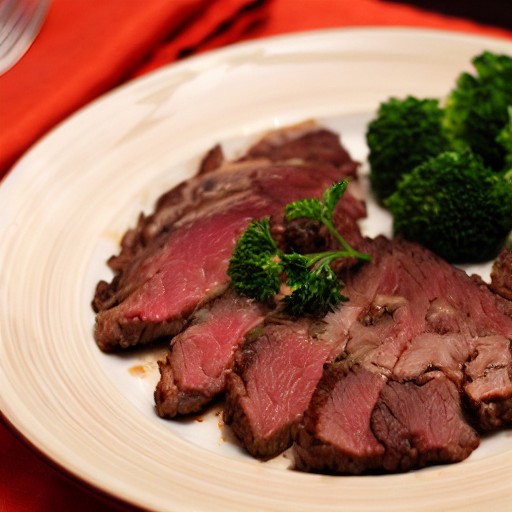} &
    \includegraphics[width=0.09\textwidth]{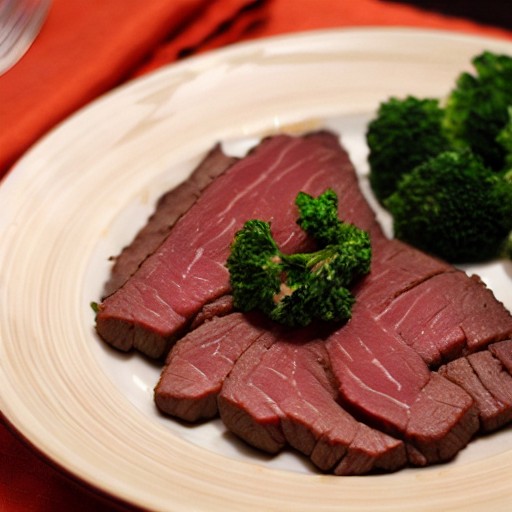} &
    \includegraphics[width=0.09\textwidth]{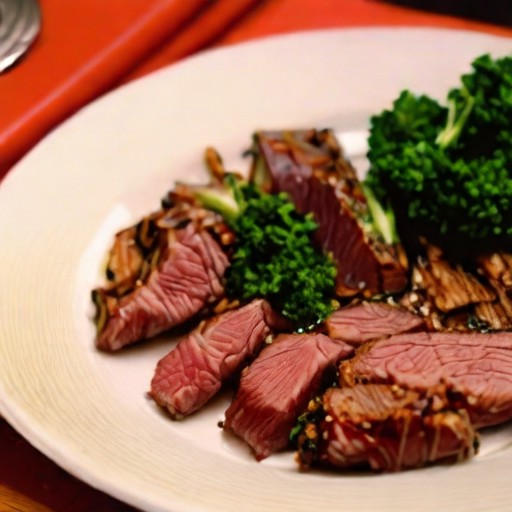} &
    \includegraphics[width=0.09\textwidth]{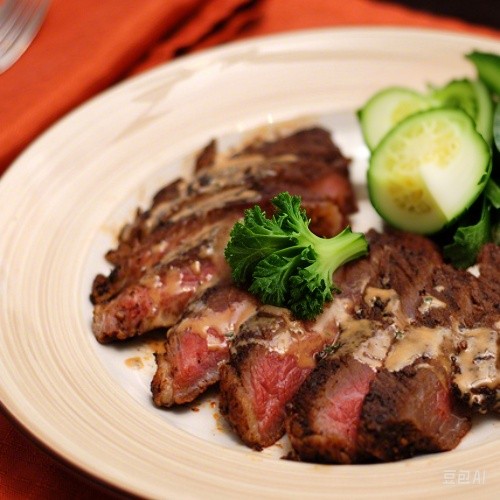} &
    \includegraphics[width=0.09\textwidth]{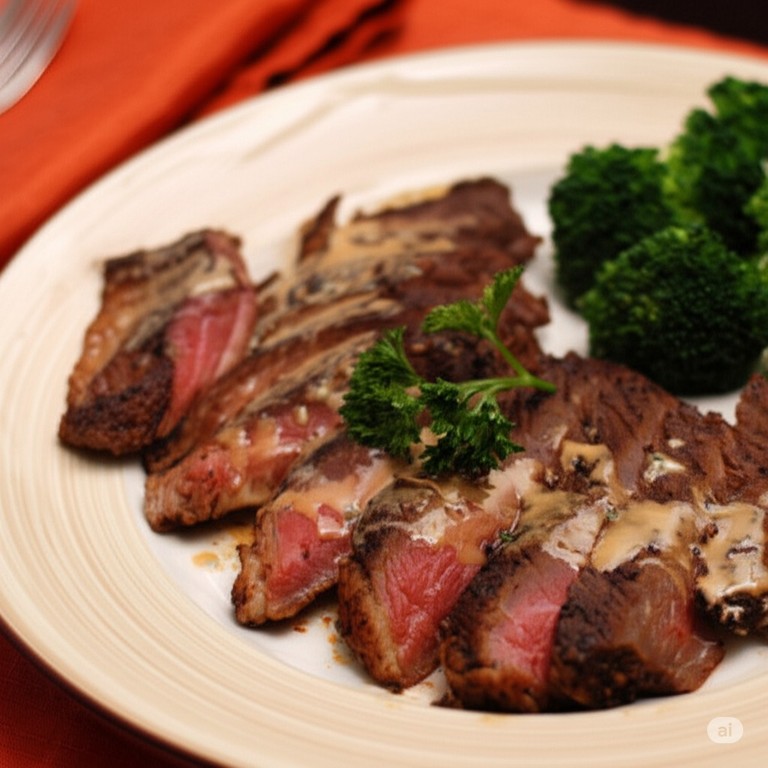} &
    \includegraphics[width=0.09\textwidth]{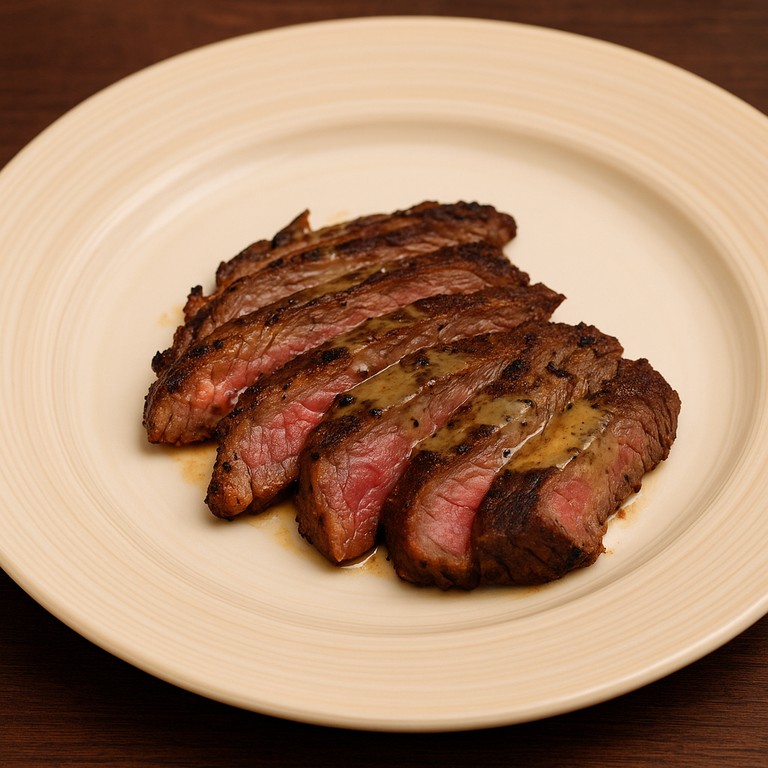} &
    \includegraphics[width=0.09\textwidth]{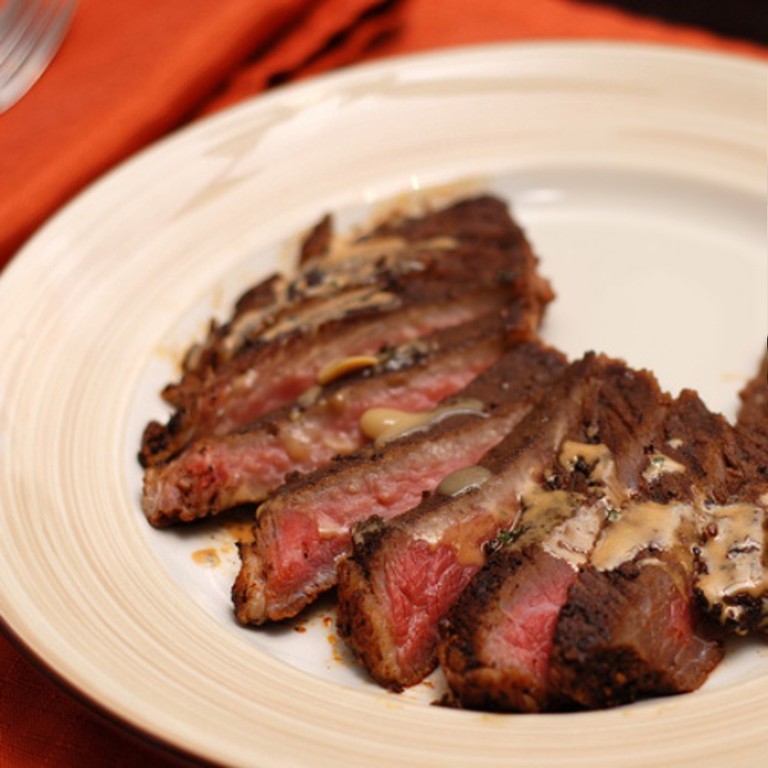} \\
    \multicolumn{9}{c}{\scriptsize\textbf{Remove:} Put just the beef on the plate.} \\
    [0.5em]
    \includegraphics[width=0.09\textwidth]{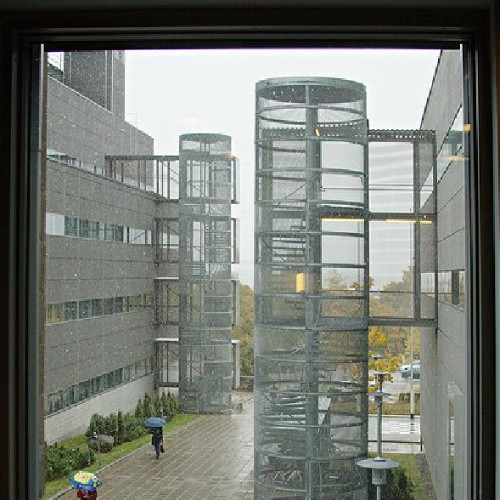} &
    \includegraphics[width=0.09\textwidth]{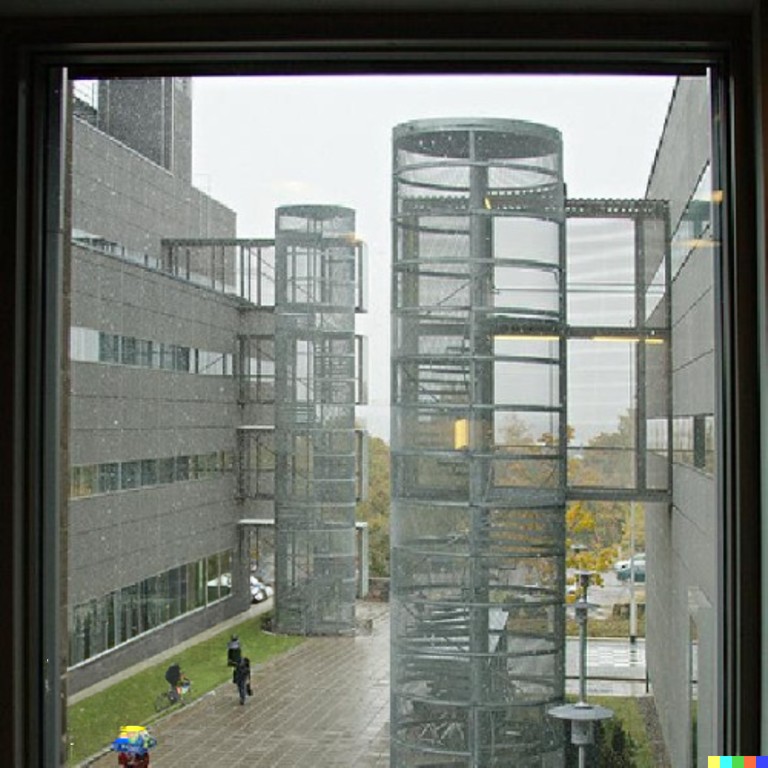} &
    \includegraphics[width=0.09\textwidth]{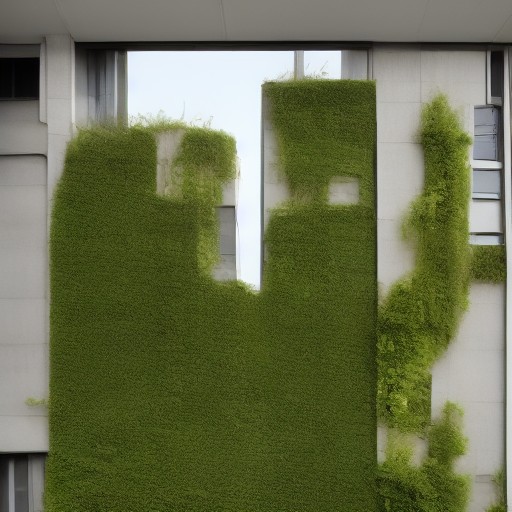} &
    \includegraphics[width=0.09\textwidth]{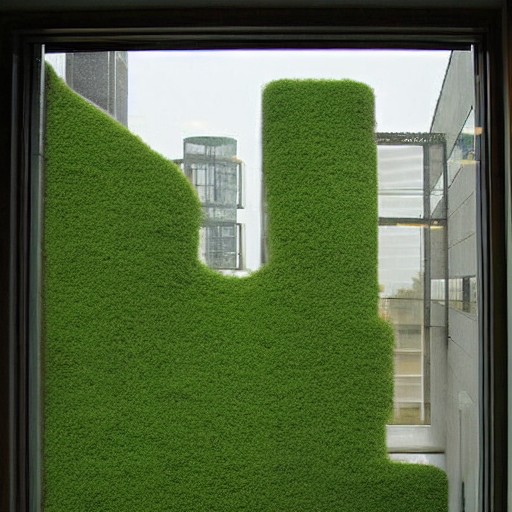} &
    \includegraphics[width=0.09\textwidth]{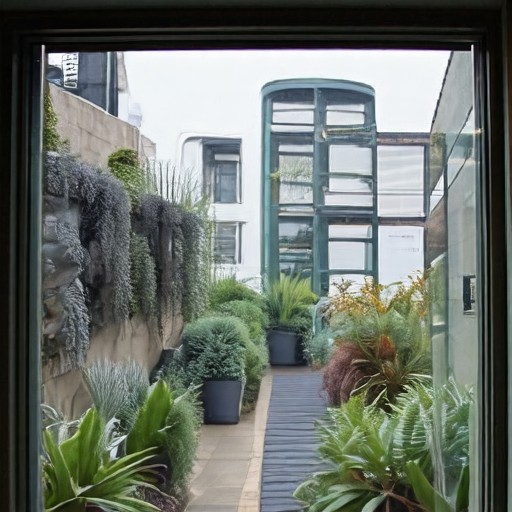} &
    \includegraphics[width=0.09\textwidth]{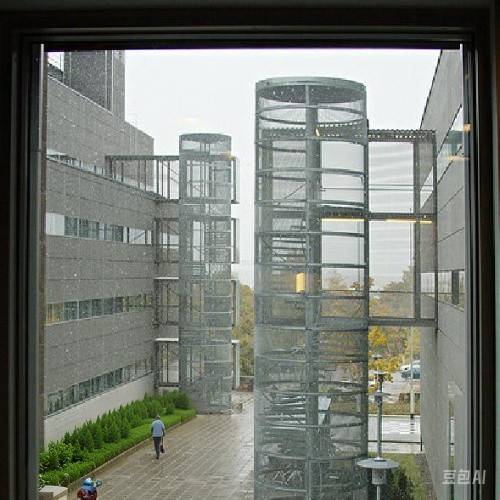} &
    \includegraphics[width=0.09\textwidth]{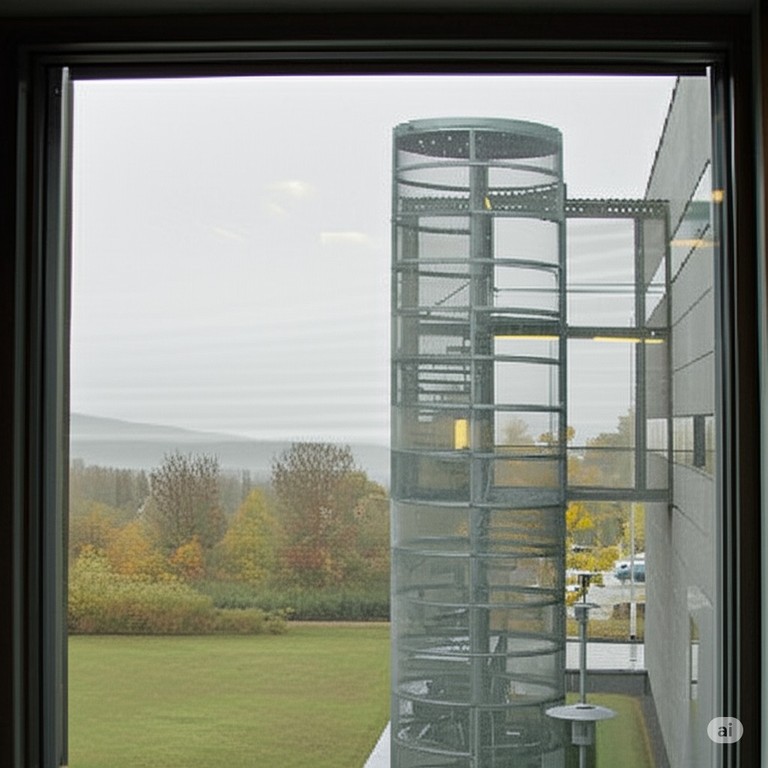} &
    \includegraphics[width=0.09\textwidth]{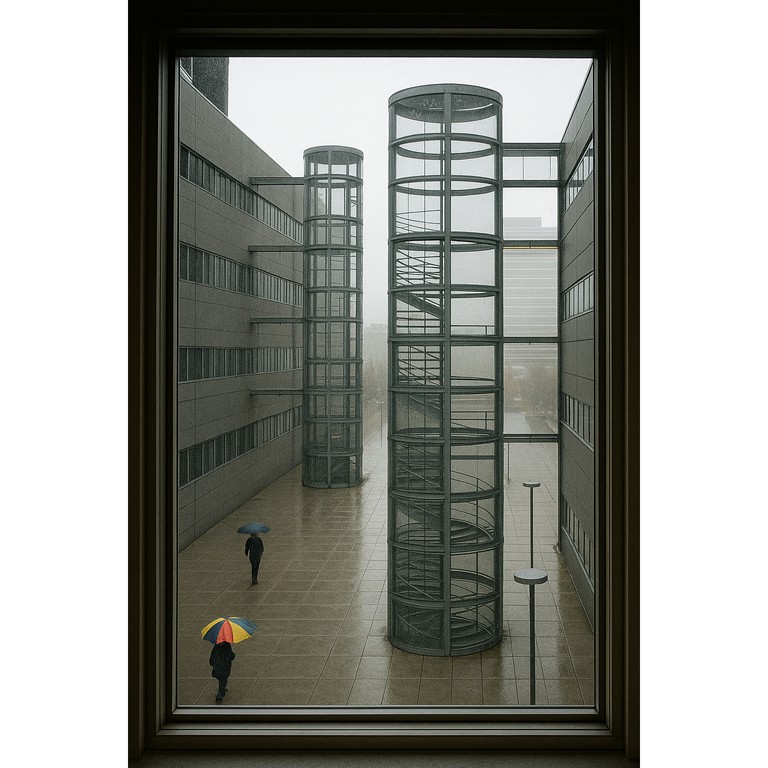} &
    \includegraphics[width=0.09\textwidth]{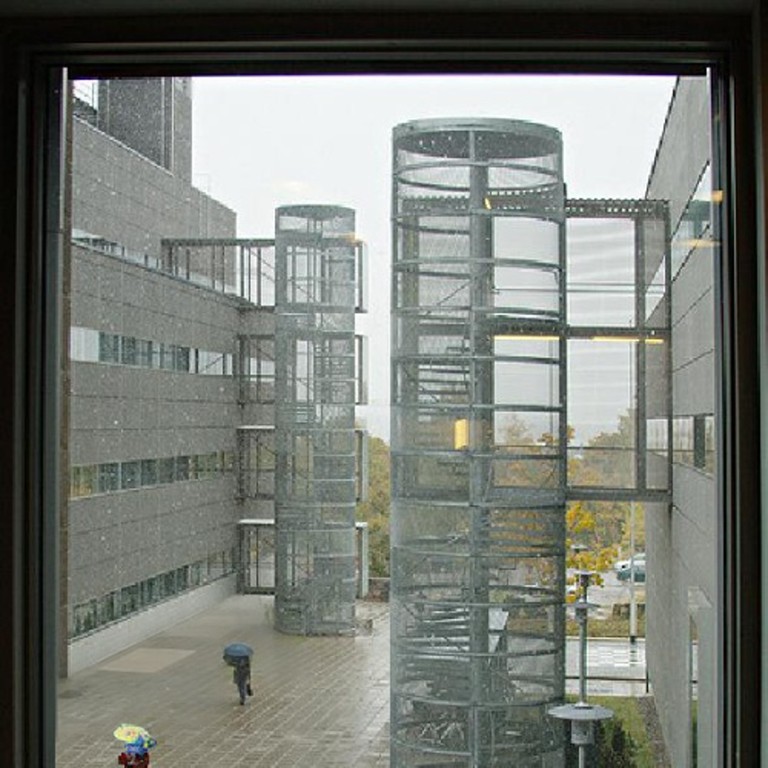} \\
    \multicolumn{9}{c}{\scriptsize\textbf{Remove:} What if there was no plant life along the building.} \\
    [0.5em]
    \includegraphics[width=0.09\textwidth]{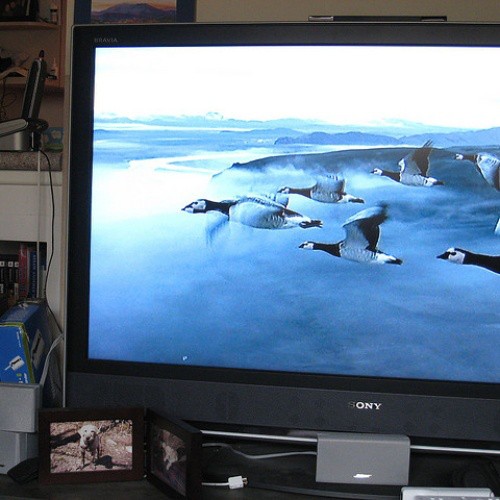} &
    \includegraphics[width=0.09\textwidth]{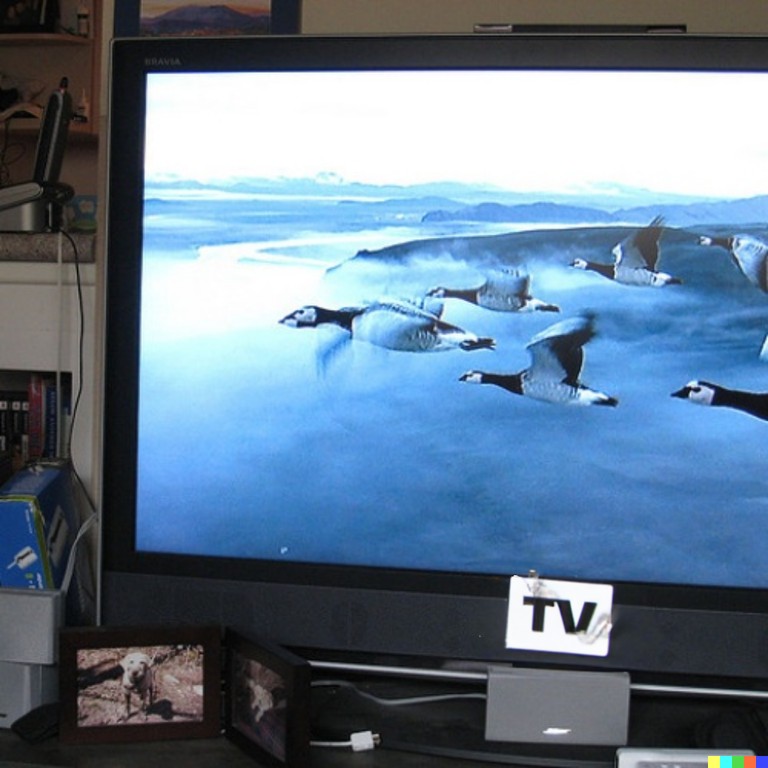} &
    \includegraphics[width=0.09\textwidth]{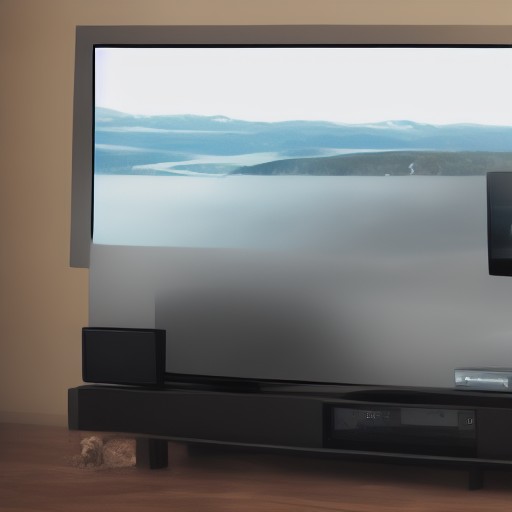} &
    \includegraphics[width=0.09\textwidth]{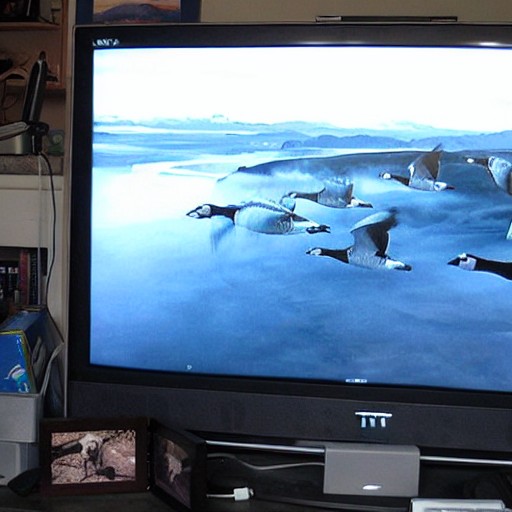} &
    \includegraphics[width=0.09\textwidth]{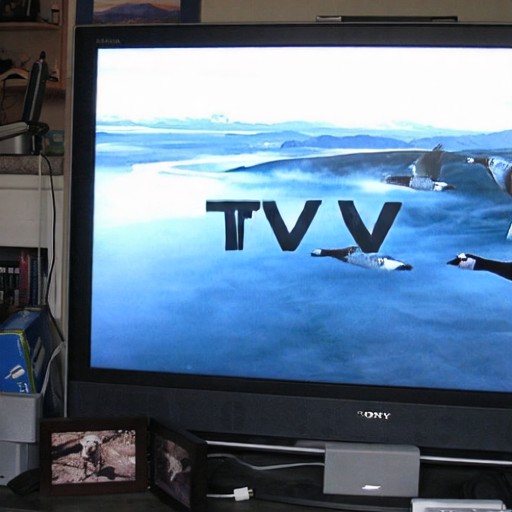} &
    \includegraphics[width=0.09\textwidth]{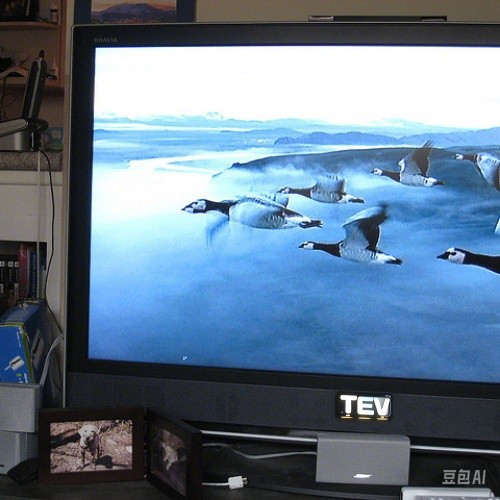} &
    \includegraphics[width=0.09\textwidth]{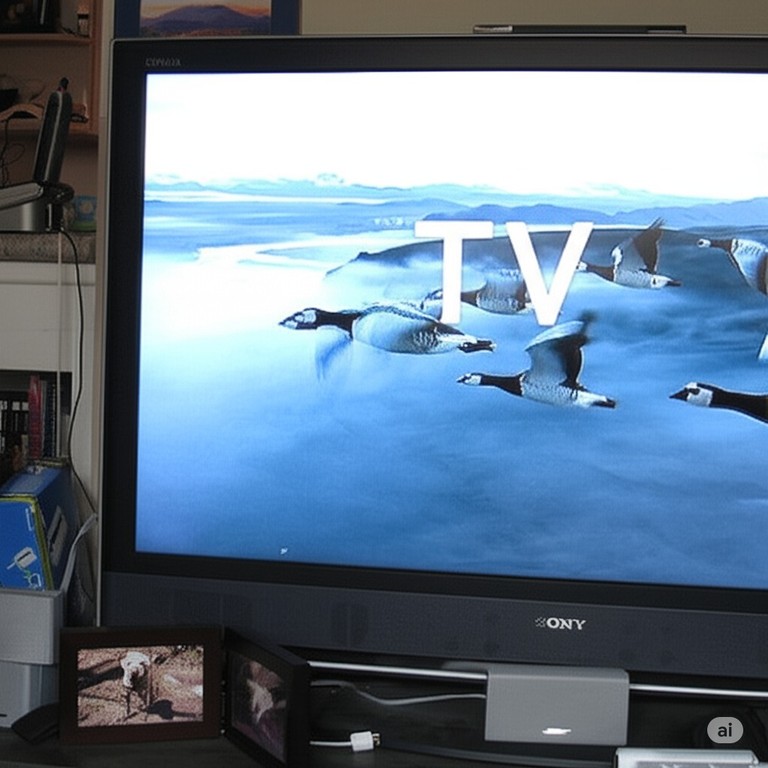} &
    \includegraphics[width=0.09\textwidth]{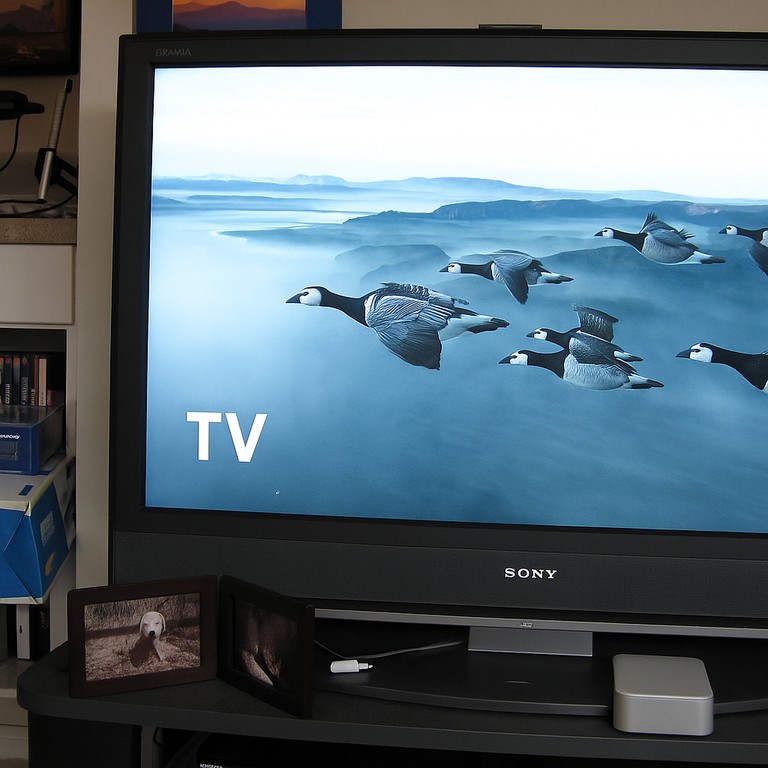} &
    \includegraphics[width=0.09\textwidth]{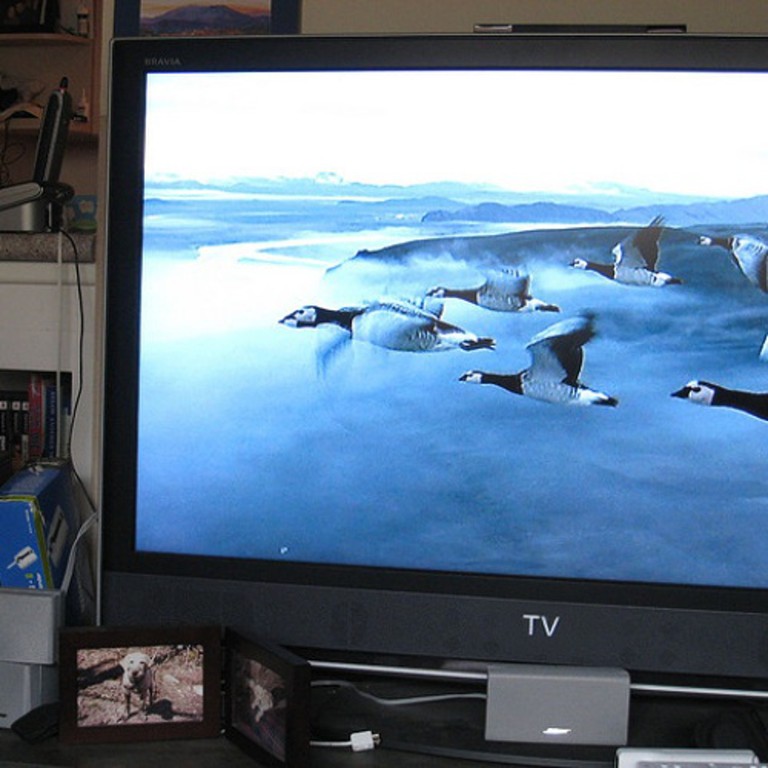} \\
    \multicolumn{9}{c}{\scriptsize\textbf{Text/Pattern:} Change the text on the television to ``TV''.} \\
\end{tabular}
\caption{\textbf{Our workflow performs the best on the first half of 14 challenging examples from MagicBrush.}}
\label{fig:challenging-magicbrush-first-half}
\end{center}

\end{figure*}

\begin{figure*}[htbp]

\begin{center}
\small
\begin{tabular}{c@{\hspace{0.3em}}c@{\hspace{0.3em}}c@{\hspace{0.3em}}c@{\hspace{0.3em}}c@{\hspace{0.3em}}c@{\hspace{0.3em}}c@{\hspace{0.3em}}c@{\hspace{0.3em}}c}
    \scriptsize\textsc{Original} & \scriptsize\textsc{Reference} & \scriptsize\textsc{IP2P} & \scriptsize\textsc{MB} & \scriptsize\textsc{UE} & \scriptsize\textsc{Doubao} & \scriptsize\textsc{Gemini} & \scriptsize\textsc{GPT-4o} & \scriptsize\textsc{\textbf{Ours}} \\
    
    \includegraphics[width=0.09\textwidth]{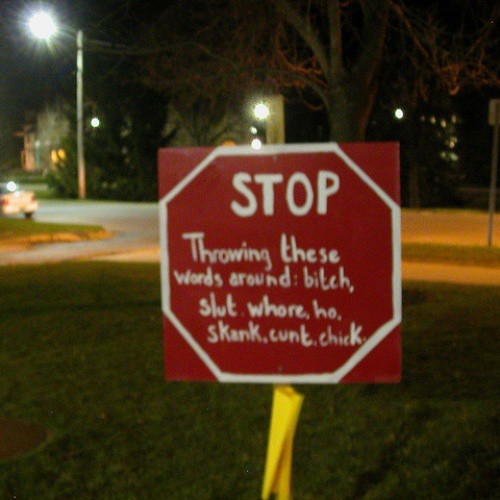} &
    \includegraphics[width=0.09\textwidth]{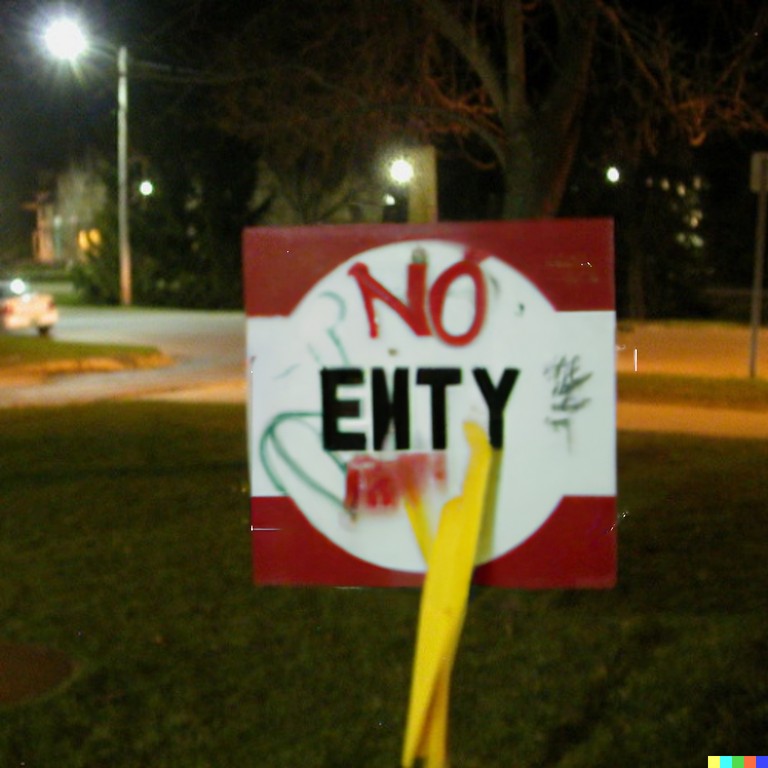} &
    \includegraphics[width=0.09\textwidth]{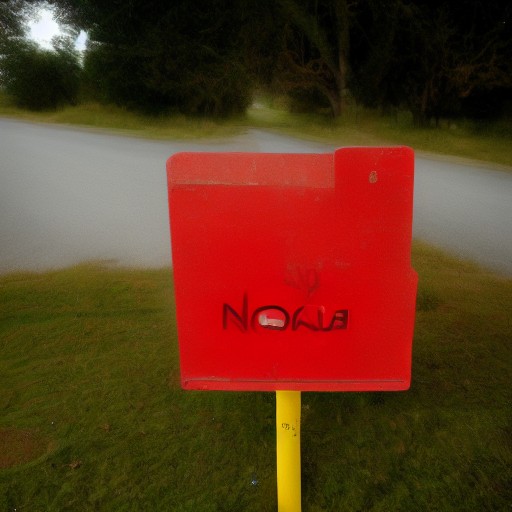} &
    \includegraphics[width=0.09\textwidth]{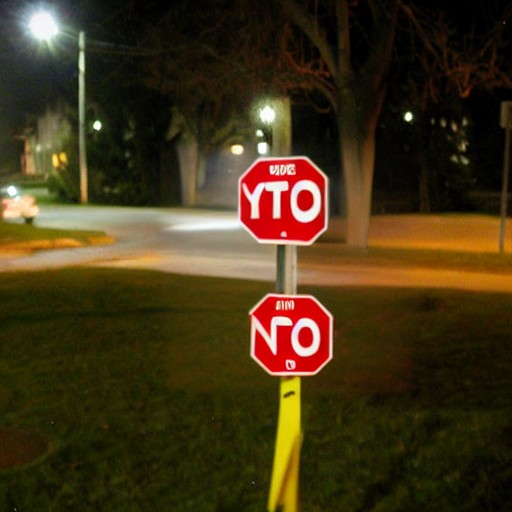} &
    \includegraphics[width=0.09\textwidth]{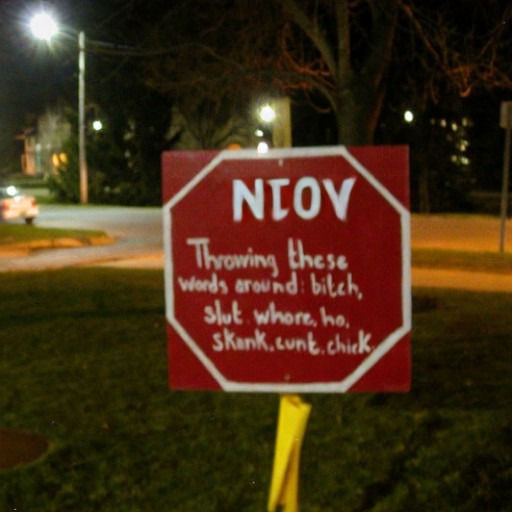} &
    \includegraphics[width=0.09\textwidth]{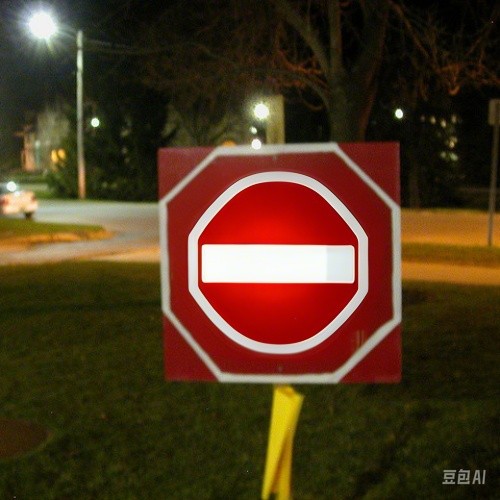} &
    \includegraphics[width=0.09\textwidth]{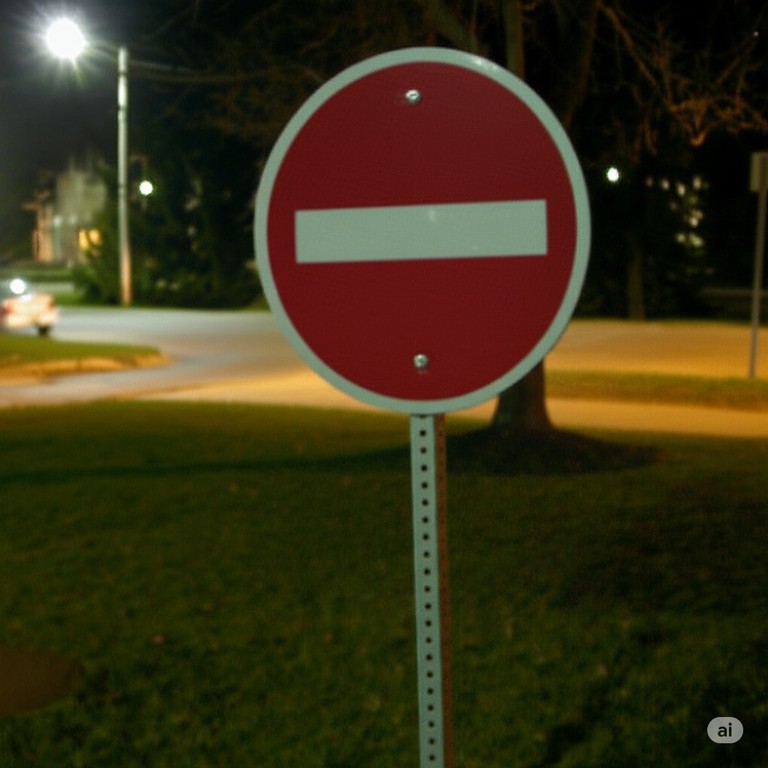} &
    \includegraphics[width=0.09\textwidth]{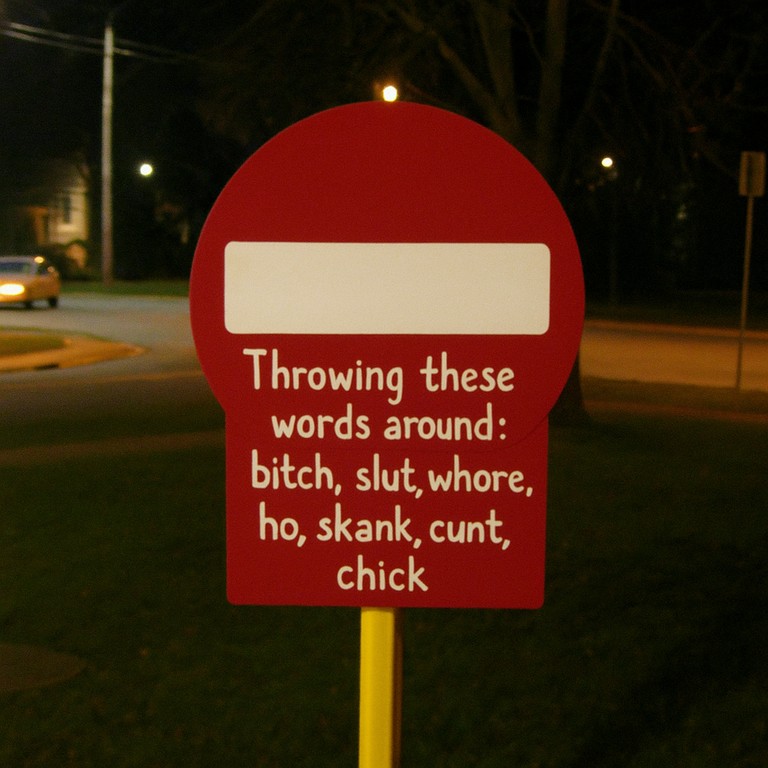} &
    \includegraphics[width=0.09\textwidth]{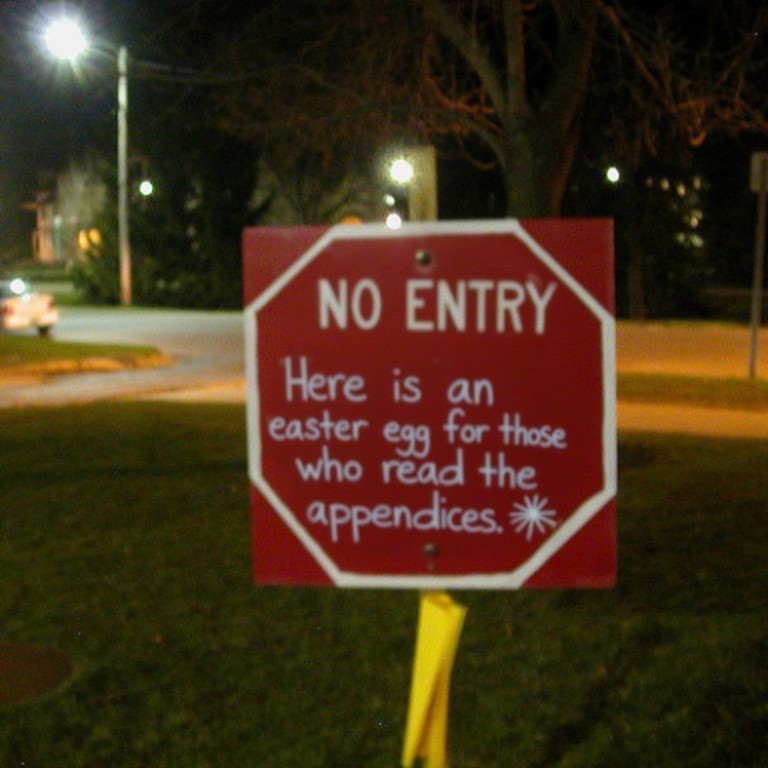} \\
    \multicolumn{9}{c}{\scriptsize\textbf{Text/Pattern:} Change the stop sign into a no entry sign.} \\
    [0.5em]
    \includegraphics[width=0.09\textwidth]{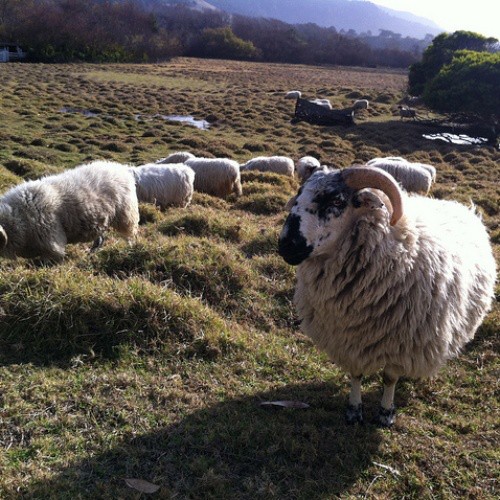} &
    \includegraphics[width=0.09\textwidth]{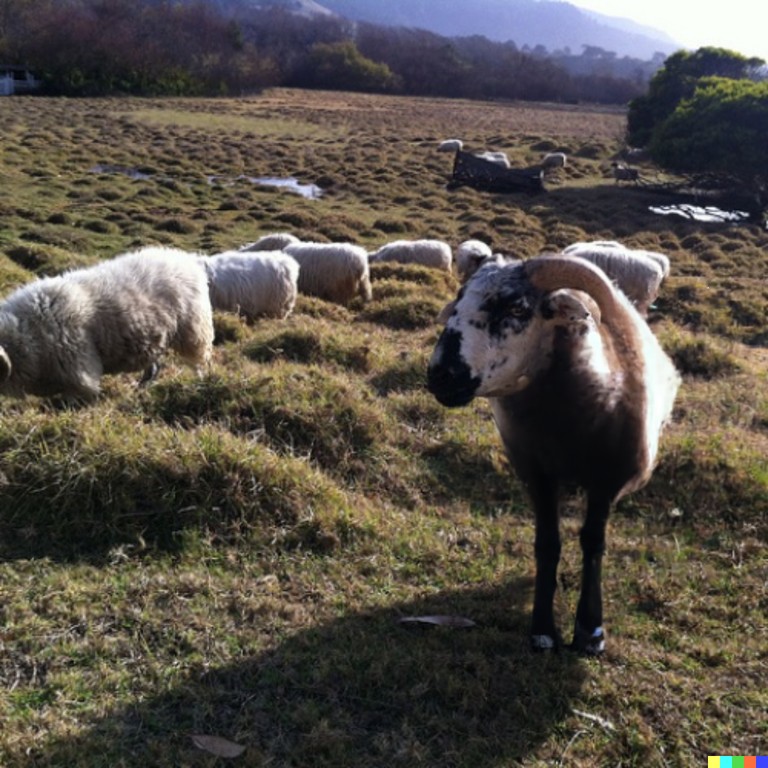} &
    \includegraphics[width=0.09\textwidth]{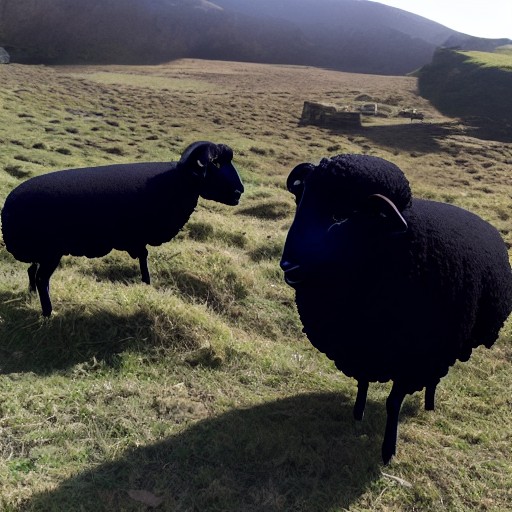} &
    \includegraphics[width=0.09\textwidth]{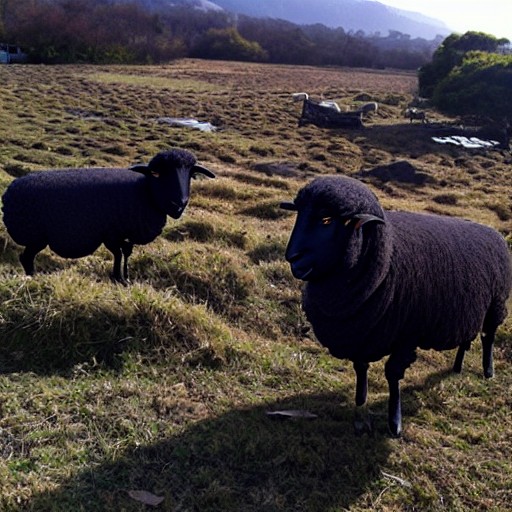} &
    \includegraphics[width=0.09\textwidth]{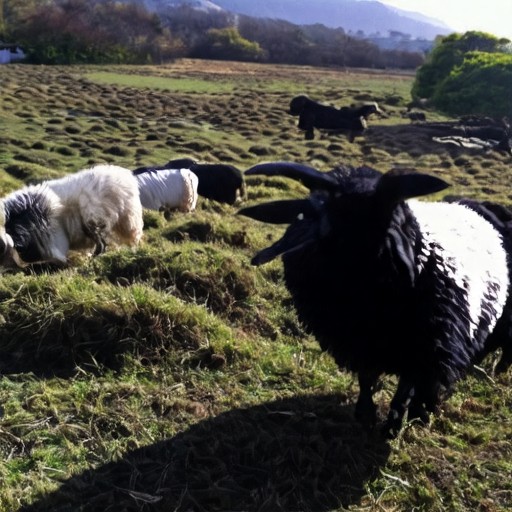} &
    \includegraphics[width=0.09\textwidth]{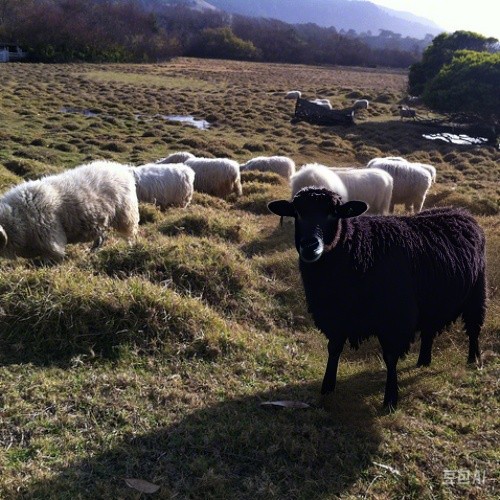} &
    \includegraphics[width=0.09\textwidth]{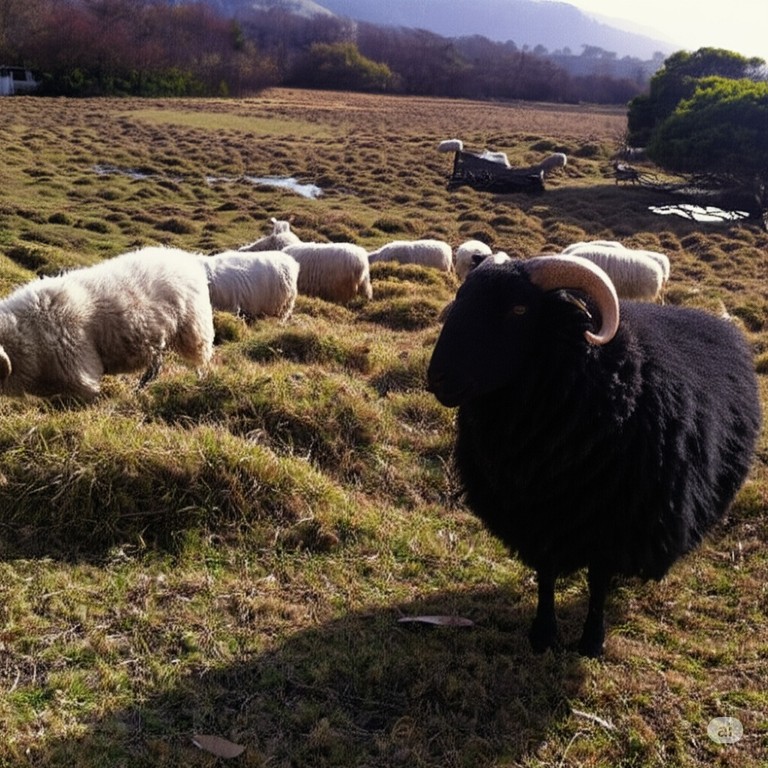} &
    \includegraphics[width=0.09\textwidth]{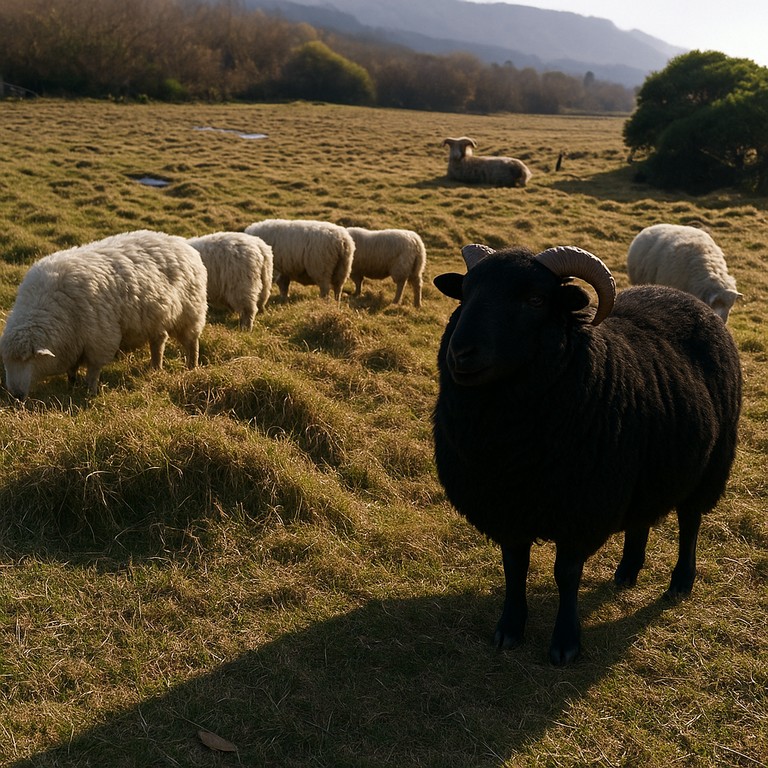} &
    \includegraphics[width=0.09\textwidth]{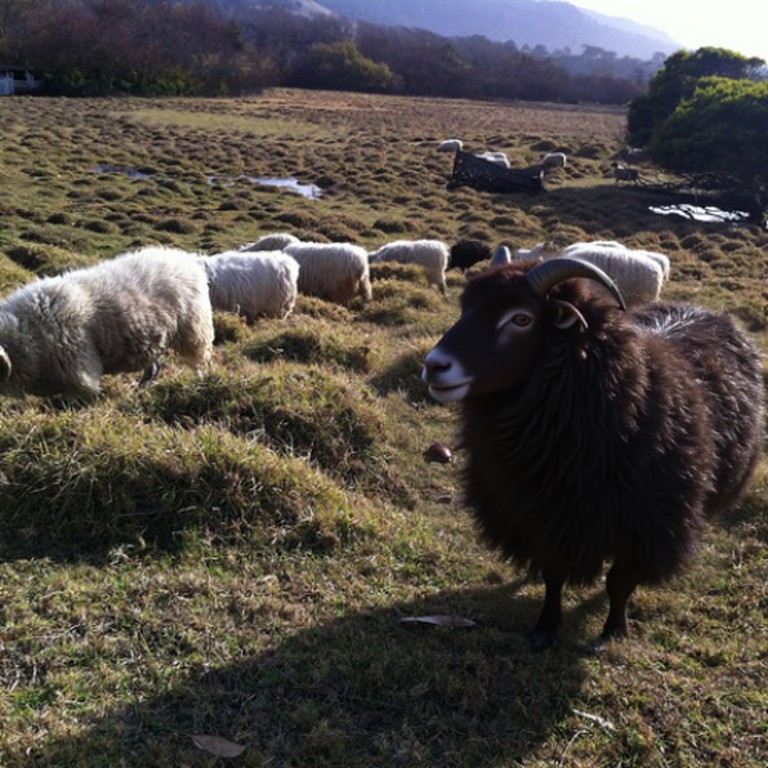} \\
    \multicolumn{9}{c}{\scriptsize\textbf{Color:} Make one of the sheep a black sheep.} \\
    [0.5em]
    \includegraphics[width=0.09\textwidth]{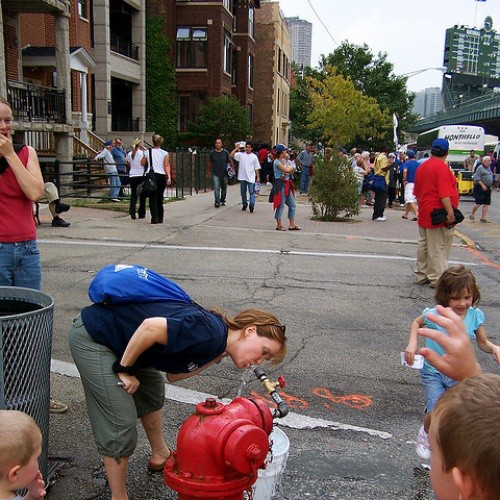} &
    \includegraphics[width=0.09\textwidth]{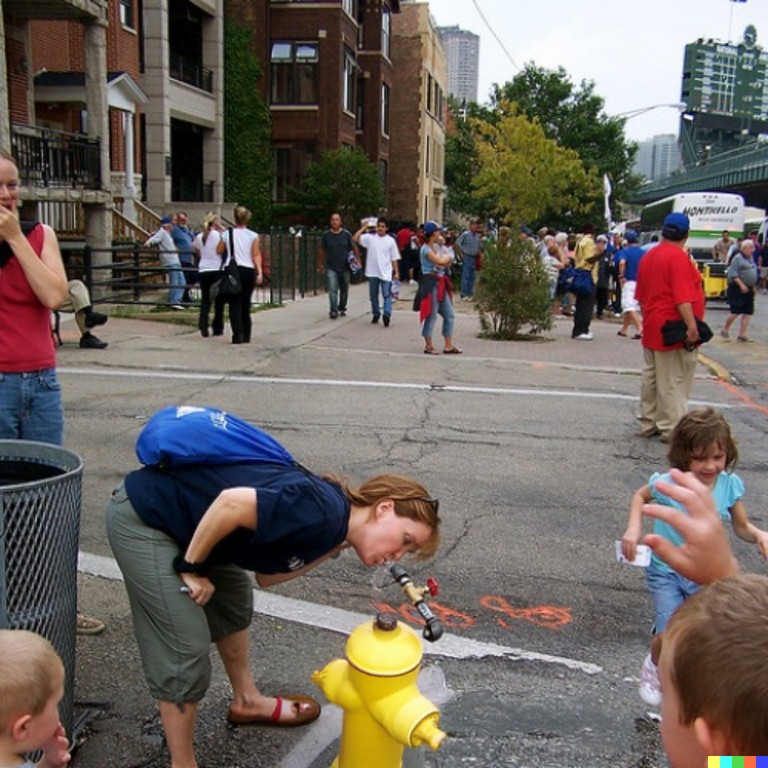} &
    \includegraphics[width=0.09\textwidth]{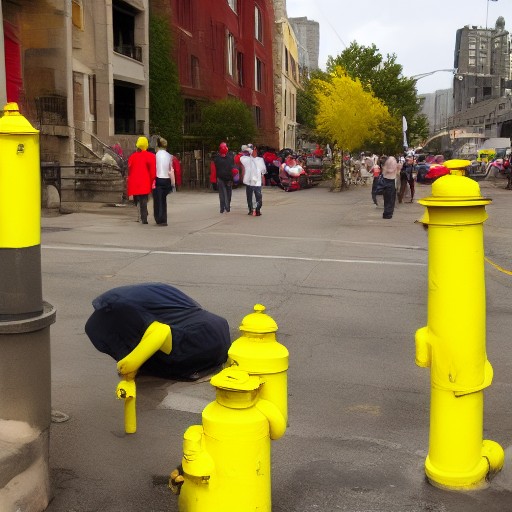} &
    \includegraphics[width=0.09\textwidth]{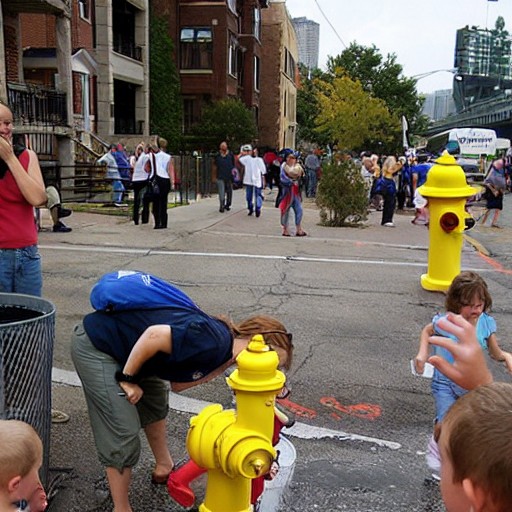} &
    \includegraphics[width=0.09\textwidth]{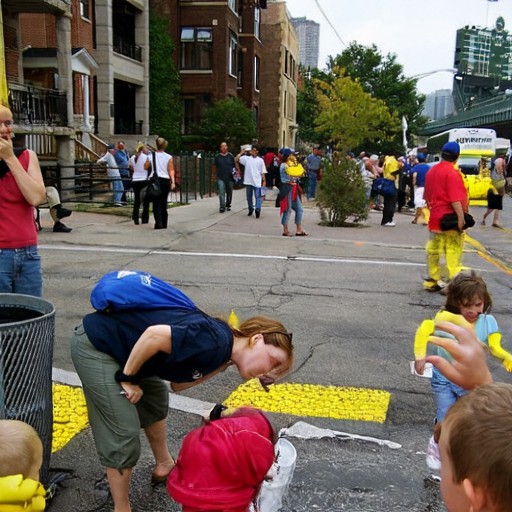} &
    \includegraphics[width=0.09\textwidth]{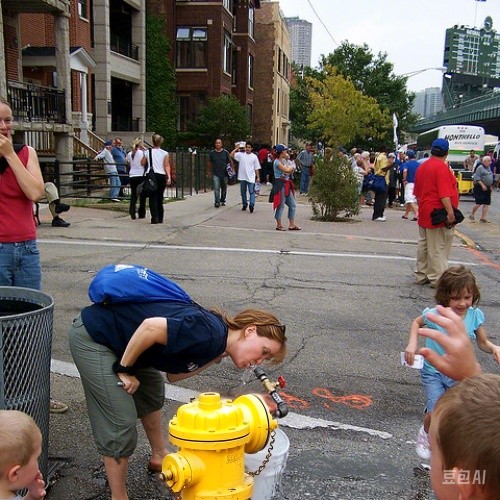} &
    \includegraphics[width=0.09\textwidth]{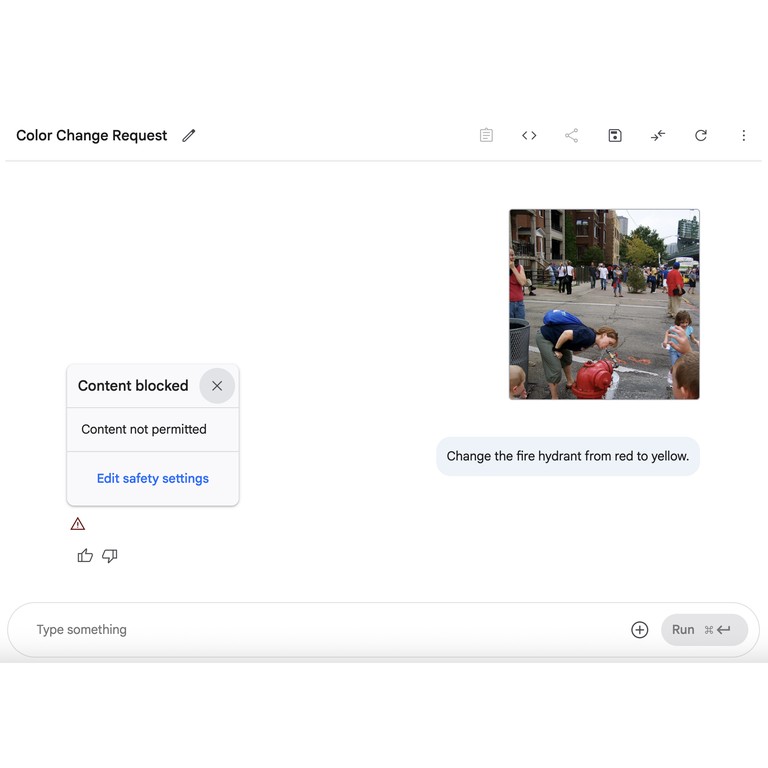} &
    \includegraphics[width=0.09\textwidth]{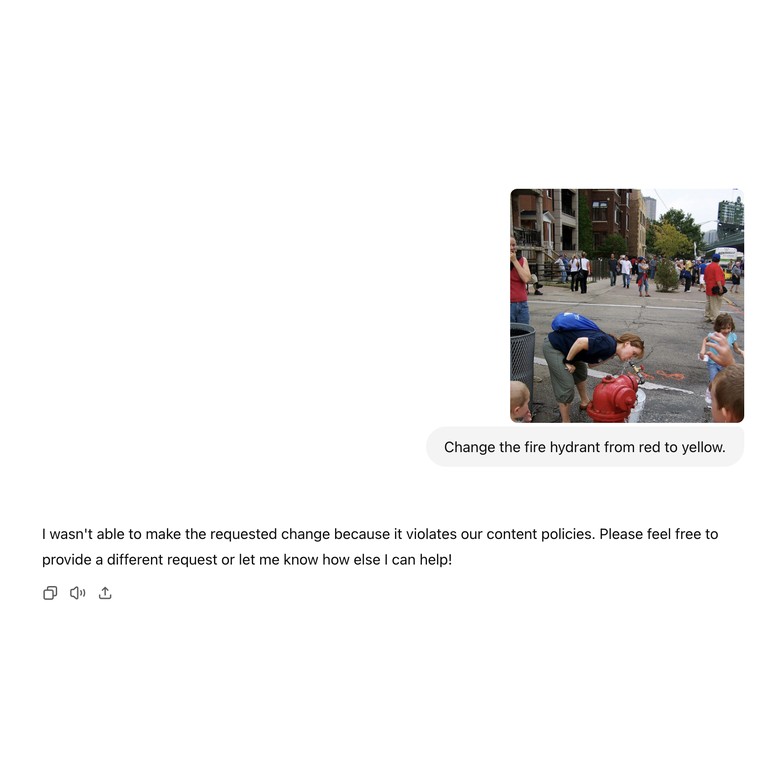} &
    \includegraphics[width=0.09\textwidth]{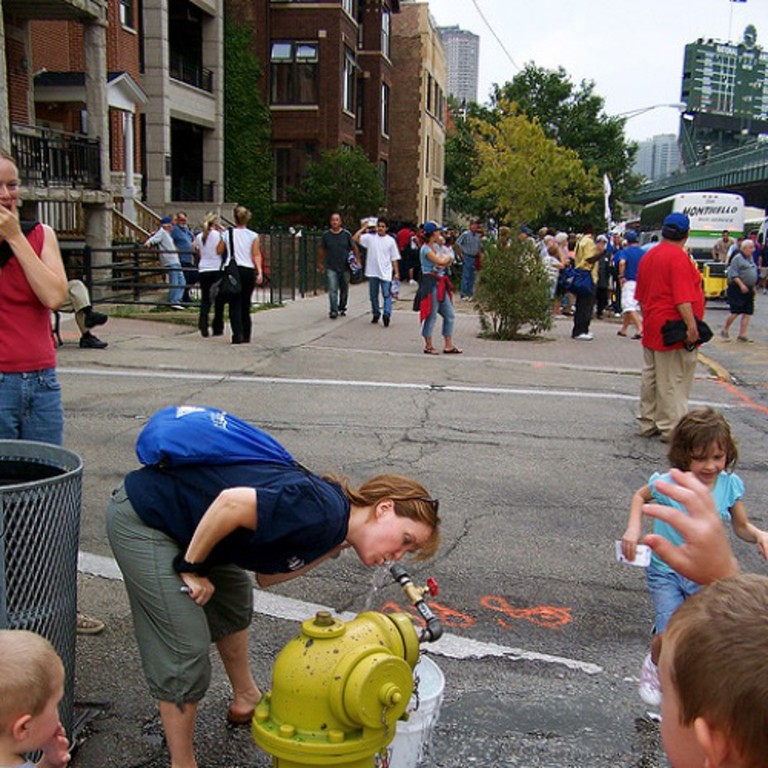} \\
    \multicolumn{9}{c}{\scriptsize\textbf{Color:} Change the fire hydrant from red to yellow.} \\
    [0.5em]
    \includegraphics[width=0.09\textwidth]{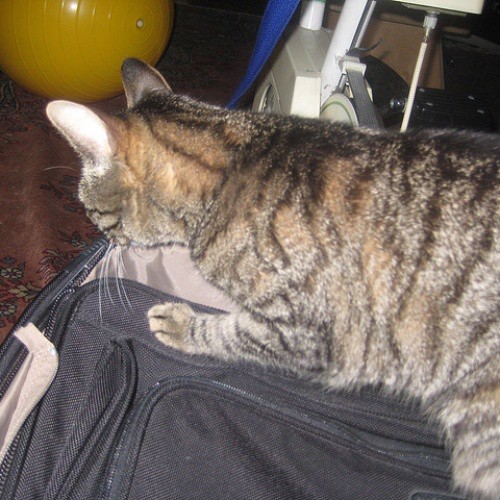} &
    \includegraphics[width=0.09\textwidth]{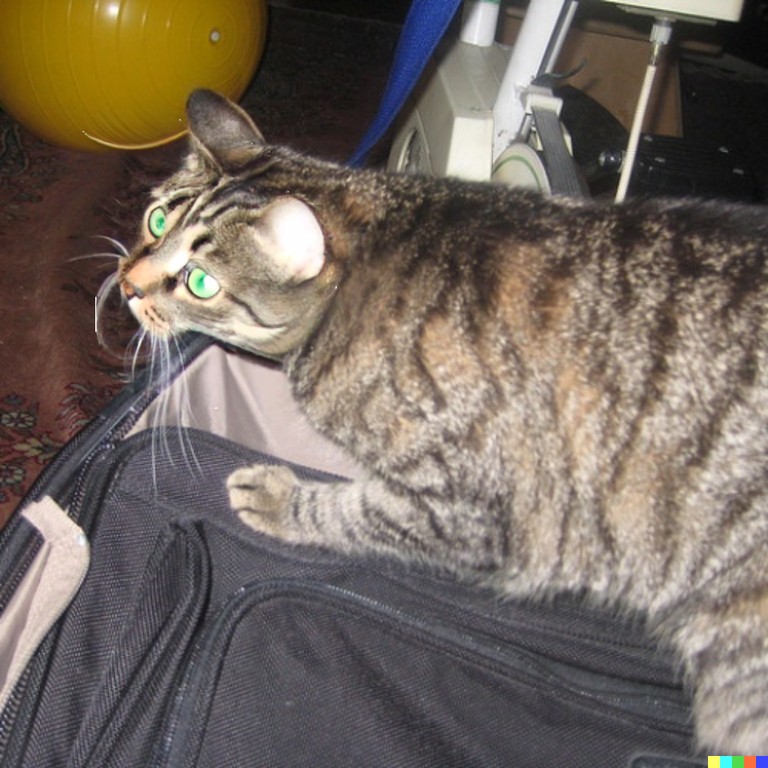} &
    \includegraphics[width=0.09\textwidth]{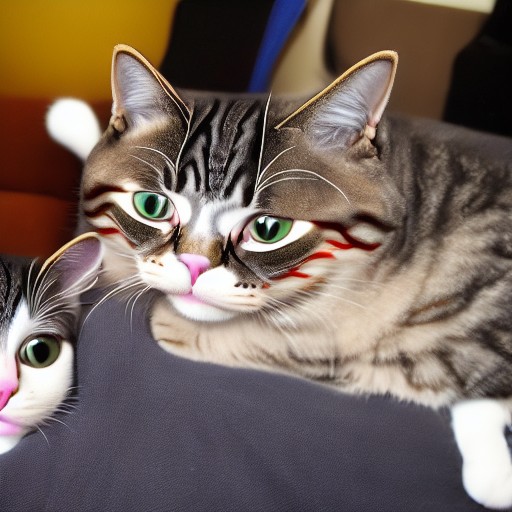} &
    \includegraphics[width=0.09\textwidth]{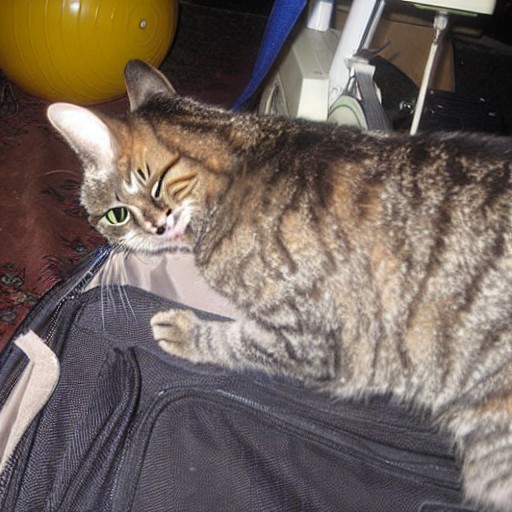} &
    \includegraphics[width=0.09\textwidth]{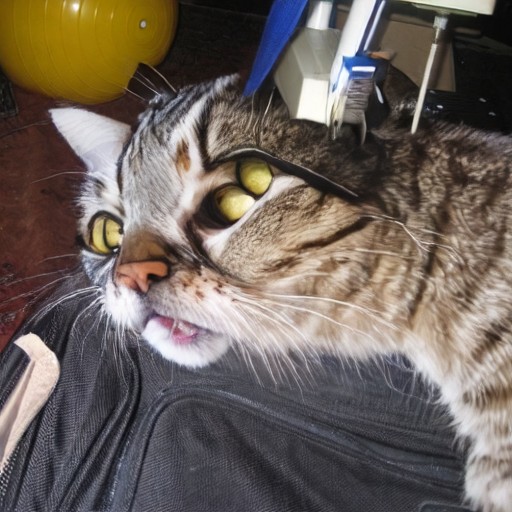} &
    \includegraphics[width=0.09\textwidth]{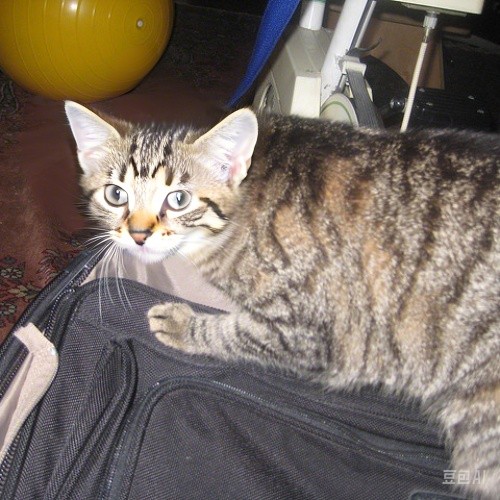} &
    \includegraphics[width=0.09\textwidth]{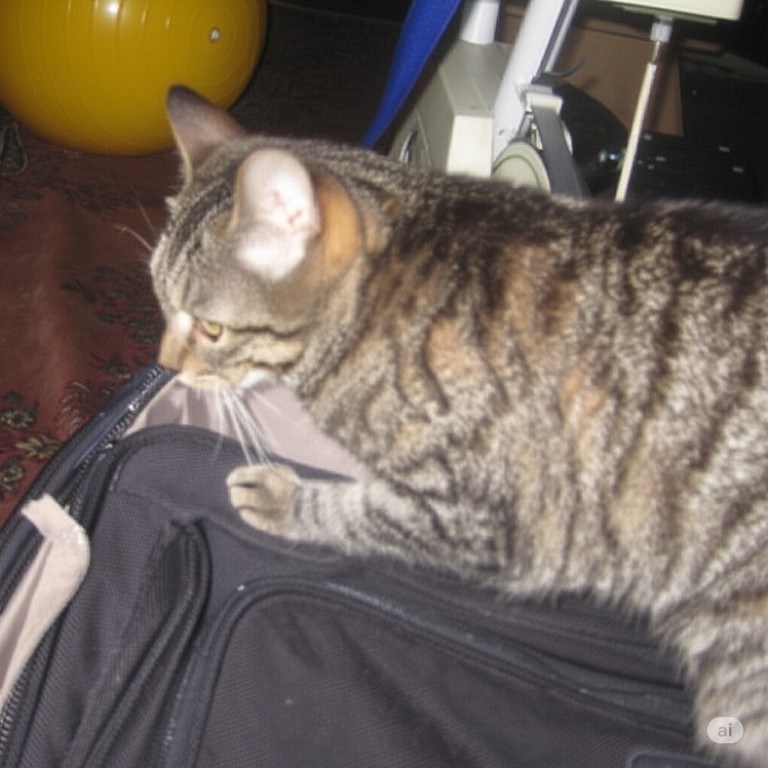} &
    \includegraphics[width=0.09\textwidth]{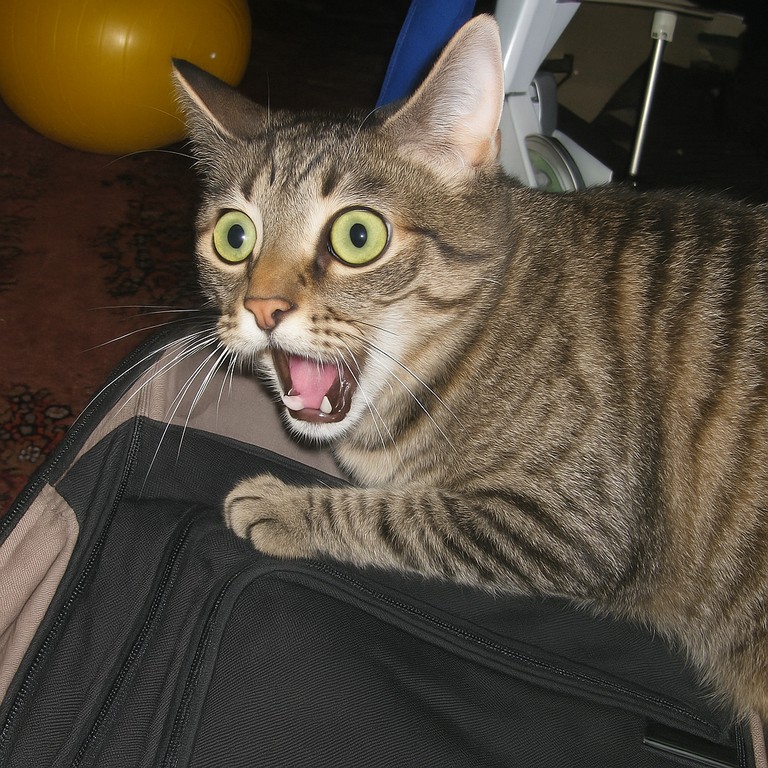} &
    \includegraphics[width=0.09\textwidth]{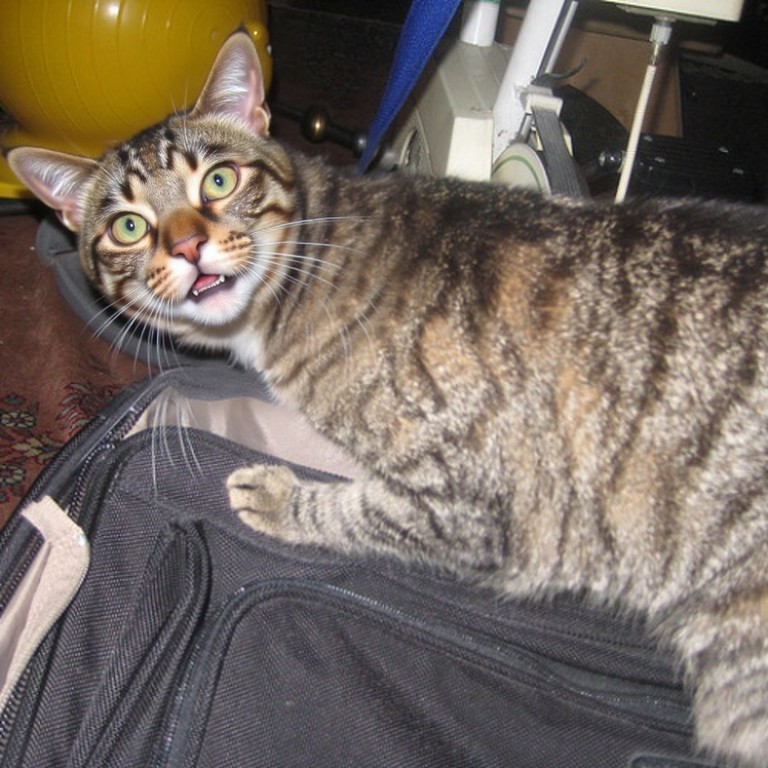} \\
    \multicolumn{9}{c}{\scriptsize\textbf{Action:} Let the cat look shocked.} \\
    [0.5em]
    \includegraphics[width=0.09\textwidth]{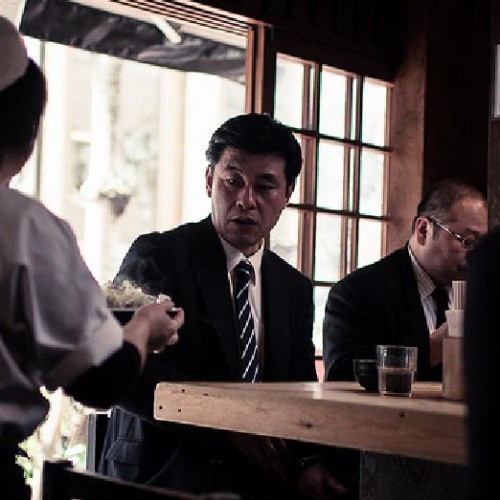} &
    \includegraphics[width=0.09\textwidth]{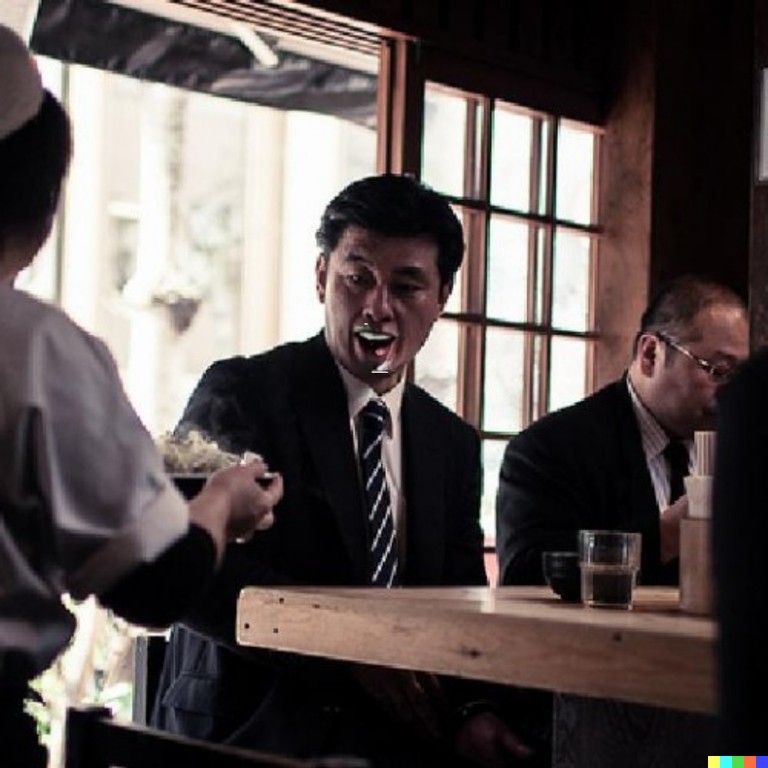} &
    \includegraphics[width=0.09\textwidth]{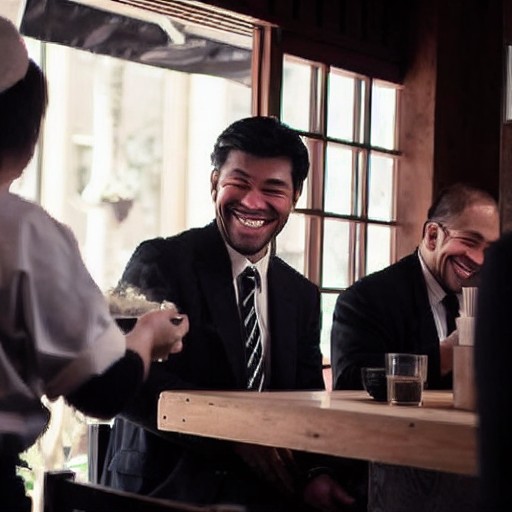} &
    \includegraphics[width=0.09\textwidth]{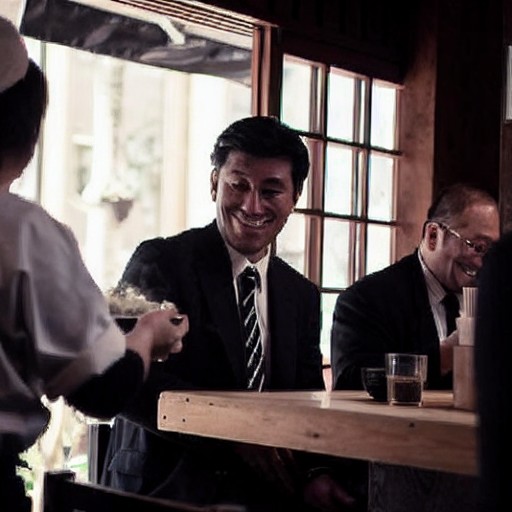} &
    \includegraphics[width=0.09\textwidth]{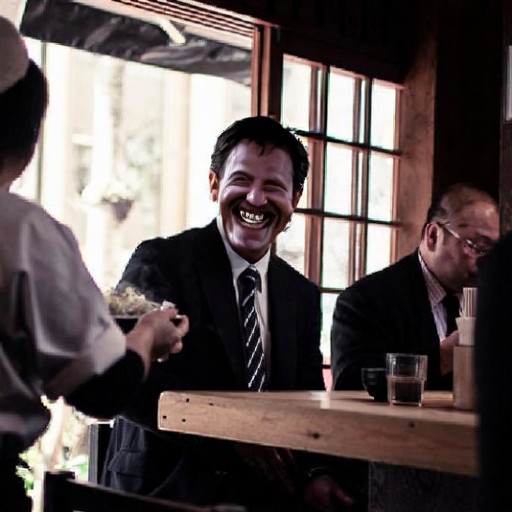} &
    \includegraphics[width=0.09\textwidth]{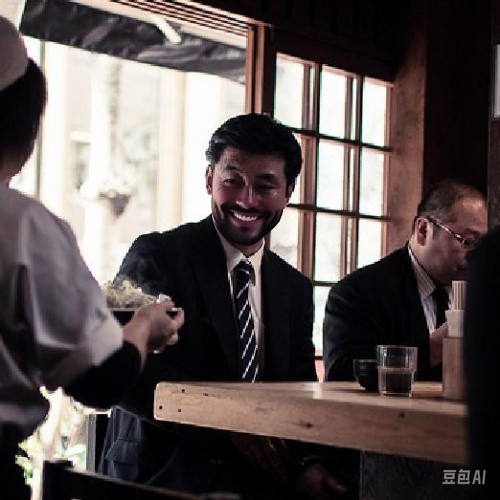} &
    \includegraphics[width=0.09\textwidth]{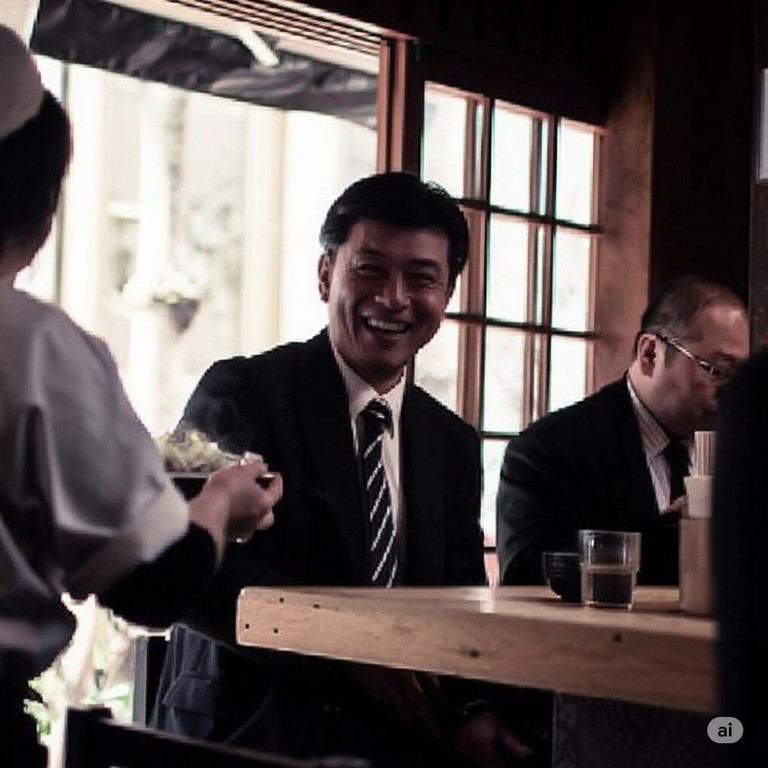} &
    \includegraphics[width=0.09\textwidth]{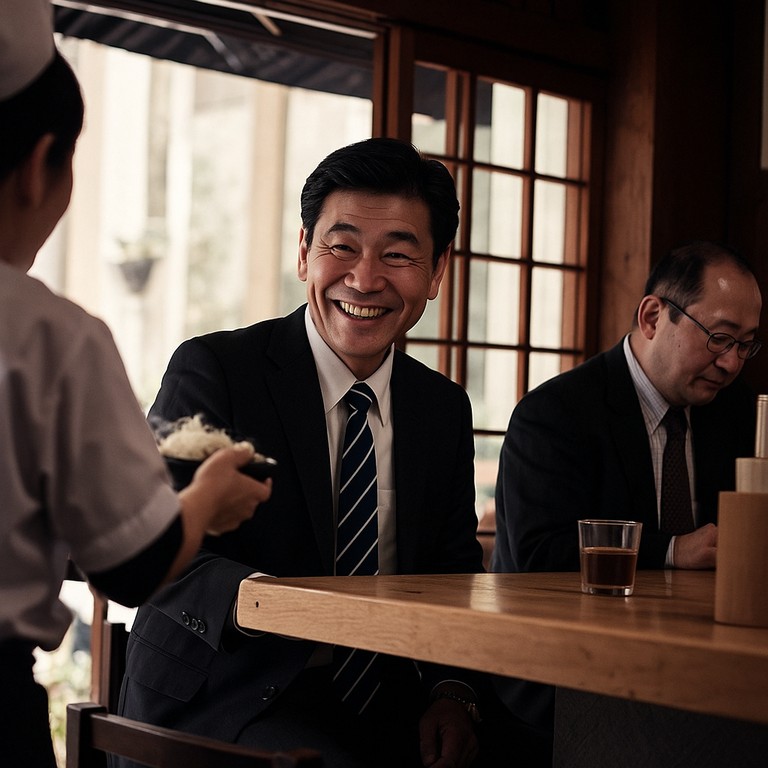} &
    \includegraphics[width=0.09\textwidth]{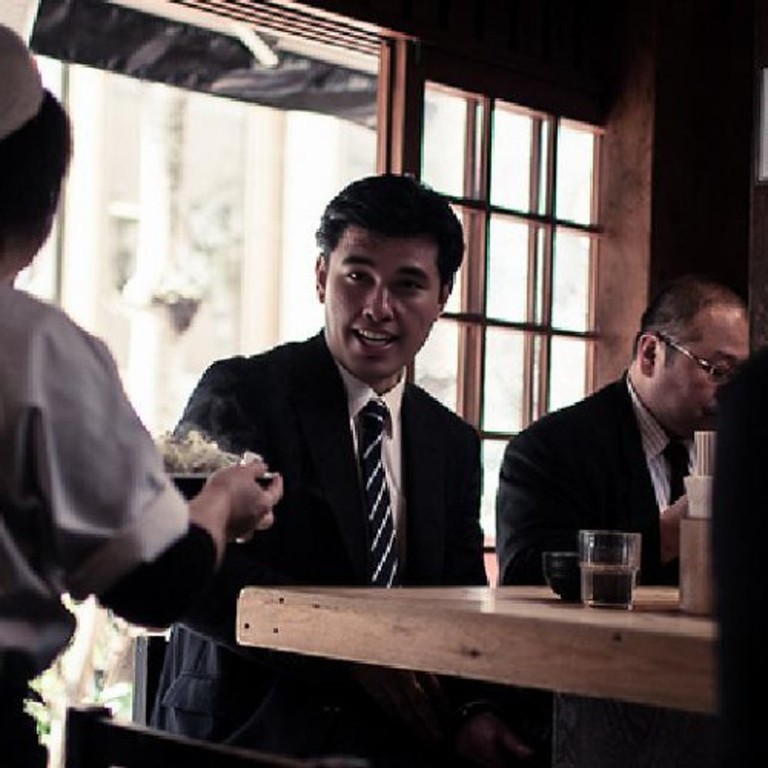} \\
    \multicolumn{9}{c}{\scriptsize\textbf{Action:} Make the man smile.} \\
    [0.5em]
    \includegraphics[width=0.09\textwidth]{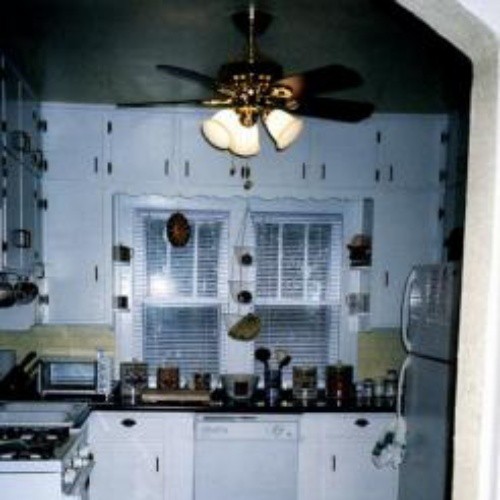} &
    \includegraphics[width=0.09\textwidth]{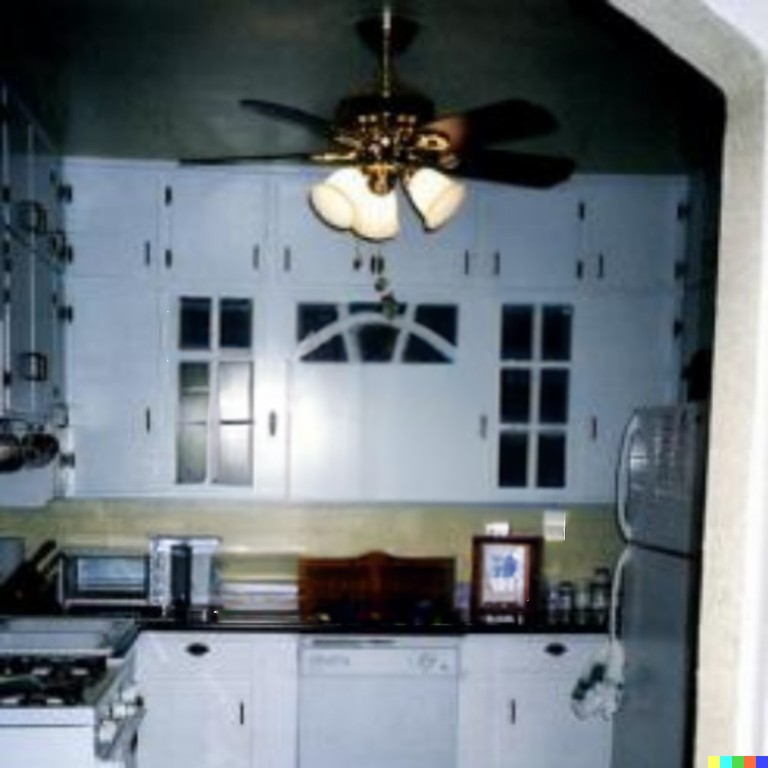} &
    \includegraphics[width=0.09\textwidth]{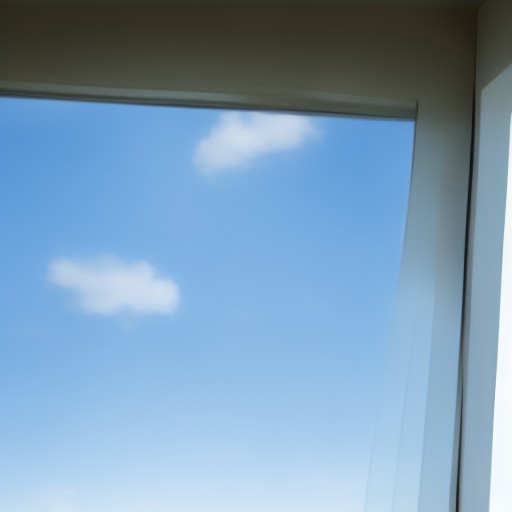} &
    \includegraphics[width=0.09\textwidth]{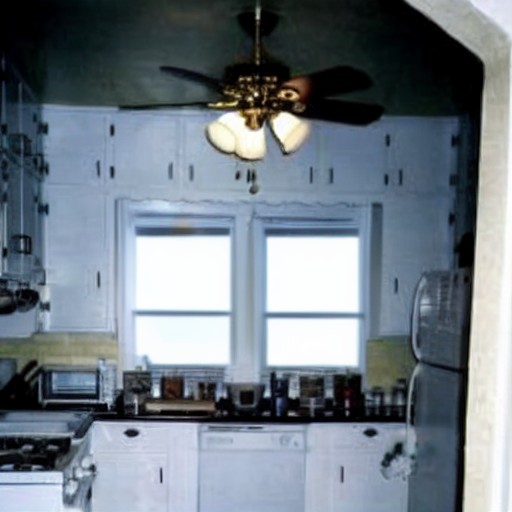} &
    \includegraphics[width=0.09\textwidth]{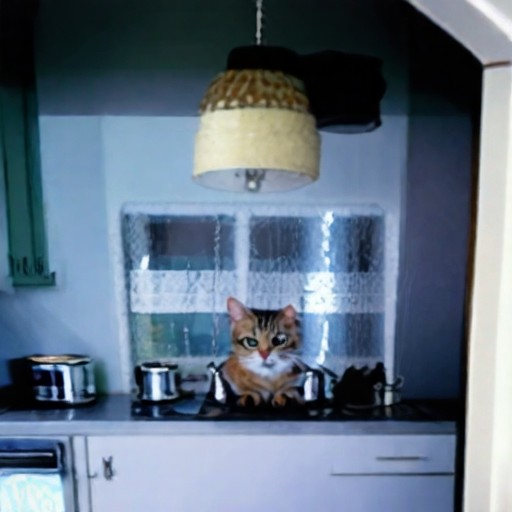} &
    \includegraphics[width=0.09\textwidth]{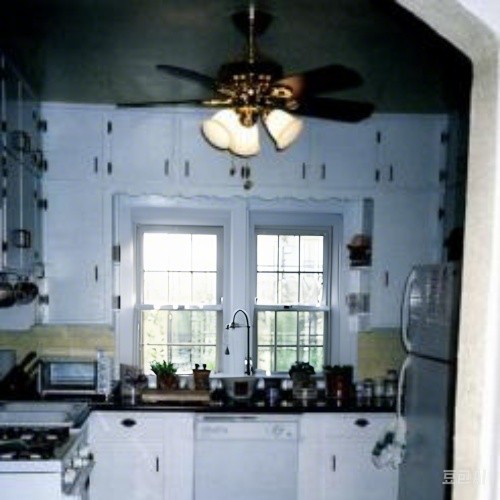} &
    \includegraphics[width=0.09\textwidth]{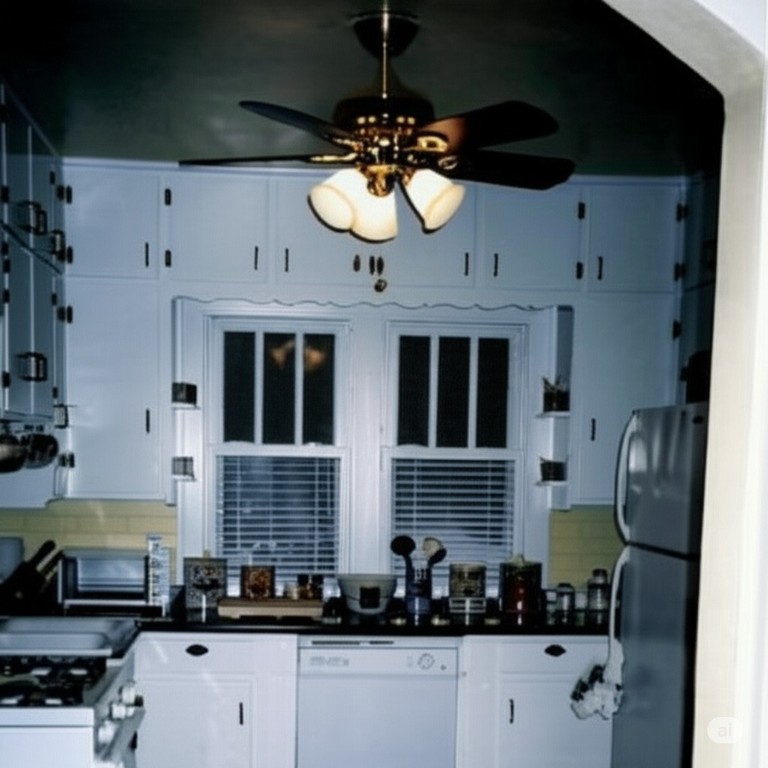} &
    \includegraphics[width=0.09\textwidth]{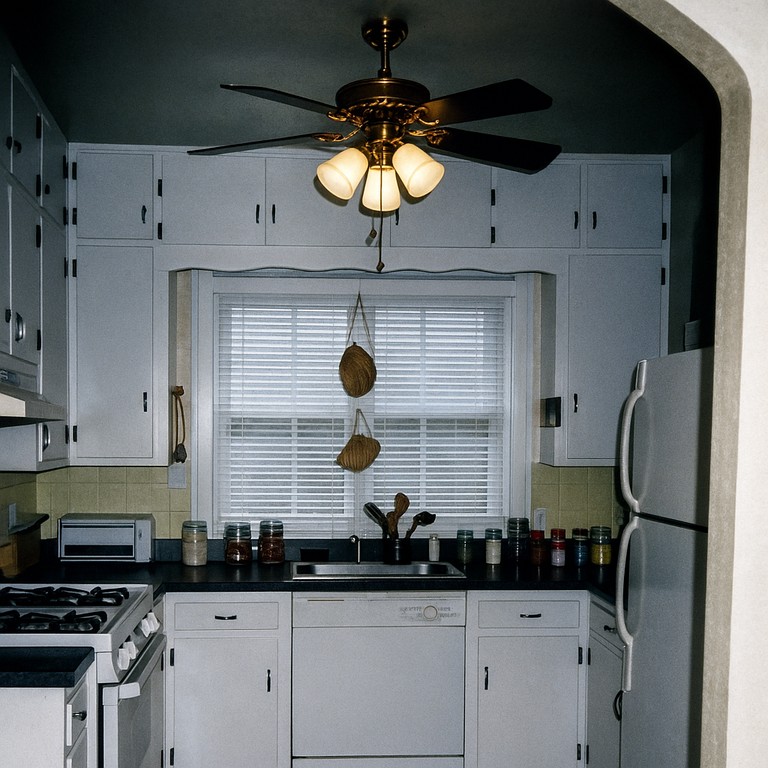} &
    \includegraphics[width=0.09\textwidth]{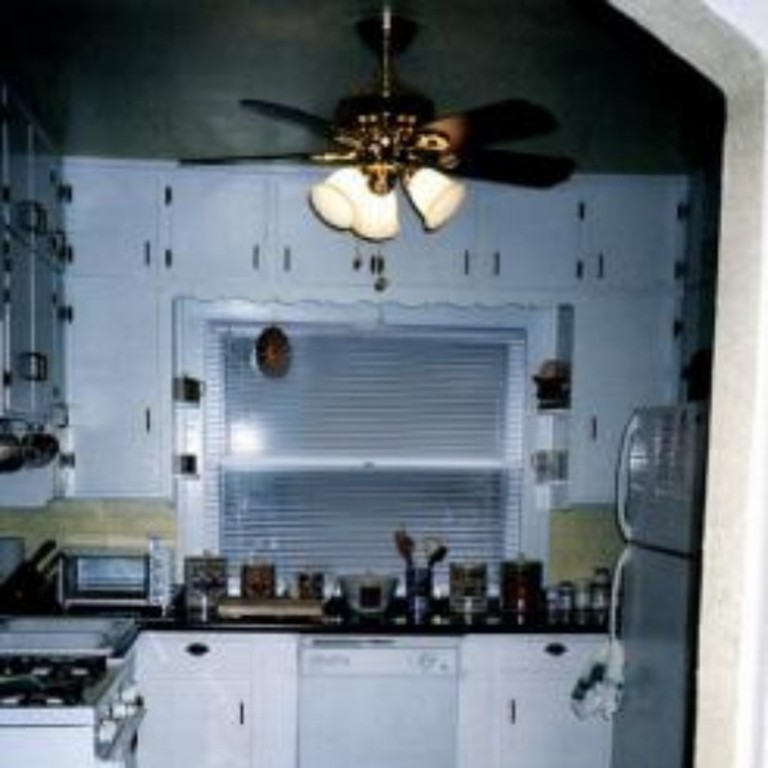} \\
    \multicolumn{9}{c}{\scriptsize\textbf{Counting:} Turn the two windows into a single window.} \\
    [0.5em]
    \includegraphics[width=0.09\textwidth]{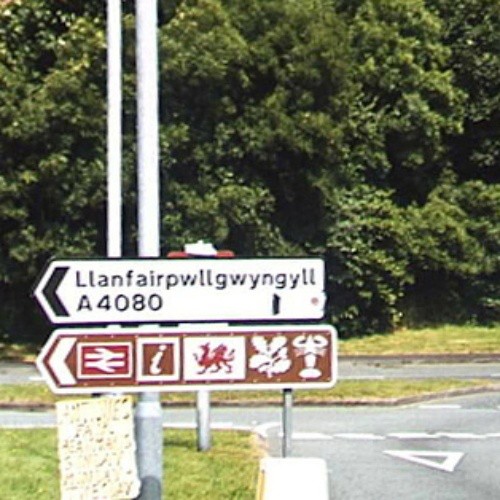} &
    \includegraphics[width=0.09\textwidth]{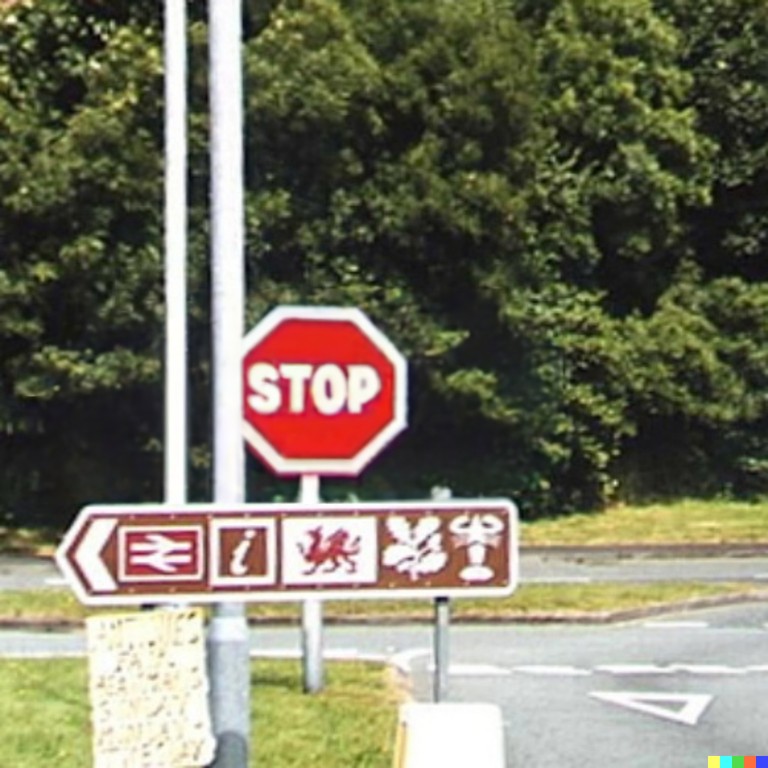} &
    \includegraphics[width=0.09\textwidth]{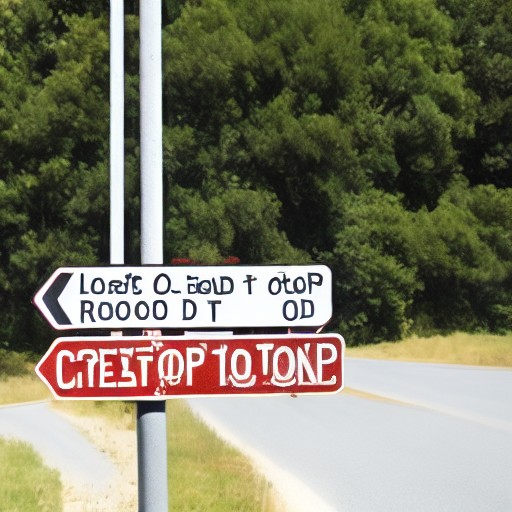} &
    \includegraphics[width=0.09\textwidth]{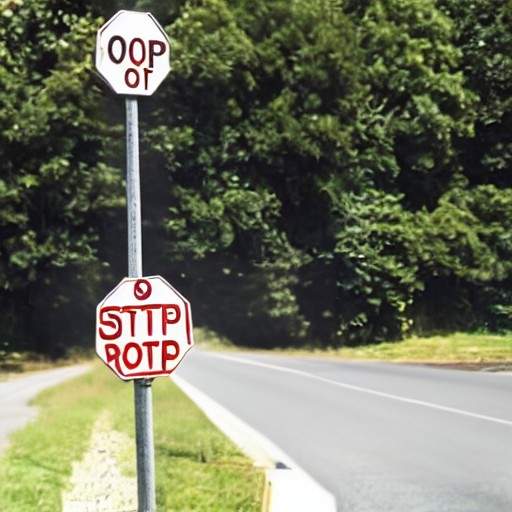} &
    \includegraphics[width=0.09\textwidth]{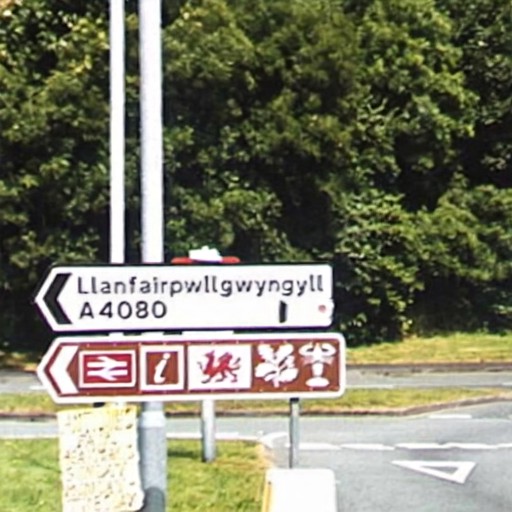} &
    \includegraphics[width=0.09\textwidth]{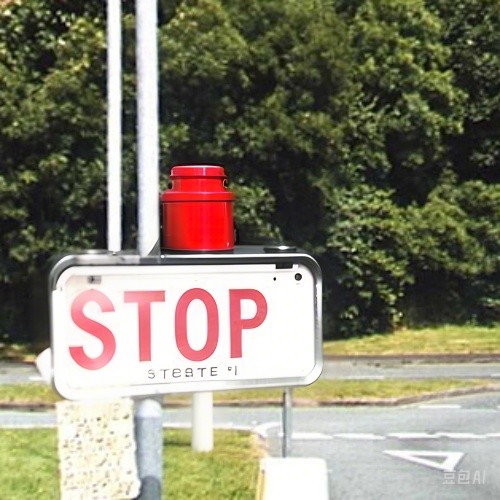} &
    \includegraphics[width=0.09\textwidth]{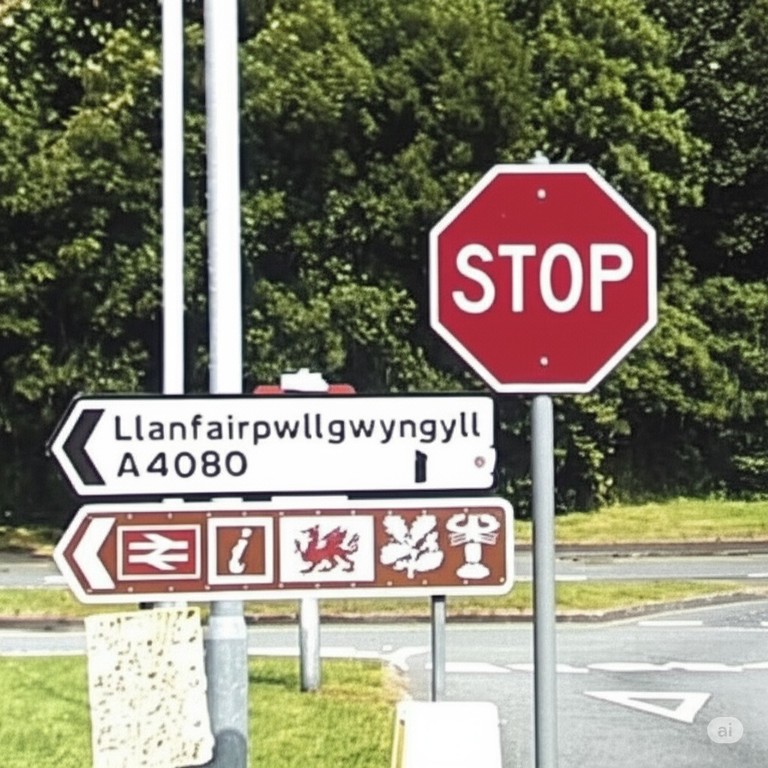} &
    \includegraphics[width=0.09\textwidth]{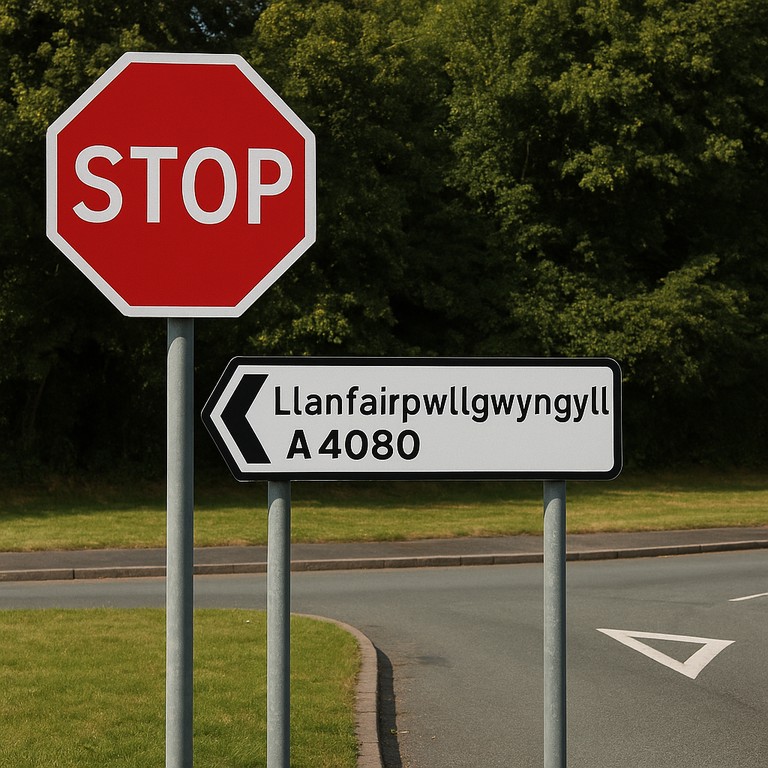} &
    \includegraphics[width=0.09\textwidth]{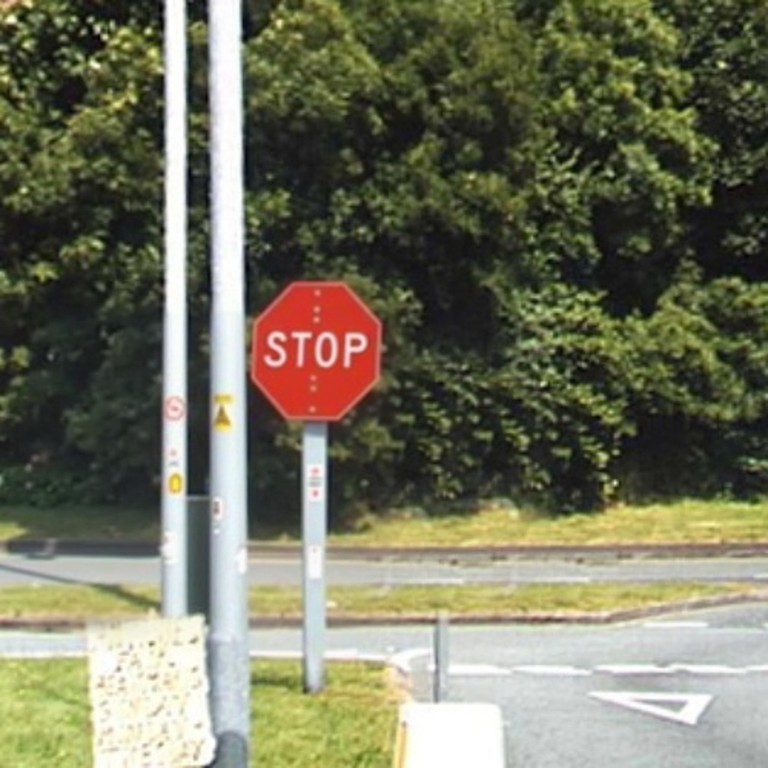} \\
    \multicolumn{9}{c}{\scriptsize\textbf{Counting:} Let there be a stop sign and only one road sign.} \\
\end{tabular}
\caption{\textbf{Our workflow performs the best on the second half of 14 challenging examples from MagicBrush.}}
\label{fig:challenging-magicbrush-second-half}
\end{center}

\end{figure*}

We further compare the performance of our workflow against baseline methods on selected challenging examples. A challenging example is an example where all baseline methods fail. We select challenging examples from the Emu Edit and MagicBrush test sets. For the Emu Edit test set, the baseline methods are Emu Edit, IP2P, MagicBrush, UltraEdit, Doubao SeedEdit, Gemini 2.0 Flash, and GPT-4o (\cref{fig:challenging-emuedit-first-half} and \cref{fig:challenging-emuedit-second-half}). For the MagicBrush test set, the baseline methods are Reference (the best DALL·E 2 generations), IP2P, MagicBrush, UltraEdit, Doubao SeedEdit, Gemini 2.0 Flash, and GPT-4o (\cref{fig:challenging-magicbrush-first-half} and \cref{fig:challenging-magicbrush-second-half}). After we exclude the style change task, both test sets cover 7 different editing categories, as reported in the original papers. We select 2 challenging examples from each category, totaling 14 for each test set. We translate the original prompts into Chinese when using Doubao SeedEdit.

Our workflow qualitatively outperforms strong baselines including Doubao SeedEdit, Gemini 2.0 Flash, and GPT-4o. Notably, our workflow strictly preserves details outside the edited region. In contrast, fine-grained texture of the image corrupt severely after a single step of editing by either Gemini 2.0 Flash or GPT-4o, which is an innate limitation of mask-free methods. Moreover, GPT-4o stretches a square input image into a rectangular image without being prompted, leading to inconsistent resolutions.

We encourage readers with more experience in the baseline methods to independently verify their performance upper-bound.

\section{Ablation Studies}
\label{app:ablation-studies}


We qualitatively demonstrate the effect of removing each component from our workflow (\cref{fig:ablation}). Then, we demonstrate that our workflow mitigates many issues of the specialized inpainting model FLUX.1 [dev] Fill (\cref{fig:flux-fill}). Together, the analysis in this section shows that our workflow improves upon the backbone image generation model.

\subsection{Necessity of Components}

\begin{figure*}[h]

\begin{center}
\scriptsize
\begin{tabular}{c@{\hspace{0.3em}}c@{\hspace{0.3em}}c@{\hspace{0.3em}}c@{\hspace{0.3em}}c@{\hspace{0.3em}}c@{\hspace{0.3em}}c}
    \textsc{Original} & \textsc{Hinted} & \textsc{No Context} & \textsc{No Blur} & \textsc{No Hints} & \textsc{No Canny} & \textsc{Full} \\
    \includegraphics[width=0.12\textwidth]{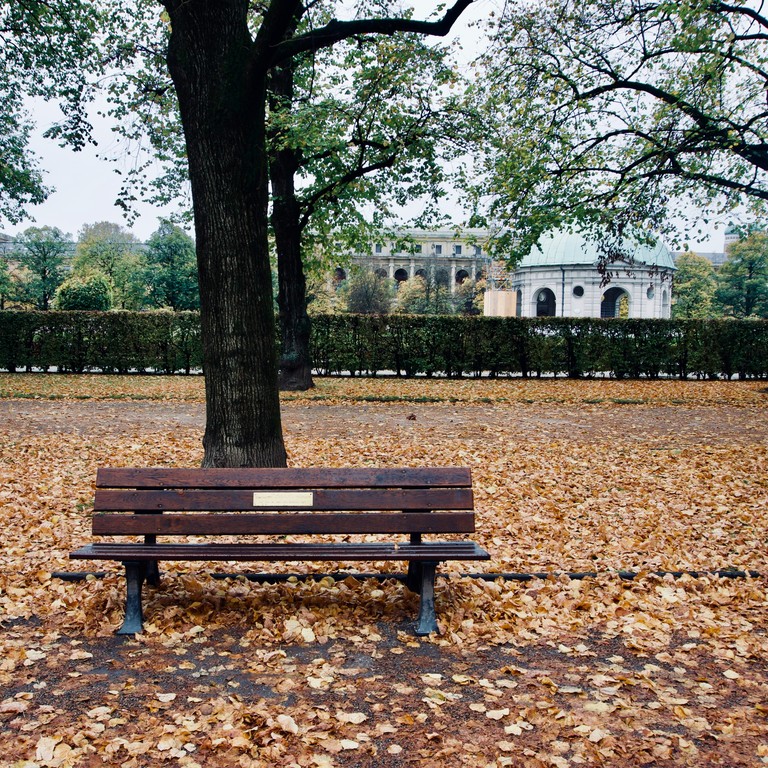} &
    \includegraphics[width=0.12\textwidth]{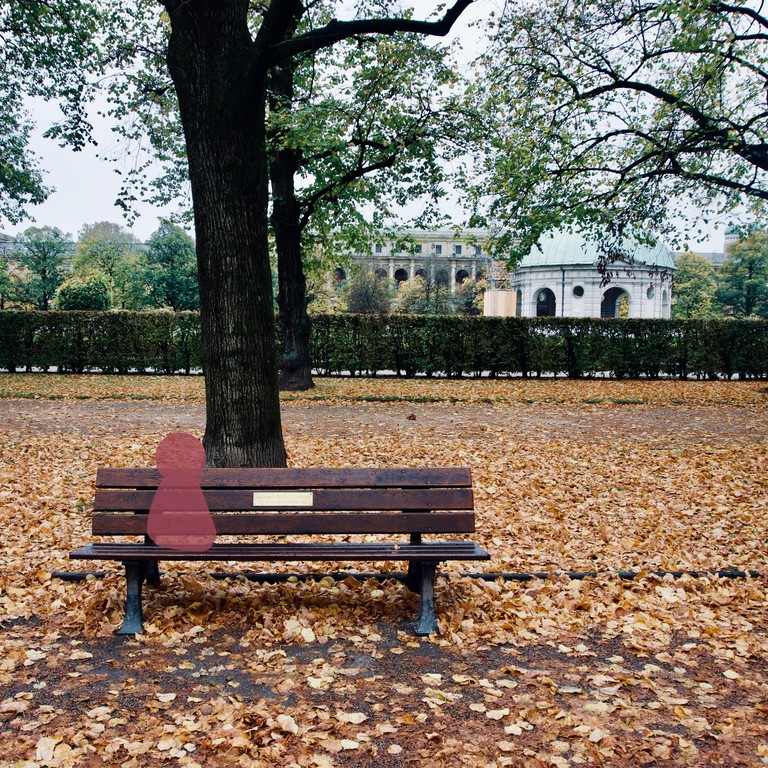} &
    \includegraphics[width=0.12\textwidth]{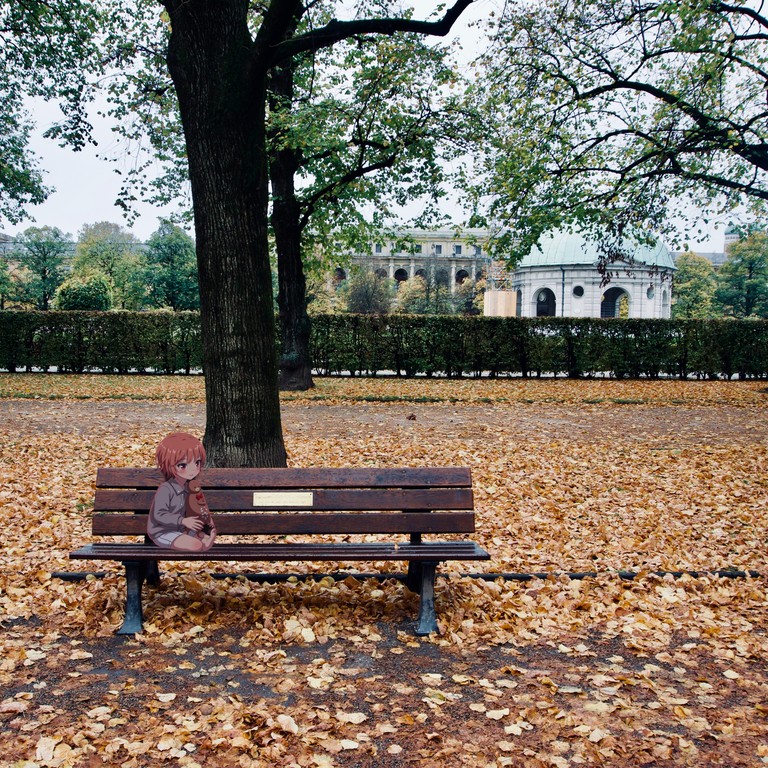} &
    \includegraphics[width=0.12\textwidth]{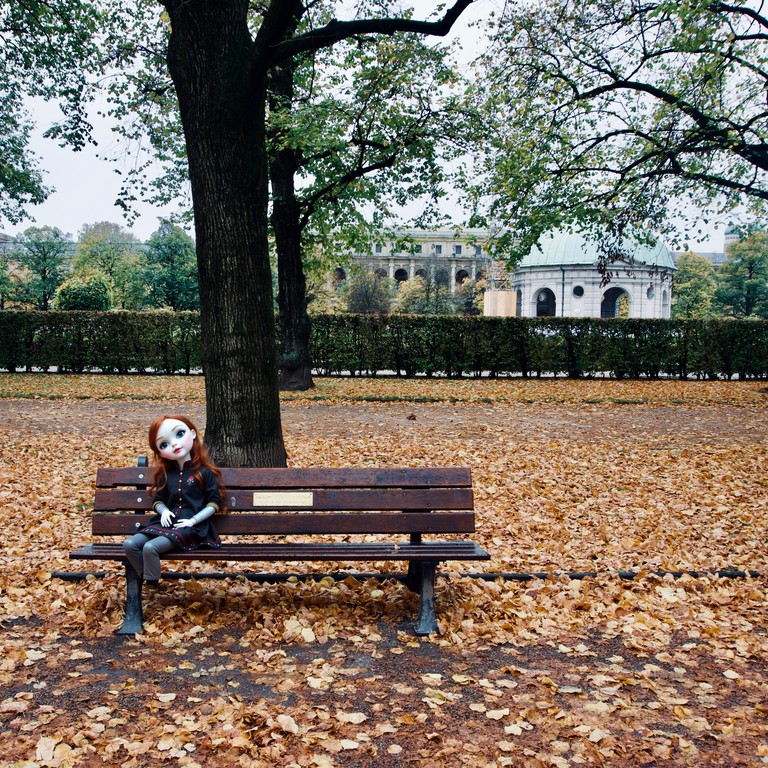} &
    \includegraphics[width=0.12\textwidth]{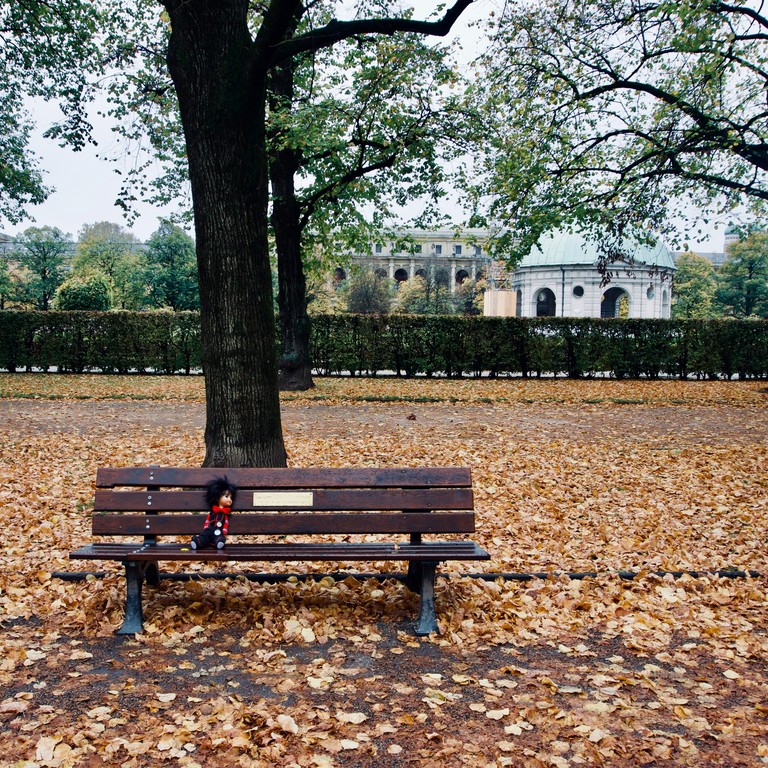} &
    \includegraphics[width=0.12\textwidth]{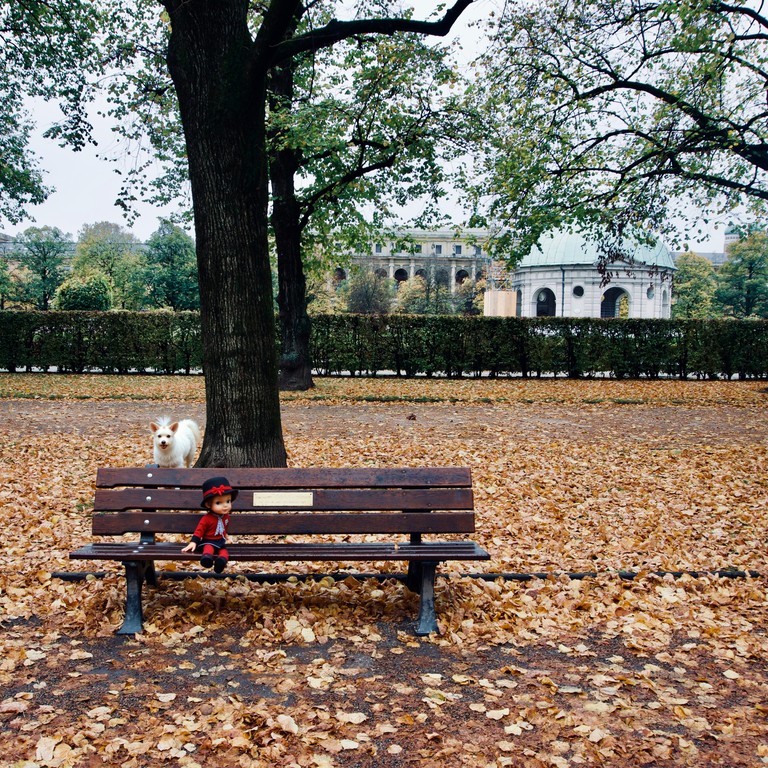} &
    \includegraphics[width=0.12\textwidth]{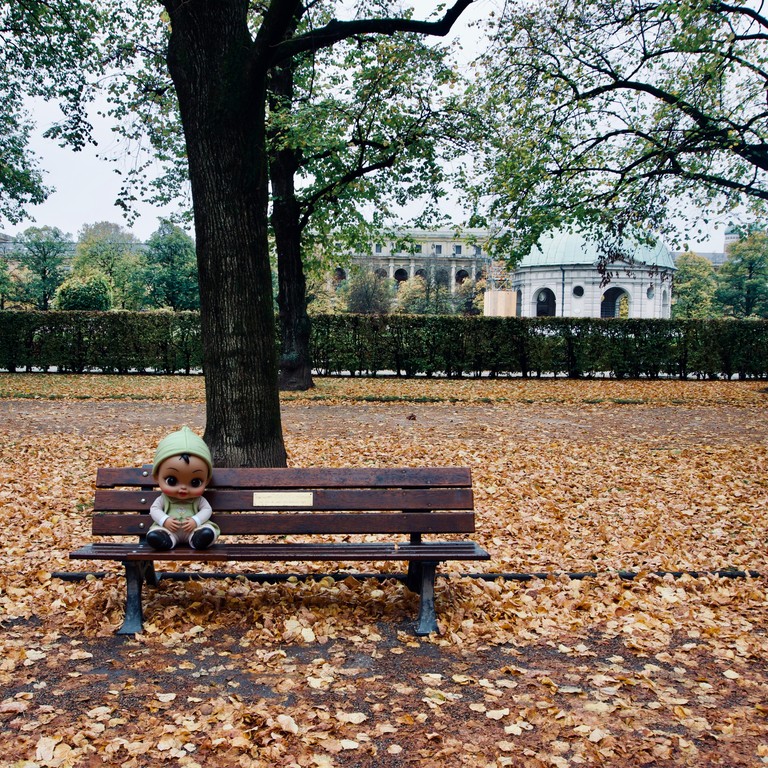} \\
\end{tabular}
\caption{\textbf{Ablation studies show that each component of our workflow is necessary.} Removing any one leads to suboptimal quality. }
\label{fig:ablation}
\end{center}

\end{figure*}

Removing any component from our workflow leads to the following negative consequences (\cref{fig:ablation}).

\paragraph{Context Selection.} Without an appropriate context, the generated region might show a different style than the overall image. Even when the style is correct, the inpainted region could still show a very different texture, an artifact that distinguishes the region from surrounding pixels. 

\paragraph{Soft Inpainting.} Without soft-inpainting, both major and minor artifacts appear. Here, the major artifact is the missing shoes from the doll. The minor artifact is the abrupt change of texture around the mask edges (on the bench backrest).

\paragraph{Color and Edge Hinting.} Without color and edge hints, the object sometimes can still be added. However, the size of the object does not follow the user's requirement. 

\paragraph{Two-Stage Denoising.} Without two-stage denoising, the generated object and the color patch have different sizes.

\subsection{Flux.1 [dev] Fill Failures}

\begin{figure}[ht!]
\begin{center}
\subfigure[\textbf{Global Editing Failures:} A person sits in a bedroom.]{
\scriptsize
\begin{tabular}{c@{\hspace{0.3em}}c@{\hspace{0.3em}}c}
    \textsc{Original} & \textsc{Flux.1 [dev] Fill} & \textsc{Ours} \\
    \includegraphics[width=0.25\textwidth]{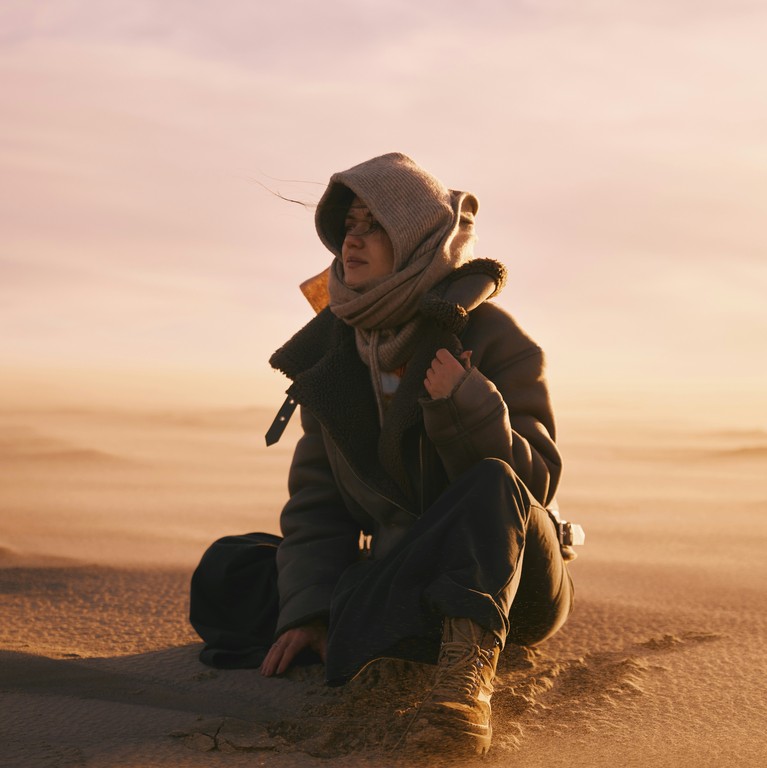} &
    \includegraphics[width=0.25\textwidth]{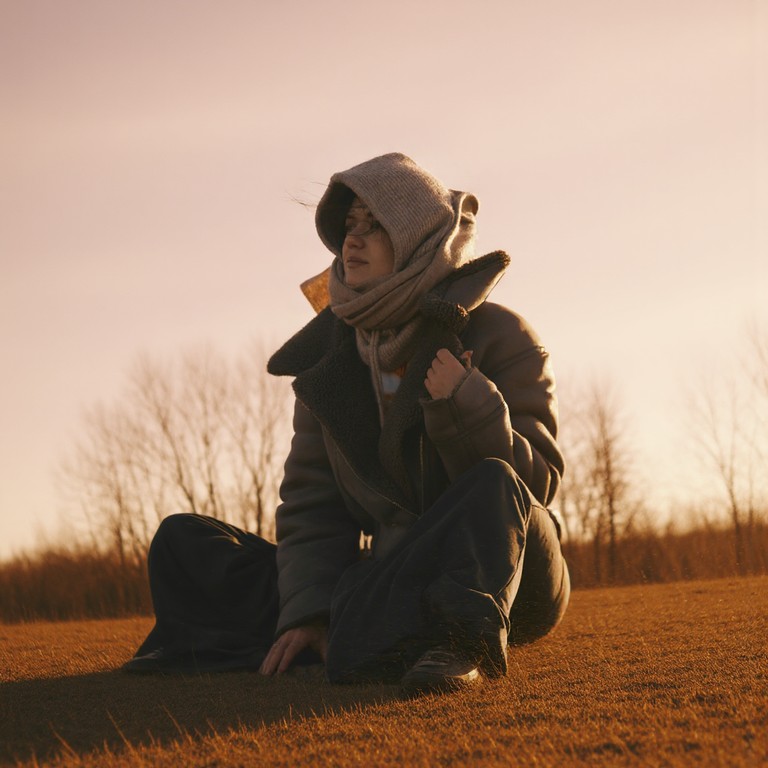} &
    \includegraphics[width=0.25\textwidth]{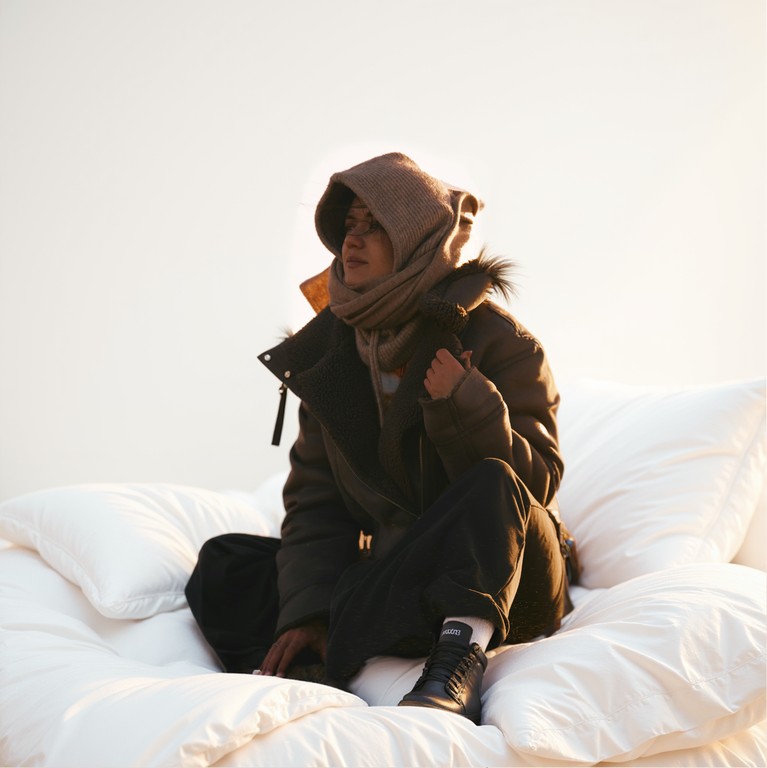} \\
\end{tabular}
}

\subfigure[\textbf{Color Drifts:} Two horses grazing on a grassy field under a clear sky.]{
\scriptsize
\begin{tabular}{c@{\hspace{0.3em}}c@{\hspace{0.3em}}c}
    \textsc{Original} & \textsc{Flux.1 [dev] Fill} & \textsc{Ours} \\
    \includegraphics[width=0.25\textwidth]{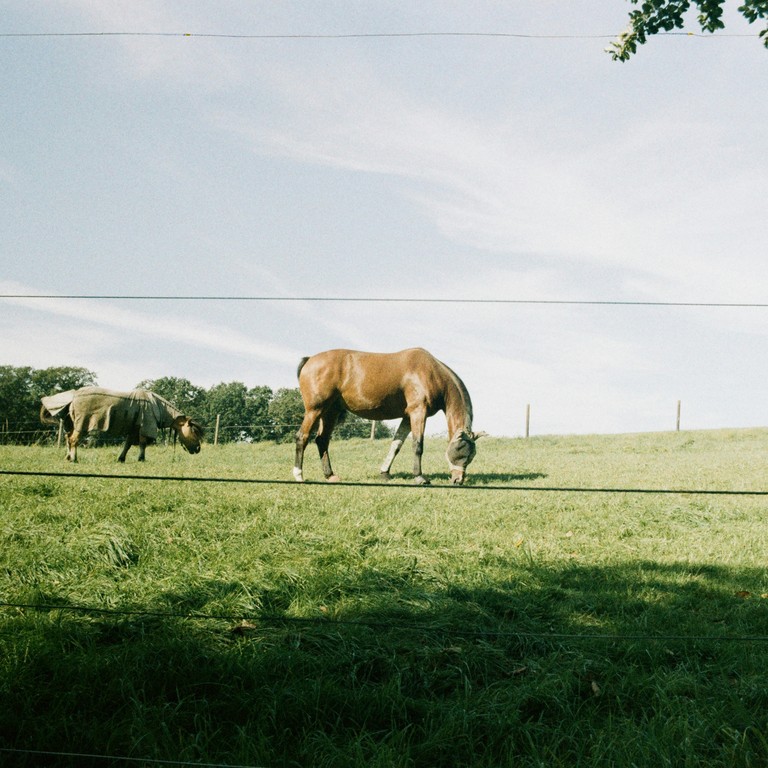} &
    \includegraphics[width=0.25\textwidth]{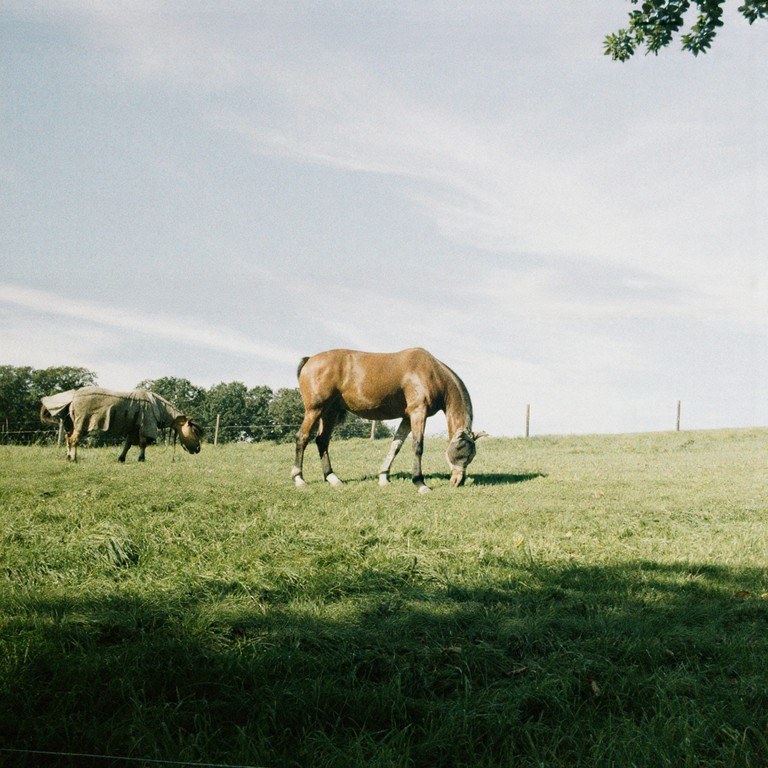} &
    \includegraphics[width=0.25\textwidth]{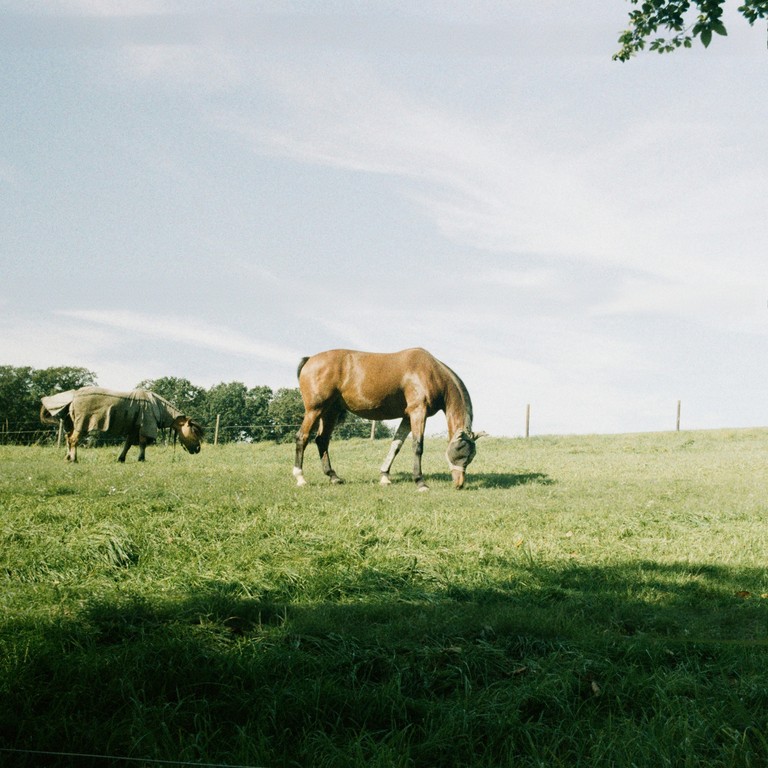} \\
\end{tabular}
}

\subfigure[\textbf{Prompt Bleeding:} A flamingo standing on a camel walking on a desert.]{
\scriptsize
\begin{tabular}{c@{\hspace{0.3em}}c@{\hspace{0.3em}}c}
    \textsc{Original} & \textsc{Flux.1 [dev] Fill} & \textsc{Ours} \\
    \includegraphics[width=0.25\textwidth]{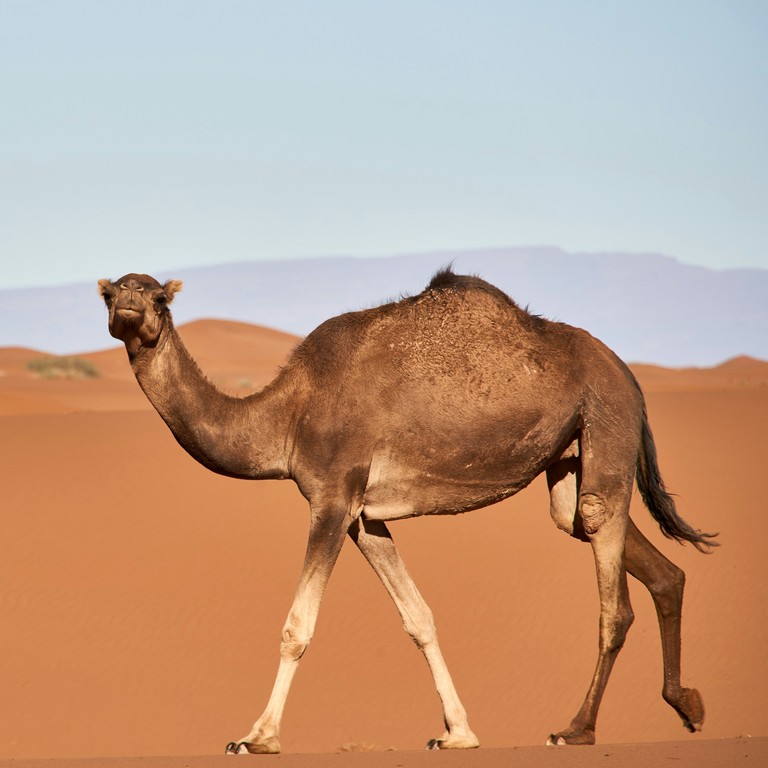} &
    \includegraphics[width=0.25\textwidth]{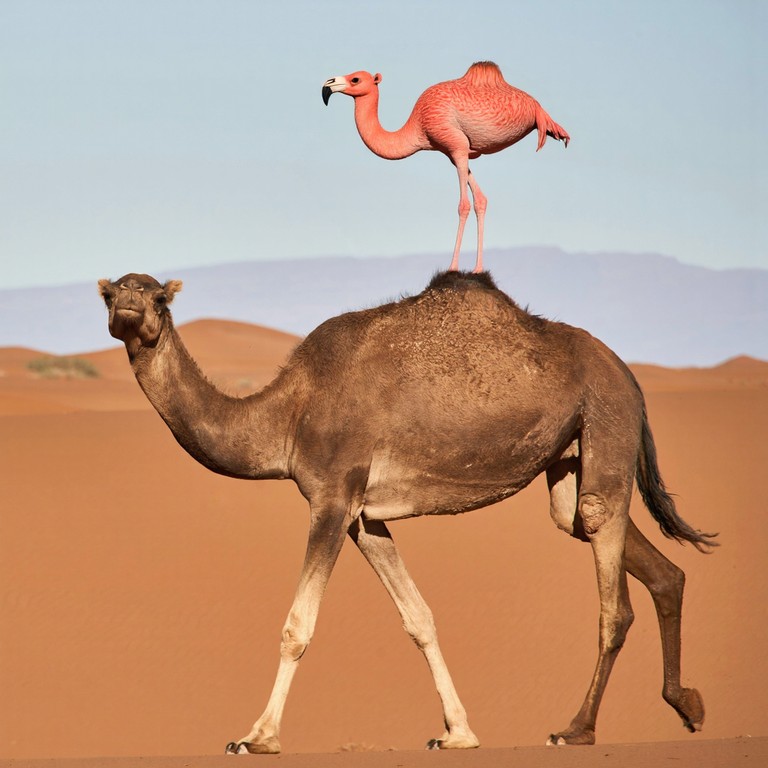} &
    \includegraphics[width=0.25\textwidth]{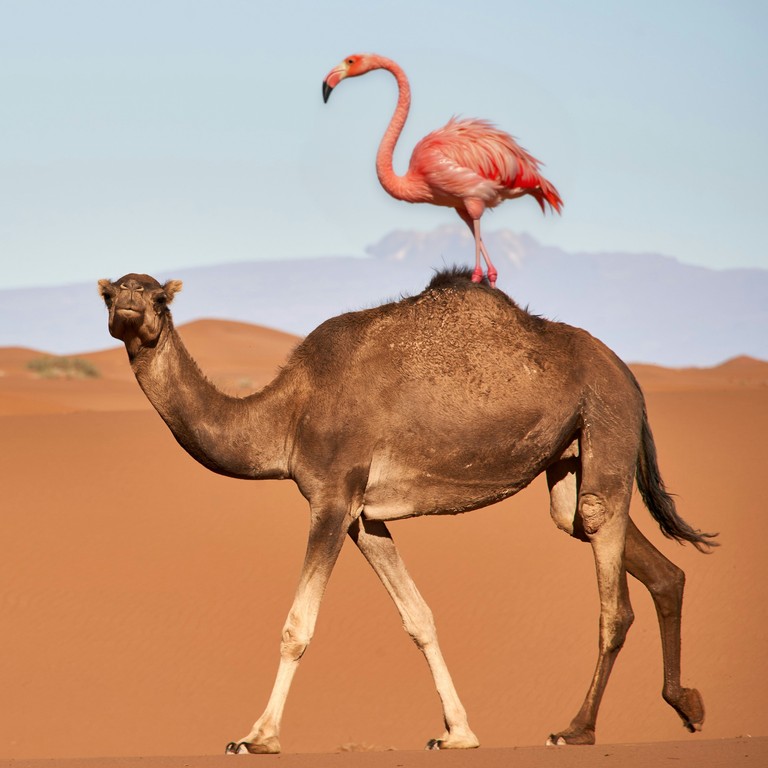} \\
\end{tabular}
}

\caption{\textbf{Our workflow mitigates issues of the strong Flux.1 [dev] Fill model.} While Flux.1 [dev] Fill is a specialized inpainting checkpoint, it suffers intrinsic limitations including failures in global editing, color drifts, and prompt bleeding. In contrast, our training-free workflow is free from these issues. In the captions, we show the description prompt for each editing task. Note that we do not use the editing prompt (``add a flamingo on top of the camel''), because these two methods are both designed to accept description prompts instead of editing prompts.}
\label{fig:flux-fill}
\end{center}

\end{figure}

Our workflow mitigates many issues of the Flux.1 [dev] Fill, a specialized inpainting model trained on the same backbone of Flux.1 [dev]. We discuss the issues below and demonstrate them in \cref{fig:flux-fill}.

\paragraph{Global Editing Failures.} While Flux.1 [dev] Fill performs comparably with our method in some local editing tasks, it has a low success rate in global (background) editing. This is probably because Flux.1 [dev] Fill was trained with more images with local edits than ones with global edits. Our method requires no further training, so no additional bias is imposed on either local or global editing.

\paragraph{Color Drifts.} Flux.1 [dev] Fill sometimes results in an obvious color drift outside the mask (saturation of the grass color drops), a weakness acknowledged by the official developers.\footnote{\url{https://huggingface.co/black-forest-labs/FLUX.1-Fill-dev\#limitations}} This is because unlike our workflow, Flux.1 [dev] Fill is not strictly a local editing model. By design, our method strictly preserves pixel values outside the mask.

\paragraph{Prompt Bleeding.} When multiple objects are mentioned in the prompt, Flux.1 [dev] Fill suffers from the well-known prompt bleeding issue, where characteristics of objects are mixed instead of being independent (a flamingo that looks like a camel). Empirically, by providing color and edge hints, our method effectively handles the prompt bleeding issue, despite using the same prompt containing multiple objects. Moreover, including multiple objects in the prompt improves model's understanding of the context in our workflow, which sometimes leads to even better results.

\section{Simple Inputs}
\label{app:simple-inputs}
\begin{figure}[h]
\begin{center}
\scriptsize
\begin{tabular}{c@{\hspace{0.3em}}c@{\hspace{0.3em}}c@{\hspace{0.3em}}c@{\hspace{0.3em}}c}
    \textsc{Original} & \textsc{Hinted} & \textsc{Silhouette} & \textsc{Mask} & \textsc{Result} \\
    \includegraphics[width=0.18\textwidth]{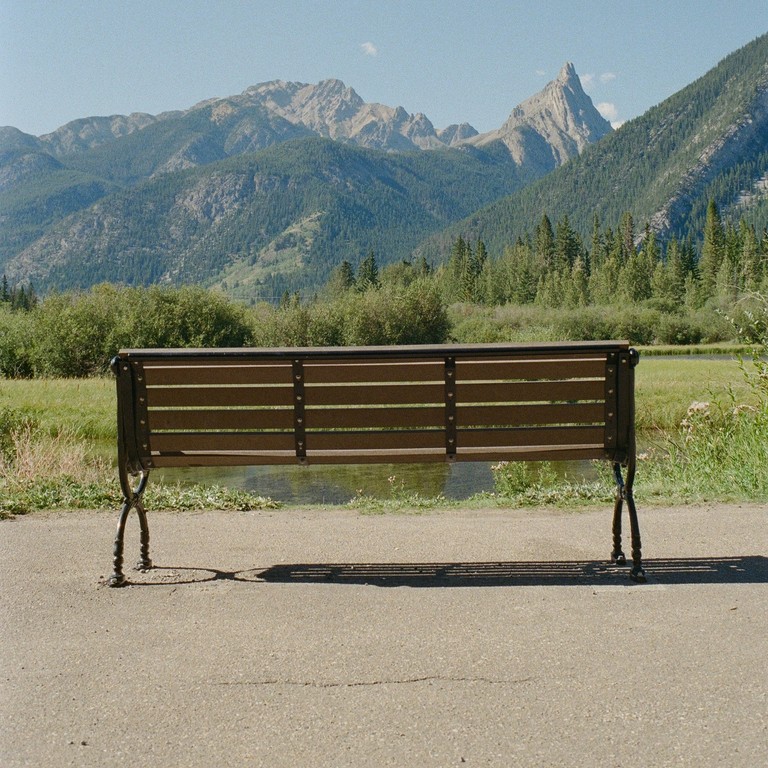} &
    \includegraphics[width=0.18\textwidth]{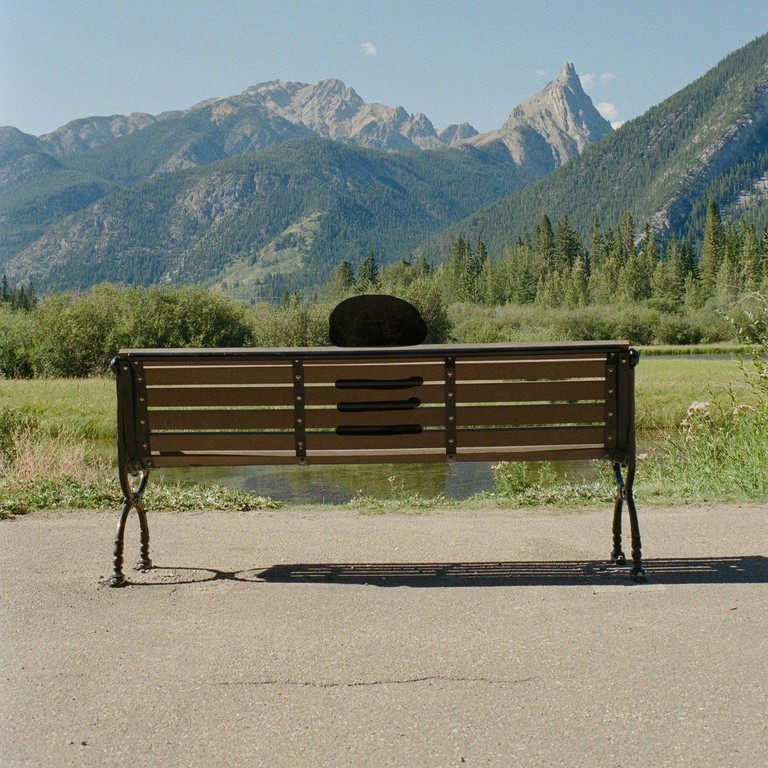} &
    \includegraphics[width=0.18\textwidth]{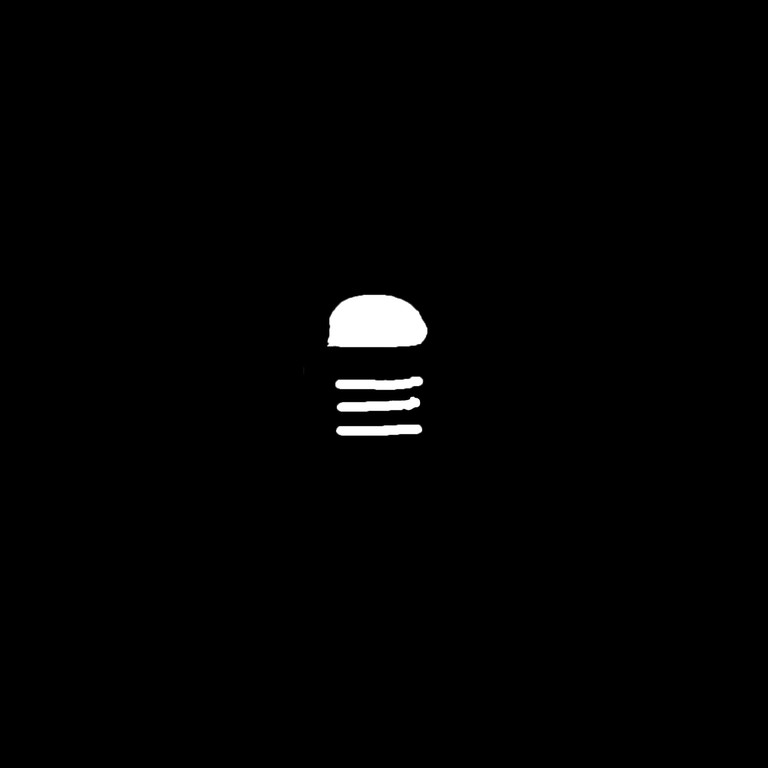} &
    \includegraphics[width=0.18\textwidth]{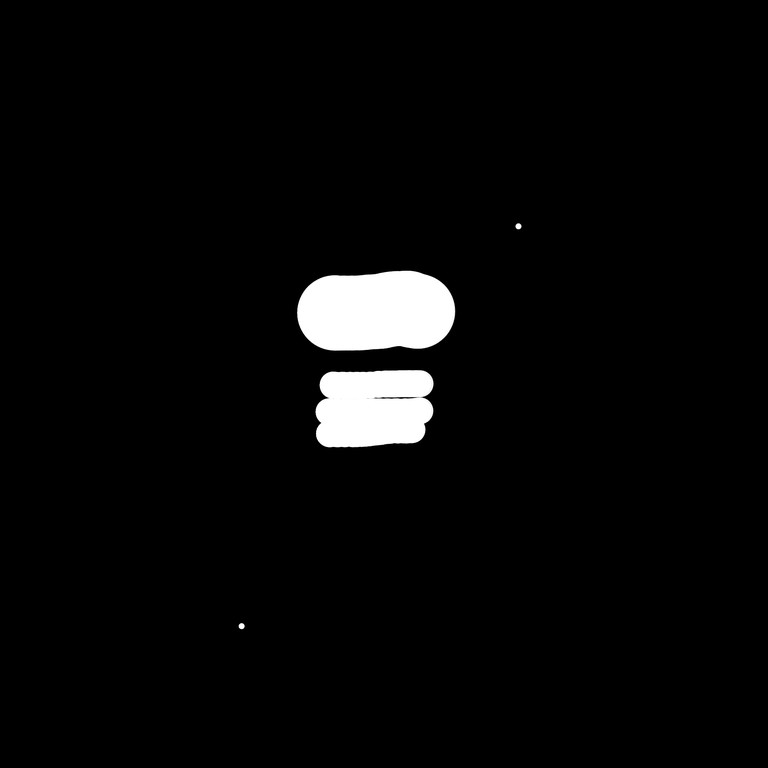} &
    \includegraphics[width=0.18\textwidth]{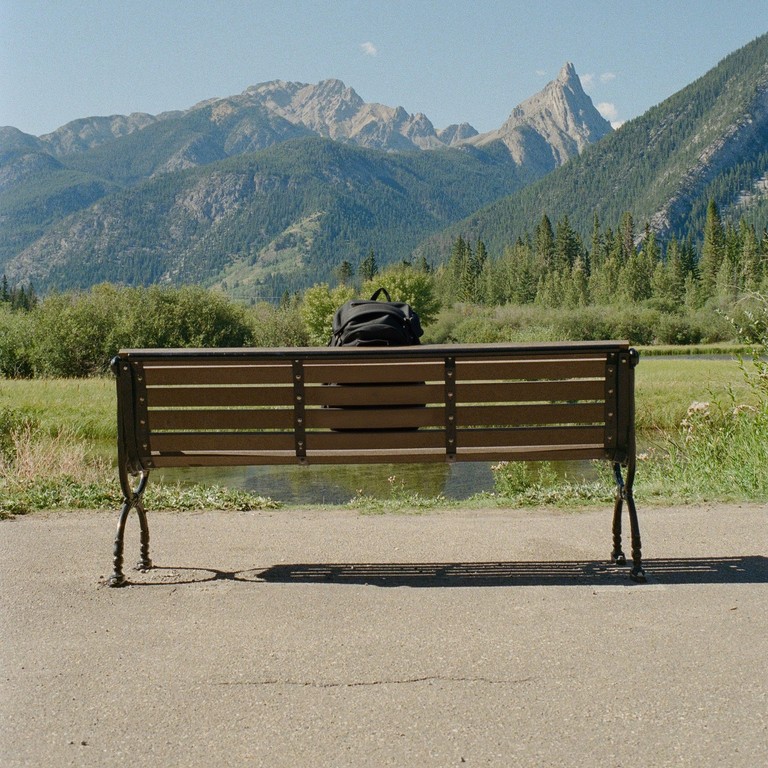} \\
    \includegraphics[width=0.18\textwidth]{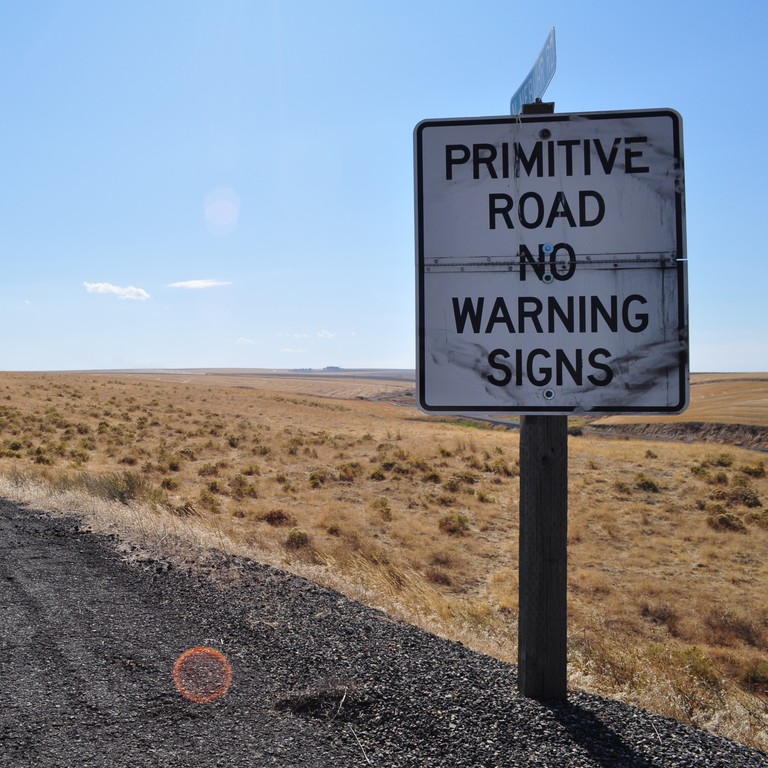} &
    \includegraphics[width=0.18\textwidth]{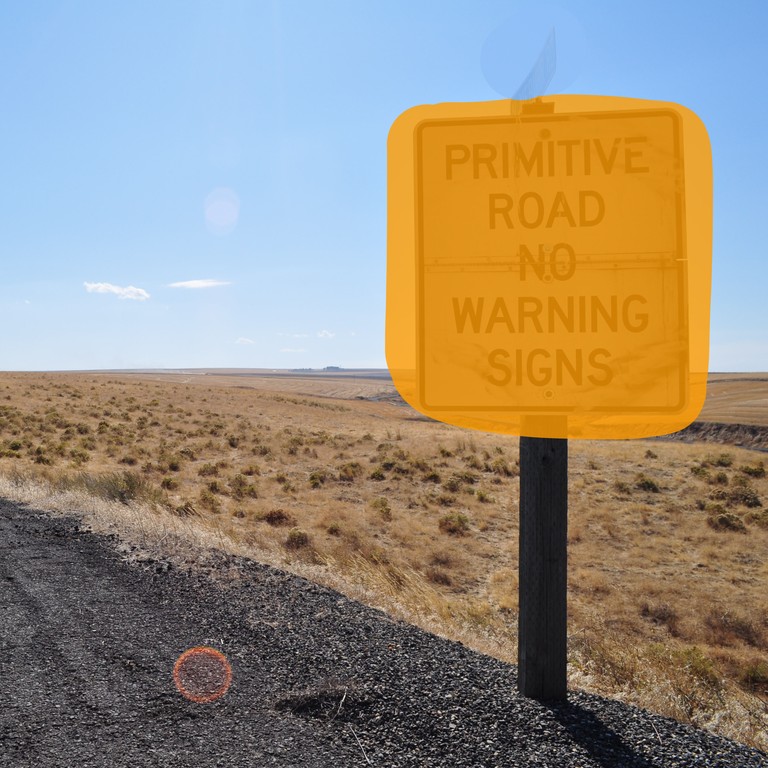} &
    \includegraphics[width=0.18\textwidth]{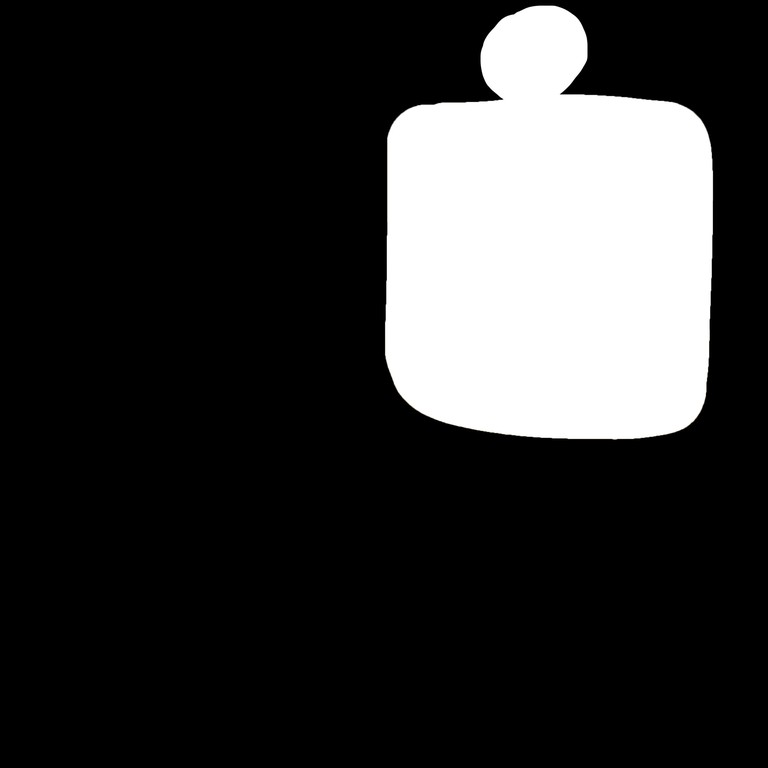} &
    \includegraphics[width=0.18\textwidth]{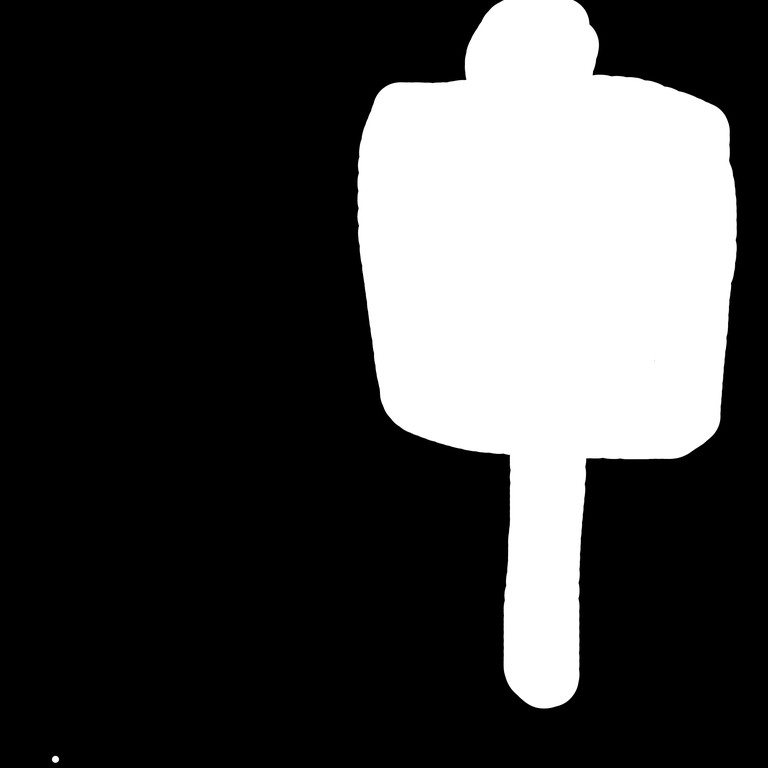} &
    \includegraphics[width=0.18\textwidth]{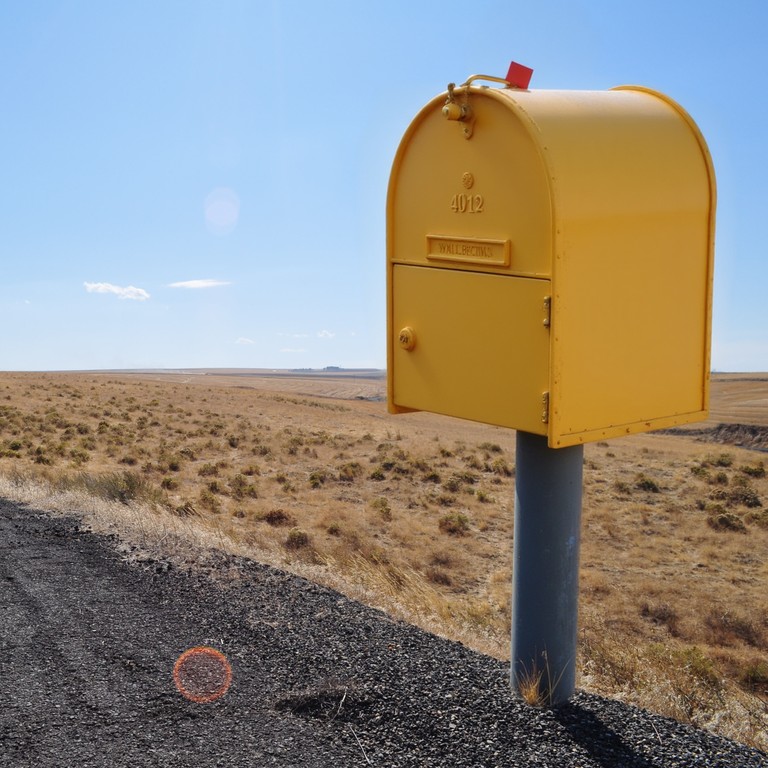} \\
    \includegraphics[width=0.18\textwidth]{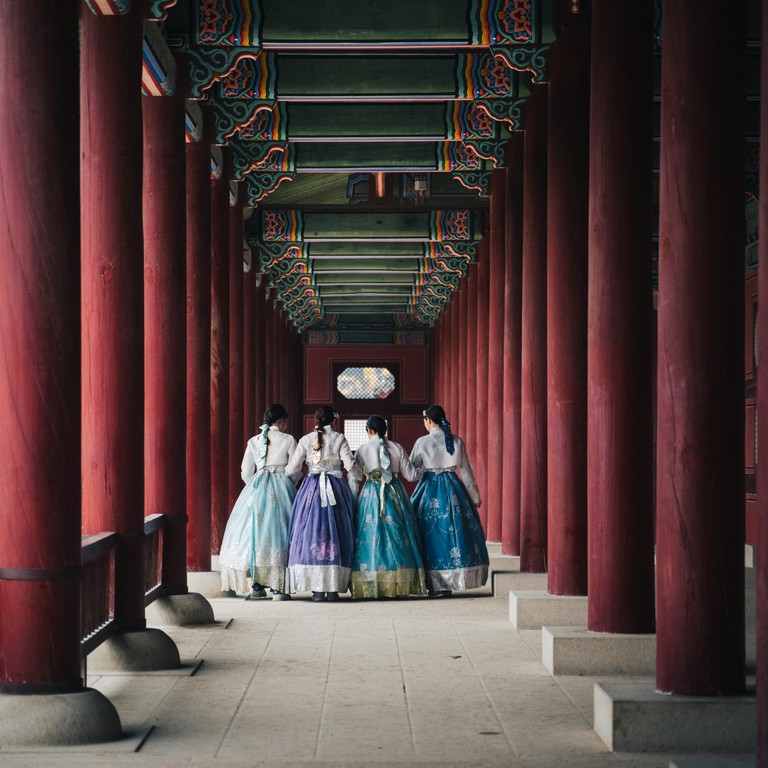} &
    \includegraphics[width=0.18\textwidth]{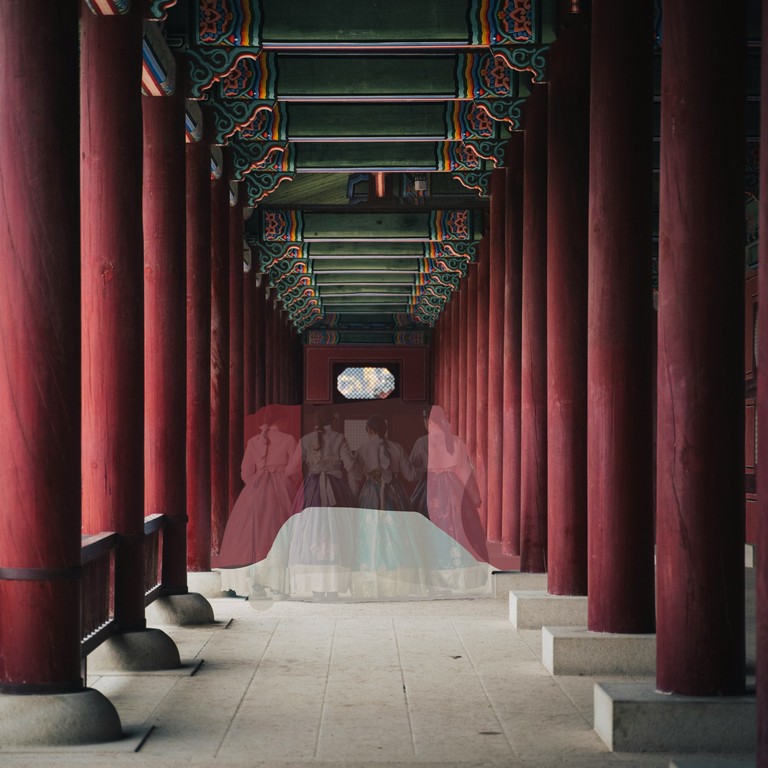} &
    \includegraphics[width=0.18\textwidth]{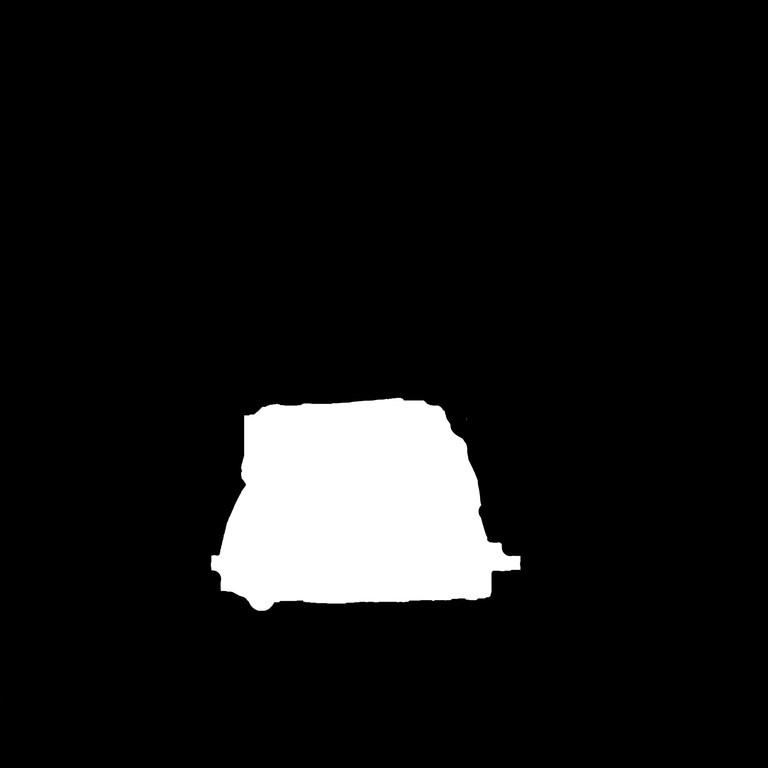} &
    \includegraphics[width=0.18\textwidth]{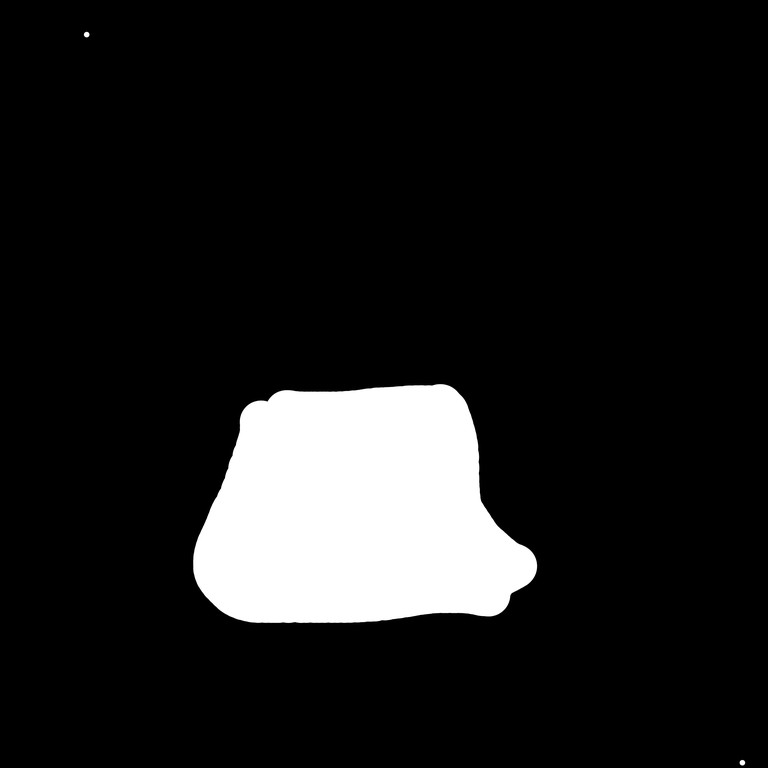} &
    \includegraphics[width=0.18\textwidth]{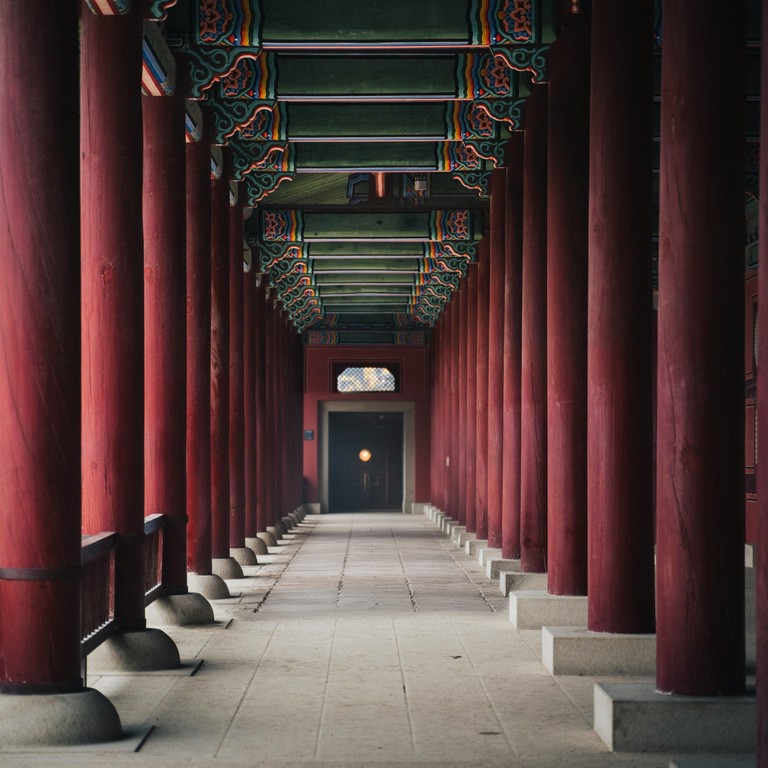} \\
    \includegraphics[width=0.18\textwidth]{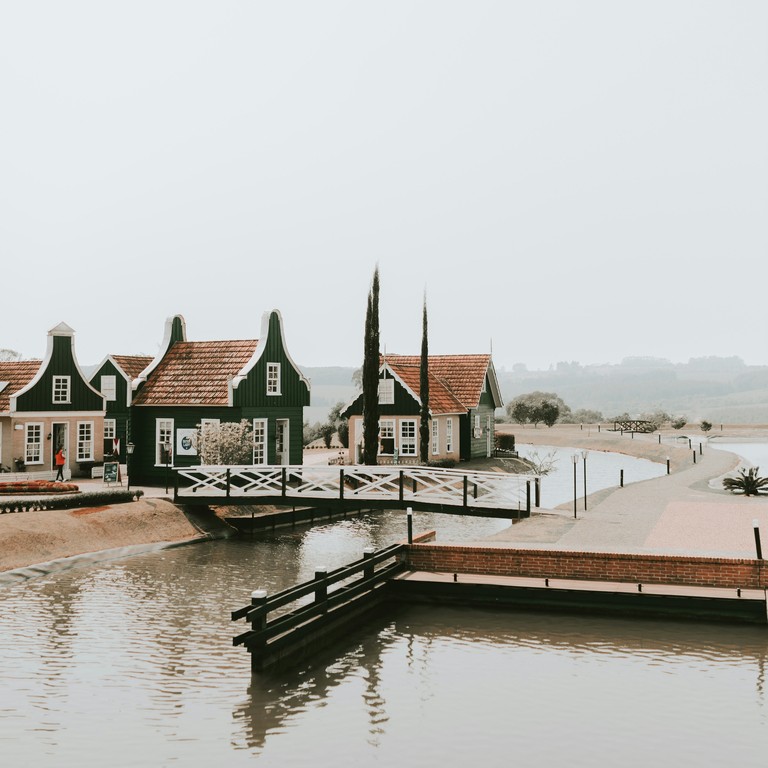} &
    \includegraphics[width=0.18\textwidth]{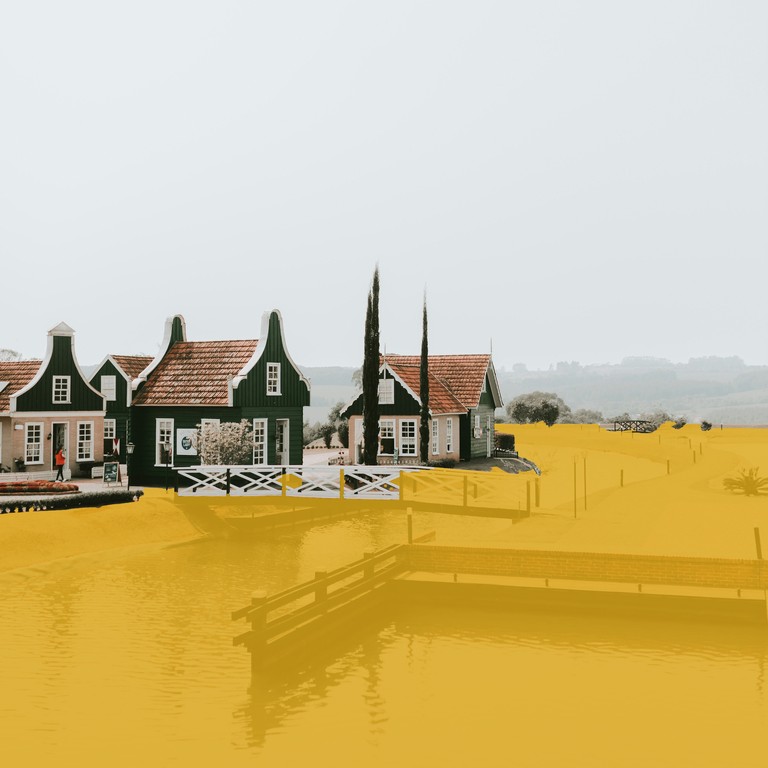} &
    \includegraphics[width=0.18\textwidth]{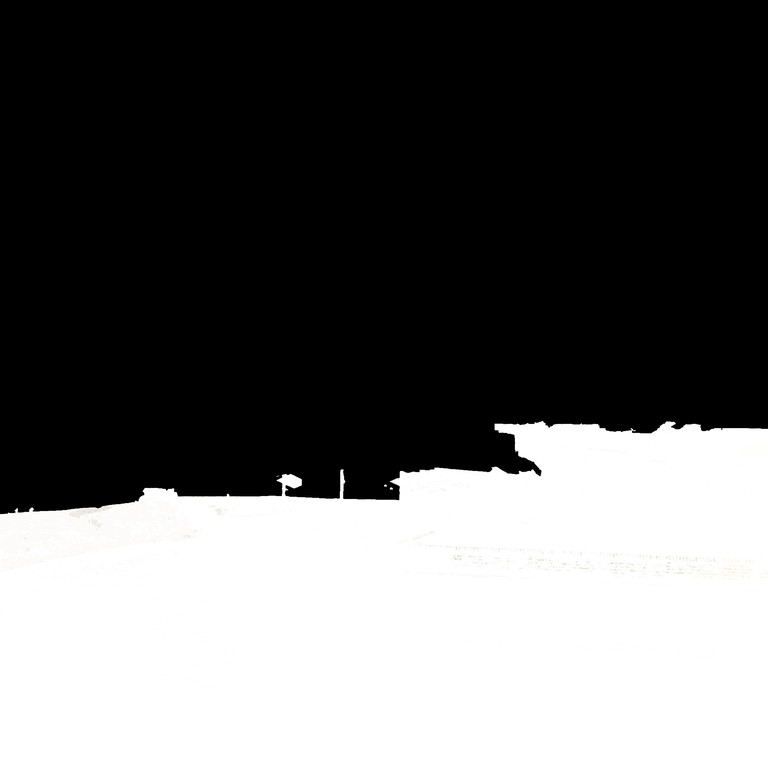} &
    \includegraphics[width=0.18\textwidth]{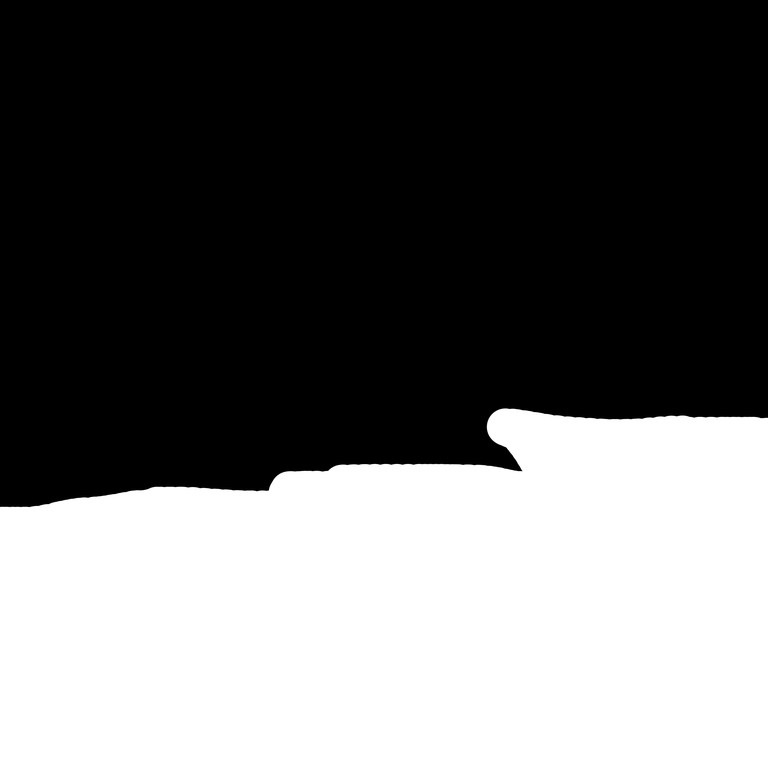} &
    \includegraphics[width=0.18\textwidth]{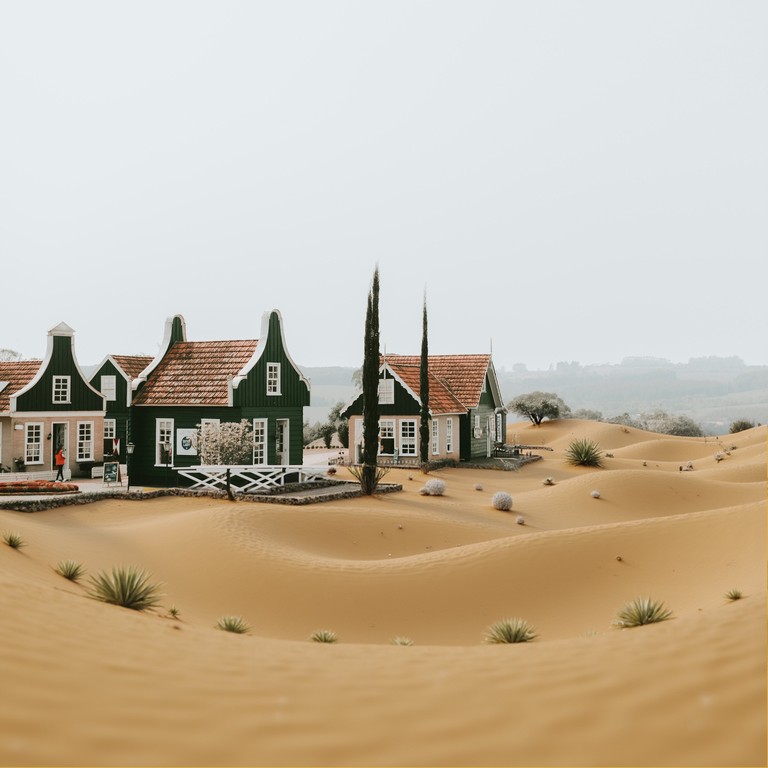} \\
    \includegraphics[width=0.18\textwidth]{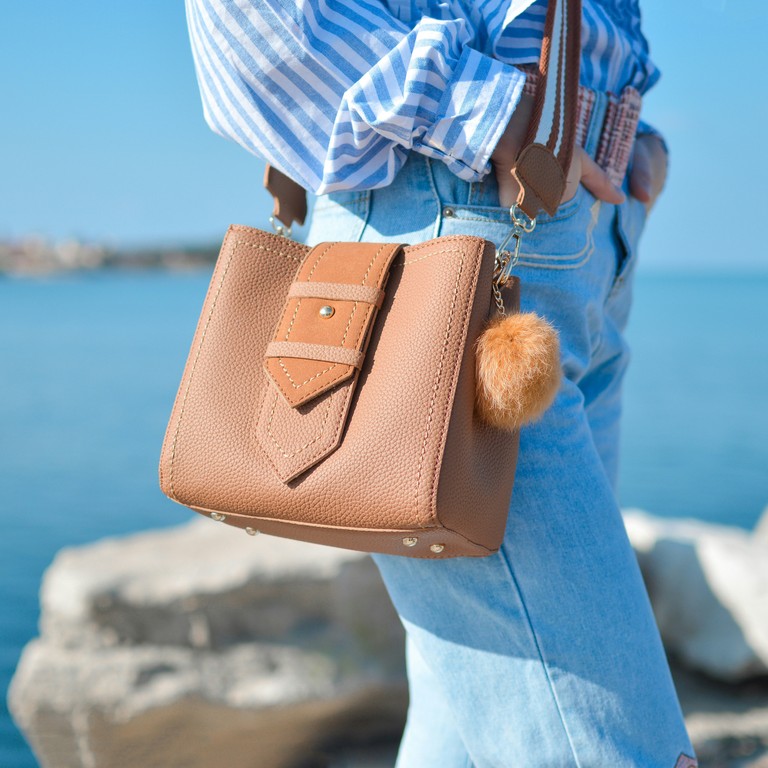} &
    \includegraphics[width=0.18\textwidth]{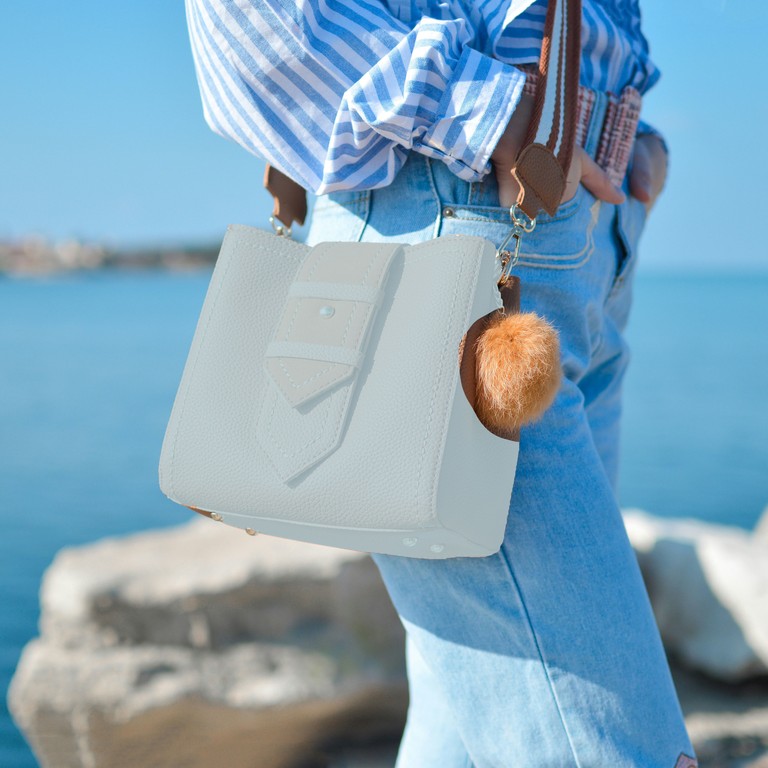} &
    \includegraphics[width=0.18\textwidth]{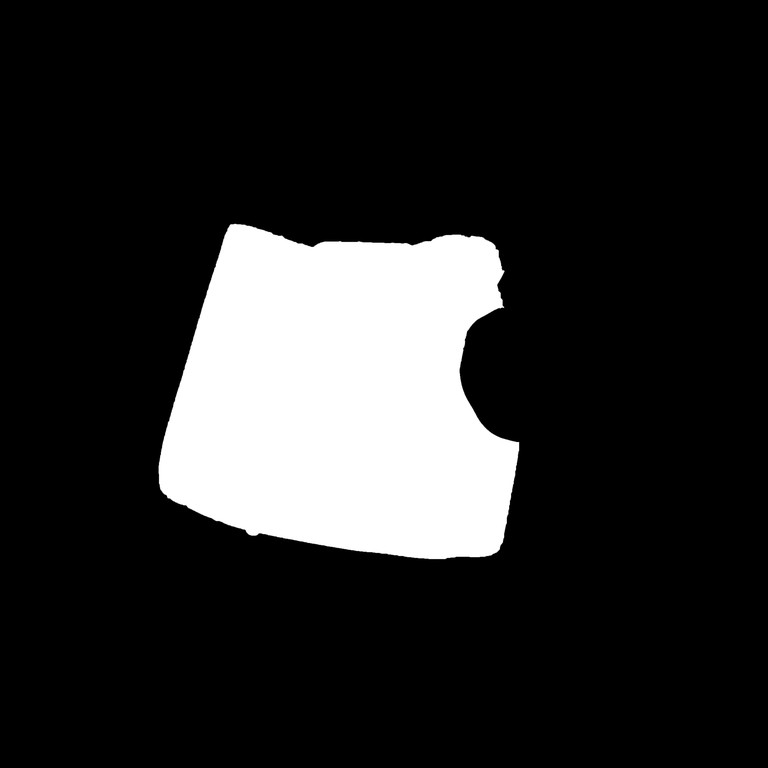} &
    \includegraphics[width=0.18\textwidth]{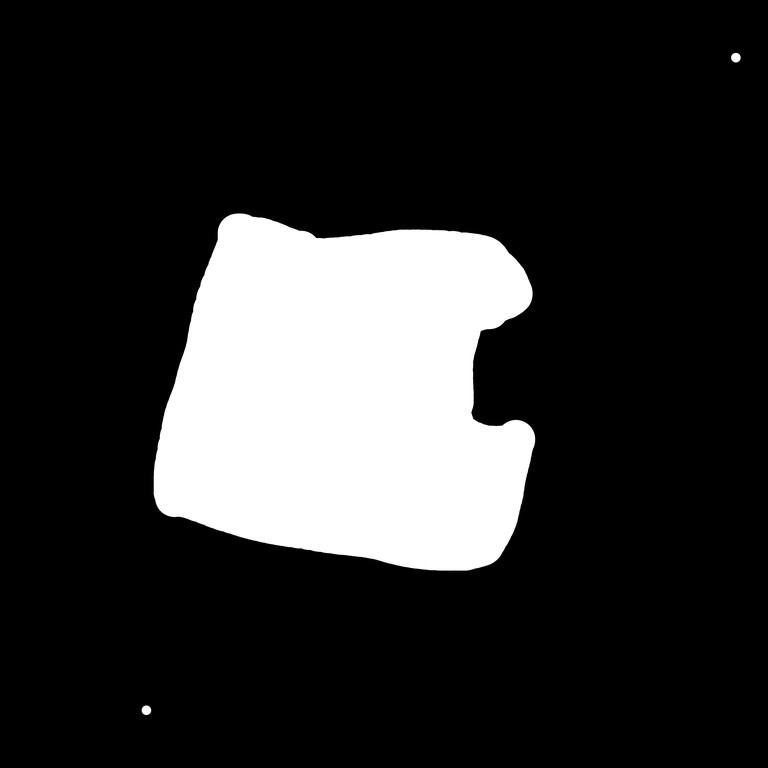} &
    \includegraphics[width=0.18\textwidth]{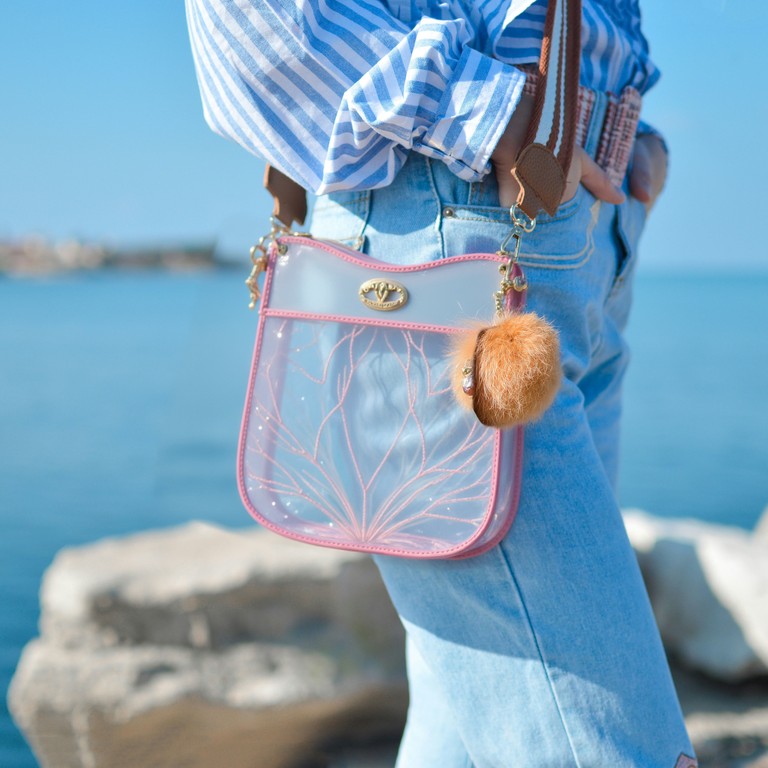} \\
    \includegraphics[width=0.18\textwidth]{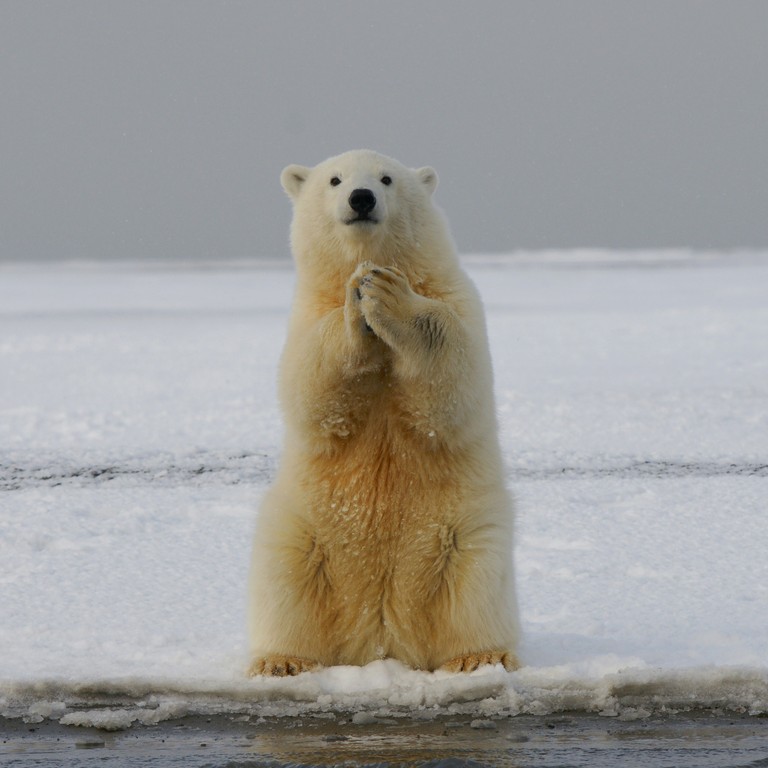} &
    \includegraphics[width=0.18\textwidth]{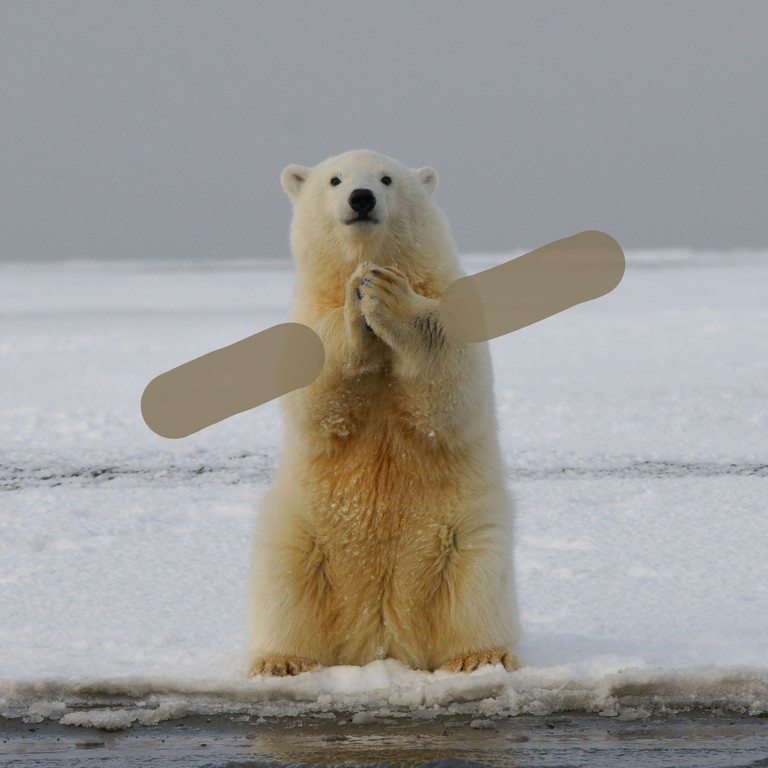} &
    \includegraphics[width=0.18\textwidth]{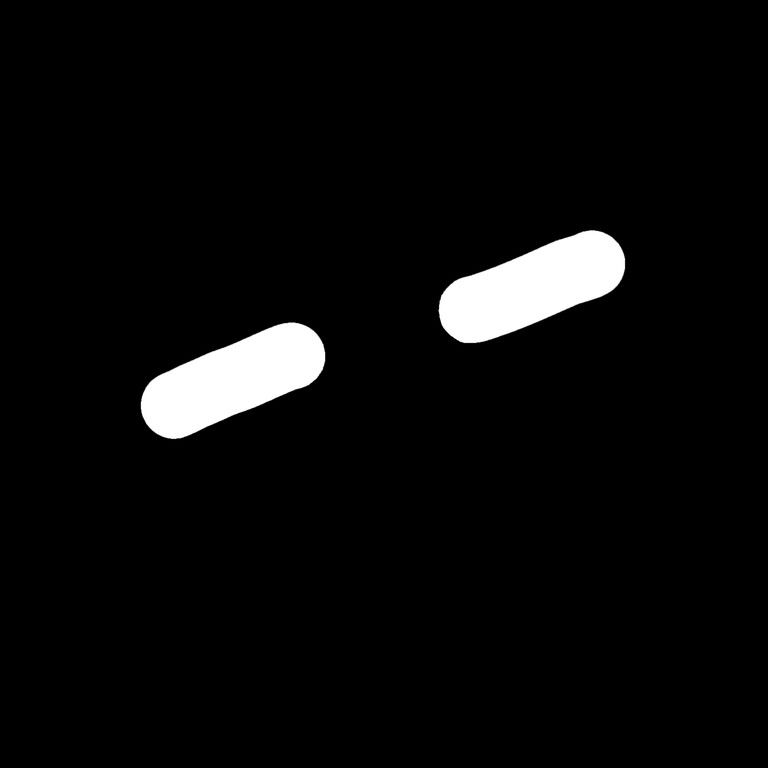} &
    \includegraphics[width=0.18\textwidth]{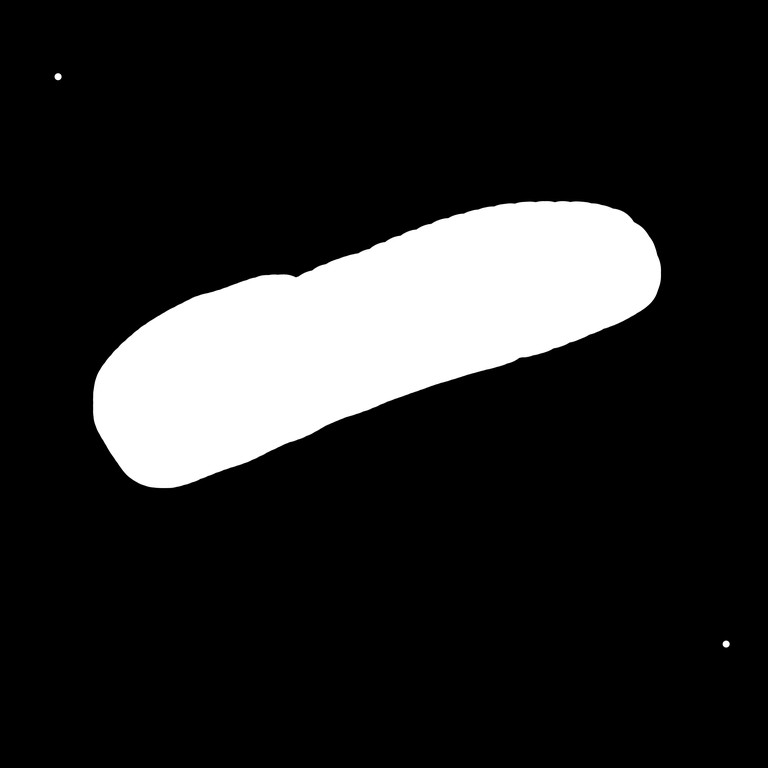} &
    \includegraphics[width=0.18\textwidth]{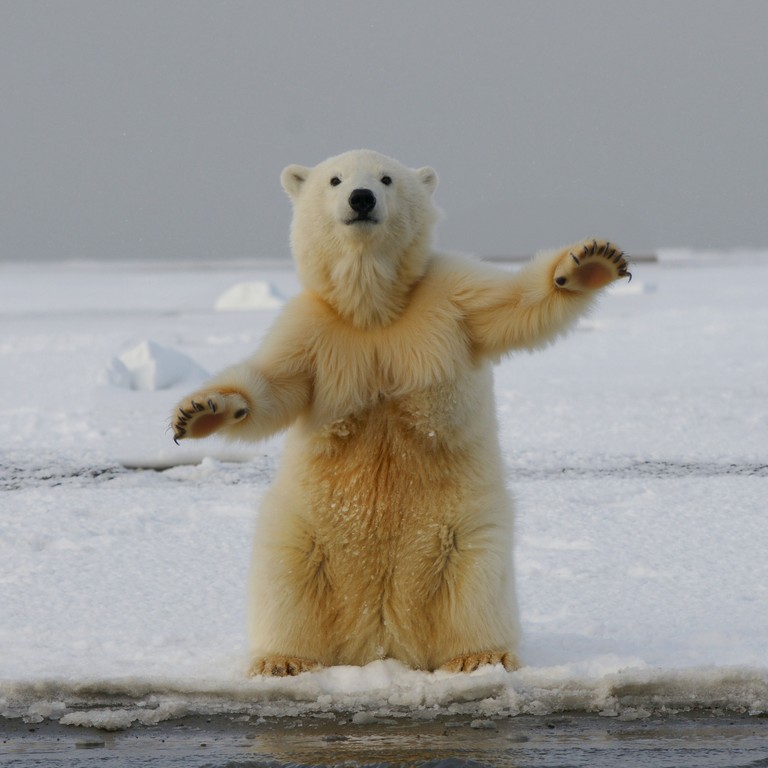} \\
\end{tabular}

\caption{\textbf{\spice~can achieve excellent editing results from simple user inputs.} To better visualize the hint, we further show the silhouette of the hint. Silhouettes are not used in the generation process, but are manually extracted after generation for visualization purpose only. Neither the hints nor the masks need to precisely match the desired shape of the edited object.}
\label{fig:simple-editing}
\end{center}

\end{figure}

While our workflow accepts extra inputs from the user, generating the extra inputs poses minimal burden. In \cref{fig:simple-editing}, we show that simple color hints (\textsc{Hinted}) and simple masks (\textsc{Mask}) can produce high-quality edits.

\section{Measuring Properties of Generated Objects}
\label{app:measuring-properties-of-generated-objects}

To quantify the precision of our workflow, we need to measure the properties of the generated objects. The properties include size, location, rotation, color, and aspect ratio. We use the Language Segment-Anything\footnote{\url{https://github.com/luca-medeiros/lang-segment-anything}} tool to identify the segmentation mask of a generated object. Here, we use segmentation mask to refer to the pixels that cover the object. We use the default model (\texttt{sam2.1\_hiera\_small}) and settings provided by the developers, which empirically performs well for our purpose. For an object, its width and height are defined as the width and height of the bounding box of the segmentation mask. The location is defined as the center of the bounding box. The rotation (for the crescent) is calculated as the direction from the center of mass of the segmentation mask to the center of the bounding box. The color is defined as the average RGB value of pixels on the generated image covered by the segmentation mask. To find a single value to represent the color, we convert the average RGB value into HSV representation and use the hue. The aspect ratio is the width divided by height.

After measuring these properties of the generated object, we calculate a percentage error between the generated property value and specified property value. For size, we vary the diameter of a red circle and compare the height and width of the apple's bounding box with the diameter. For location, we vary the center coordinates of a red circle and compare the center coordinates of the cherry's bounding box with the center coordinates of the circle. For rotation, we vary the rotation angle of a crescent-shaped color patch and compare the orientation of the crescent with the orientation of the color patch. For color, we vary the hue of a color patch and compare the average hue of the plastic chair to the hue of the patch. For aspect ratio, we vary the width-to-height ratio of a rectangle and compare the width-to-height ratio of the painting with the width-to-height ratio of the rectangle. The definition of percentage errors are listed below:
\begin{align*}
    \text{Percentage Error of Width} & = \frac{\text{Generated Width} - \text{Specified Width}}{\text{Specified Width}}\times 100\%,\\
    \text{Percentage Error of Height} &= \frac{\text{Generated Height} - \text{Specified Height}}{\text{Specified Height}}\times 100\%,\\
    \text{Percentage Error of X}&= \frac{\text{Generated X} - \text{Specified X}}{\text{Specified X}}\times 100\%,\\
    \text{Percentage Error of Y}&= \frac{\text{Generated Y} - \text{Specified Y}}{\text{Specified Y}}\times 100\%,\\
    \text{Percentage Error of Rotation}&= \frac{\text{Generated Rotation} - \text{Specified Rotation}}{360^\circ}\times 100\%,\\
    \text{Percentage Error of Color}&= \frac{\text{Generated Hue} - \text{Specified Hue}}{1.0}\times 100\%,\\
    \text{Percentage Error of Aspect Ratio}&= \frac{\text{Generated Aspect Ratio} - \text{Specified Aspect Ratio}}{\text{Specified Aspect Ratio}}\times 100\%.\\
\end{align*}
To account for the periodic nature of rotation and hue, we use 360$^\circ$ for rotation and 1.0 for hue instead of the specified property value.

\section{Changing the Random Seed}
\label{app:changing-the-random-seed}

\begin{figure*}[htbp]

\begin{center}
\small
\subfigure[IP2P]{
\begin{tabular}{c@{\hspace{0.3em}}c@{\hspace{0.3em}}c@{\hspace{0.3em}}c@{\hspace{0.3em}}c@{\hspace{0.3em}}c@{\hspace{0.3em}}c@{\hspace{0.3em}}c@{\hspace{0.3em}}c@{\hspace{0.3em}}c}
    \scriptsize\textsc{Seed 0} & \scriptsize\textsc{Seed 1} & \scriptsize\textsc{Seed 2} & \scriptsize\textsc{Seed 3} & \scriptsize\textsc{Seed 4} & \scriptsize\textsc{Seed 5} & \scriptsize\textsc{Seed 6} & \scriptsize\textsc{Seed 7} & \scriptsize\textsc{Seed 8} & \scriptsize\textsc{Seed 9} \\
    
    \includegraphics[width=0.09\textwidth]{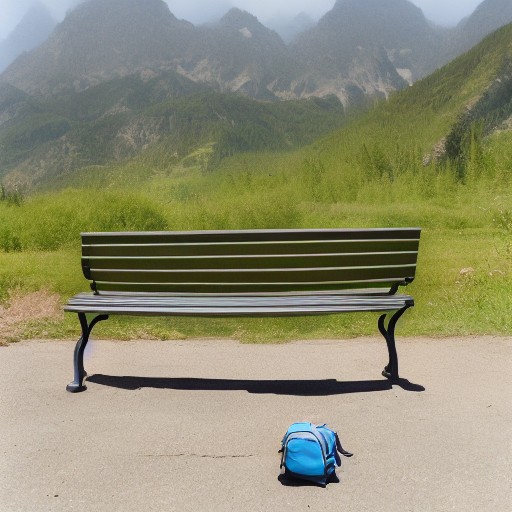} &
    \includegraphics[width=0.09\textwidth]{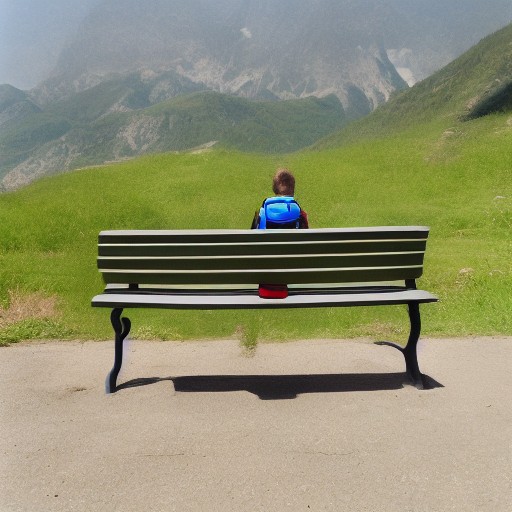} &
    \includegraphics[width=0.09\textwidth]{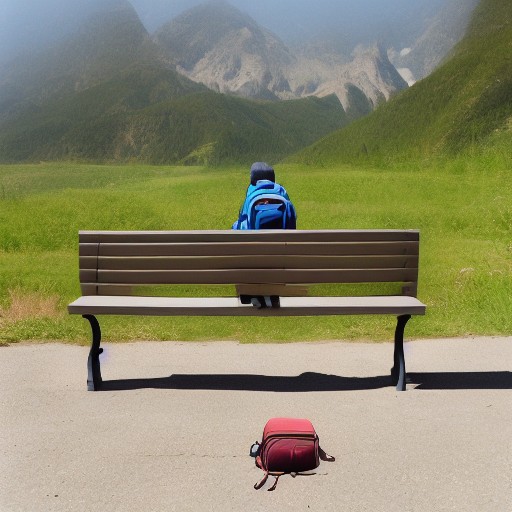} &
    \includegraphics[width=0.09\textwidth]{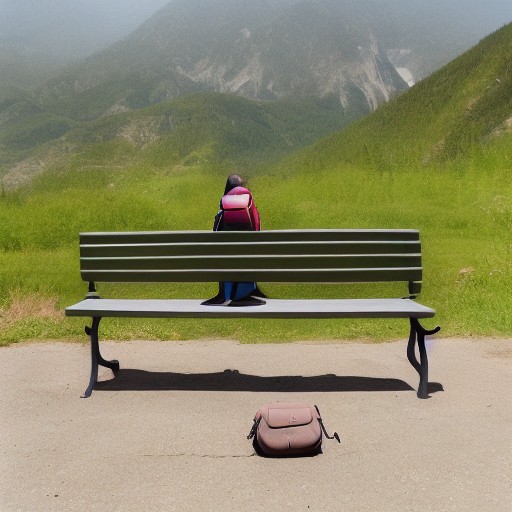} &
    \includegraphics[width=0.09\textwidth]{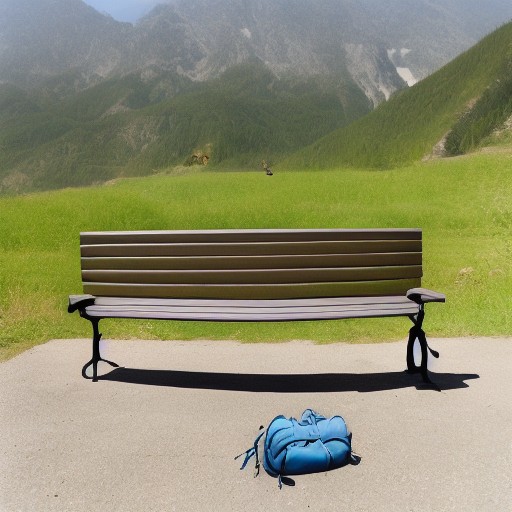} &
    \includegraphics[width=0.09\textwidth]{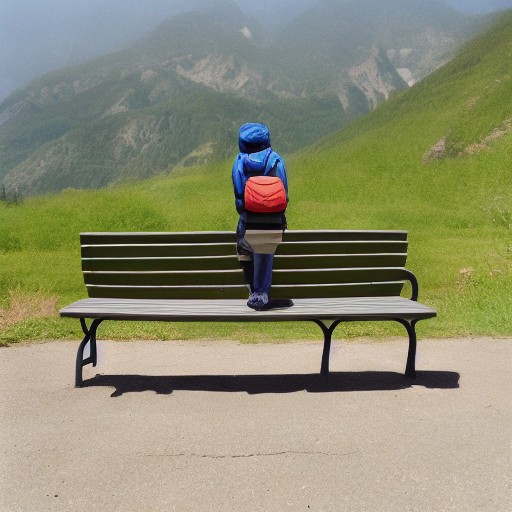} &
    \includegraphics[width=0.09\textwidth]{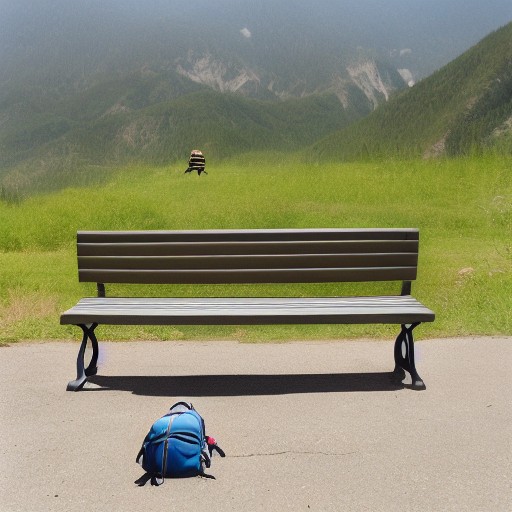} &
    \includegraphics[width=0.09\textwidth]{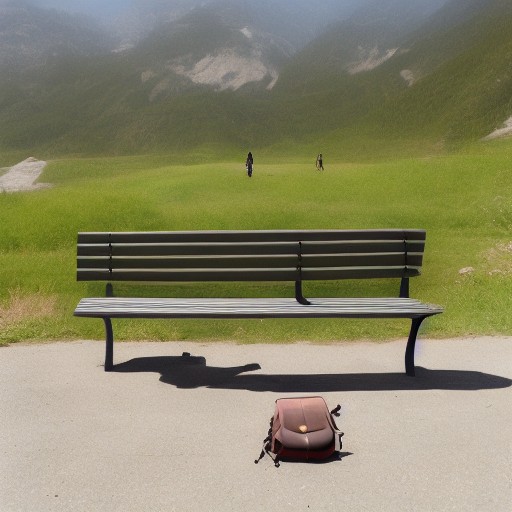} &
    \includegraphics[width=0.09\textwidth]{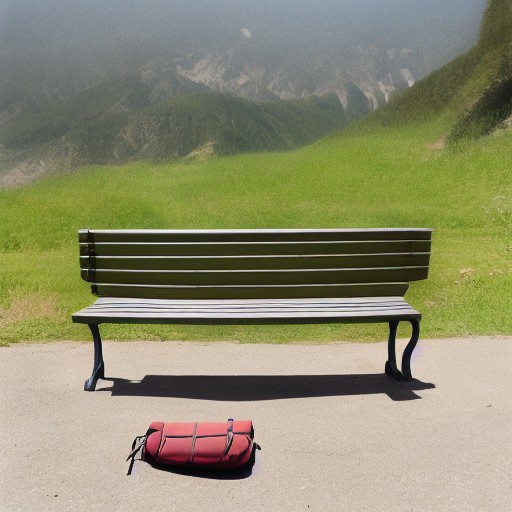} &
    \includegraphics[width=0.09\textwidth]{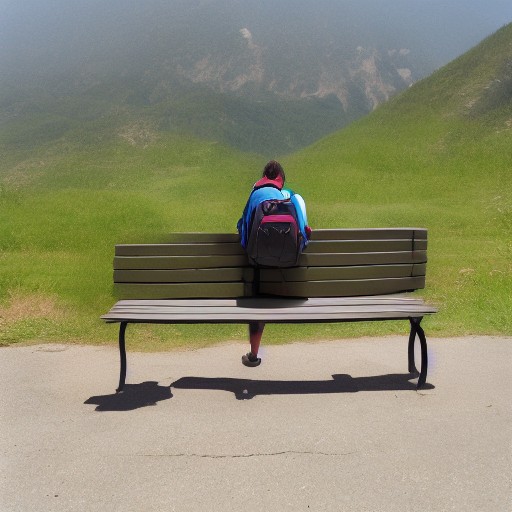} \\
\end{tabular}
}
\subfigure[MagicBrush]{
\begin{tabular}{c@{\hspace{0.3em}}c@{\hspace{0.3em}}c@{\hspace{0.3em}}c@{\hspace{0.3em}}c@{\hspace{0.3em}}c@{\hspace{0.3em}}c@{\hspace{0.3em}}c@{\hspace{0.3em}}c@{\hspace{0.3em}}c}  
    \includegraphics[width=0.09\textwidth]{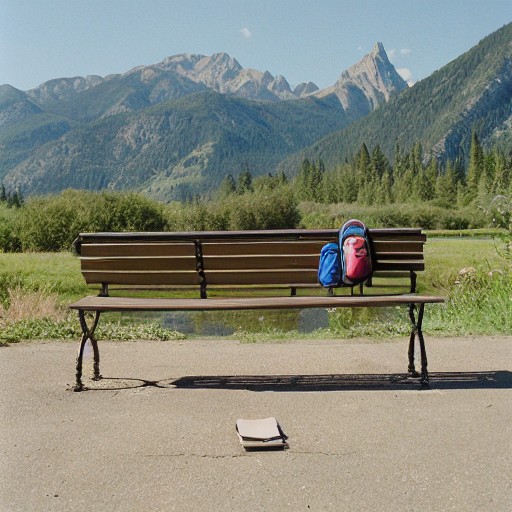} &
    \includegraphics[width=0.09\textwidth]{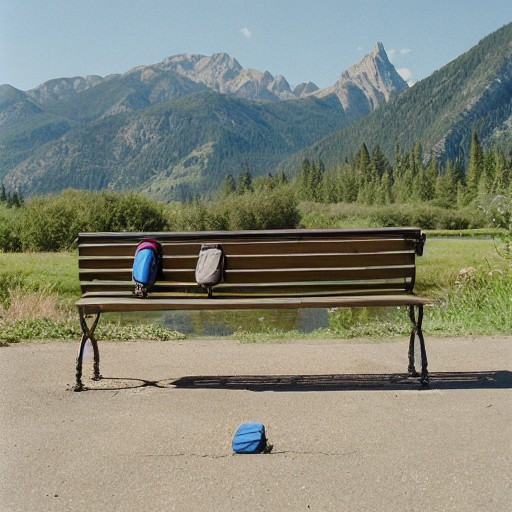} &
    \includegraphics[width=0.09\textwidth]{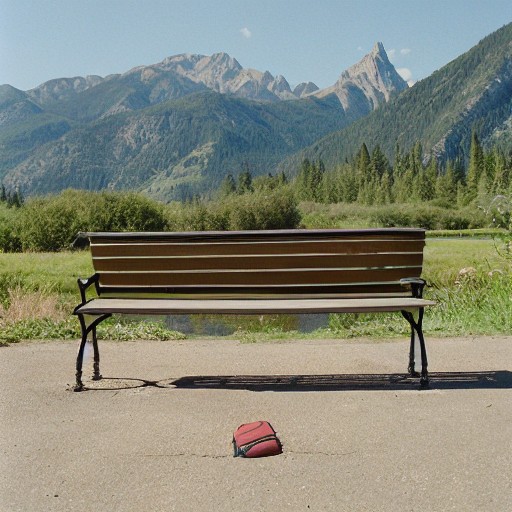} &
    \includegraphics[width=0.09\textwidth]{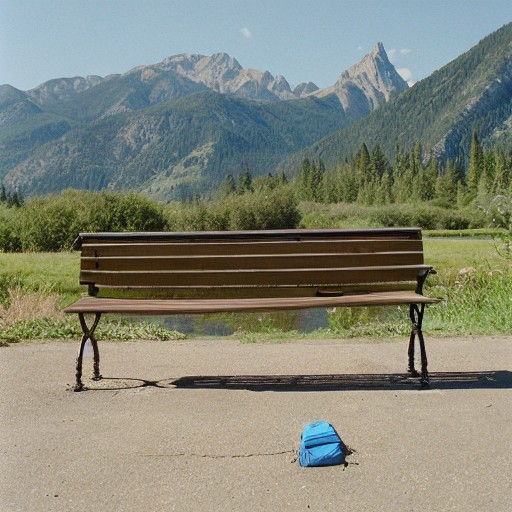} &
    \includegraphics[width=0.09\textwidth]{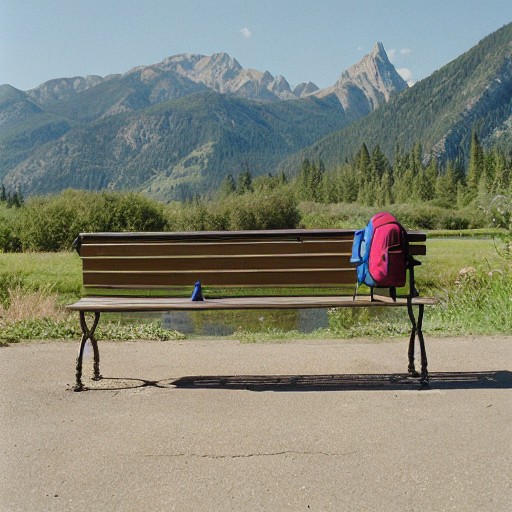} &
    \includegraphics[width=0.09\textwidth]{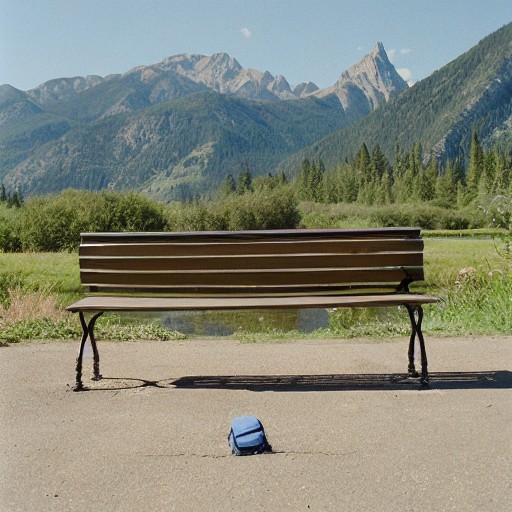} &
    \includegraphics[width=0.09\textwidth]{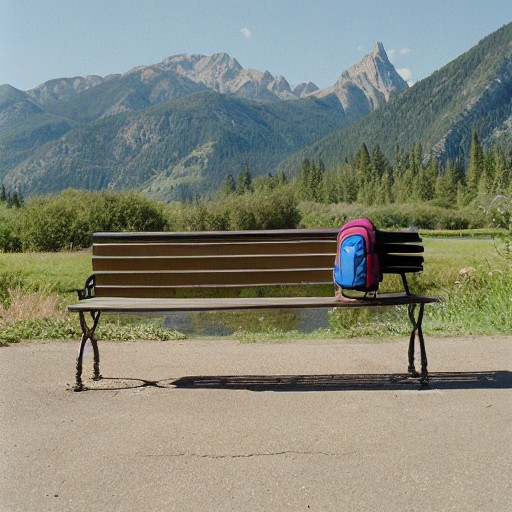} &
    \includegraphics[width=0.09\textwidth]{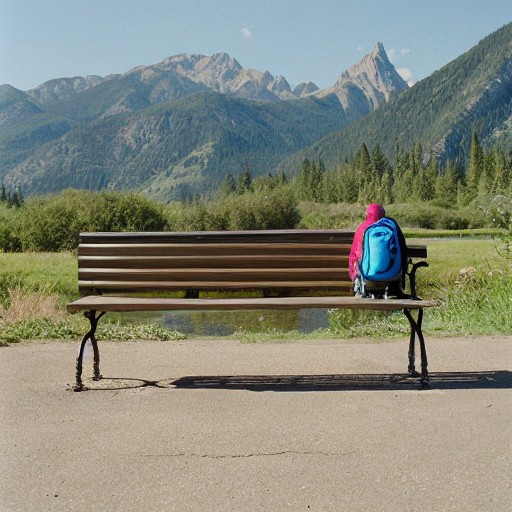} &
    \includegraphics[width=0.09\textwidth]{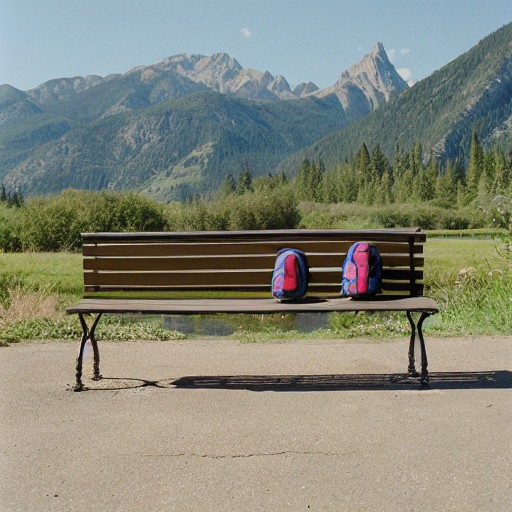} &
    \includegraphics[width=0.09\textwidth]{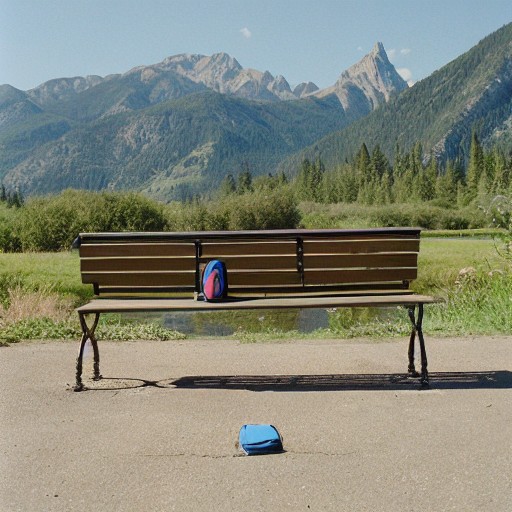} \\
\end{tabular}
}
\subfigure[UltraEdit]{
\begin{tabular}{c@{\hspace{0.3em}}c@{\hspace{0.3em}}c@{\hspace{0.3em}}c@{\hspace{0.3em}}c@{\hspace{0.3em}}c@{\hspace{0.3em}}c@{\hspace{0.3em}}c@{\hspace{0.3em}}c@{\hspace{0.3em}}c}  
    \includegraphics[width=0.09\textwidth]{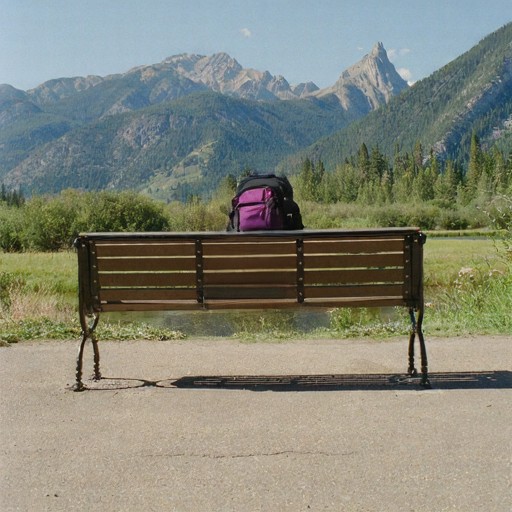} &
    \includegraphics[width=0.09\textwidth]{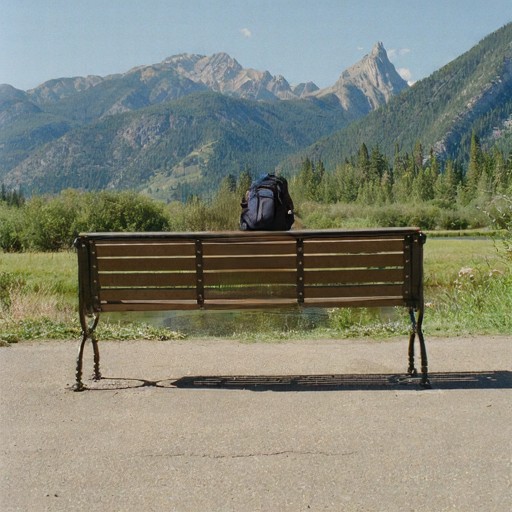} &
    \includegraphics[width=0.09\textwidth]{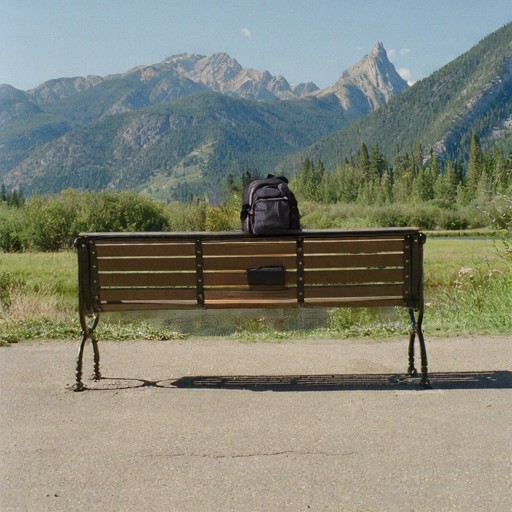} &
    \includegraphics[width=0.09\textwidth]{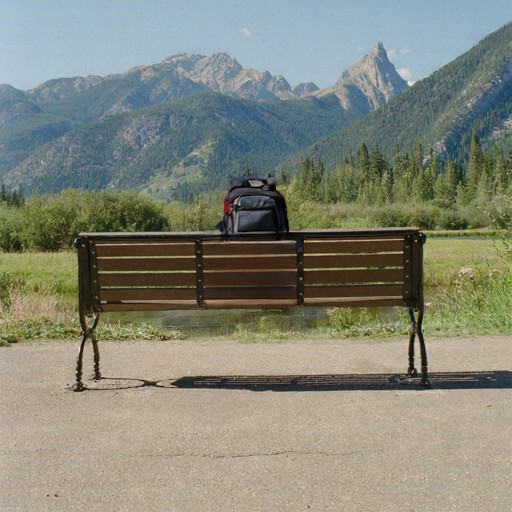} &
    \includegraphics[width=0.09\textwidth]{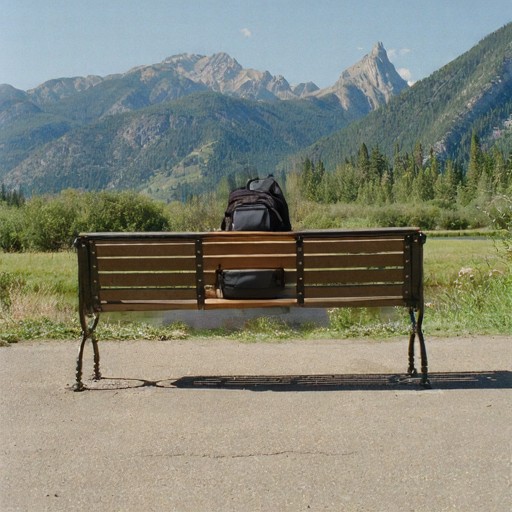} &
    \includegraphics[width=0.09\textwidth]{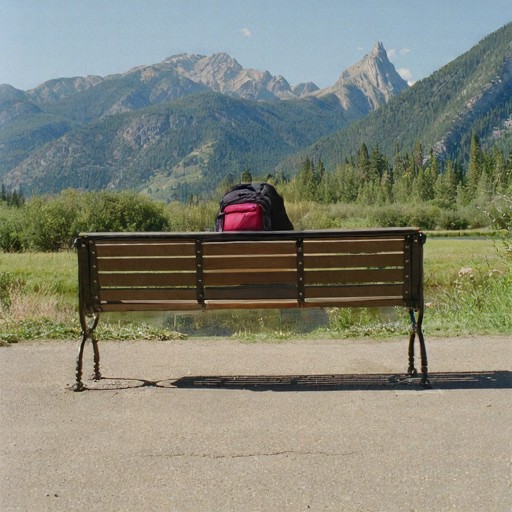} &
    \includegraphics[width=0.09\textwidth]{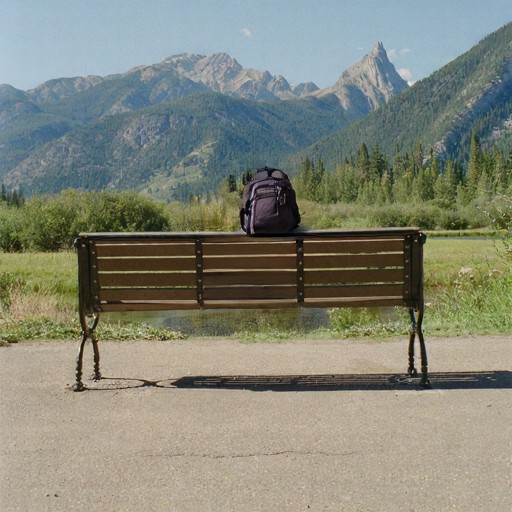} &
    \includegraphics[width=0.09\textwidth]{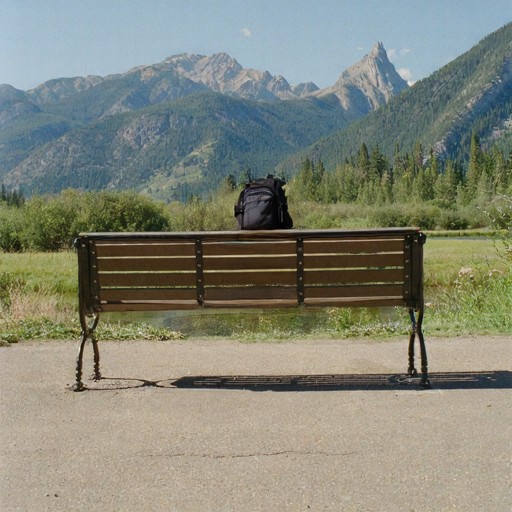} &
    \includegraphics[width=0.09\textwidth]{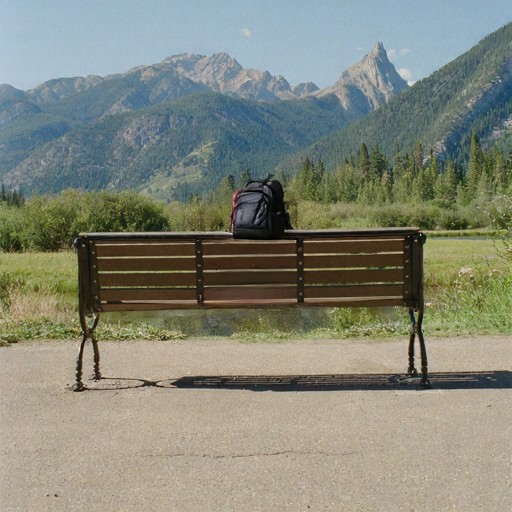} &
    \includegraphics[width=0.09\textwidth]{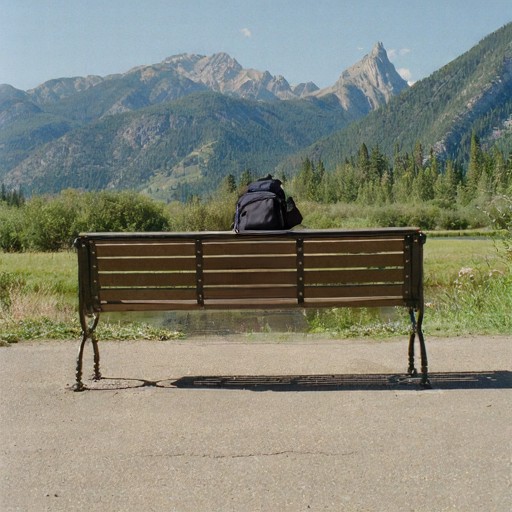} \\
\end{tabular}
}
\caption{\textbf{For baseline methods, changing the random seed does not help.} The random seeds are shown on the top. With 10 times the original compute, baseline methods still fail on the challenging editing task from EditEval of adding a bag to the bench. Our workflow succeeds in the first edit and will succeed in fewer than 10 edits should the first edit fail. The original image and our result are in \cref{fig:benchmark-results}.}
\label{fig:change-seed}
\end{center}

\end{figure*}

With baseline methods, a user can change the random seed to get multiple results and select the best one. However, even with more compute, baseline models will consistently fail on the same challenging example (\cref{fig:change-seed}). In contrast, with our workflow, a user can better utilize the increased compute if the user performs editing step by step.

\section{Iterative Construction}
\label{app:iterative-construction}

\cref{fig:iterative-construction} shows all 40 steps we use to iteratively construct \cref{subfig:iterative-heart}. Some steps involve subtle but important edits that are best visualized at a high resolution. We will release a video to better illustrate the editing operation done at each step.

We further evaluate iterative editing on recent (October 2025) state-of-the-art models and qualitatively demonstrate how image quality worsens as edits accumulate. The three models we evaluate are \textbf{FLUX.1 Kontext [pro]}, \textbf{Sora (GPT)}, and \textbf{Gemini 2.5 Flash Image (Nano Banana)}. Access to FLUX.1 Kontext [Pro] and Sora was provided by their respective official websites, and access to Gemini 2.5 Flash Image (Nano Banana) was provided via Google AI Studio.\footnote{\url{https://flux1.ai} and \url{https://sora.chatgpt.com}} For each model, we apply a sequence of 12 iterative edits to the same base image of a man sitting on a chair, wearing a yellow jacket.\footnote{\url{https://www.pexels.com/photo/man-in-yellow-long-sleeve-shirt-sitting-on-black-chair-smiling-7562191/}} The editing prompt at each step was: ``Change the man's jacket color to x'' where x cycles twice through the colors red, orange, yellow, green, blue, and purple, for a total of 12 editing steps. All models failed in this iterative editing task, with image quality dropping dramatically after 12 steps. Below, we further explain and visualize the failure patterns of each model.

\begin{figure}[!htbp]
  \centering
  \stepsgrid{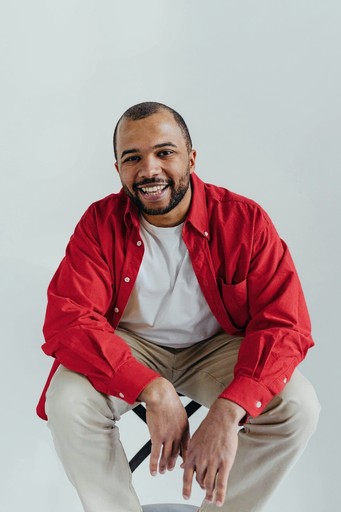}
  \caption{\textbf{Noise accumulates in the iterative editing results by FLUX.1 Kontext [pro].} Obvious artifacts appear at around 7 steps.}
  \label{fig:flux-steps}
\end{figure}

\paragraph{FLUX.1 Kontext [pro].}
This model is a diffusion-based image editor with strong local inpainting and instruction-following capabilities \citep{BFL2025FluxKontext}. However, repeated edits introduce obvious artifacts, as shown in \cref{fig:flux-steps}.

\begin{figure}[!htbp]
  \centering
  \stepsgrid{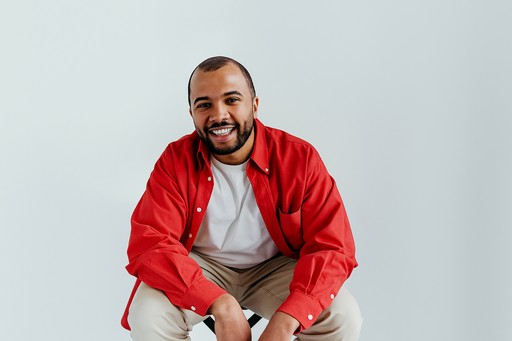}
  \caption{\textbf{Sora does not preserve details over 12 iterative editing steps.} The background color changes from gray to light blue. The facial structure of the man also changes.}
  \label{fig:sora-steps}
\end{figure}

\paragraph{Sora.}
Sora preserves global structure well in early steps but changes the aspect ratio from the first step. Moreover, degradation appears as gradual over-smoothing and inconsistency in fine detail, as shown in \cref{fig:sora-steps}. It also struggles with preserving facial structure as noted in practitioner reports.\footnote{\url{https://davidharris.uk/2025/04/11/kling-vs-runway-vs-sora-vs-lumalabs-a-deep-dive-into-ai-video-generation-platforms/}}

\begin{figure}[!htbp]
  \centering
  \stepsgrid{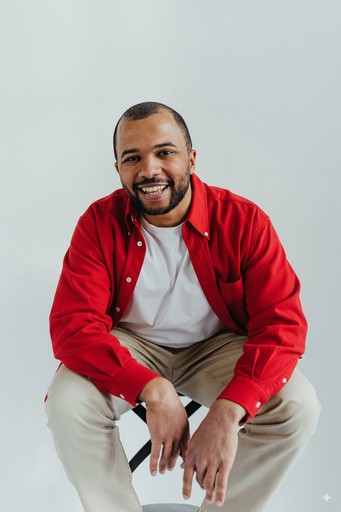}
  \caption{\textbf{Gemini 2.5 Flash Image fails to maintain character consistency over 12 editing steps.} The background, skin tone, and details on the jacket all change.}
  \label{fig:gemini-steps}
\end{figure}

\begin{figure}[p] 
  \centering
  \includegraphics[width=0.95\textwidth,height=0.95\textheight,keepaspectratio]{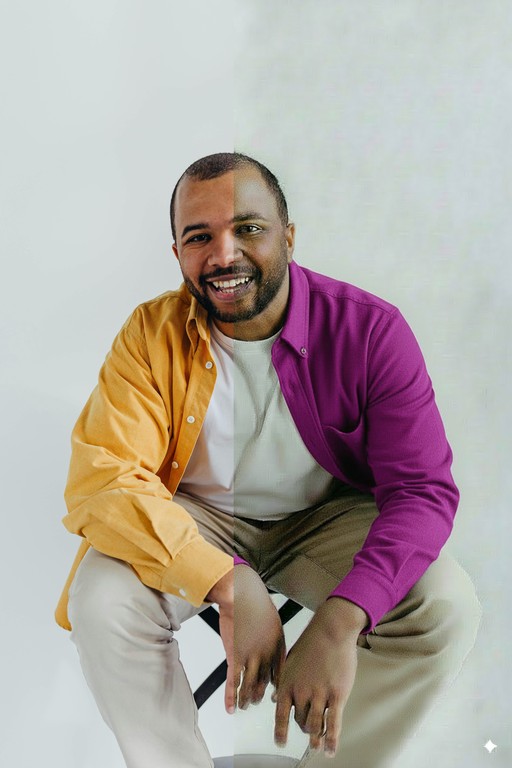}
  \caption{\textbf{For clarity, we present a full-page comparison between the initial image and the final output from Gemini 2.5 Flash Image.} The skin tone shifts noticeably, and the background becomes darker and more textured. The buttons on the man's jacket disappear in the edited version, and fine details (eyes, teeth, and fingernails) morph and differ from the original. Additional indentations and fine lines also appear around the mouth.}
  \label{fig:gemini-fullpage}
\end{figure}

\begin{figure*}[htbp]

\begin{center}
\small
\begin{tabular}{c@{\hspace{0.3em}}c@{\hspace{0.3em}}c@{\hspace{0.3em}}c@{\hspace{0.3em}}c@{\hspace{0.3em}}c@{\hspace{0.3em}}c@{\hspace{0.3em}}c@{\hspace{0.3em}}}
    \scriptsize\textsc{Step 1} & \scriptsize\textsc{Step 2} & \scriptsize\textsc{Step 3} & \scriptsize\textsc{Step 4} & \scriptsize\textsc{Step 5} & \scriptsize\textsc{Step 6} & \scriptsize\textsc{Step 7} & \scriptsize\textsc{Step 8} \\
    
    \includegraphics[width=0.11\textwidth]{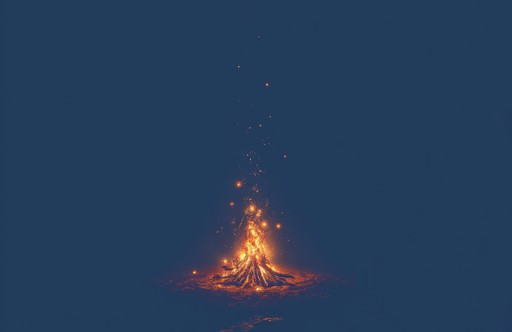} &
    \includegraphics[width=0.11\textwidth]{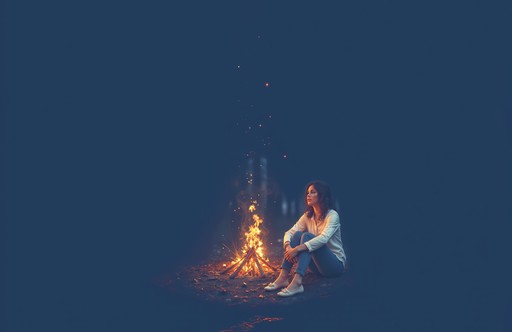} &
    \includegraphics[width=0.11\textwidth]{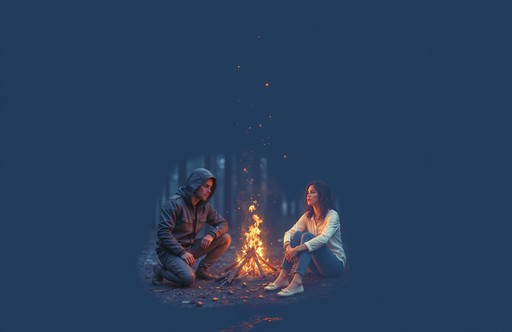} &
    \includegraphics[width=0.11\textwidth]{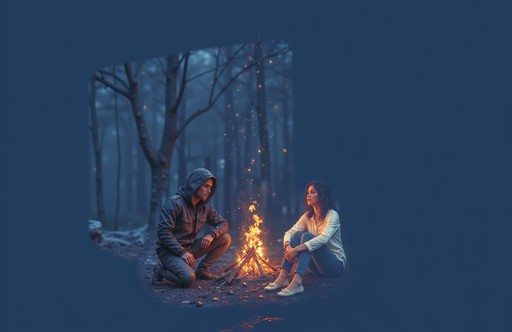} &
    \includegraphics[width=0.11\textwidth]{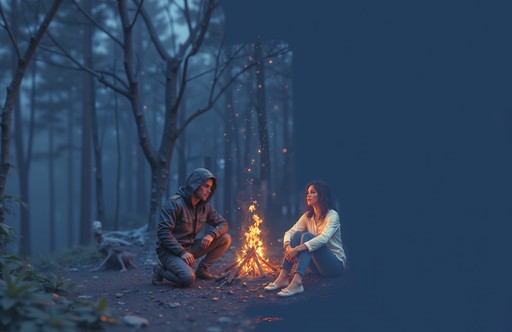} &
    \includegraphics[width=0.11\textwidth]{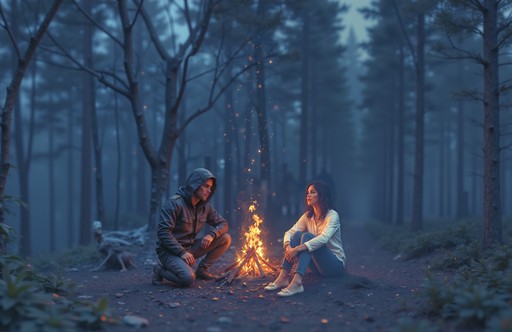} &
    \includegraphics[width=0.11\textwidth]{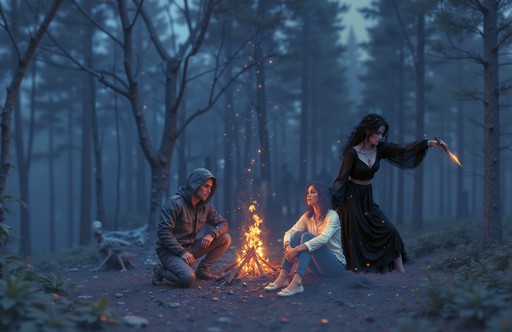} &
    \includegraphics[width=0.11\textwidth]{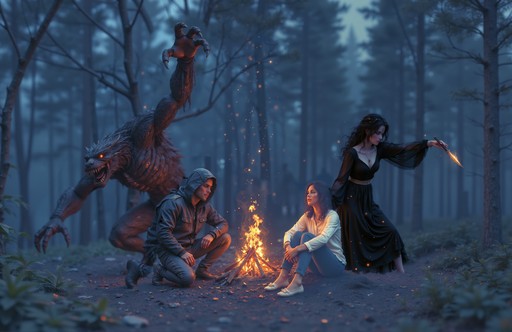} \\

    \scriptsize\textsc{Step 9} & \scriptsize\textsc{Step 10} & \scriptsize\textsc{Step 11} & \scriptsize\textsc{Step 12} & \scriptsize\textsc{Step 13} & \scriptsize\textsc{Step 14} & \scriptsize\textsc{Step 15} & \scriptsize\textsc{Step 16} \\
    
    \includegraphics[width=0.11\textwidth]{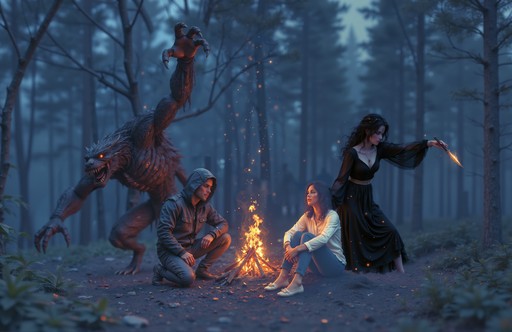} &
    \includegraphics[width=0.11\textwidth]{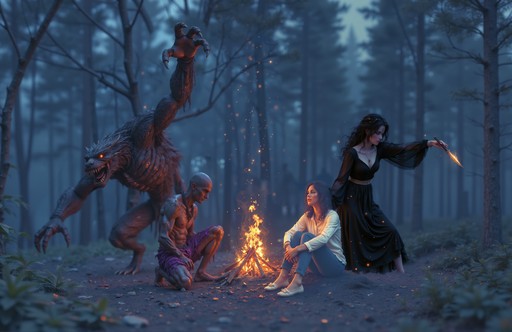} &
    \includegraphics[width=0.11\textwidth]{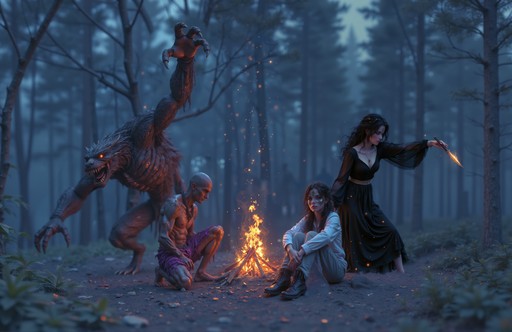} &
    \includegraphics[width=0.11\textwidth]{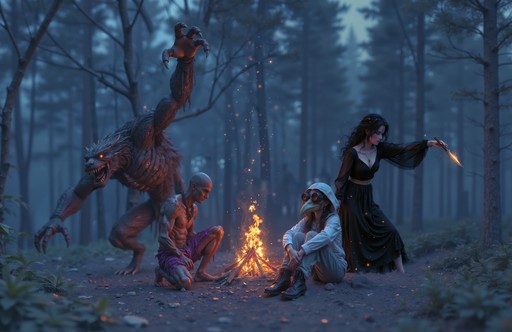} &
    \includegraphics[width=0.11\textwidth]{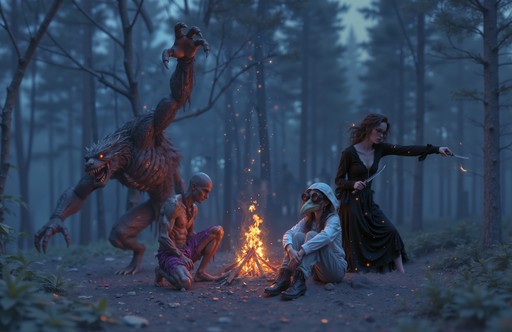} &
    \includegraphics[width=0.11\textwidth]{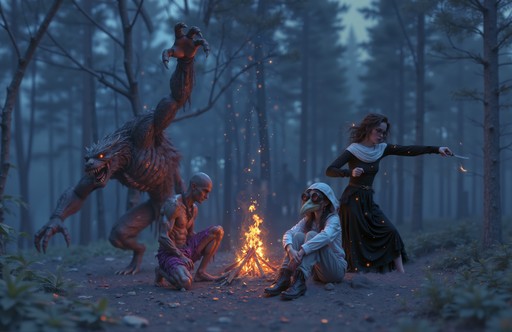} &
    \includegraphics[width=0.11\textwidth]{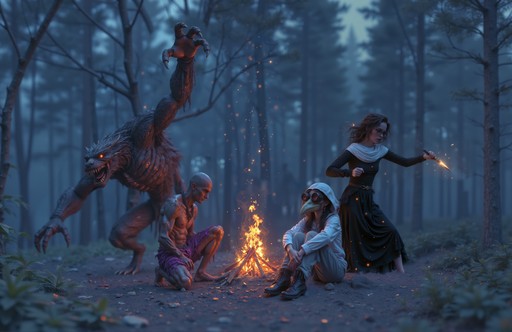} &
    \includegraphics[width=0.11\textwidth]{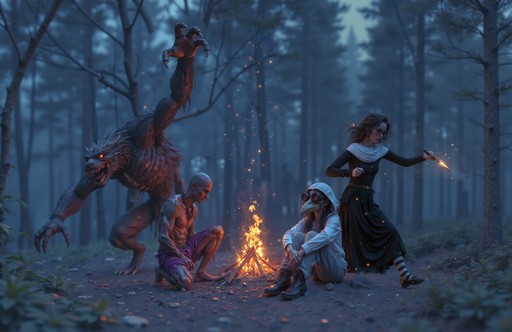} \\

    \scriptsize\textsc{Step 17} & \scriptsize\textsc{Step 18} & \scriptsize\textsc{Step 19} & \scriptsize\textsc{Step 20} & \scriptsize\textsc{Step 21} & \scriptsize\textsc{Step 22} & \scriptsize\textsc{Step 23} & \scriptsize\textsc{Step 24} \\
    
    \includegraphics[width=0.11\textwidth]{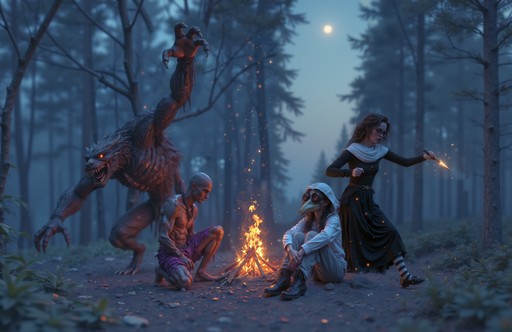} &
    \includegraphics[width=0.11\textwidth]{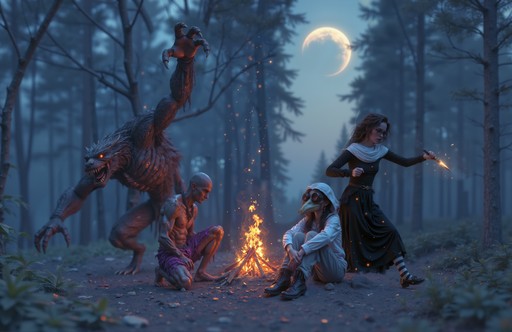} &
    \includegraphics[width=0.11\textwidth]{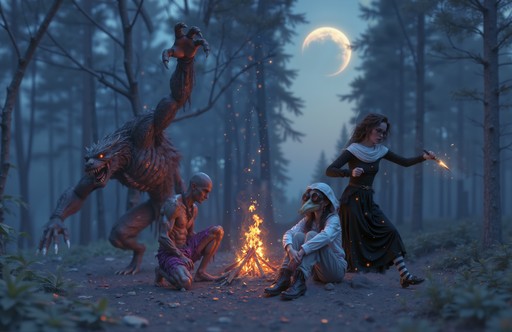} &
    \includegraphics[width=0.11\textwidth]{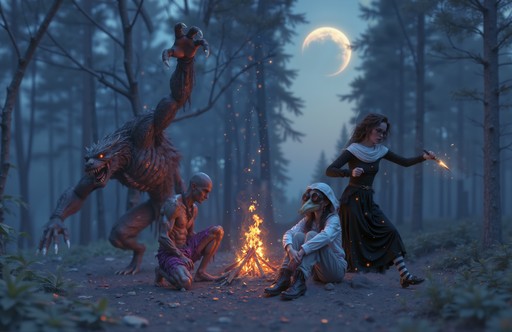} &
    \includegraphics[width=0.11\textwidth]{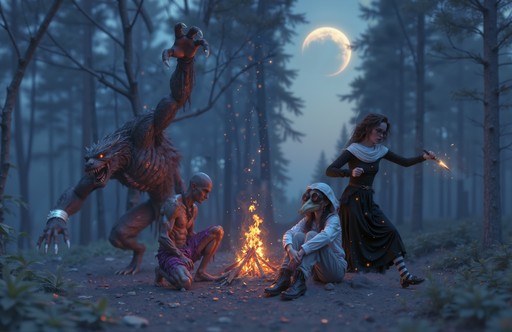} &
    \includegraphics[width=0.11\textwidth]{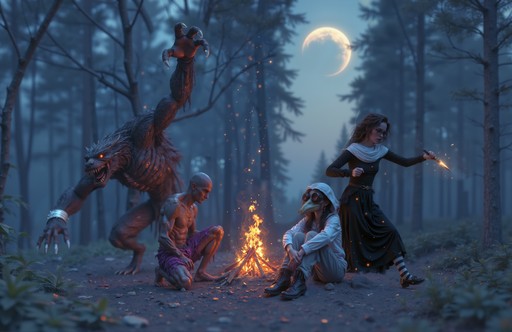} &
    \includegraphics[width=0.11\textwidth]{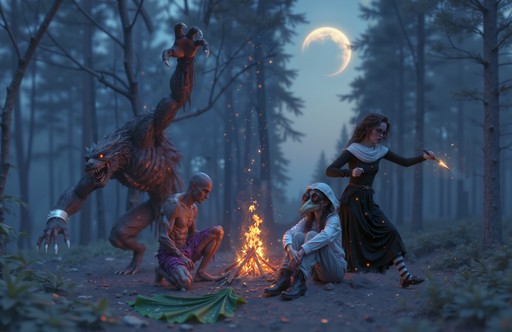} &
    \includegraphics[width=0.11\textwidth]{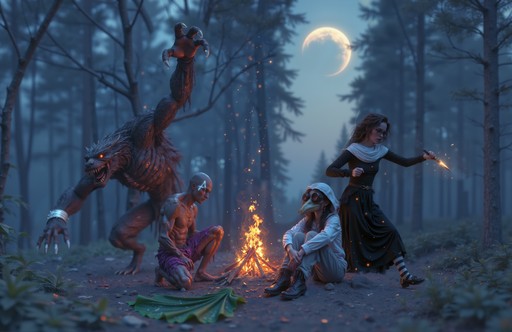} \\

    \scriptsize\textsc{Step 25} & \scriptsize\textsc{Step 26} & \scriptsize\textsc{Step 27} & \scriptsize\textsc{Step 28} & \scriptsize\textsc{Step 29} & \scriptsize\textsc{Step 30} & \scriptsize\textsc{Step 31} & \scriptsize\textsc{Step 32} \\
    
    \includegraphics[width=0.11\textwidth]{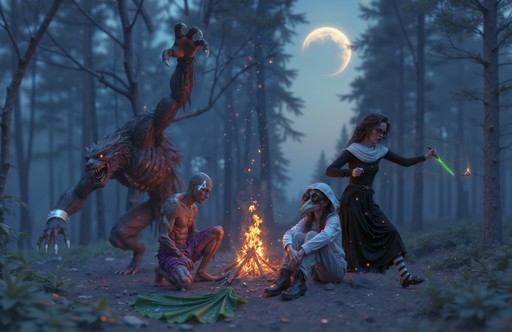} &
    \includegraphics[width=0.11\textwidth]{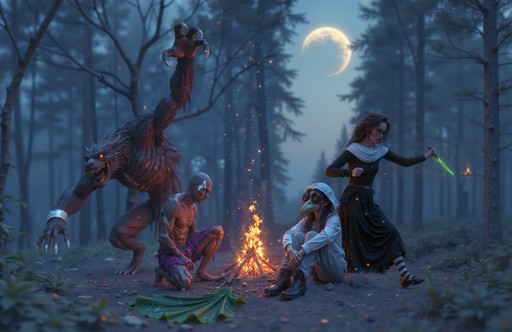} &
    \includegraphics[width=0.11\textwidth]{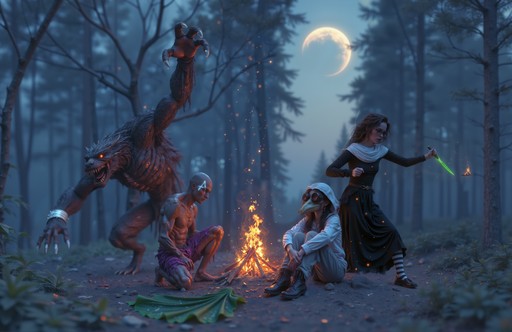} &
    \includegraphics[width=0.11\textwidth]{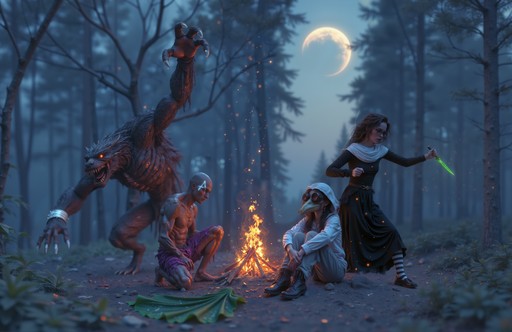} &
    \includegraphics[width=0.11\textwidth]{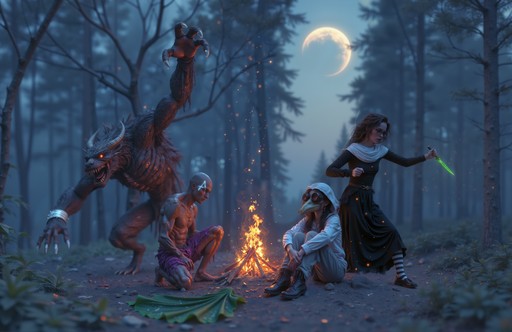} &
    \includegraphics[width=0.11\textwidth]{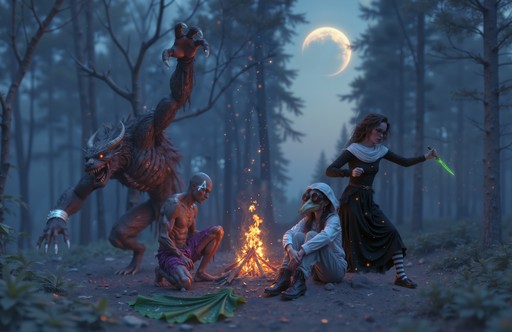} &
    \includegraphics[width=0.11\textwidth]{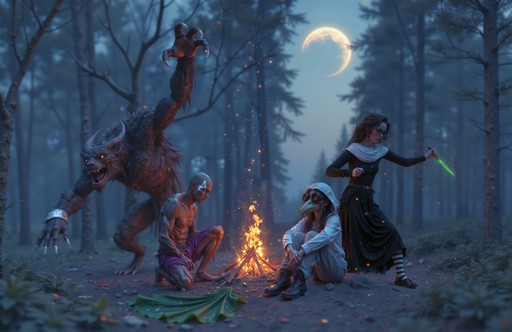} &
    \includegraphics[width=0.11\textwidth]{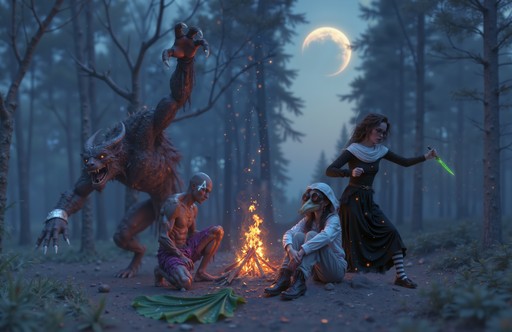} \\

    \scriptsize\textsc{Step 33} & \scriptsize\textsc{Step 34} & \scriptsize\textsc{Step 35} & \scriptsize\textsc{Step 36} & \scriptsize\textsc{Step 37} & \scriptsize\textsc{Step 38} & \scriptsize\textsc{Step 39} & \scriptsize\textsc{Step 40} \\
    
    \includegraphics[width=0.11\textwidth]{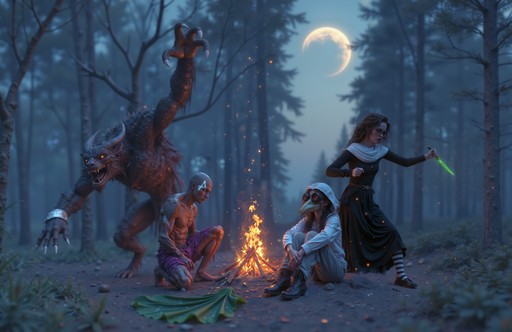} &
    \includegraphics[width=0.11\textwidth]{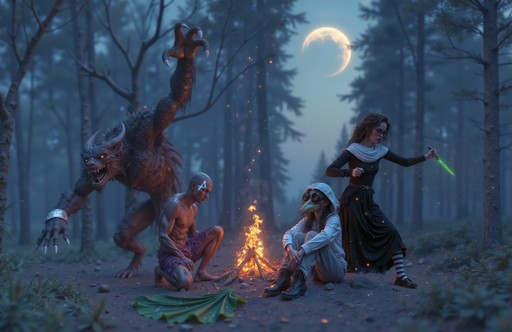} &
    \includegraphics[width=0.11\textwidth]{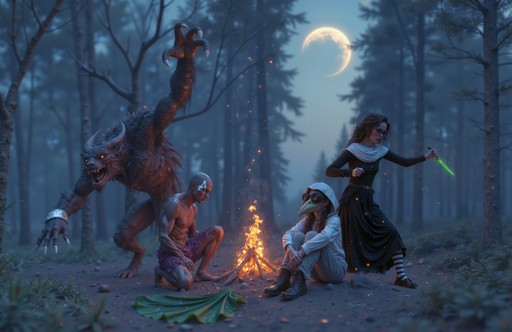} &
    \includegraphics[width=0.11\textwidth]{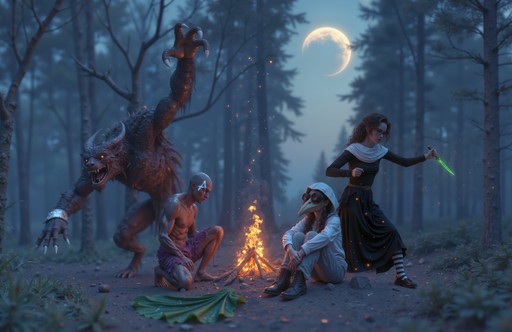} &
    \includegraphics[width=0.11\textwidth]{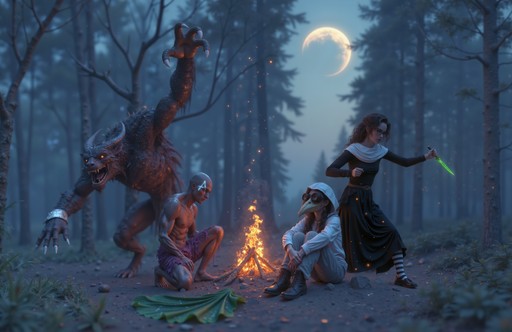} &
    \includegraphics[width=0.11\textwidth]{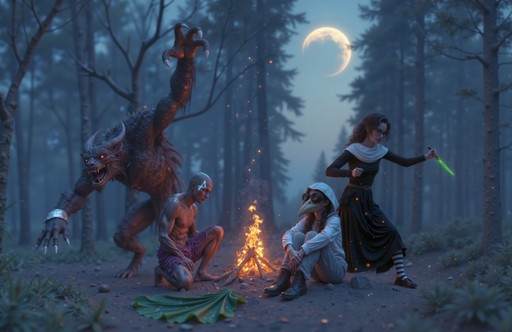} &
    \includegraphics[width=0.11\textwidth]{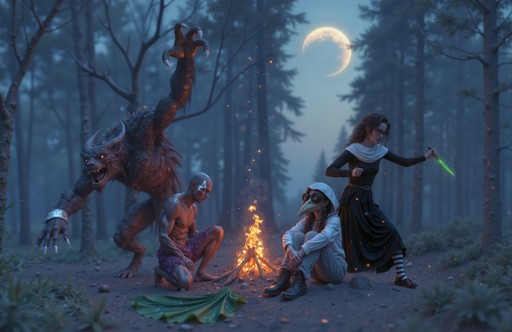} &
    \includegraphics[width=0.11\textwidth]{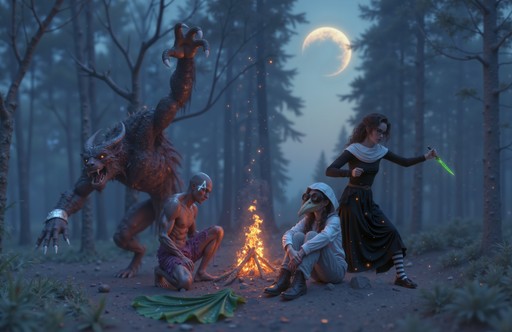} \\
\end{tabular}
\caption{\textbf{We use 40 steps to iteratively construct the image and fix detail errors.} With our workflow, high frequency details do not deteriorate over the steps. With baseline methods, the deterioration happens at the first step.}
\label{fig:iterative-construction}
\end{center}

\end{figure*}

\paragraph{Gemini 2.5 Flash Image.}
During iterative editing, Gemini 2.5 Flash Image preserves global structure reliably, but texture and details still degrade, as shown in both \cref{fig:gemini-steps} and \cref{fig:gemini-fullpage}.

\section{Results of Different Art Styles}
\label{app:art-styles}

\begin{figure}[t!]
\begin{center}

\scriptsize
\begin{tabular}{c@{\hspace{0.3em}}c}
    \textsc{BoleroMix (Pony) v1.41} & \textsc{Flux.1 [dev]} \\
    \includegraphics[width=0.48\textwidth]{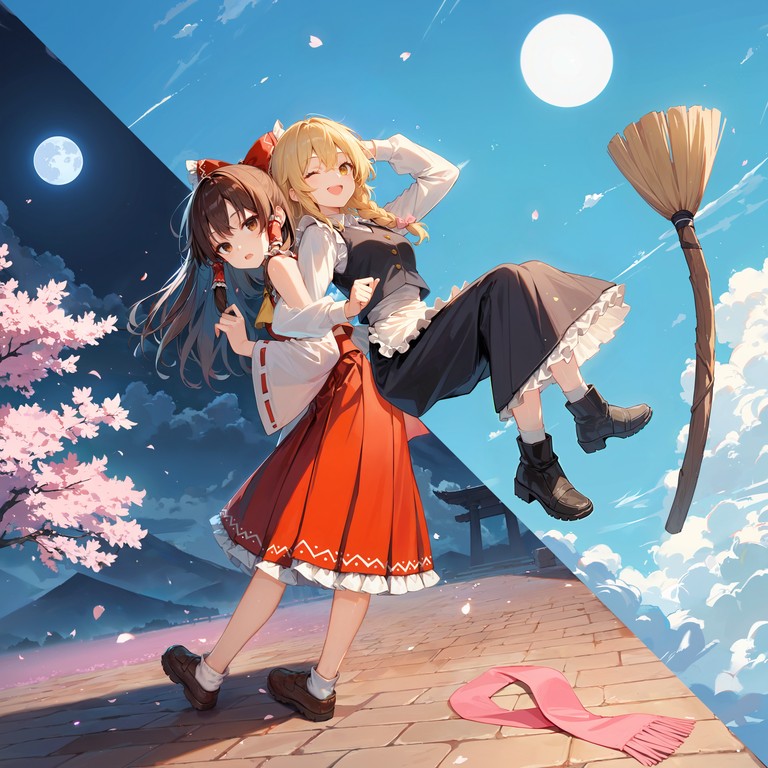} &
    \includegraphics[width=0.48\textwidth]{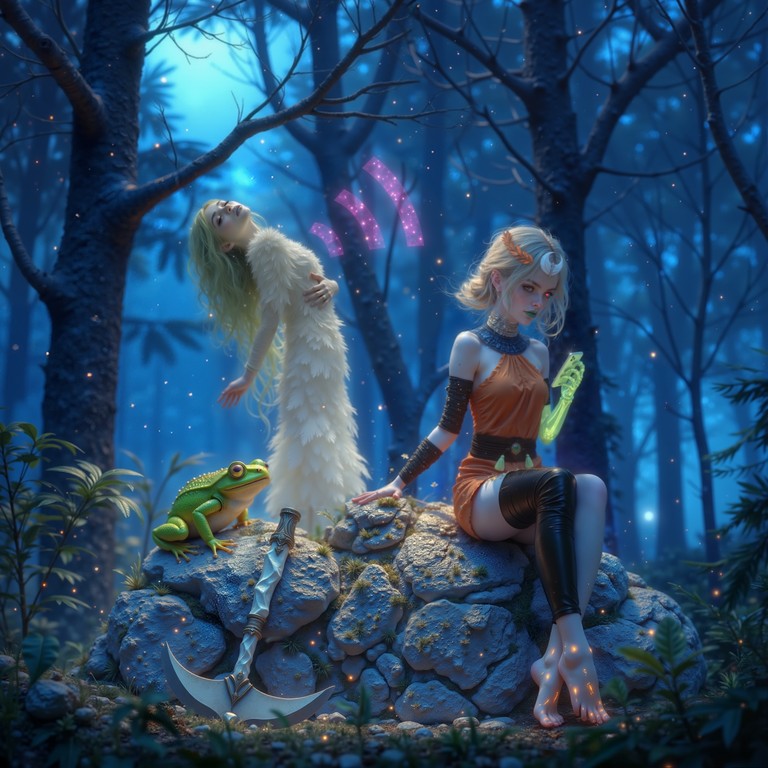} \\
\end{tabular}

\caption{\textbf{Our workflow excels at editing and iteratively generating images of different art styles.} Both images are generated using around 100 to 200 editing steps, within 4 hours in one sitting. The characters are from \textit{Touhou Project} and \textit{Hades 2}. Character LoRAs are not used, so many editing steps are spent on correcting details of the character's clothes. Note that due to the complex composition and high resolution (2048$\times$2048), it is extremely hard, if not impossible, to generate these images with a single text-to-image step.}
\label{fig:art-styles}
\end{center}

\end{figure}

\cref{fig:art-styles} shows two additional results. One is generated with BoleroMix (Pony) v1.41 as the backbone, and the other is generated with the same Flux.1 [dev] as the backbone. These additional results illustrate the flexibility of our workflow in editing and iteratively generating images of different art styles. Note that these are fan-made images, where details may differ from those of the original characters.

\section{Hyperparameter Recommendations}
\label{app:hyperparameter-recommendations}

\begin{figure}[ht!]
\begin{center}
\subfigure[Context Size]{
\scriptsize
\begin{tabular}{c@{\hspace{0.3em}}c@{\hspace{0.3em}}c@{\hspace{0.3em}}c@{\hspace{0.3em}}c@{\hspace{0.3em}}c}
    \textsc{Hinted} & \textsc{Small} & \textsc{Medium} & \textsc{Large} & \textsc{Top Left} & \textsc{Top Right} \\
    \includegraphics[width=0.15\textwidth]{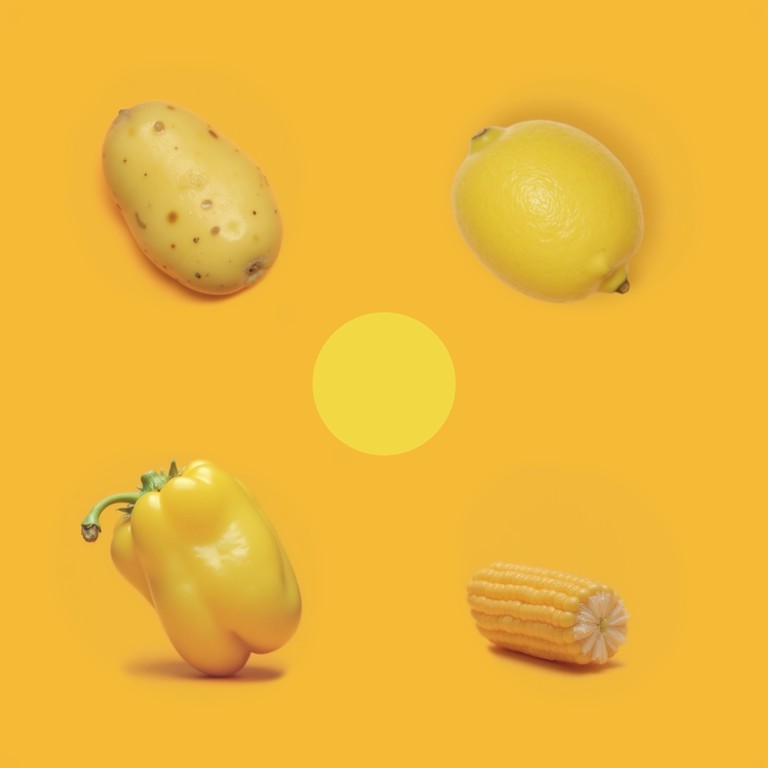} &
    \includegraphics[width=0.15\textwidth]{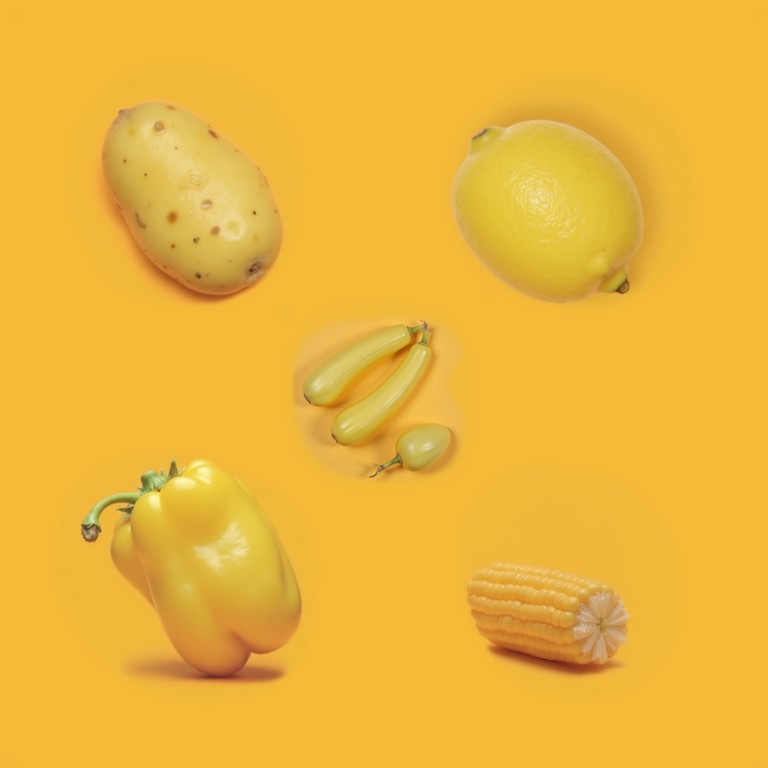} &
    \includegraphics[width=0.15\textwidth]{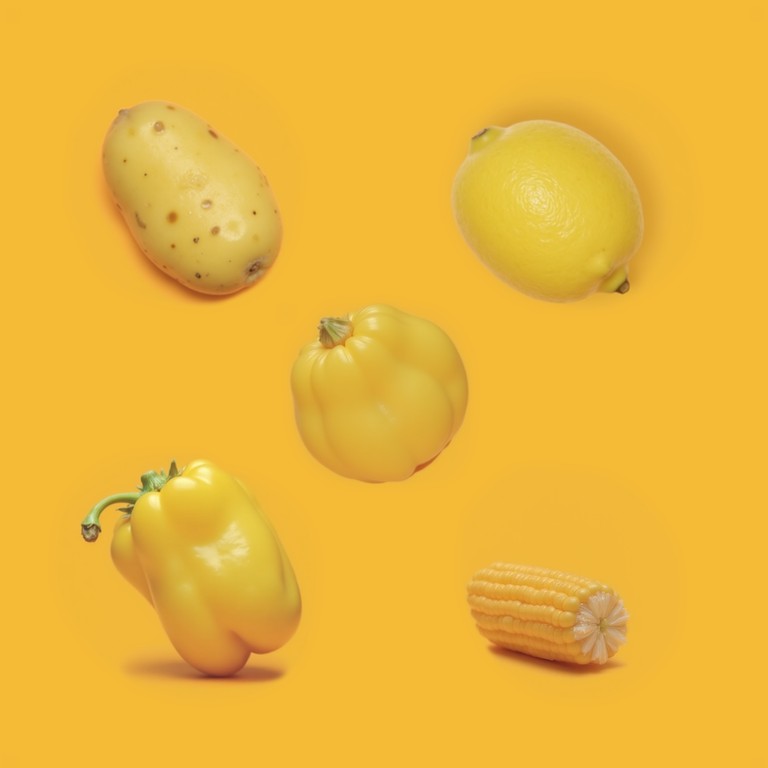} &
    \includegraphics[width=0.15\textwidth]{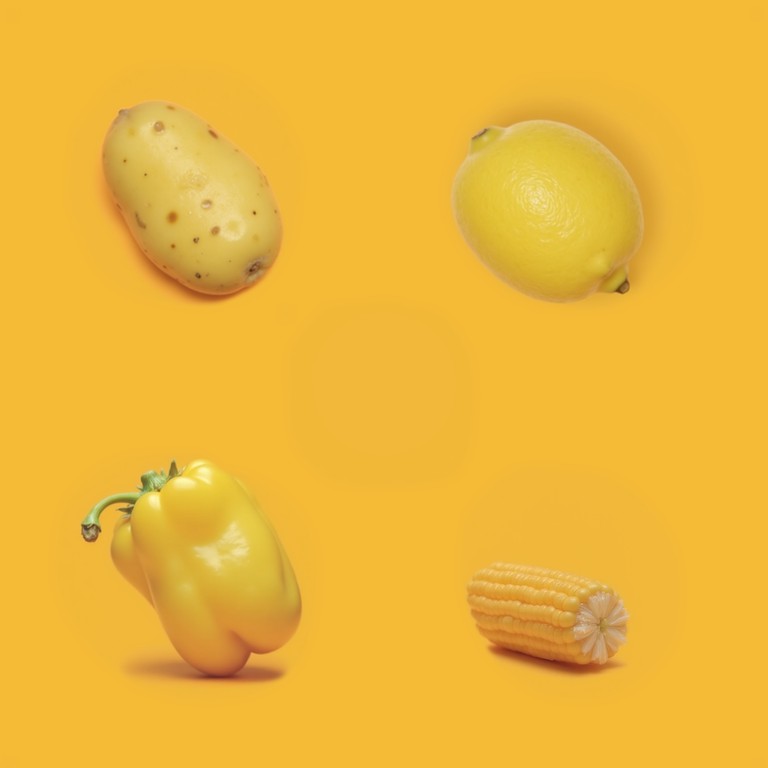} &
    \includegraphics[width=0.15\textwidth]{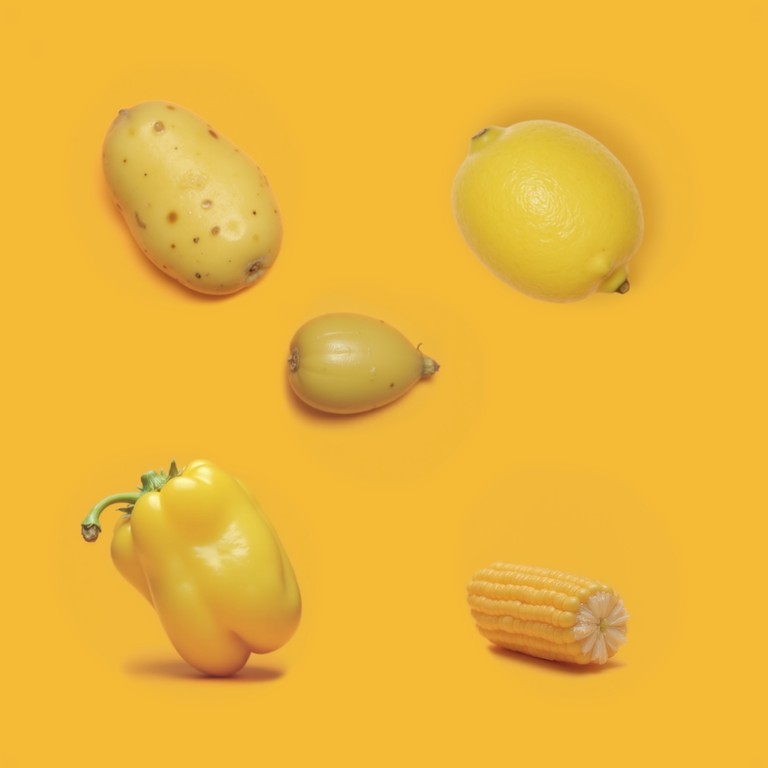} &
    \includegraphics[width=0.15\textwidth]{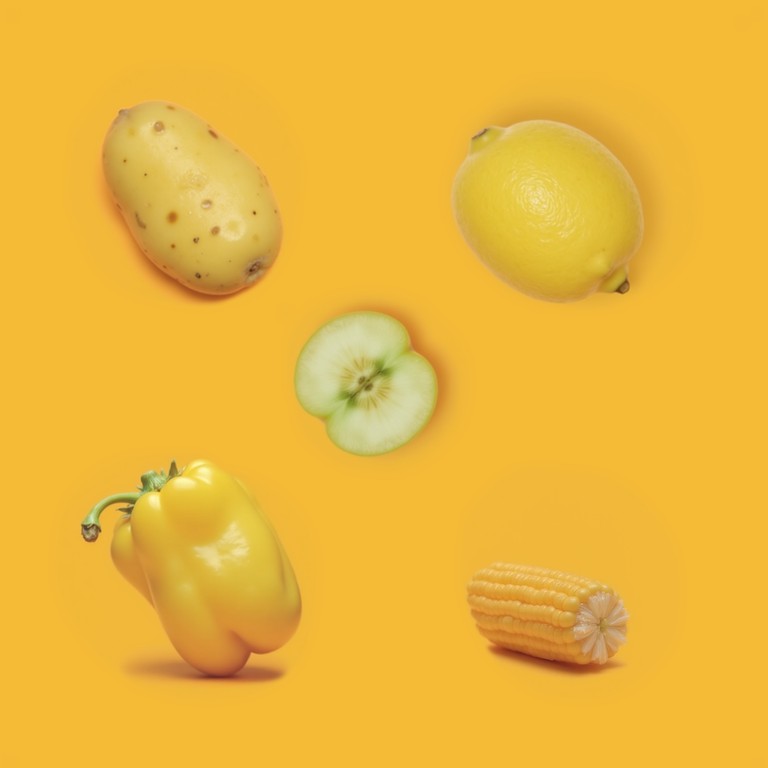} \\
\end{tabular}
\label{subfig:context-size}
}

\subfigure[Denoising Strength]{
\scriptsize
\begin{tabular}{c@{\hspace{0.3em}}c@{\hspace{0.3em}}c@{\hspace{0.3em}}c@{\hspace{0.3em}}c@{\hspace{0.3em}}c}
    \textsc{Hinted} & \textsc{0.1} & \textsc{0.3} & \textsc{0.5} & \textsc{0.7} & \textsc{0.9} \\
    \includegraphics[width=0.15\textwidth]{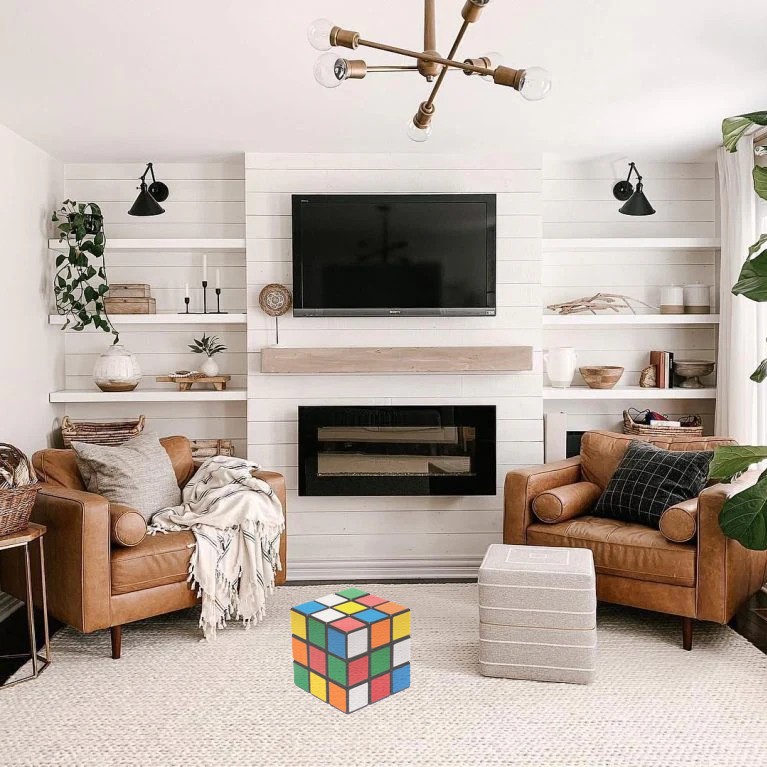} &
    \includegraphics[width=0.15\textwidth]{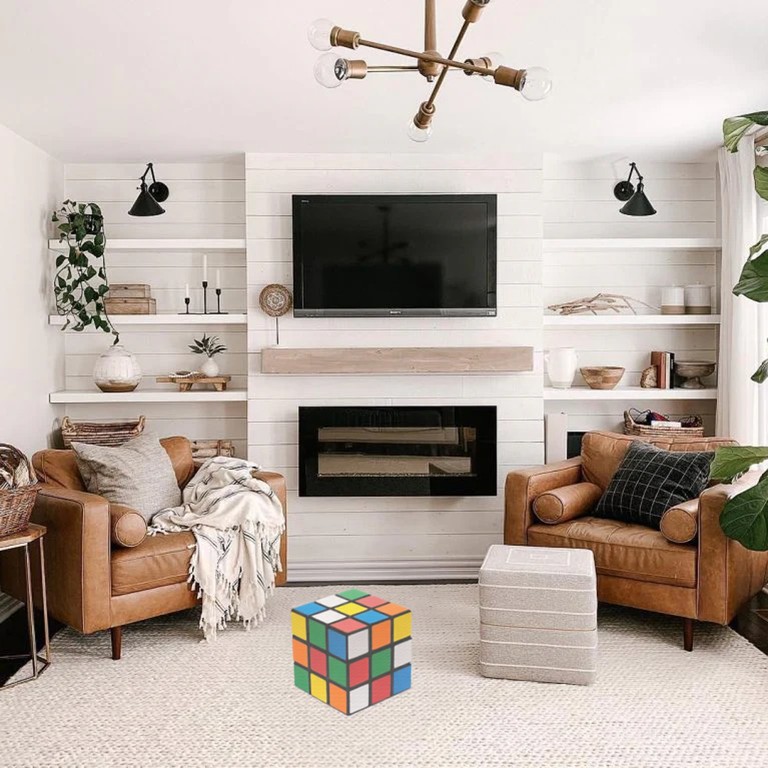} &
    \includegraphics[width=0.15\textwidth]{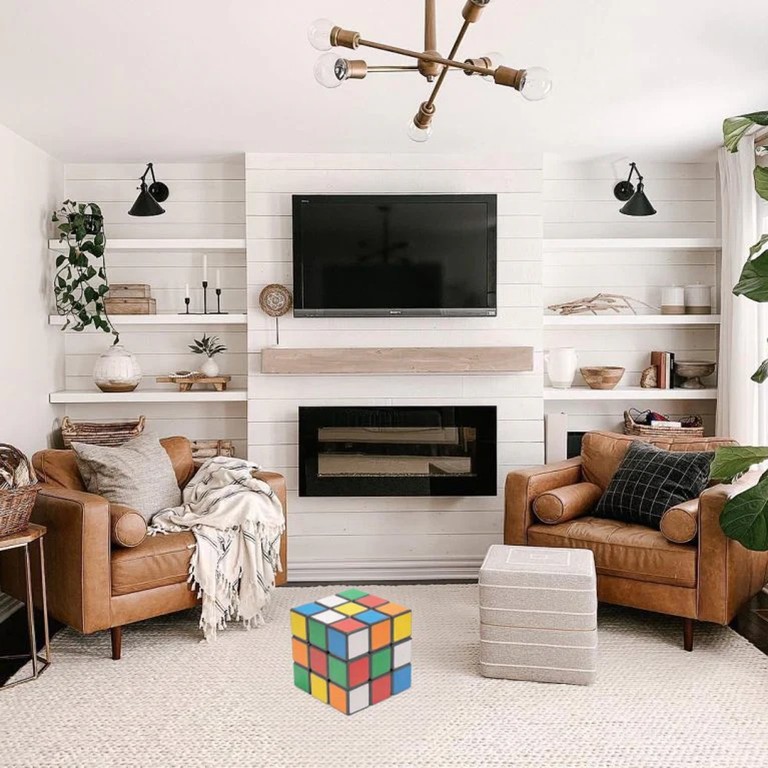} &
    \includegraphics[width=0.15\textwidth]{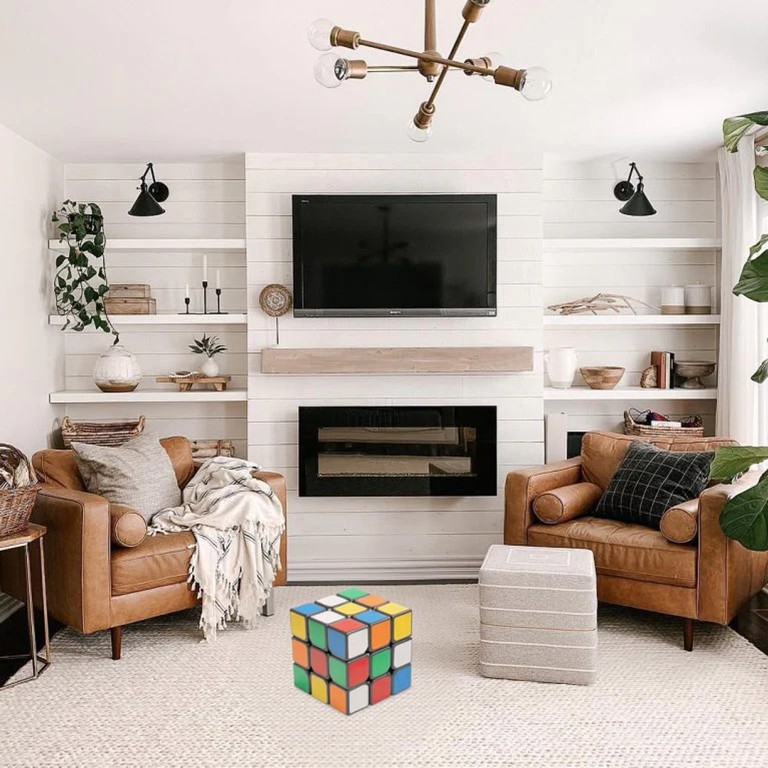} &
    \includegraphics[width=0.15\textwidth]{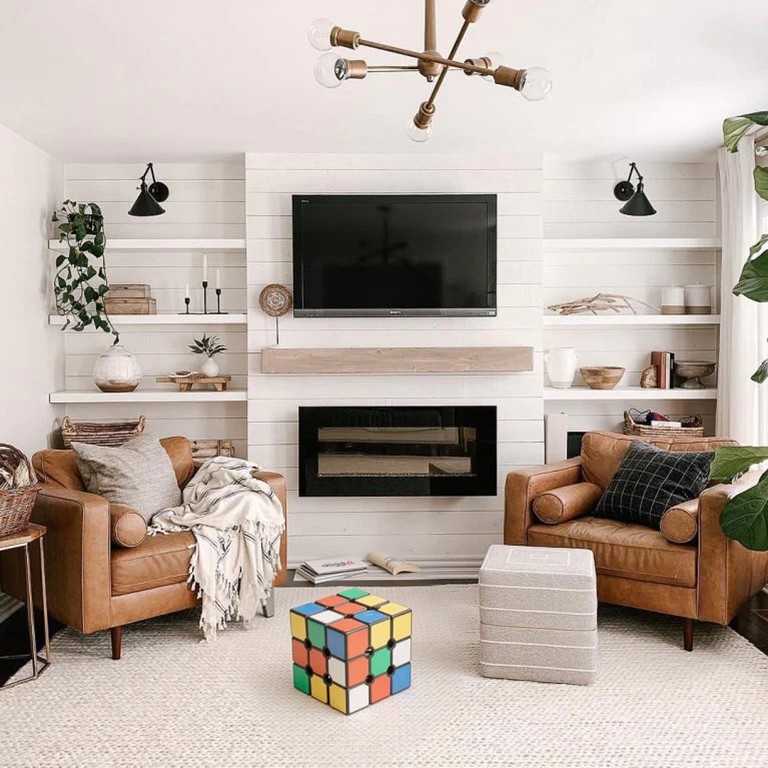} &
    \includegraphics[width=0.15\textwidth]{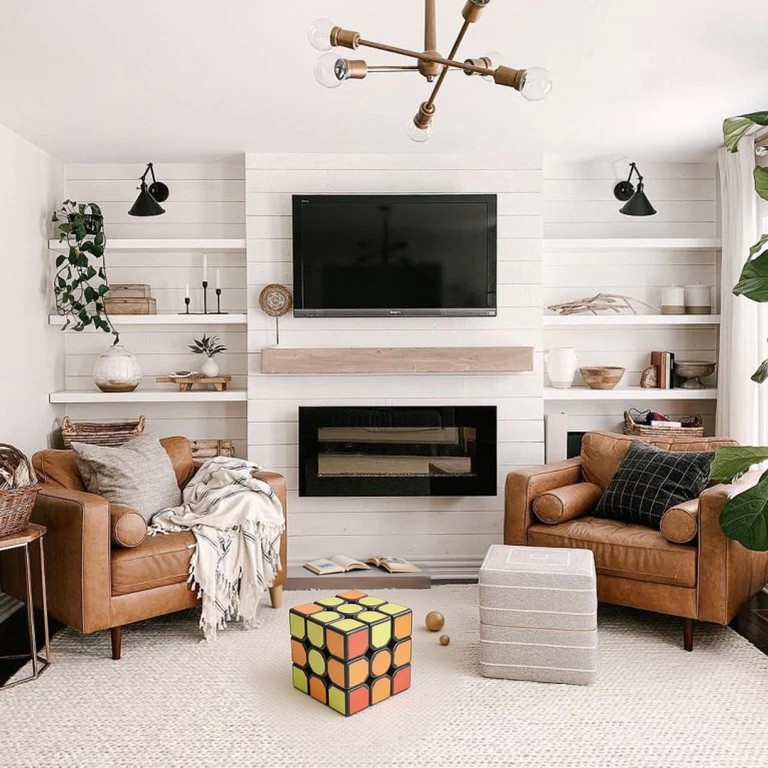} \\
\end{tabular}
\label{subfig:denoising-strength}
}

\subfigure[Canny Model Steps]{
\scriptsize
\begin{tabular}{c@{\hspace{0.3em}}c@{\hspace{0.3em}}c@{\hspace{0.3em}}c@{\hspace{0.3em}}c@{\hspace{0.3em}}c}
    \textsc{Hinted} & \textsc{5} & \textsc{10} & \textsc{15} & \textsc{20} & \textsc{25} \\
    \includegraphics[width=0.15\textwidth]{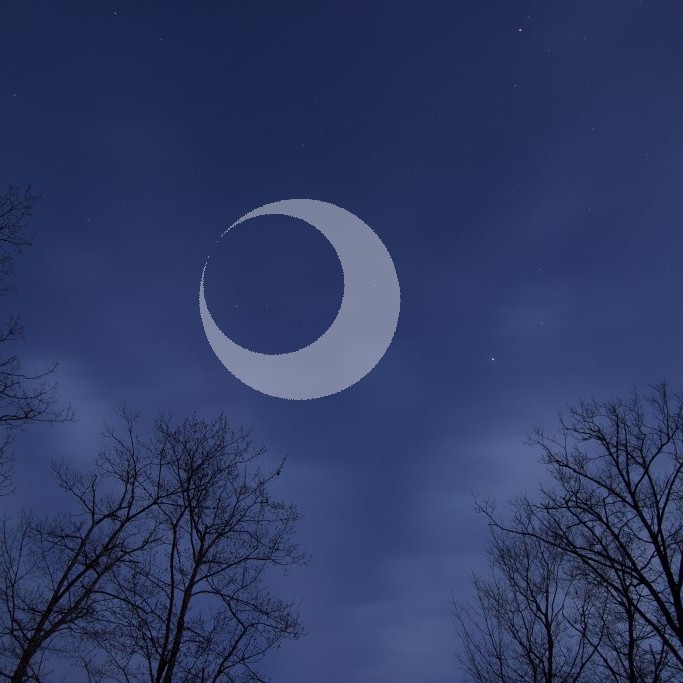} &
    \includegraphics[width=0.15\textwidth]{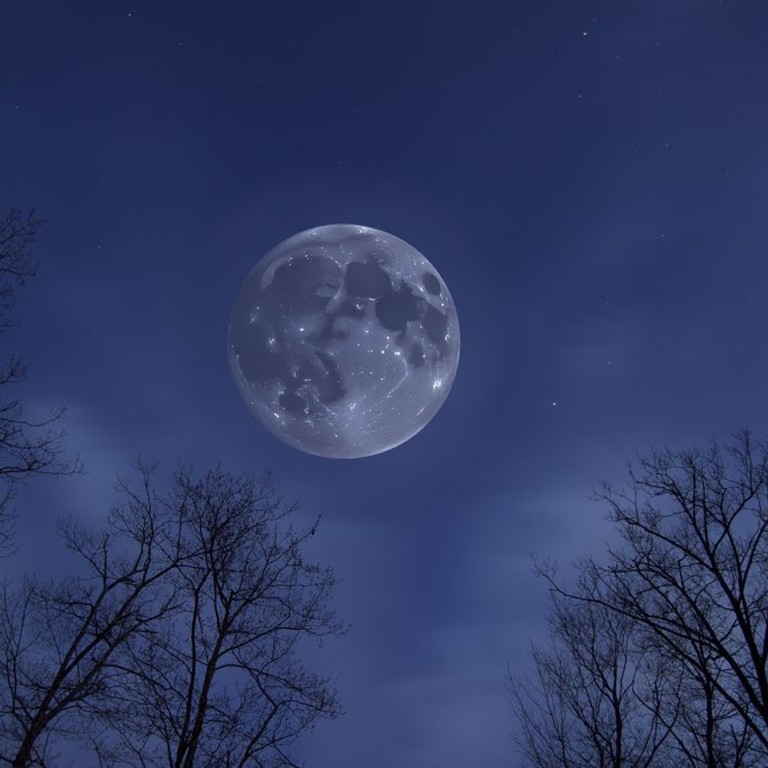} &
    \includegraphics[width=0.15\textwidth]{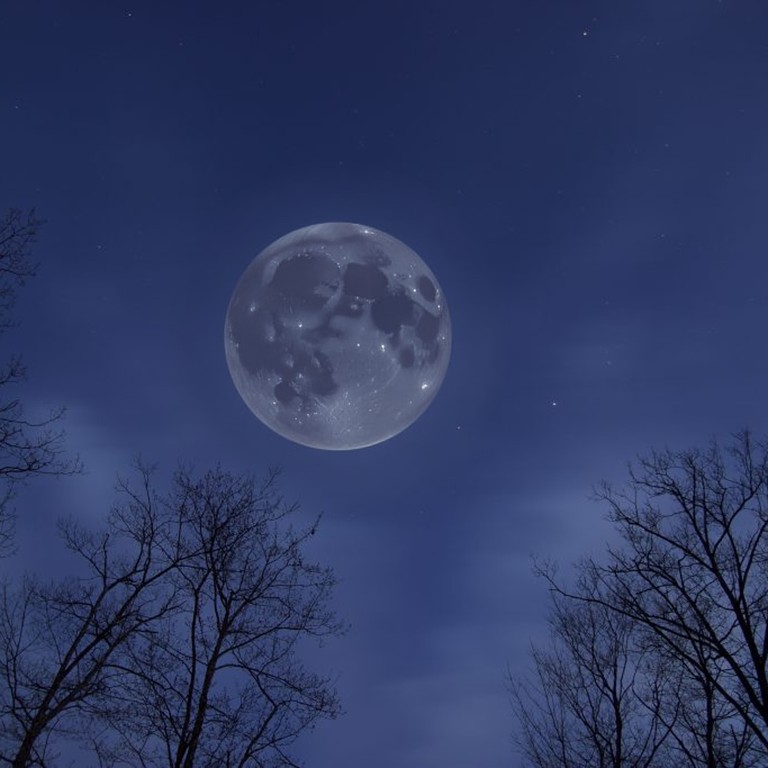} &
    \includegraphics[width=0.15\textwidth]{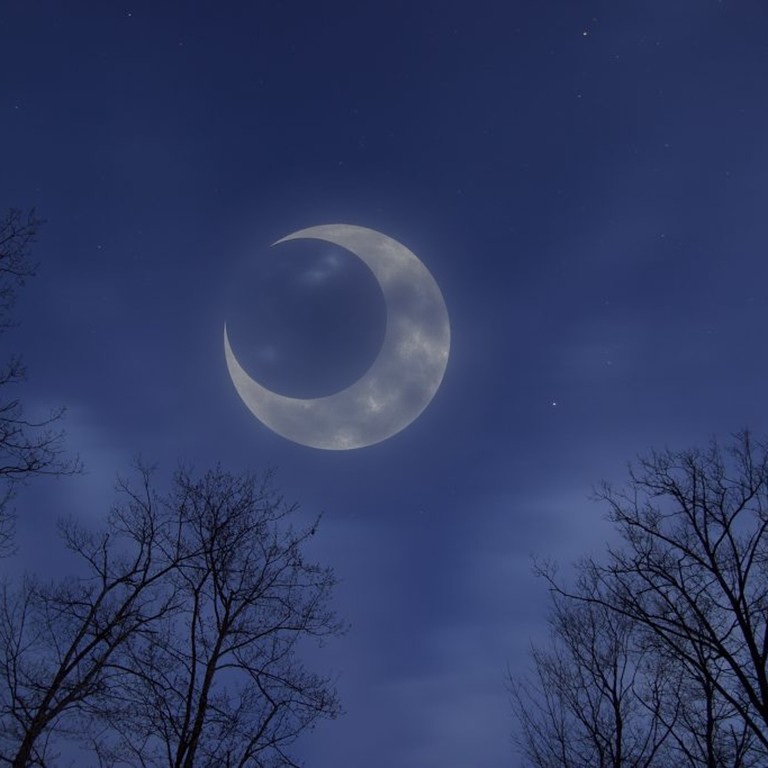} &
    \includegraphics[width=0.15\textwidth]{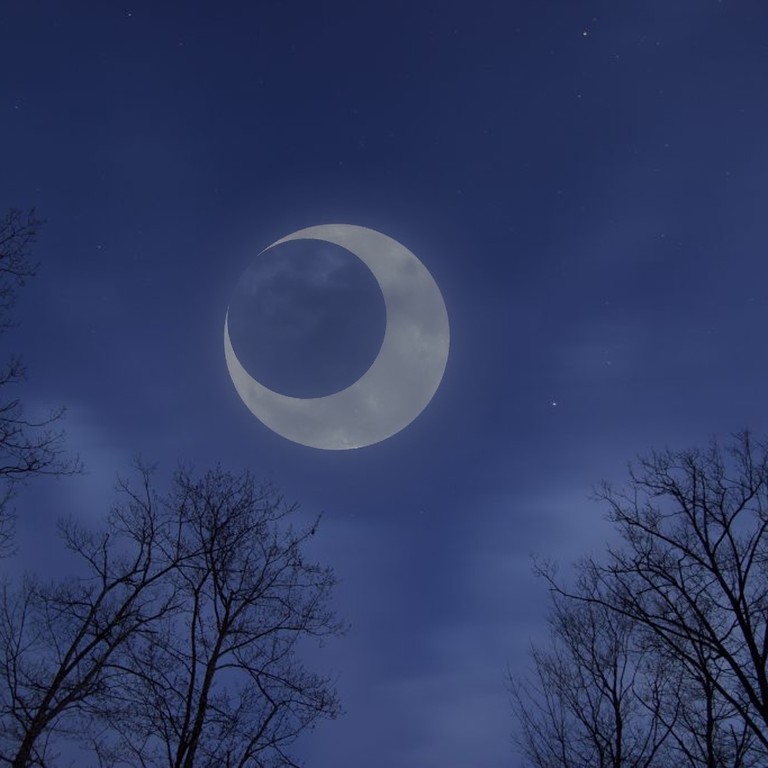} &
    \includegraphics[width=0.15\textwidth]{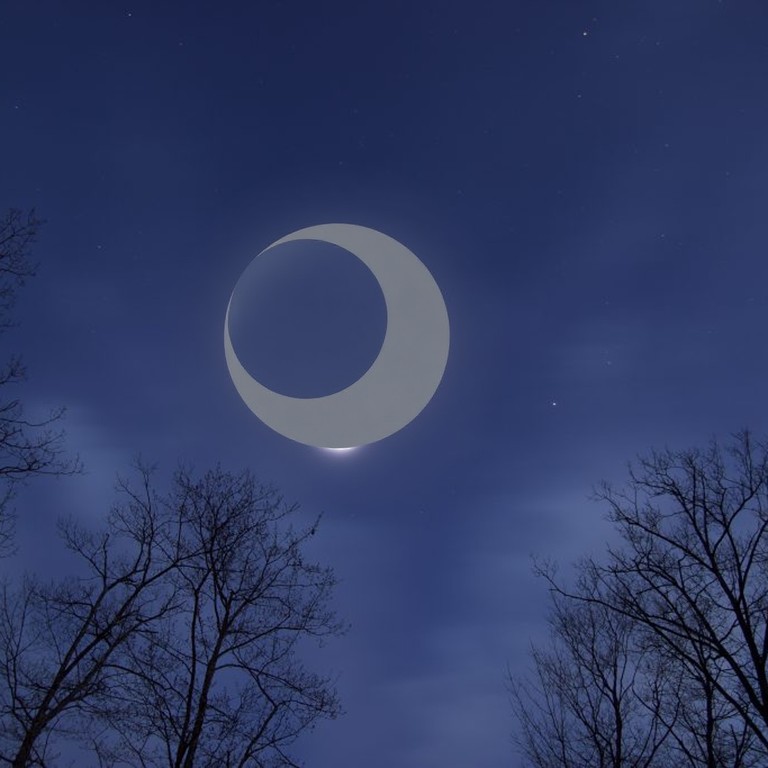} \\
\end{tabular}
\label{subfig:canny-model-steps}
}

\caption{\textbf{Users can achieve desired effects by customizing the value of 3 hyperparameters.} (a) To improve the consistency of edited region with a certain part of the image, the user can cover this part in the context. (b) Lower denoising strength preserves the color, whereas higher denoising strength inserts realistic shadows. (c) Fewer Canny model steps produce high-frequency details, whereas more steps allows the generated image to faithfully follow the hinted shape. Some examples look sub-optimal, because they show the results of only varying one hyperparameter. In practice, users can adjust all three parameters together for better edits.}
\label{fig:hyperparameters}
\end{center}

\end{figure}

We first demonstrate the effect of each hyperparameter (\cref{fig:hyperparameters}). Then, we recommend several combination of hyperparameters can help the user reliably generate an image that DALL·E 3 and GPT-4o cannot (\cref{fig:rabbit}).

\paragraph{Context Size.} During inpainting, a diffusion model generates content consistent with what is already in the context. In \cref{subfig:context-size}, the user requires another yellow vegetable to be added to the center of the image. The prompt is ``yellow vegetables on an orange background''. The color hint is the yellow circle in the center. The model's interpretation of the color hint depends on the selected context. When a small context includes none of the other vegetables, the model generates multiple vegetables from the single yellow circle. When a medium context includes corners of the other vegetables, the model generates one vegetable. When a large context includes all other vegetables, the model fails to generate a new vegetable. When the context includes one vegetable either on the top left or the top right corner, the model generates a vegetable with the same direction of shade. While we do not theoretically explain the failure when using large context sizes, we observe it frequently in other editing tasks. We also observe the failure to add objects using the ``whole image'' mode (\cref{sec:mask-generation}), an observation that motivated our design of context dots.

\paragraph{Denoising Strength.} The model balances hinted colors and realistic shadows at a medium denoising strength. In \cref{subfig:denoising-strength}, the user requires a Rubik's cube with certain face colors to be added to the image. Because the colors cannot be succinctly described by words, the user inserts a reference image as the hint. With a low denoising strength, the colors are preserved, but the shadows are not generated. We observe the opposite with a high denoising strength.

\paragraph{Canny Model Steps.} The model balances realistic details and hinted edges at a medium number of Canny model steps. In \cref{subfig:canny-model-steps}, the user requires a crescent to be added to the sky, but the prompt is simply ``a moon''. With few Canny model steps, the moon looks realistic, but its shape is wrong. We observe the opposite with many Canny model steps.

To generate \cref{subfig:customizable}, the recommended hyperparameter ranges are shown in \cref{tab:hyperparamemter-recommendation}. Note that several consecutive editing steps are necessary in a certain editing region to achieve the shown effect. 

\begin{table}[ht]
\caption{\textbf{We recommend different hyperparameter combinations according to specific editing tasks.} Empirically, these combinations have a higher success rate than others.}
\label{tab:hyperparamemter-recommendation}

\begin{center}
\begin{small}
\begin{tabular}{llll}
\toprule
\textsc{Object} & \textsc{Context} & \textsc{Denoising} & \textsc{Canny}\\
\midrule
Ears   & Head   & Low      & High     \\
Bridge & Violin & Moderate & Moderate \\
Potato & Arm    & High     & Moderate \\
Fish   & Dress  & Moderate & Low      \\
\bottomrule
\end{tabular}
\end{small}
\end{center}

\end{table}

\section{Comparison with MagicQuill}
\label{app:magicquill}

MagicQuill \cite{liu2024magicquill} is a method very similar to ours, because it also accepts edge and color information from the user as its input. However, the following differences in design choices lead to a better performance of our method. 

\paragraph{Fewer ControlNets.} MagicQuill uses three ControlNets, namely inpainting, edge, and color ControlNets. The control branch also needs to be further trained. Meanwhile, our workflow uses only one ControlNet, which is the Canny-edge ControlNet, and no further training is required. 

\paragraph{Higher Flexibility.} As our workflow uses different models for different stages, it can be adapted even when the ControlNets are not released as a module on the base model but as a full model (for example, the official Flux.1 [dev] Canny model by Black Forest Labs). Hence, the higher flexibility allows us to adapt our workflow to the Flux.1 [dev] base model within one week of the release of Flux.1 [dev] Canny in November 2024, whereas MagicQuill still supports neither SDXL or FLUX models by April 2025.\footnote{\url{https://github.com/ant-research/MagicQuill/issues/89}}

\paragraph{Better Detail Preservation.} Our workflow does not downsample the color information. Hence, high frequency details are preserved from user input. In contrast, MagicQuill downsamples user color input to a 32$\times$32 resolution during pre-processing, an information bottleneck that limits the ability of users to provide detailed color hints.

\begin{figure}[t!]
\begin{center}

\scriptsize
\begin{tabular}{c@{\hspace{0.3em}}c@{\hspace{0.3em}}c}
    \textsc{Our Hinted Image} & \textsc{Our Blurred Mask} & \textsc{Our Result (SD 1.5)} \\
    \includegraphics[width=0.3\textwidth]{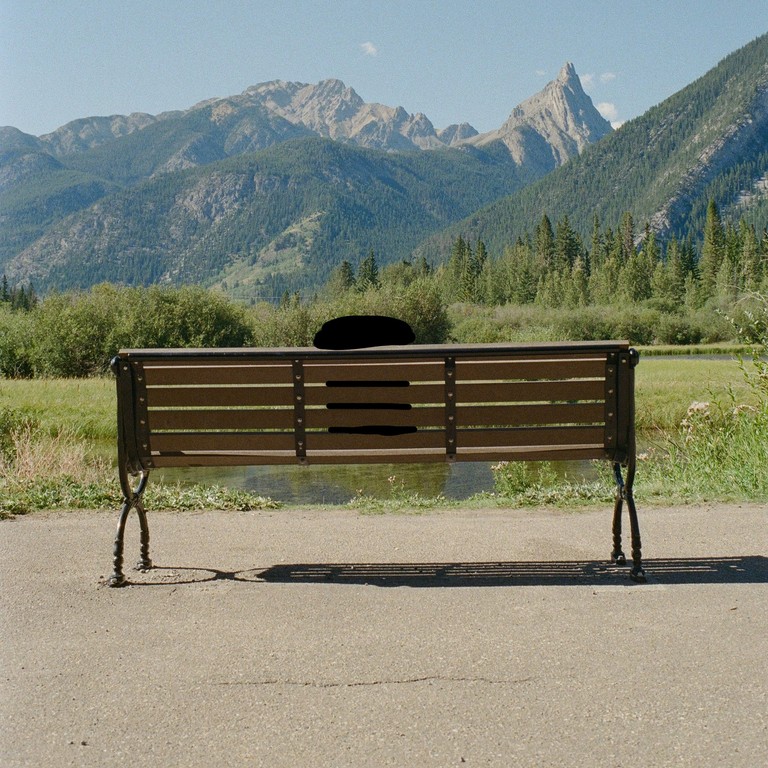} &
    \includegraphics[width=0.3\textwidth]{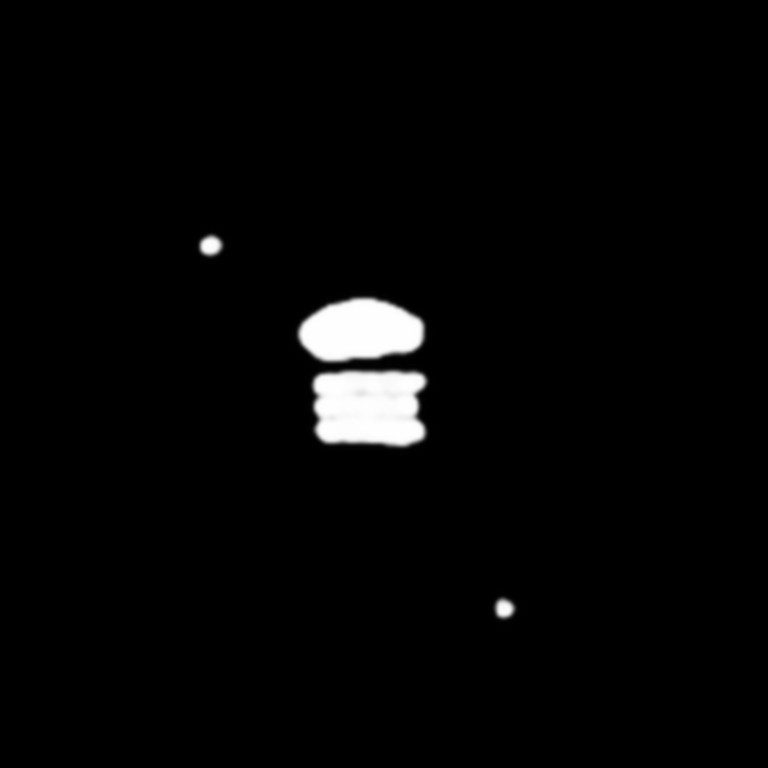} &
    \includegraphics[width=0.3\textwidth]{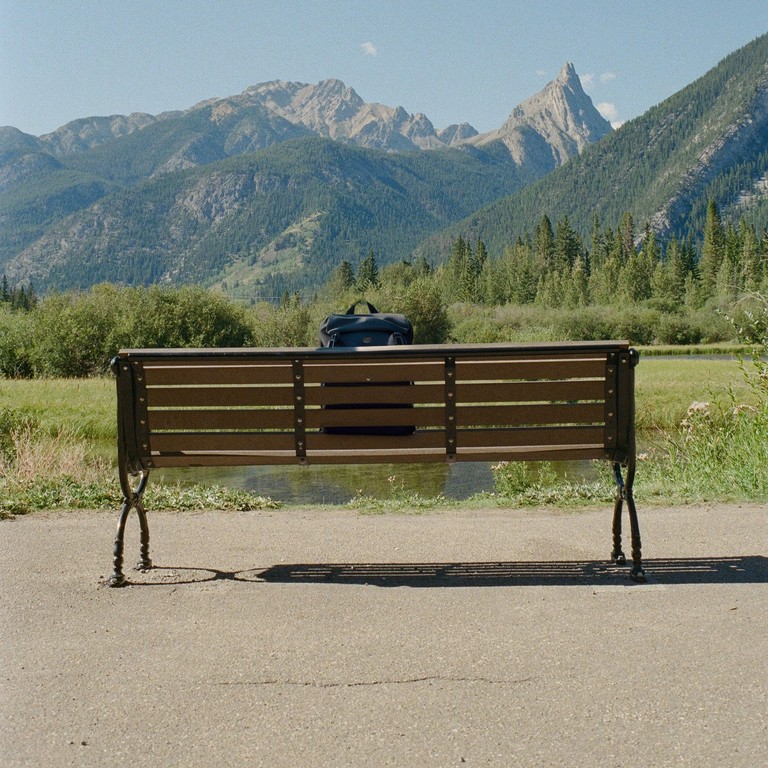} \\
    \textsc{MagicQuill} & \textsc{Gemini 2.0 Flash} & \textsc{GPT-4o} \\
    \includegraphics[width=0.3\textwidth]{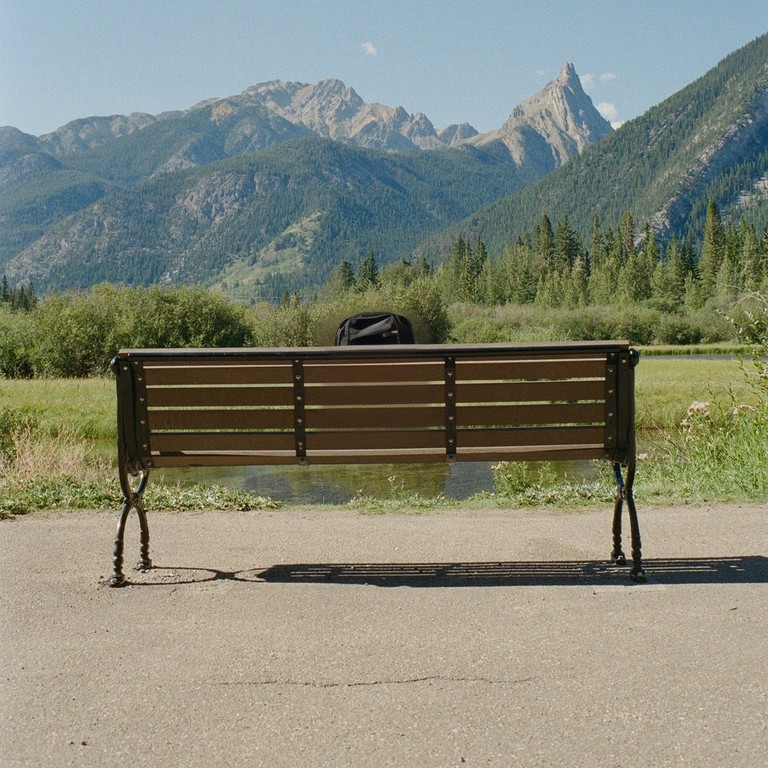} &
    \includegraphics[width=0.3\textwidth]{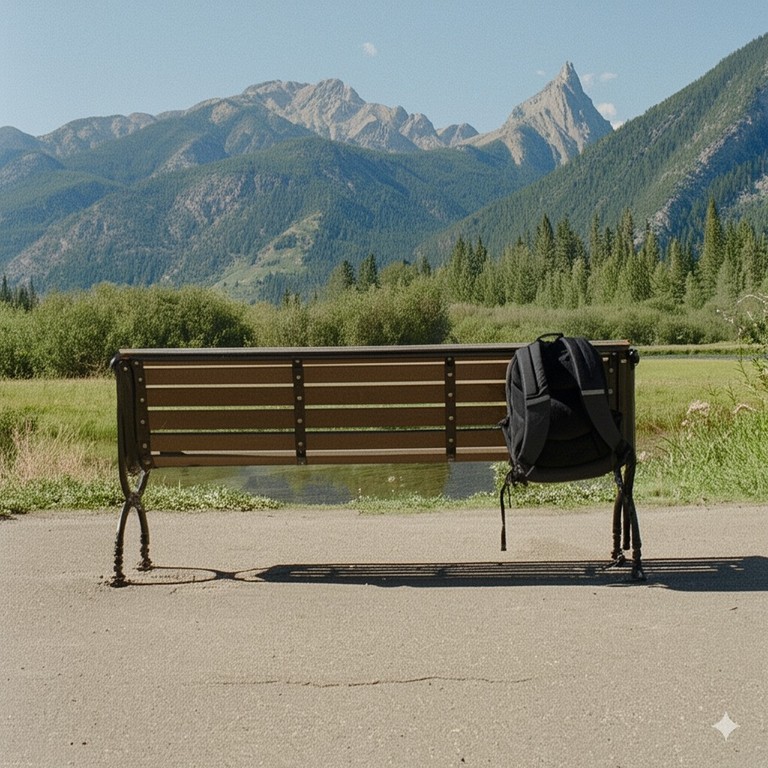} &
    \includegraphics[width=0.3\textwidth]{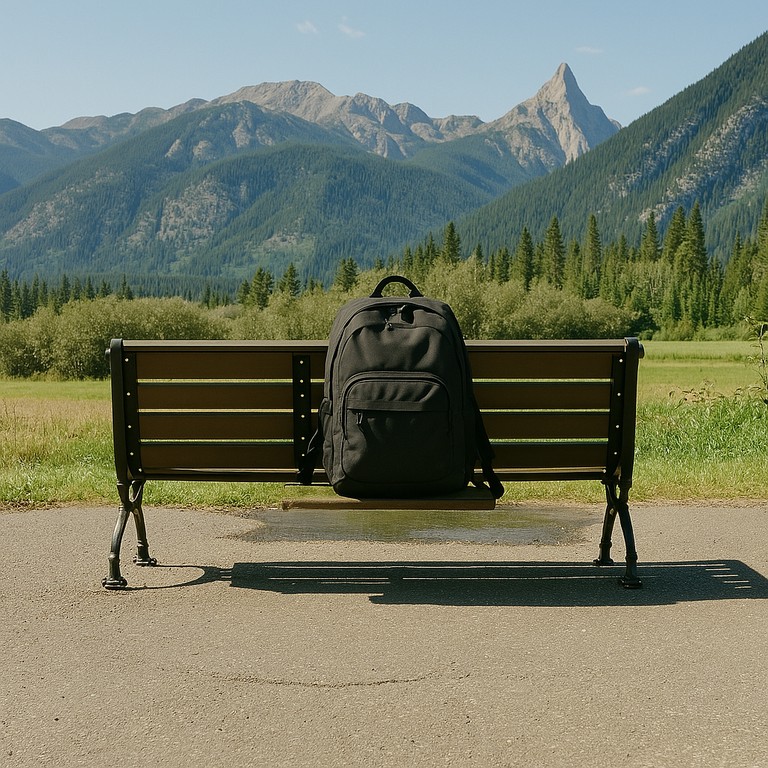} \\
\end{tabular}

\caption{\textbf{Using Realistic Vision V6.0 B1, an SD-1.5-based model released in December 2023, as the backbone, our workflow outperforms both MagicQuill (with the same backbone) and latest commercial VLMs (released in March 2025).} For our workflow and MagicQuill, we used the same description prompt ``black backpack on a bench''. For Gemini 2.0 Flash and GPT-4o, we used the same editing prompt ``add a black backpack onto this bench''. Only our method results in almost perfect spatial relationships, with only a minor error at the bottom of the backpack. MagicQuill is unable to generate the lower part of the backpack, which is supposed to be visible through slits on the backrest. Gemini 2.0 Flash and GPT-4o both generate a floating backpack in front of the backrest. The original image and our better result of using Flux.1 [dev] as the backbone are shown in \cref{fig:benchmark-results}.}
\label{fig:magic-quill-comparison}
\end{center}

\end{figure}

\paragraph{Disentangled Mask and Colors} MagicQuill assumes that the user only wants to edit the region around locations where the color patch is provided. The mask is uniformly expanded in all directions from the color patch. However, consider the example of adding a backpack on a bench (\cref{fig:magic-quill-comparison}). In this case, the mask should be expanded more in the horizontal direction than the vertical direction, because the mask should not fully cover the backrest of the bench. To address this difficulty, the disentangled design in our workflow returns the freedom to the user.

\paragraph{Streamlined Components.} First, our workflow does not uses an MLLM to guess user inputs. While MLLM provides convenience in MagicQuill for common objects, it heavily hallucinates when adding an object that is partially occluded by other objects, because the provided color patch does not have the shape of the full un-occluded object. The hallucination leads to negative user experience when challenging edits are required. Second, our workflow does not use an additional CNN to extract edges. Empirically, the Canny edges provides satisfactory performance and can be quickly extracted in our workflow.

\paragraph{Unlimited Resolution.} By default, MagicQuill generates an image with the resolution of the whole image and replaces the masked region with generated content. This is troublesome when the user wants to iteratively perform global and local edits on the same image. Our workflow mitigates the multi-scale editing issue by proposing context dots, an intuitive and user-friendly method for specifying generation resolutions. The existing ``Resolution Adjustement'' functionality in MagicQuill (updated December 6, 2024) is not designed for multi-scale editing.

\paragraph{Multi-Purpose.} By changing parameters and resolutions, our workflow smoothly transitions from an image editor to a detail enhancer, or from an inpainter to an outpainter. Examples of both transitions can be found in \cref{fig:iterative-construction}. Neither detail enhancing nor outpainting is supported by MagicQuill.

Due to the above advantages, our method outperforms MagicQuill on a challenging example (\cref{fig:magic-quill-comparison}), with the same Realistic Vision V6.0 B1 (based on SD 1.5) model\footnote{\url{https://civitai.com/models/4201?modelVersionId=245598}} as the backbone. To ensure fair comparison, we take the first-shot, single-step result from our method, but try our best to tune the hyperparameters and to draw accurate shapes for MagicQuill. After over 20 attempts, we are still unable to correctly add the backpack onto the bench using MagicQuill. 

Here, we additionally provide comparison with GPT-4o (March 2025) and Gemini 2.0 Flash (March 2025). The fact that an SD-1.5-based model (December 2023) outperforms latest commercial VLMs suggests intrinsic limitations of text-only image editing frameworks. 

We encourage readers with more experience in the baseline methods to independently verify their performance upper-bound.

\section{Limitations}
\label{app:limitations}

Our workflow has the following limitations.

\paragraph{Texture.} Our workflow performs seamless inpainting when the original image is either (1) the output of the workflow or (2) the text-to-image output of the same base model used in the workflow. If the goal is to edit a real photo or an image generated by another base model, the distribution mismatch will cause minor texture mismatches that are impossible to fix perfectly. While such artifacts are barely noticeable with human eyes, they would not escape heuristics-based editing artifact detectors.

\paragraph{Reference Images.} Our workflow does not allow copying all details of a reference image in a single editing step. If the user requires editing complicated apparel, including larger pieces of clothing and smaller accessories, we recommend using multiple editing steps, varying the context size according to the item size, and adding one item at a time.

\paragraph{Local Minima.}
If the base model has some strong local minima, our workflow is unable to help the user generate content out of those local minima. Two examples from the base model (Flux.1 [dev] with a LoRA) we use are (1) a fixed hairstyle regardless of the edge hints and (2) two patellas on one knee regardless of the color hints. To mitigate these artifacts, a convenient and successful workaround is to increase the Canny model steps (correspondingly reducing the base model steps), since the Canny model and the base model seldom suffer from the same local minima. The resulting image after successful edits is shown in the right panel of \cref{fig:art-styles}.

\paragraph{User Input.}
To achieve a high editing quality, a user needs to provide mask and color guidance and to tune hyperparameters. This is harder for the user than baseline methods, where a user only needs to select a better-looking image, if there exists such an image in the outputs. We have not thoroughly tested our workflow in fully automated settings. 

\section{Workflow Implementation}
\label{app:workflow-implementation}

The steps in our workflow can be easily implemented (or have already been separately implemented) in the most popular diffusion model Web UIs (Stable Diffusion Web UI Automatic1111, ComfyUI, and Stable Diffusion Web UI Forge). While there are myriad other methods which perform similar functionalities, we narrow down the necessary components to the set of our choice. The novelty of our workflow is that we achieve greatly improved image editing quality with this small set of simple, user-friendly steps.

For Flux.1 [dev], our workflow is best implemented in ComfyUI. To implement our workflow in ComfyUI, we refer to the following YouTube and Reddit tutorials:
\begin{enumerate}
    \item \url{https://www.youtube.com/watch?v=mI0UWm7BNtQ}
    \item \url{https://www.youtube.com/watch?v=GvZXK4phJmE}
    \item \url{https://www.reddit.com/r/comfyui/comments/1d1x5ek/dual_prompting_with_split_sigmas}
\end{enumerate}

Other than Flux.1 [dev], we thoroughly test our workflow with BoleroMix (Pony) v1.41,\footnote{\url{https://civitai.com/models/448716?modelVersionId=629179}} a Japanese animation-style base model derived from SDXL \cite{podell2024sdxl}. For this one and other SDXL-derived models, the user can implement the workflow by simply enabling relevant add-ons in Stable Diffusion Web UI Automatic1111. We also test our workflow in other models at a smaller scale, where all steps in our workflow similarly work. 

We have tested our workflow on one NVIDIA RTX A6000, NVIDIA A100 (80 GB), or NVIDIA GeForce RTX 4090. On A6000, each editing step using Flux.1 [dev] (1024$\times$1024 resolution and 30 diffusion steps) only takes around 30 seconds. The price of a server with a single A6000 can be as low as 0.5\$ per hour,\footnote{\url{https://cloud.vast.ai/} (January 2025)} equivalent to 0.004\$ per editing step. This is one tenth of the cost of DALL·E 3, which is 0.04\$ per image.\footnote{\url{https://openai.com/api/pricing/}} Due to the low success rate of DALL·E 3 on hard editing tasks (\cref{sec:customizable-editing}), the cost gap is only larger in real scenarios.

The workflow, together with the installation instructions for popular Web UIs, has been released in a GitHub repository. Please see the abstract of the paper for the link.

\section{Societal Impact}
\label{app:societal-impact}

For the first time, our workflow unlocks the ability to perform photo-realistic edits with perfect details using diffusion models. Since the workflow is fully open source, malicious users may misuse the workflow, leading to negative societal consequences. However, as there exist various techniques for detecting AI-generated content, we do not believe the edited images will pose an immediate threat to the public. Still, we strongly encourage the users to adhere to all applicable laws and respect moral standards when generating images with our workflow and using them.


\newpage
\section*{NeurIPS Paper Checklist}


\begin{enumerate}

\item {\bf Claims}
    \item[] Question: Do the main claims made in the abstract and introduction accurately reflect the paper's contributions and scope?
    \item[] Answer: \answerYes{}
    \item[] Justification: We clearly state the task and multiple contributions of our proposed method. The contributions are thoroughly discussed in corresponding sections in the main paper and appendices.

\item {\bf Limitations}
    \item[] Question: Does the paper discuss the limitations of the work performed by the authors?
    \item[] Answer: \answerYes{}
    \item[] Justification: Please see \cref{app:limitations}.

\item {\bf Theory assumptions and proofs}
    \item[] Question: For each theoretical result, does the paper provide the full set of assumptions and a complete (and correct) proof?
    \item[] Answer: \answerNA{}
    \item[] Justification: We do not provide any theoretical results in the paper.

    \item {\bf Experimental result reproducibility}
    \item[] Question: Does the paper fully disclose all the information needed to reproduce the main experimental results of the paper to the extent that it affects the main claims and/or conclusions of the paper (regardless of whether the code and data are provided or not)?
    \item[] Answer: \answerYes{}
    \item[] Justification: We release the code and our own inputs and outputs in the supplemental material. The reader should be able to fully reproduce our results.

\item {\bf Open access to data and code}
    \item[] Question: Does the paper provide open access to the data and code, with sufficient instructions to faithfully reproduce the main experimental results, as described in supplemental material?
    \item[] Answer: \answerYes{}
    \item[] Justification: Open access is our top priority. The data and code is included in the supplemental material. 

\item {\bf Experimental setting/details}
    \item[] Question: Does the paper specify all the training and test details (e.g., data splits, hyperparameters, how they were chosen, type of optimizer, etc.) necessary to understand the results?
    \item[] Answer: \answerYes{}
    \item[] Justification: We clearly specify all hyperparameters and the way they are chosen (\cref{app:evaluation-details}). 

\item {\bf Experiment statistical significance}
    \item[] Question: Does the paper report error bars suitably and correctly defined or other appropriate information about the statistical significance of the experiments?
    \item[] Answer: \answerYes{}
    \item[] Justification: In \cref{tab:editeval-quantitative}, we report standard deviation of the metrics. Because the results are close, some comparisons are not statistically significant. Hence, we further conduct extensive qualitative study to compare the models.

\item {\bf Experiments compute resources}
    \item[] Question: For each experiment, does the paper provide sufficient information on the computer resources (type of compute workers, memory, time of execution) needed to reproduce the experiments?
    \item[] Answer: \answerYes{}
    \item[] Justification: For the whole project, we used a single local NVIDIA RTX A6000, including preliminary and failed experiments. We also estimate the cost of real application scenarios in \cref{app:workflow-implementation}. 
    
\item {\bf Code of ethics}
    \item[] Question: Does the research conducted in the paper conform, in every respect, with the NeurIPS Code of Ethics \url{https://neurips.cc/public/EthicsGuidelines}?
    \item[] Answer: \answerYes{}
    \item[] Justification: Our work does not cause any potential harm.

\item {\bf Broader impacts}
    \item[] Question: Does the paper discuss both potential positive societal impacts and negative societal impacts of the work performed?
    \item[] Answer: \answerYes{}
    \item[] Justification: Please see \cref{app:societal-impact}.
    
\item {\bf Safeguards}
    \item[] Question: Does the paper describe safeguards that have been put in place for responsible release of data or models that have a high risk for misuse (e.g., pretrained language models, image generators, or scraped datasets)?
    \item[] Answer: \answerYes{}
    \item[] Justification: We require that the users adhere to usage guidelines (\cref{app:societal-impact}). 

\item {\bf Licenses for existing assets}
    \item[] Question: Are the creators or original owners of assets (e.g., code, data, models), used in the paper, properly credited and are the license and terms of use explicitly mentioned and properly respected?
    \item[] Answer: \answerYes{}
    \item[] Justification: All assets are properly credited.

\item {\bf New assets}
    \item[] Question: Are new assets introduced in the paper well documented and is the documentation provided alongside the assets?
    \item[] Answer: \answerYes{}
    \item[] Justification: We include the code in the supplemental material. The code will be open source after the review period. We will also release tutorial and documentation for ease-of-use.

\item {\bf Crowdsourcing and research with human subjects}
    \item[] Question: For crowdsourcing experiments and research with human subjects, does the paper include the full text of instructions given to participants and screenshots, if applicable, as well as details about compensation (if any)? 
    \item[] Answer: \answerYes{}
    \item[] Justification: Please see \cref{app:human-evaluation-details}.

\item {\bf Institutional review board (IRB) approvals or equivalent for research with human subjects}
    \item[] Question: Does the paper describe potential risks incurred by study participants, whether such risks were disclosed to the subjects, and whether Institutional Review Board (IRB) approvals (or an equivalent approval/review based on the requirements of your country or institution) were obtained?
    \item[] Answer: \answerNA{}
    \item[] Justification: We only conduct a small-scale human preference study with 9 annotators. The annotation takes about 1 hour for each annotator, so there are no potential risks.

\item {\bf Declaration of LLM usage}
    \item[] Question: Does the paper describe the usage of LLMs if it is an important, original, or non-standard component of the core methods in this research? Note that if the LLM is used only for writing, editing, or formatting purposes and does not impact the core methodology, scientific rigorousness, or originality of the research, declaration is not required.
    \item[] Answer: \answerNA{}.
    \item[] Justification: Our proposed method does not use any LLM as a component.

\end{enumerate}

\end{document}